\documentclass[iop,apj]{emulateapj}
\usepackage{amsmath,amssymb,amstext,float}

\usepackage[citecolor=black,linkcolor=black]{hyperref}
\usepackage{color}

\usepackage[all]{hypcap} 
\usepackage{aas_macros}
\usepackage{mathtools}
\usepackage{xspace}
\usepackage{morefloats}
\extrafloats{100}
\usepackage{perpage} 
\MakePerPage{footnote} 
\shorttitle{The Role Of Twist in Kinked Flux Rope Emergence}
\shortauthors{Knizhnik et al.}
\newcommand{\beg}[1]{\begin{equation}\label{#1}}
\newcommand{\done}{\end{equation}}
\newcommand{\pd}[2]{\frac{\partial #1}{\partial #2}}

\newcommand{\vecB}{\textbf{B}}
\newcommand{\vecJ}{\textbf{J}}

\newcommand{\vecv}{\textbf{v}}

\newcommand{\unit}[1]{\hat{\textbf{#1}}}

\newcommand{\curl}[1]{\nabla\times{#1}}
\newcommand{\divv}[1]{\nabla\cdot{#1}}
\newcommand{\dss}{$\delta$-spots\xspace}
\newcommand{\ds}{$\delta$-spot\xspace}
\DeclareMathOperator{\sech}{sech}
\numberwithin{equation}{section}
\begin{document}

\title{The Role of Twist in Kinked Flux Rope Emergence and Delta-Spot 
Formation}
\author{K. J. Knizhnik\altaffilmark{1}, M. G. Linton\altaffilmark{2}, and C. R. DeVore\altaffilmark{3}}
\altaffiltext{1}{National Research Council Research Associate Residing At Naval Research Laboratory, 4555 Overlook Ave SW, Washington, DC 20375, USA}
\altaffiltext{2}{Naval Research Laboratory, 4555 Overlook Ave SW, Washington, DC 20375}
\altaffiltext{3}{Heliophysics Science Division, NASA Goddard Space Flight Center, 8800 Greenbelt Rd, Greenbelt MD 20771, USA}

\begin{abstract}
It has been observationally well established that the magnetic configurations most favorable for producing energetic flaring events reside in \dss, a class of sunspots defined as having opposite polarity umbrae sharing a common penumbra. They are frequently characterized by extreme compactness, strong rotation and anti-Hale orientation. Numerous studies have shown that nearly all of the largest solar flares originate in \dss, making the understanding of these structures a fundamental step in predicting space weather. Despite their important influence on the space environment, surprisingly little is understood about the origin and behavior of \dss. {\color{black} In this paper, we perform a systematic study of the behavior of emerging flux ropes to test a theoretical model for the formation of \dss: the kink instability of emerging flux ropes. We simulated the emergence of highly twisted, kink-unstable flux ropes from the convection zone into the corona, and compared their photospheric properties to those of emerged weakly twisted, kink-stable flux ropes. We show that the photospheric manifestations of the emergence of highly twisted flux ropes closely match the observed properties of \dss, and we discuss the resulting implications for observations. Our results strongly support and extend previous theoretical work that suggested} that the kink instability of emerging flux ropes is a promising candidate to explain \ds formation, as it reproduces their key characteristics very well. 
\end{abstract}

\maketitle

\section{introduction}\label{sec:intro}
One of the most important features of the solar atmosphere, from the standpoint of space weather forecasting, is the subset of sunspots known as \dss. Classified as having opposite polarity umbrae that share a common penumbra \citep{Kunzel65}, \dss exhibit, by definition, extremely compact photospheric flux distributions. They are often observed to rotate rapidly after emergence \citep{Kurokawa87,Kurokawa91,Tanaka91,LF00,LF03}, contain highly twisted magnetic fields \citep{LF00,LF03,Holder04}, and violate Hale's law \citep{Smith68,Zirin87}. Hale's law states that for a given 11-year solar cycle, the majority of sunspot pairs observed in the Northern (Southern) solar hemisphere will be oriented with the negative (positive) polarity spot to the East (West). This ordering flips with each successive solar cycle. \dss are overwhelmingly associated with explosive flares, being responsible for $>90\%$ of X-class flares during the last two solar cycles \citep{Tanaka91,Shi94,Sammis00,Guo14}. Conversely, over 90\% of the \dss studied by \citet{Tanaka91} were flare active.
The details of why \dss manifest their unique properties, however, are poorly understood. 
Understanding the observed behavior of \dss is critical for predicting their explosive flares. \par
Numerical studies explaining \ds formation have typically come in four varieties: 1) The kink instability of emerging flux ropes \citep{Linton98,Linton99,Fan99,Takasao15,Toriumi17,Toriumi17b}, in which a highly twisted flux rope emerges with a deformed axis, producing a complex active region (AR) that manifests a strongly sheared polarity inversion line across the entire AR. \citet{Takasao15} simulated one such region, NOAA 11429, and found that it was likely formed through the emergence of a kink-unstable flux rope. 
2) The splitting off of a portion of a subsurface flux rope to create small parasitic polarities \citep{Jaeggli16,Toriumi17,Toriumi17b}. 3) The emergence of multiple buoyant sections along a single emerging flux rope \citep{Toriumi14, Fang15, Prior16c,Toriumi17b}, and 4) the interaction of two emerging flux ropes, each containing a single buoyant section {\color{black}\citep{Toriumi14, Jouve18}}. \par 
In this paper, we will focus on the first mechanism, i.e., the generation of \dss via the kink instability of emerging flux ropes. Much of the theoretical underpinning of the theory of the kink instability of twisted flux ropes was developed for applications in tokamak physics by \citet{Shafranov57} and \citet{Kruskal58}, and then in solar physics \citep{Gold60,Anzer68,Hood80, Mikic90}. In highly twisted flux ropes, the kink instability converts twist (rotation of field lines about the flux rope's central axis) into writhe \citep[a deformation of the axis itself;][]{Moffatt92}. The kink instability sets in when the destabilizing magnetic pressure of the azimuthal field encircling the flux rope overwhelms the stabilizing magnetic tension in the central axial field. This occurs when flux ropes are highly twisted \citep[e.g.,][]{Shafranov57,Kruskal58}, and, crucially for interpreting observations and predictive modeling, the helical sense of the kink will have the same sense as that of the flux rope twist \citep{Linton96}.\par 
\citet{Linton99} modeled the kinking of a twisted flux rope and argued that, if such a flux rope were to emerge through the photosphere, the kink would cause it to emerge with a significantly tilted configuration (suggestive of anti-Haleness), then rotate and reorient itself into a Hale configuration, in agreement with observed evolution of \dss \citep{LF03}. Taken together, simulations of highly twisted flux ropes and the analytical work of \citet{Linton96,Linton98,Linton99} suggest that the kink instability of emerging flux ropes is a promising mechanism to explain the formation and behavior of a significant proportion of \dss.    \par
{\color{black} An open question is: what is the initial twist on these emerging flux ropes? One way to estimate flux rope twist is by measuring the twist parameter $\alpha=(\nabla\times\vecB)/\vecB$. At the photospheric level, $\alpha$ can either be measured directly through its normal component, $\alpha_n=J_n/B_n$, with $n$ denoting the component normal to the photosphere, $J$ the current density and $B$ the magnetic field \citep[e.g.,][]{Leka96} or indirectly by fitting a coronal constant-$\alpha$ force free model to the observations \citep[often denoted $\alpha_{best}$, e.g.,][]{Pevtsov94,Pevtsov95,Longcope98,Longcope99,Holder04}. Observations of $\alpha_n$ averaged over each individual AR area find that $\alpha_n$ ranges from $\pm 1\times10^{-10}$ to$\pm1\times10^{-7} \;\mathrm{m}^{-1}$ \citep{Leka96,Otsuji15}, while linear force free fits of $\alpha_{best}$ find a similar range of $0$ to $5\times10^{-8}\;\mathrm{m}^{-1}$ \citep{Pevtsov94,Pevtsov95}. \par 
Analytical arguments suggest that the twist must be large enough for the flux ropes to maintain coherence inside the convection zone \citep{Emonet98}. \citet{Longcope99} found that if this minimum twist were transferred from the convection zone directly to the photosphere, the corresponding peak value of $\alpha$ should be larger than the observed average ranges of $\alpha_n$ or $\alpha_{best}$. 
Numerical models of emerging flux ropes have typically used initial twists larger than that required to survive the convection zone, and the twist needed for a kink is yet higher still. The question, therefore, is how to explain the relatively low observed values of photospheric $\alpha_n$. \par 
Several theoretical arguments can help resolve this difficulty. First, both $\alpha$ and $\alpha_n$ vary significantly across the initial subsurface flux rope, with $\alpha$ peaking at the center and vanishing at the edges, and $\alpha_n$ even reversing sign beyond a certain radius. Thus, taking the average of either quantity will lead to values much smaller than the peak value. 
Second, the assumption that $\alpha$ (or $\alpha_n$) is transferred directly from the convection zone to the photosphere is unlikely to hold up during the dynamic emergence of the flux rope, as significant reconnection, diffusion, and kinking are expected to reduce the twist on the emerging flux rope. It has also been demonstrated in several flux emergence simulations \citep[e.g.,][]{Fan09,Leake13}, that a significant fraction of the twist, in the form of $\alpha$, is trapped below the photosphere, so that the observed local $\alpha$ at the photosphere is about an order of magnitude lower than the local $\alpha$ in the convection zone portion of that same field line. \par 
One consequence of this reasoning is that it implies that it may not be possible to infer the flux rope's initial twist directly from $\alpha_n$ or $\alpha_{best}$. Nevertheless, it is clear that some ARs do have significantly larger values of $\alpha_{best}$ or $\alpha_n$ than others \citep{Pevtsov94,Pevtsov95,Longcope98,Holder04}. The ARs showing the largest values of either $\alpha_n$ or $\alpha_{best}$ have not been systematically studied in observations, but an individual study found $\alpha_{best}=2.5\times10^{-8}\;\mathrm{m}^{-1}$ in \dss as compared to $1.1\times10^{-8}\;\mathrm{m}^{-1}$ in typical ARs \citep{Holder04}. This suggests that the flux ropes that form \dss are more twisted than typical emerging flux ropes, and may explain these higher values of $\alpha_{best}$. 
In addition, due to the kink instability, highly twisted flux ropes are expected to manifest several other features at the photospheric level that suggest they may be the source of \dss on the photosphere.} \par  
The kink-instability model of \ds formation predicts four testable consequences for the photospheric and coronal properties of the resulting ARs.  \par 
First, since the kink instability occurs only for highly twisted flux ropes, the emerging AR should exhibit high twist. Thus, the kink-instability model naturally explains the high twists observed in \dss \citep{Holder04}.\par
Second, the conversion of twist into writhe manifests as a strong rotation of the flux distribution on the photosphere. Crucially, the sense of this rotation will be determined by the sign of the flux rope twist \citep{Linton99}, wherein a right (left) handed twist will give a right (left) handed kink, which, upon passing through the photosphere, will translate into a counter-clockwise (clockwise) rotation. This provides a testable prediction for \ds formation. Thus, the kink of the emerging flux rope is a promising mechanism for explaining the observed rotation of \dss on the photosphere \citep{Kazachenko10,Vemareddy16,Wang16}.\par
Third, the kinking motion should break the cylindrical symmetry of the flux rope about its initial axis, moving the flux rope legs off of the initial axis. In its nonlinear phase, the kink instability should cause the rising loop to fold over on itself, causing the legs of the rising flux rope to be held very close to each other, or even, in extreme cases, to form a knot \citep{Linton98}. The predicted result is that, on the photospheric level, opposite polarity regions of the emerging AR would not separate to any appreciable distance. As a result, the `compactness' of the photospheric flux distribution in \dss may readily explained by the kink instability model of \ds formation \citep{Howard96,LF00,LF03}. \par
Finally, at the coronal level, the highly twisted field in the kinked structure should contain a tremendous amount of free magnetic energy, making it susceptible to intense energy release. This would then explain the association of \dss with X-class flares \citep{Tanaka91,Shi94,Sammis00,Guo14}. \par

The picture that emerges, therefore, is that at the photosphere, kinked flux ropes are expected to be compact (if they have kinked enough to bring the legs of the flux rope close together), to be twisted (since kinking requires high twist), to rotate (with the kinking motion manifesting itself as a rotation on the photosphere) and, at the coronal level, to flare (due to high levels of twist, which corresponds to free energy, and knots, which represents topological complexity). This is consistent with the observed properties of \dss, which are known to be compact, twisted, and strongly rotating and flaring. Thus, several predictions of the kinking flux rope model are eminently testable via numerical simulations. But although this is strong theoretical support for the kink instability model of \ds formation, very few numerical studies have tested this idea. With the exception of a few simulations \citep{Matsumoto98, Fan99, Takasao15, Toriumi17b}, the overwhelming majority of flux emergence studies have focused on non-kinking, weakly twisted flux ropes. Furthermore, very few authors \citep{Fan99,Murray06,Sturrock16} have performed parameter studies to investigate how varying the initial flux rope parameters affects the photospheric flux distribution in emerging flux ropes. Of these, only one \citep{Fan99} used a kink-unstable flux rope, and this study did not model the actual emergence. Thus, although significant theoretical work has argued that \dss are the photospheric manifestations of the emergence of kinked flux ropes rising from the convection zone into the corona, no systematic numerical study of the emergence of such kinked flux ropes has been undertaken.\par 
In this paper, we perform a parameter study of the emergence of twisted flux ropes from the convection zone into the corona. We vary only a single parameter: the flux rope's initial dimensionless twist, which determines the stability of the flux rope. We study twists ranging from kink-stable to marginally stable to unstable and explore the consequences of the emergence of these flux ropes at the photospheric level.

\section{Numerical Model}\label{sec:model}
Our simulations solve the equations of magnetohydrodynamics (MHD) using the Adaptively Refined Magnetohydrodynamics Solver \citep[ARMS;][]{DeVore08} in three Cartesian dimensions. The equations have the form
\beg{cont}
\pd{\rho}{t}+\divv{\rho\vecv}=0,
\done
\beg{momentum}
\pd{\rho\vecv}{t} + \divv{\left( \rho\vecv\vecv \right)} = - \nabla P + \frac{1}{4\pi}(\curl{\vecB})\times\vecB + \rho g,
\done
\beg{energy}
\pd{T}{t}+\divv{(T\vecv)}=(2-\gamma)T\divv{\vecv},
\done
\beg{induction}
\pd{\vecB}{t} = \curl{ \left( \vecv \times \vecB \right)}+\frac{1}{4\pi}\eta(\curl{\vecB}).
\done

In these equations, $\rho$ is mass density, $T$ is temperature, $P$ is thermal pressure, $\gamma$ is the ratio of specific heats, $\vecv$ is velocity, $\vecB$ is magnetic field, $g$ is the gravitational acceleration, $\eta$ is the magnetic diffusivity and $t$ is time. We close the equations via the ideal gas equation,
\beg{ideal}
P = \rho RT,
\done
where $R=8.26\times10^7 \mathrm{dyn\;cm\;K^{-1}\;g^{-1}}$ is the gas constant. In all of the simulations presented below, we choose $g=2.7\times10^4\;\mathrm{cm\;s^{-1}}$ and $\eta=10^{10}\mathrm{cm^2\;s^{-1}}$. \par

\begin{figure*}
\centering\includegraphics[scale=0.5, trim=0.0cm 0.0cm 0.0cm 0.0cm,clip=true]{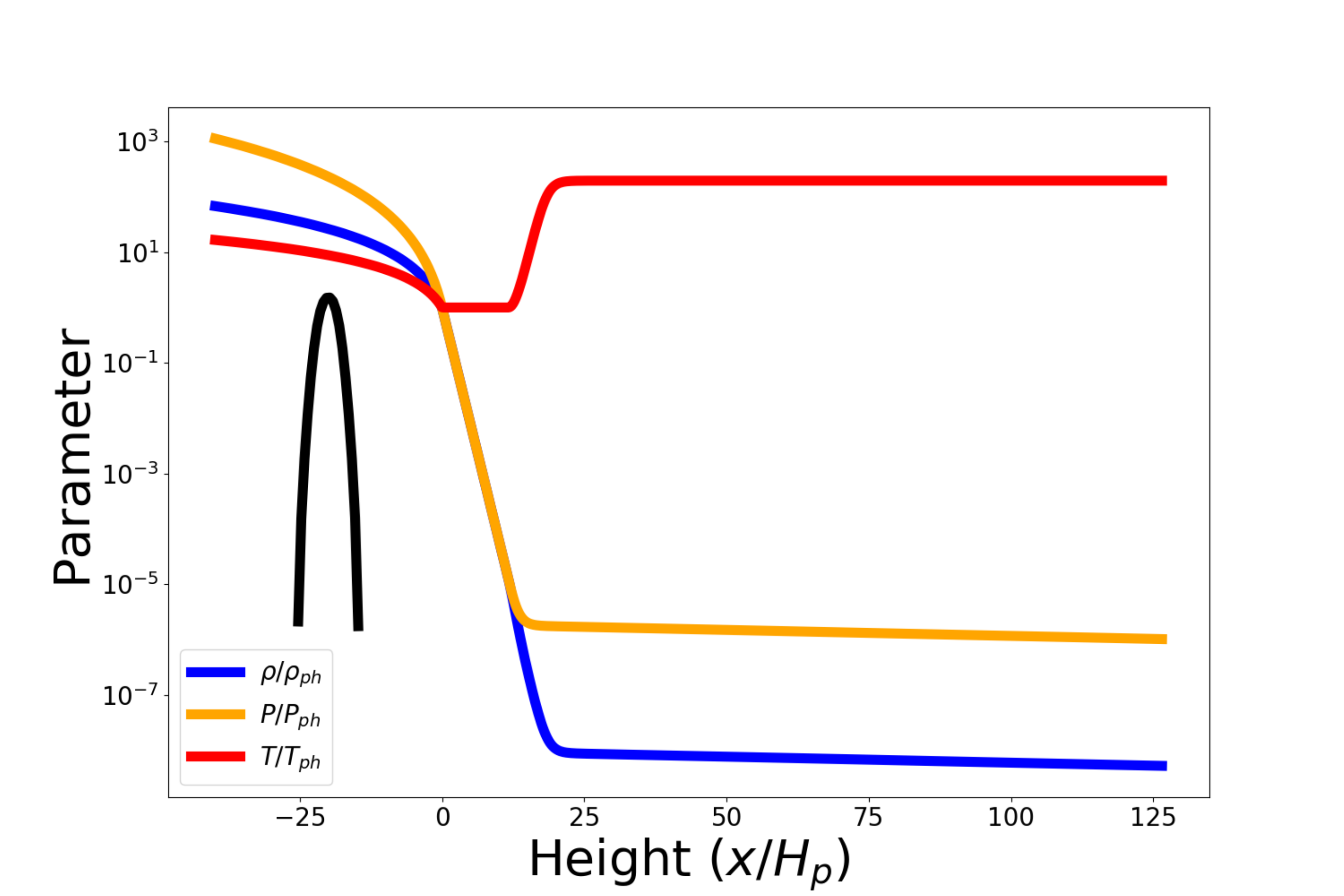}
\caption{Stratified atmosphere used in flux emergence simulations. The mass density (blue), pressure (orange) and temperature (red) are scaled by their photospheric ($x = 0$) values. Shown in black is the magnetic pressure profile for the $\zeta=1$ flux rope. The convection zone is located at $x/H_p<0$, the temperature minimum region at $0<x/H_p<12$, and the corona begins at $x/H_p=25$, being separated from the temperature minimum portion of the atmosphere by a short transition region.}
\label{fig:atmosphere}
\end{figure*}

We set up a stratified atmosphere, shown in Figure \ref{fig:atmosphere}, with a temperature profile of the form
\beg{Tcz}
T(x) = T_{ph}\left(1-\frac{x-x_{ph}}{\ell_{cz}}\right),
\done
in the convection zone, with $x$ the vertical direction. The temperature gradient $dT/dx$ has the constant value $-T_{ph}/\ell_{cz}$, and the temperature $T$ varies on the characteristic length scale $\ell_{cz}$. We use the exact solution for an adiabatically stratified Cartesian atmosphere: 
\beg{lc}
\ell_{cz} = \frac{\gamma R}{(\gamma-1)g} T_{ph}.
\done 
This profile attaches to the bottom of a constant temperature region having temperature $T_{ph}$ starting at height $x_{ph}$, which we take to be the photosphere. At the top of the constant (minimum) temperature region, at height $x_{tr}$, the transition region attaches, having a profile of the form
\beg{Ttr}
\frac{1}{T(x)} = \frac{1}{T_{cor}}+\left(\frac{1}{T_{ph}}-\frac{1}{T_{cor}}\right)\sech^2{\left(\frac{x-x_{tr}}{\ell_{tr}}\right)},
\done
where $T_{cor}$ is the asymptotic temperature of the corona ($x-x_{tr} \gg \ell_{tr}$)
for a characteristic length scale $\ell_{tr}$. Background magnetic-field free hydrostatic density and pressure profiles are then calculated from Equations \ref{momentum}, \ref{ideal}, \ref{Tcz} and \ref{Ttr}, with $\vecv=\vecB=0$ initially and the photospheric temperature and pressure pinned at  $T_{ph}=5.1\times10^3\;\mathrm{K}$, and $P_{ph}=1.14\times10^5 \;\mathrm{dyn\;cm^{-2}}$ (cf. Figure \ref{fig:atmosphere}). These choices set the photospheric density, $\rho_{ph} = 2.7\times10^{-7}\;\mathrm{g\;cm^{-3}}$, pressure scale height $H_p = RT_{ph}/g= 150\;\mathrm{km}$, sound speed $C_s=\sqrt{\gamma P/\rho}=8.4\;\mathrm{km\;s^{-1}}$, and convection zone length scale $\ell_{cz}=0.383\;\mathrm{Mm}$,. We choose $x_{ph}=0.0\;\mathrm{Mm}$, $x_{tr}=1.75\;\mathrm{Mm}$, $\ell_{tr} = 0.3\;\mathrm{Mm}$ and $T_{cor}=10^6\;\mathrm{K}$. 
The simulation box has size $L_x\times L_y\times L_z = 25\times20\times20\;\mathrm{Mm}$, with $x$ representing the vertical direction and $y$ the direction parallel to the initial flux rope axis. {\color{black} Our boundaries are periodic in the horizontal directions, closed at the bottom and open at the top. Our grid is specified such that the smallest grid cell is $0.039\;\mathrm{Mm}$ in each direction.}
We placed a flux rope with magnetic field profile
\beg{fluxrope}
\begin{split}
B_y(r) &= B_0e^{-r^2/a^2} \\
B_\phi(r) &= \zeta \frac{r}{a}B_y(r)
\end{split}
\done
in pressure balance at a depth $x=-d$. Here $r$ is the radial distance from the initial flux rope axis, $\phi$ is the azimuthal coordinate, $d = 3 \; \mathrm{Mm}$, $B_0 = 6.5\;\mathrm{kG}$ is the flux rope's initial magnetic field strength, and $a = 0.3 \; \mathrm{Mm}$ is its radius, so that the total axial flux in the flux rope is $\Phi_0 = 1.8\times10^{19}\;\mathrm{Mx}$. The parameter $\zeta$ is the dimensionless twist of the flux rope.
Twisted flux ropes are kink unstable when their twist $\zeta$ exceeds the critical twist $\zeta_c$. For the profile given in Equation \ref{fluxrope}, \citet{Linton96} showed that $\zeta_c=1$.\par

To ensure hydrostatic equilibrium is maintained when this flux rope is added to the atmosphere, the field-free atmospheric pressure $p_0(x)$ is modified by an amount
\beg{new_pres}
p_1(r) = -\frac{B_0^2}{8\pi}e^{-2r^2/a^2}\bigg[1-\frac{1}{2}\zeta^2\left(1-\frac{2r^2}{a^2}\right)\bigg],
\done
so that $\nabla p_1 = (\vecJ\times\vecB)/c$, where \beg{defJ}
\vecJ = \frac{c}{4\pi}\nabla\times\vecB,
\done 
with $c=3\times10^{10}\;\mathrm{cm\;s^{-1}}$ the speed of light. The equilibrium pressure is then
\beg{peq}
p_{eq}(x) = p_0(x) + p_1(x).
\done 
To initialize the emergence, we impose a density perturbation of the form \citep{Fan09}:
\beg{dens_perturb}
\rho_1(x,y,z) = -\rho_0(x)\frac{B^2_y(r)}{8\pi p_0(x)}e^{-y^2/\lambda^2}
\done
where $\lambda=1.2\;\mathrm{Mm}$, so that
\beg{rhoeq}
\rho_{eq}(x,y,z) = \rho_0(x) + \rho_1(x,y,z).
\done 
This generates an initial buoyant section of the flux rope, centered at $y=0$ and with an extent of $\Delta y \approx 2\lambda$. \par

We perform five simulations with these parameters, varying the value of the dimensionless twist $\zeta$, while keeping the radius fixed. We set $\zeta=\{0.5,1,1.5,2,4\}$, varying its value from kink-stable to marginally stable to kink-unstable. \par
\section{Results}\label{sec:results}
The photospheric properties of each of the simulations are both qualitatively and quantitatively different. We first present qualitative results from each simulation individually before comparing the different cases quantitatively. 

\subsection{$\zeta=0.5$}\label{sec:0.5}
\subsubsection{Subsurface Evolution}\label{sec:sub0.5}
\begin{figure*}
\centering\includegraphics[scale=0.29, trim=0.0cm 1.0cm 4.0cm 0.0cm,clip=true]{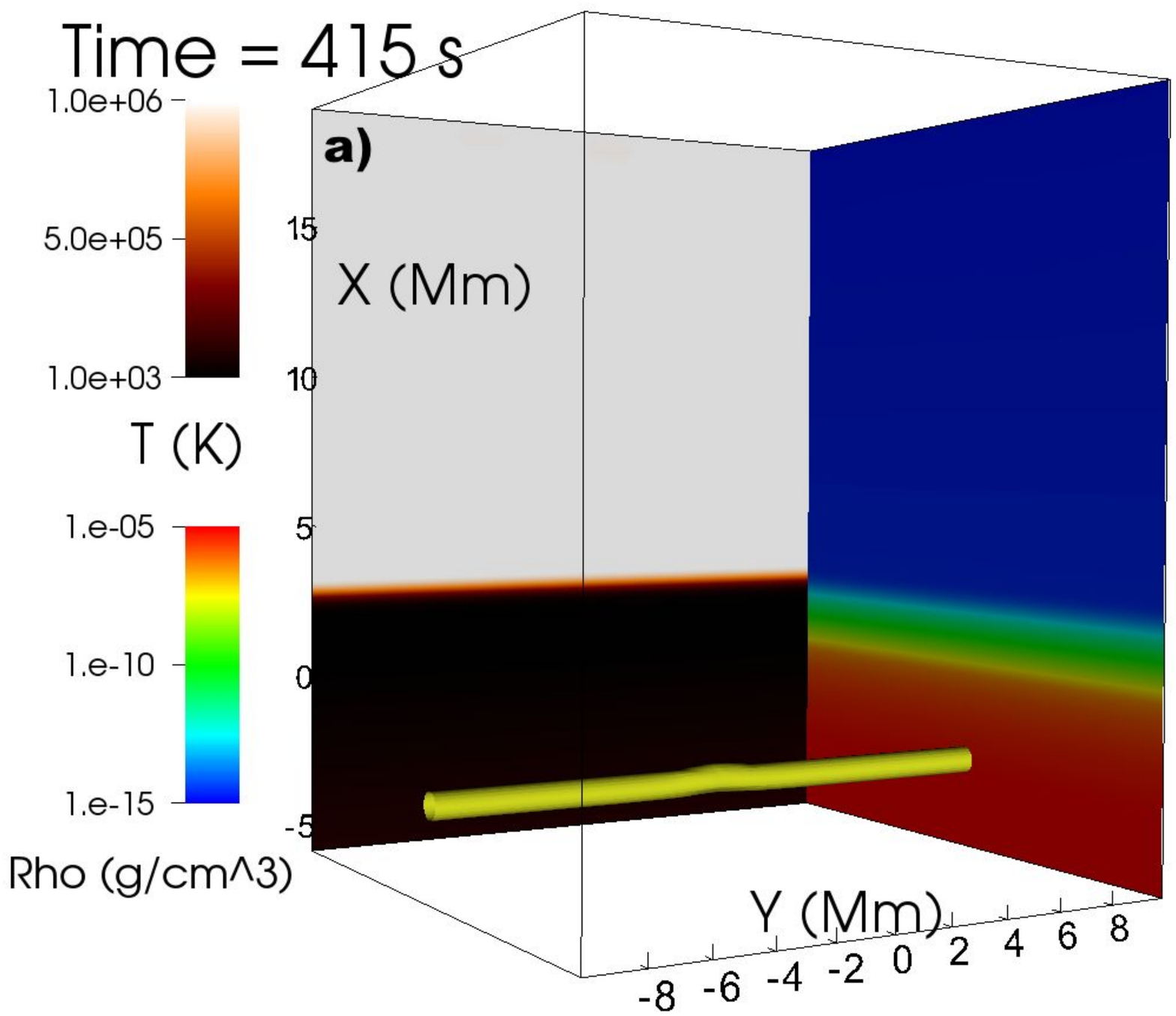}
\centering\includegraphics[scale=0.32, trim=4.0cm 1.0cm 5.0cm 3.0cm,clip=true]{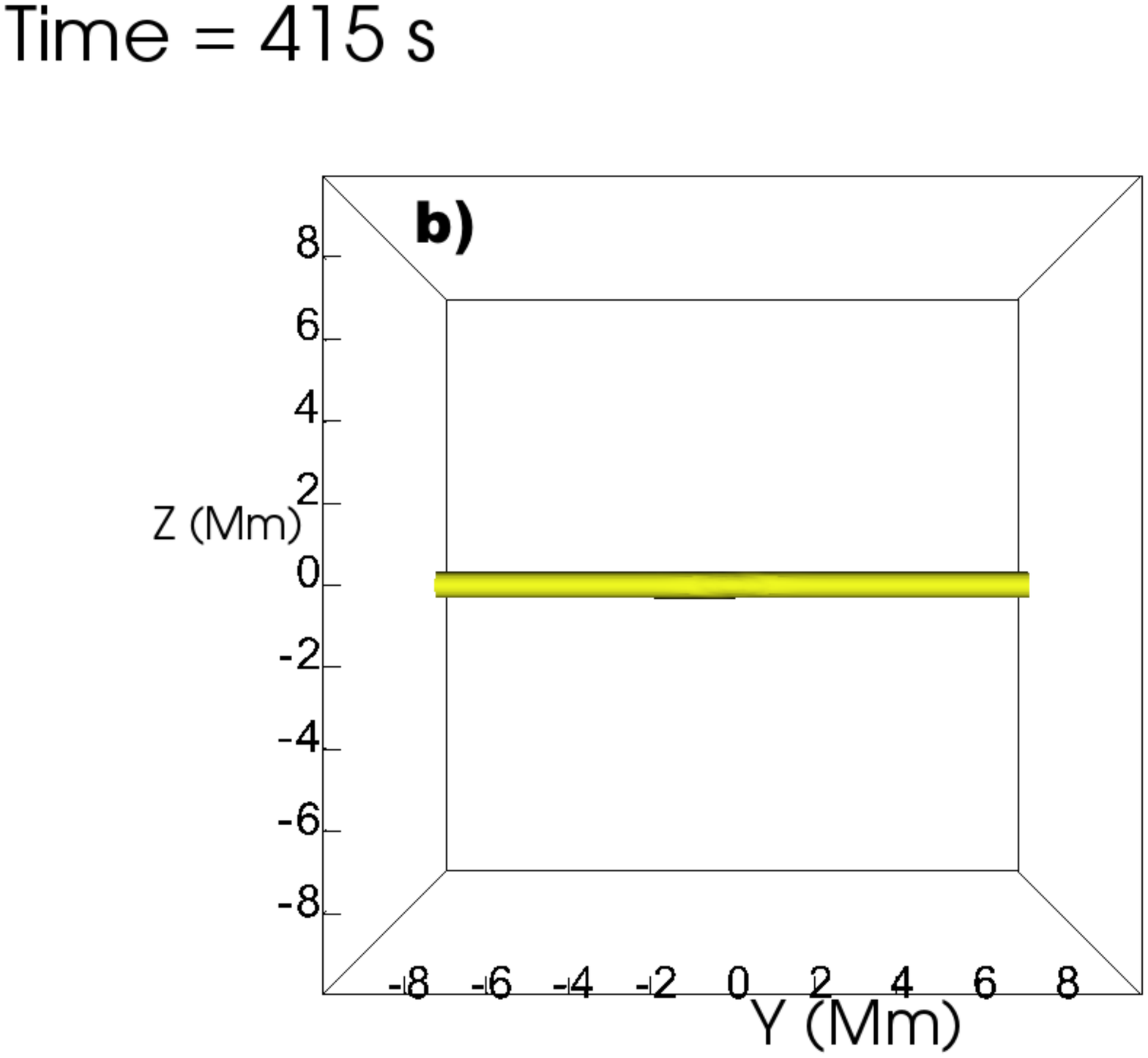}
\centering\includegraphics[scale=0.29, trim=0.0cm 1.0cm 4.0cm 0.0cm,clip=true]{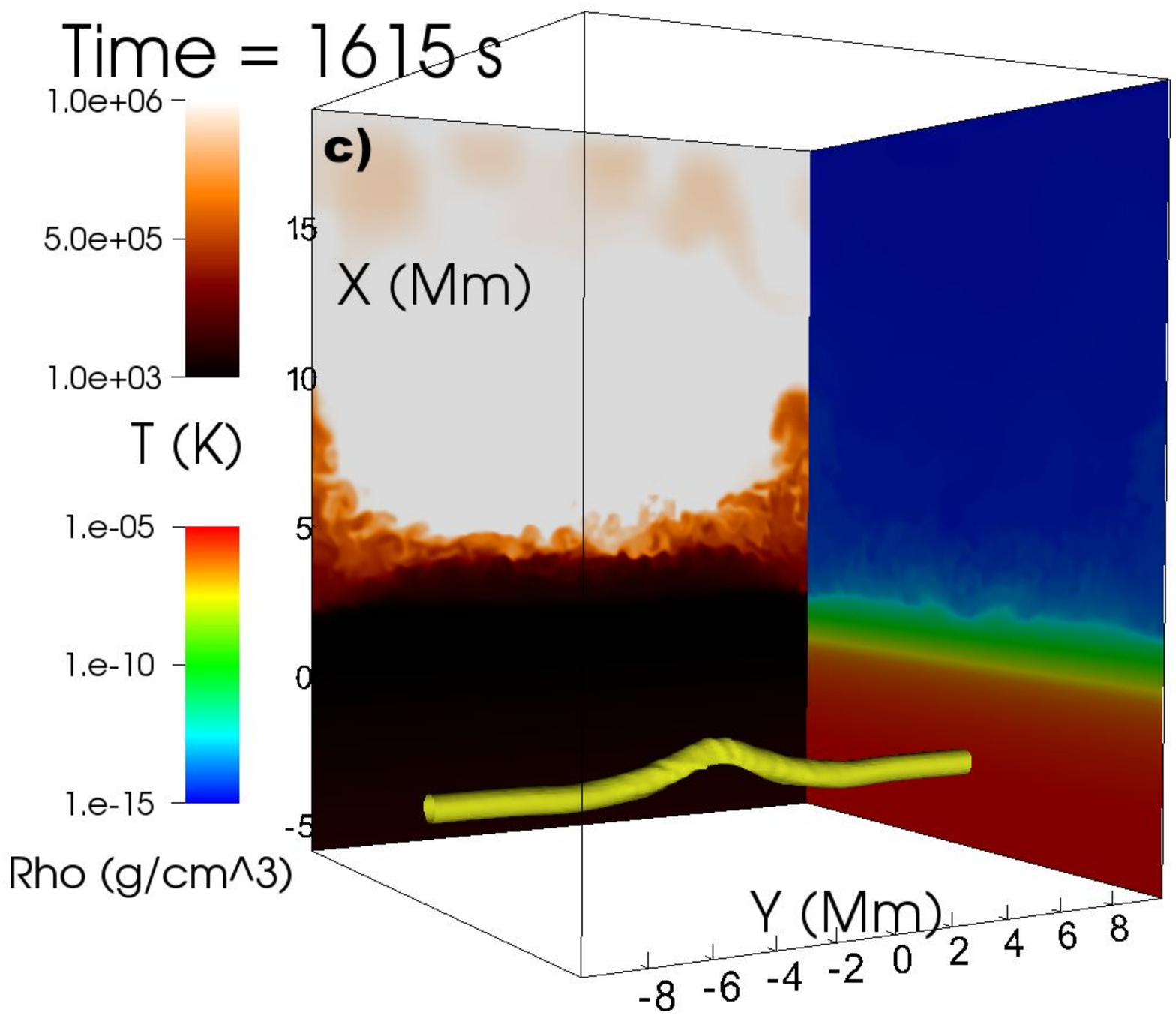}
\centering\includegraphics[scale=0.32, trim=4.0cm 1.0cm 5.0cm 3.0cm,clip=true]{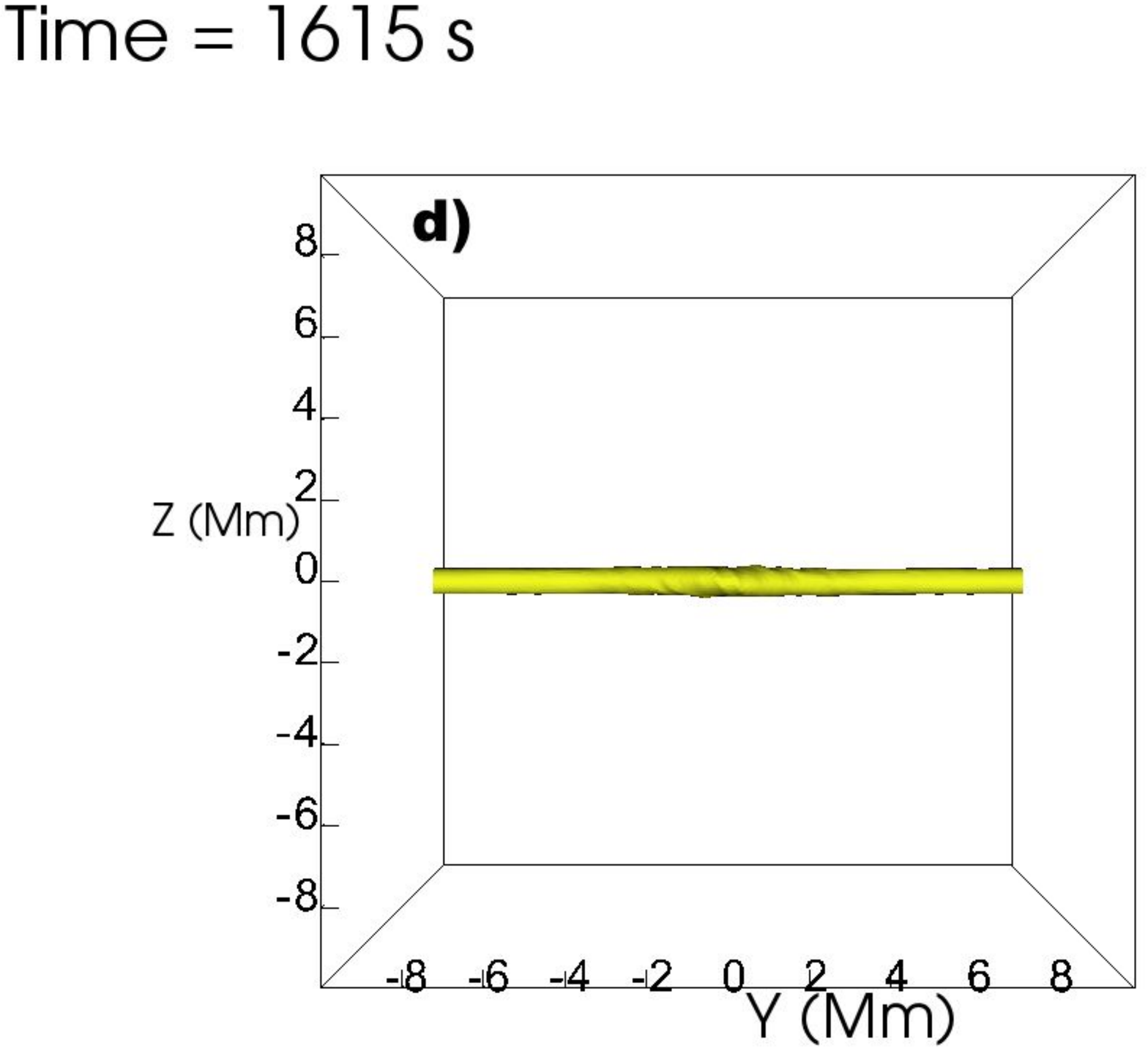}
\centering\includegraphics[scale=0.29, trim=0.0cm 1.0cm 4.0cm 0.0cm,clip=true]{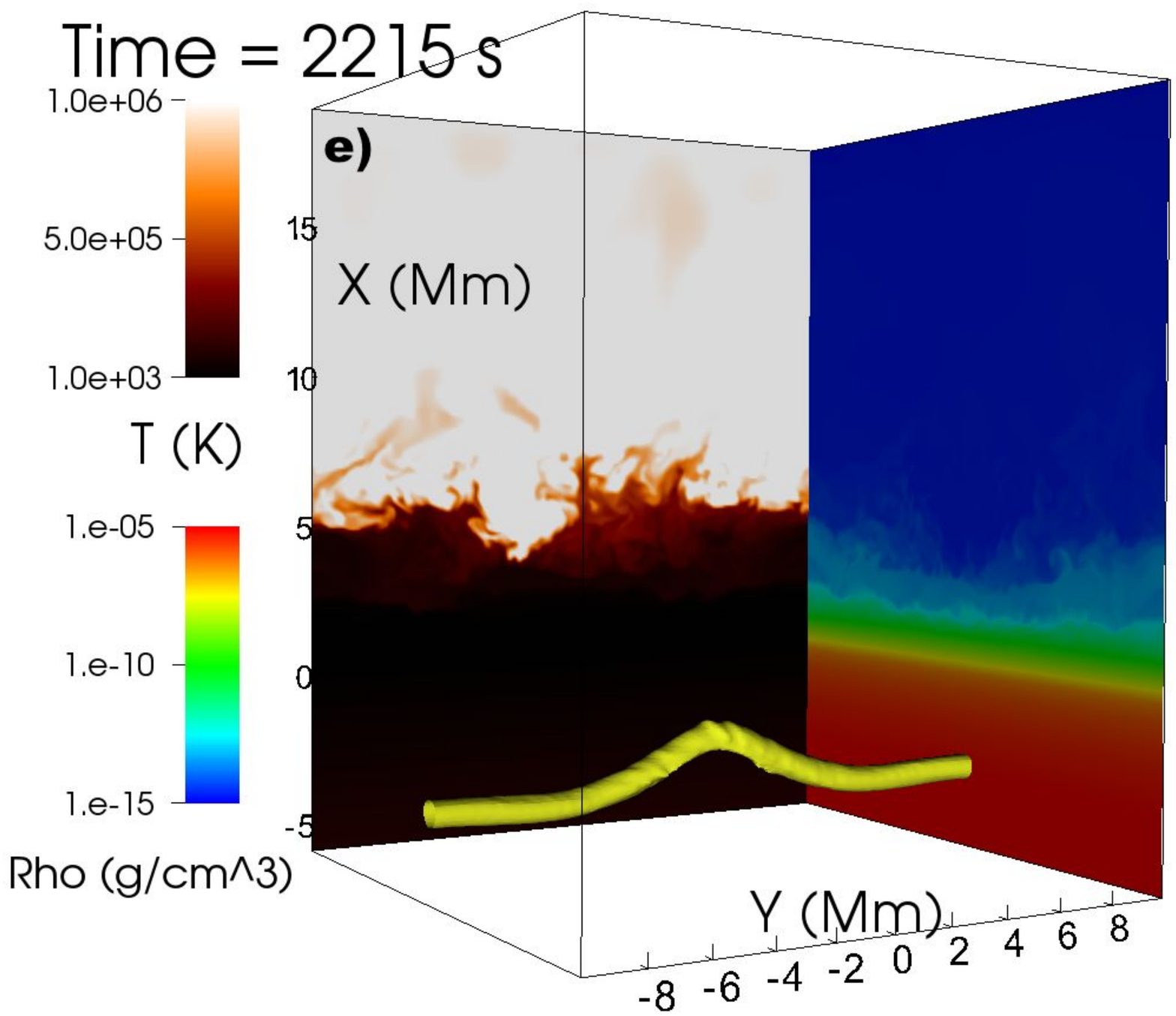}
\centering\includegraphics[scale=0.32, trim=4.0cm 1.0cm 5.0cm 3.0cm,clip=true]{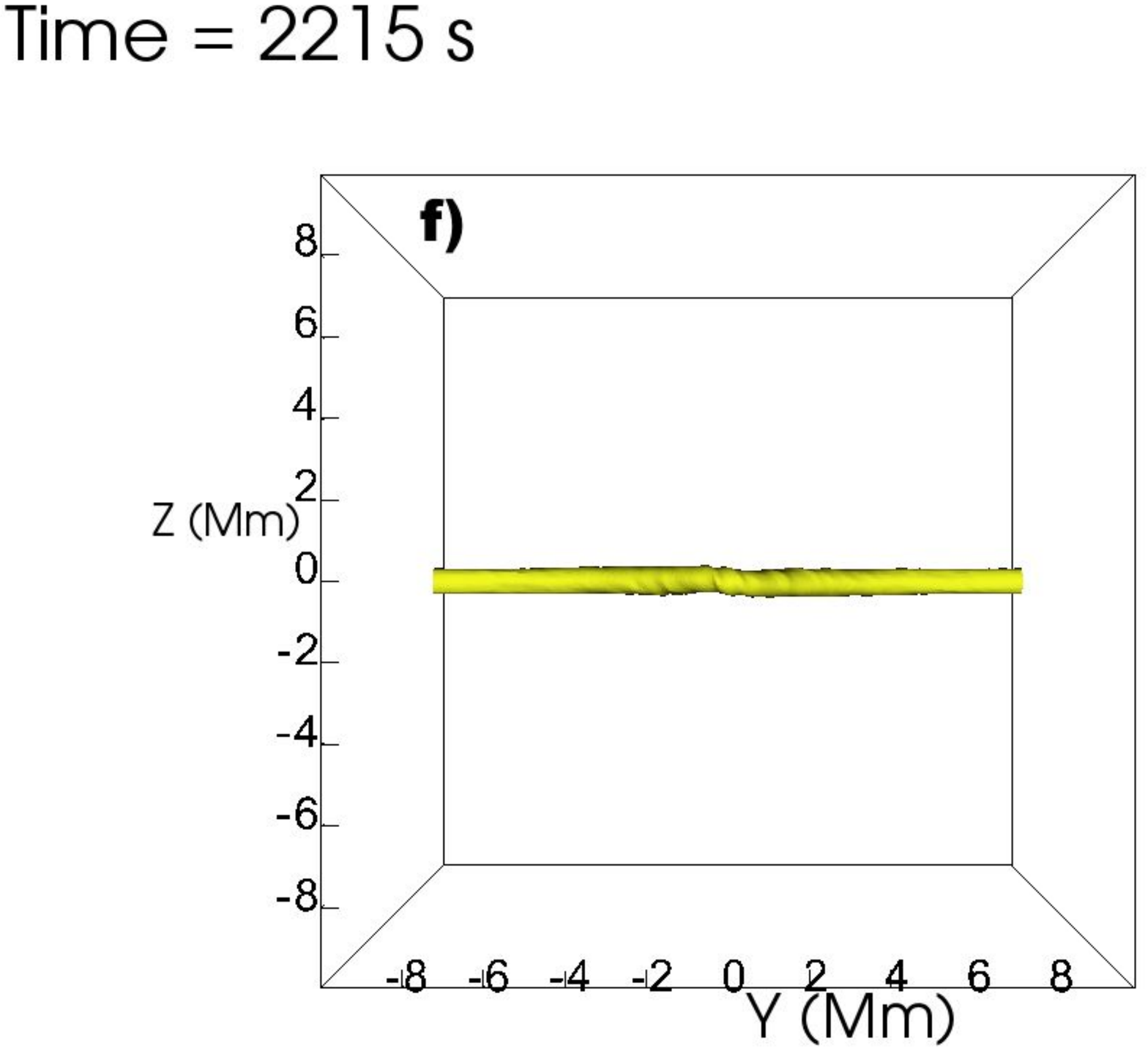}
\centering\includegraphics[scale=0.29, trim=0.0cm 1.0cm 4.0cm 0.0cm,clip=true]{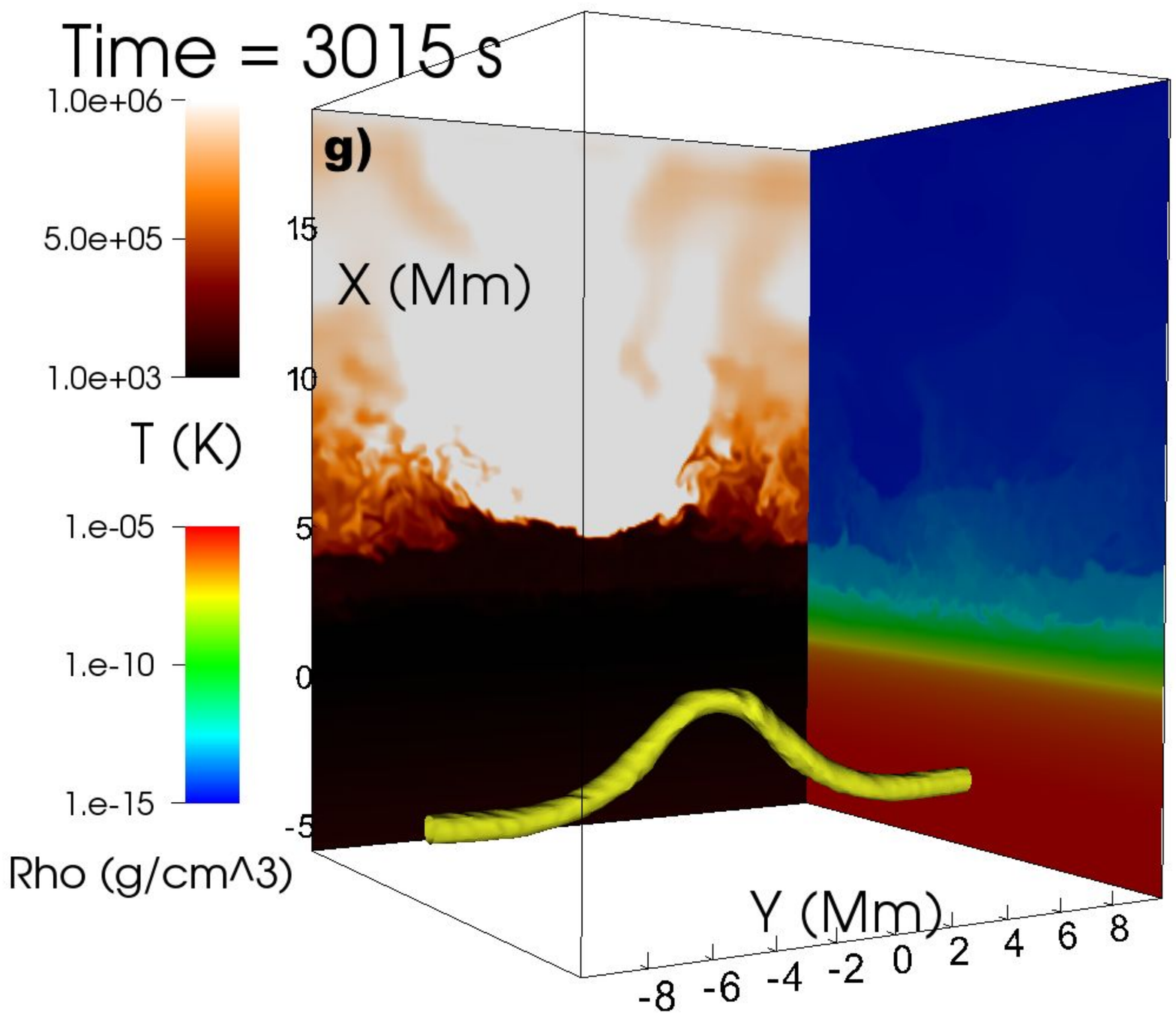}
\centering\includegraphics[scale=0.32, trim=4.0cm 1.0cm 5.0cm 3.0cm,clip=true]{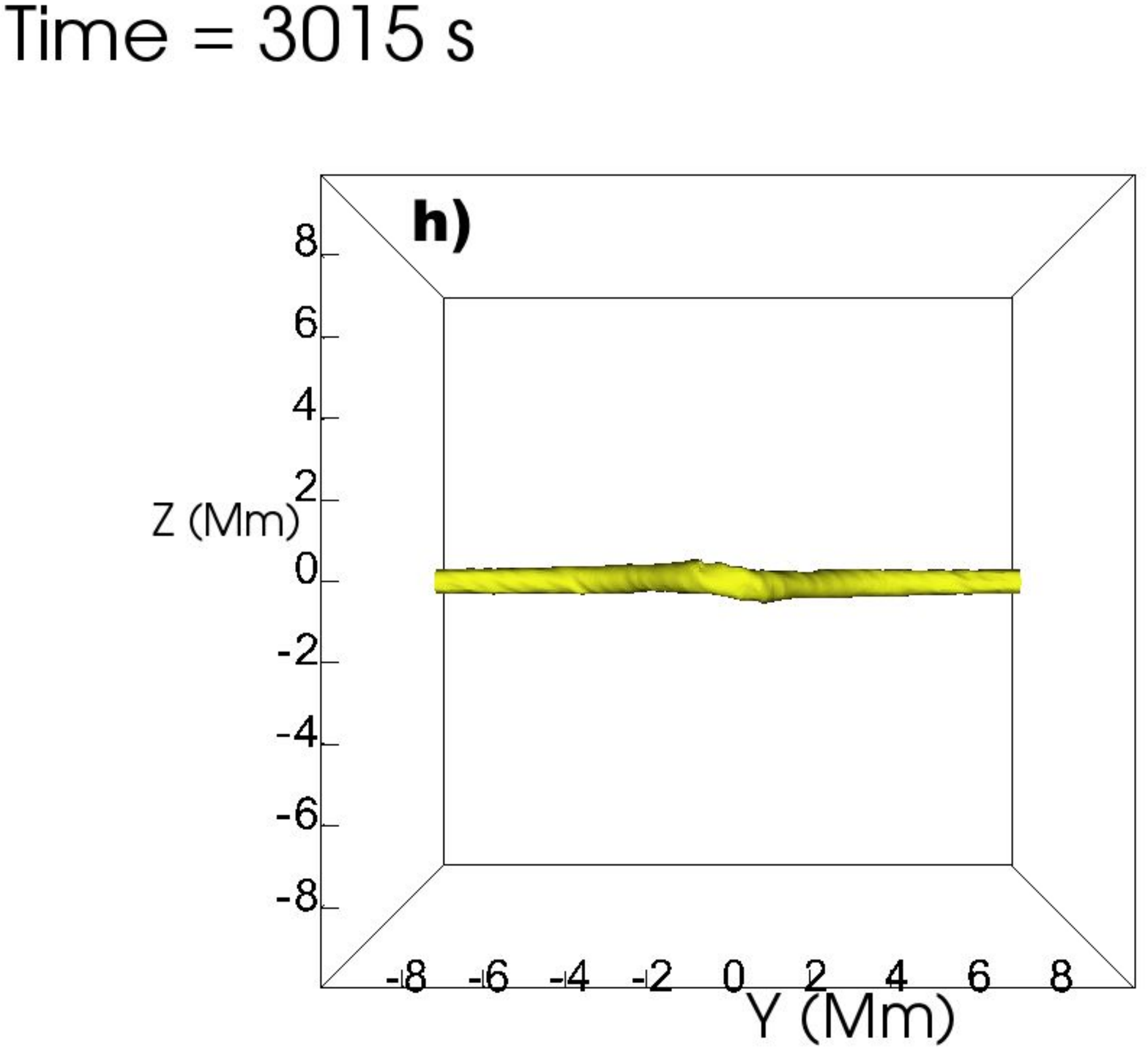}
\caption{Isosurfaces (yellow) of $B=1\;\mathrm{kG}$ during the $\zeta=0.5$ flux rope's rise through the convection zone. The photosphere is located at $x=0$. Left: Seen from the side, with color shading representing temperature, plotted on a linear scale, and mass density, plotted on a log scale. Right: Seen from above.}
\label{fig:isosurfaces05}
\end{figure*}

\begin{figure*}
\centering\includegraphics[scale=0.5, trim=0.0cm 0.0cm 0.0cm 0.0cm,clip=true]{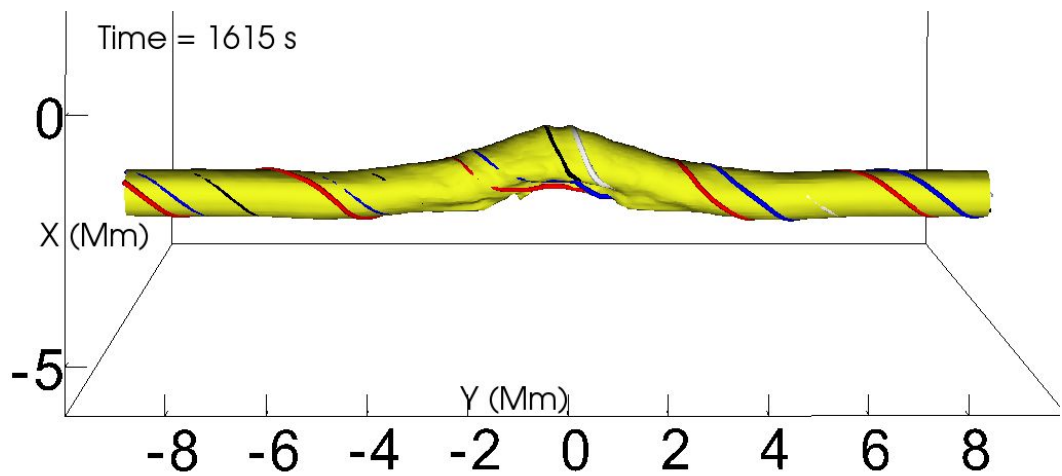}
\caption{Close-up of isosurface shown in Figures \ref{fig:isosurfaces05}c,d with several field lines overplotted on the striations visible along the isosurface.}
\label{fig:striations}
\end{figure*}

The first simulation we present is the initially kink stable flux rope, which starts with $\zeta=0.5$. The density perturbation creates a buoyant section in the middle of the flux rope, causing it to rise in an $\Omega$-shaped loop, seen in the left panels of Figure \ref{fig:isosurfaces05}. Striations, shown in close-up in Figure \ref{fig:striations}, form initially along the rising apex, but later extend down into the legs of the flux rope. These striations appear to form when field lines move apart to allow plasma to pass through, suggestive of the magnetic Rayleigh-Taylor instability \citep{Duan18}. The top-down view shown in the right panels of Figure \ref{fig:isosurfaces05} demonstrates that there is a slight rotation of the apex of the rising loop out of its initial plane, suggestive of kinking behavior. This is consistent with the finding that even flux ropes with $\zeta<1$ initially may, during the course of their rise, become kink-unstable \citep{Fan99}, though this is expected to occur due to an increase in the flux rope's radius, which is not observed here.  

\subsubsection{Photospheric Evolution}\label{sec:phot0.5}
\begin{figure*}
\centering\includegraphics[scale=0.32, trim=1.0cm 1.0cm 5.5cm 1.0cm,clip=true]{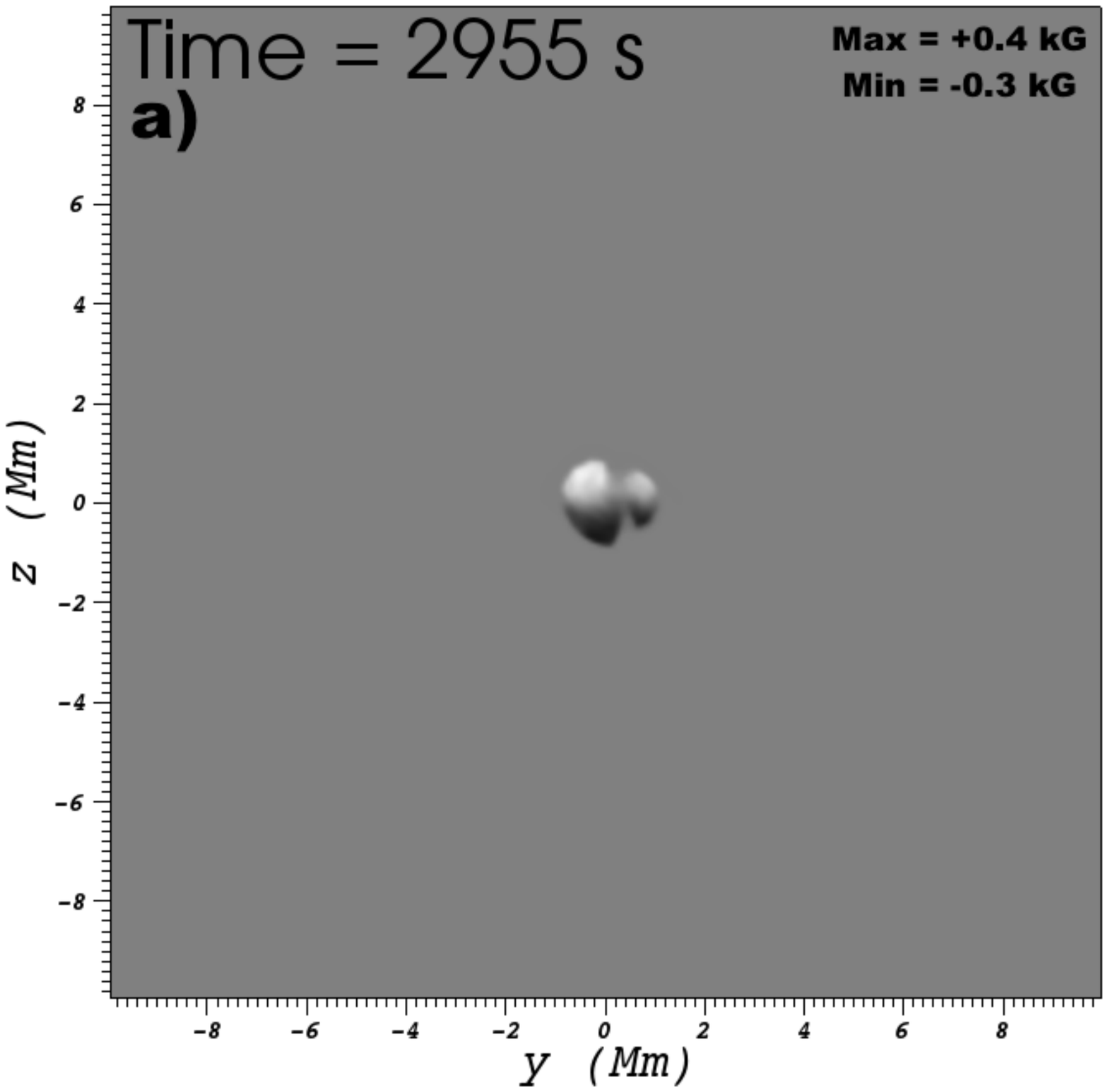}
\centering\includegraphics[scale=0.32, trim=3.0cm 1.0cm 5.5cm 1.0cm,clip=true]{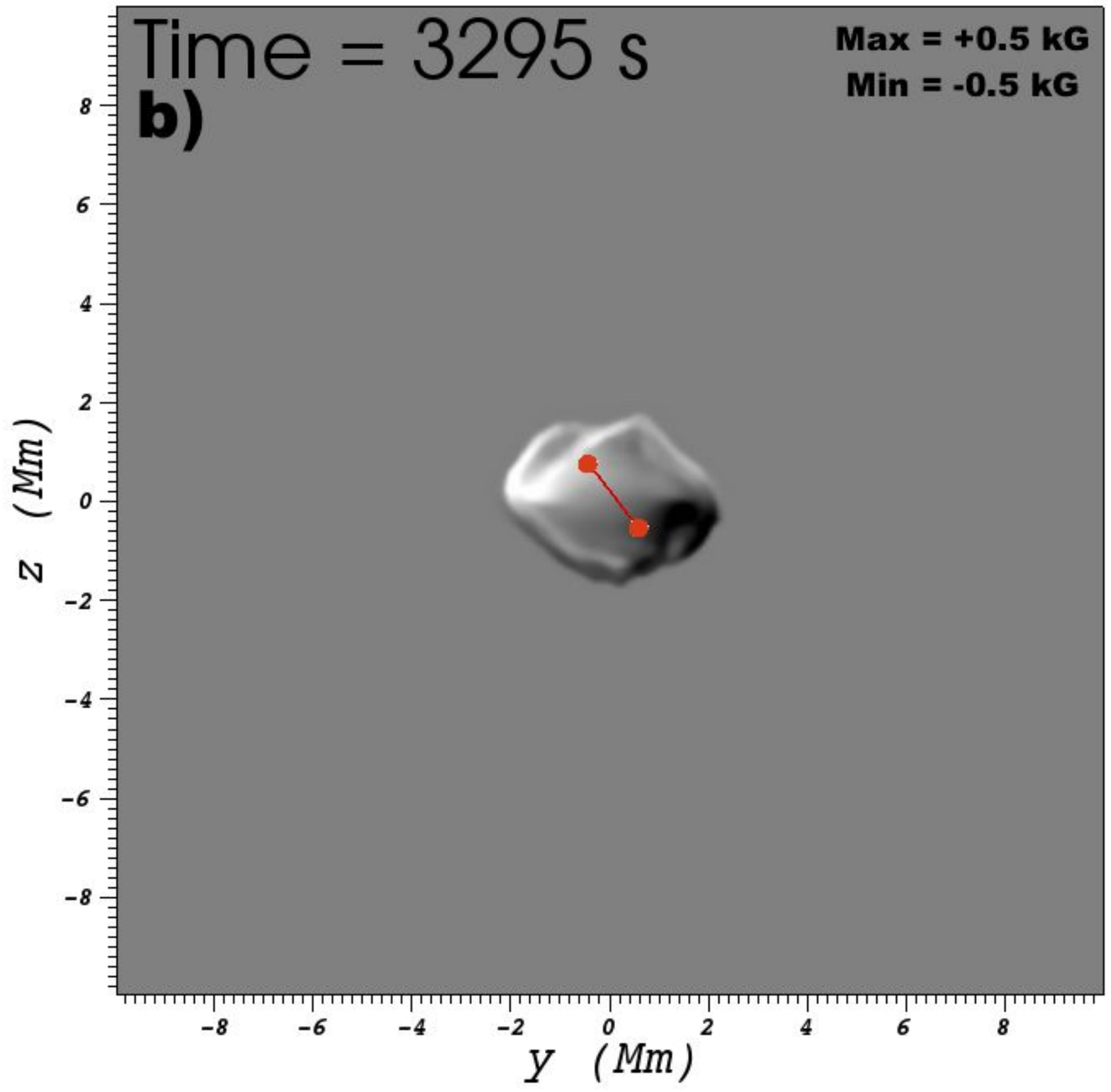}
\centering\includegraphics[scale=0.32, trim=1.0cm 1.0cm 5.5cm 1.0cm,clip=true]{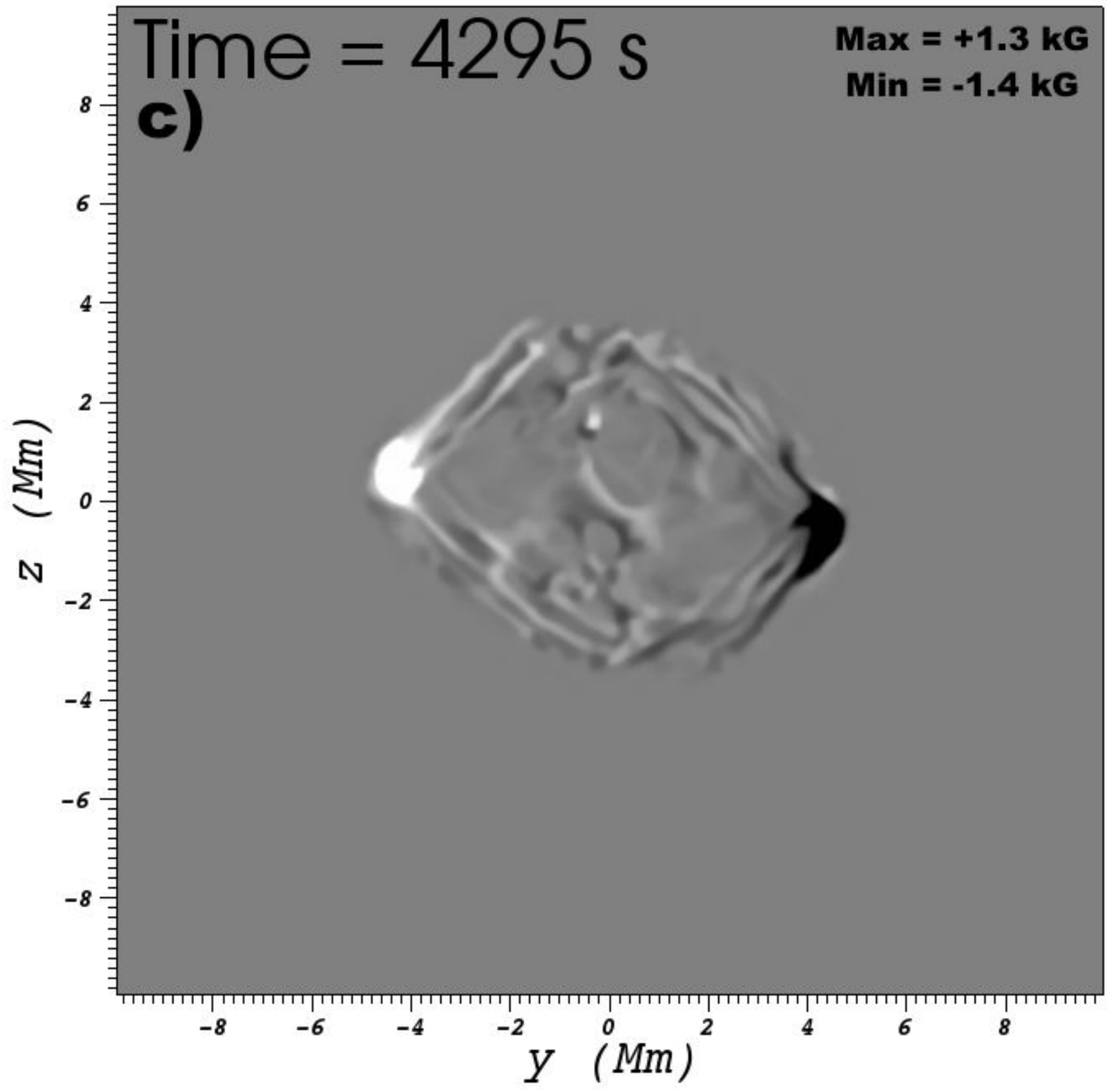}
\centering\includegraphics[scale=0.32, trim=3.0cm 1.0cm 5.5cm 1.0cm,clip=true]{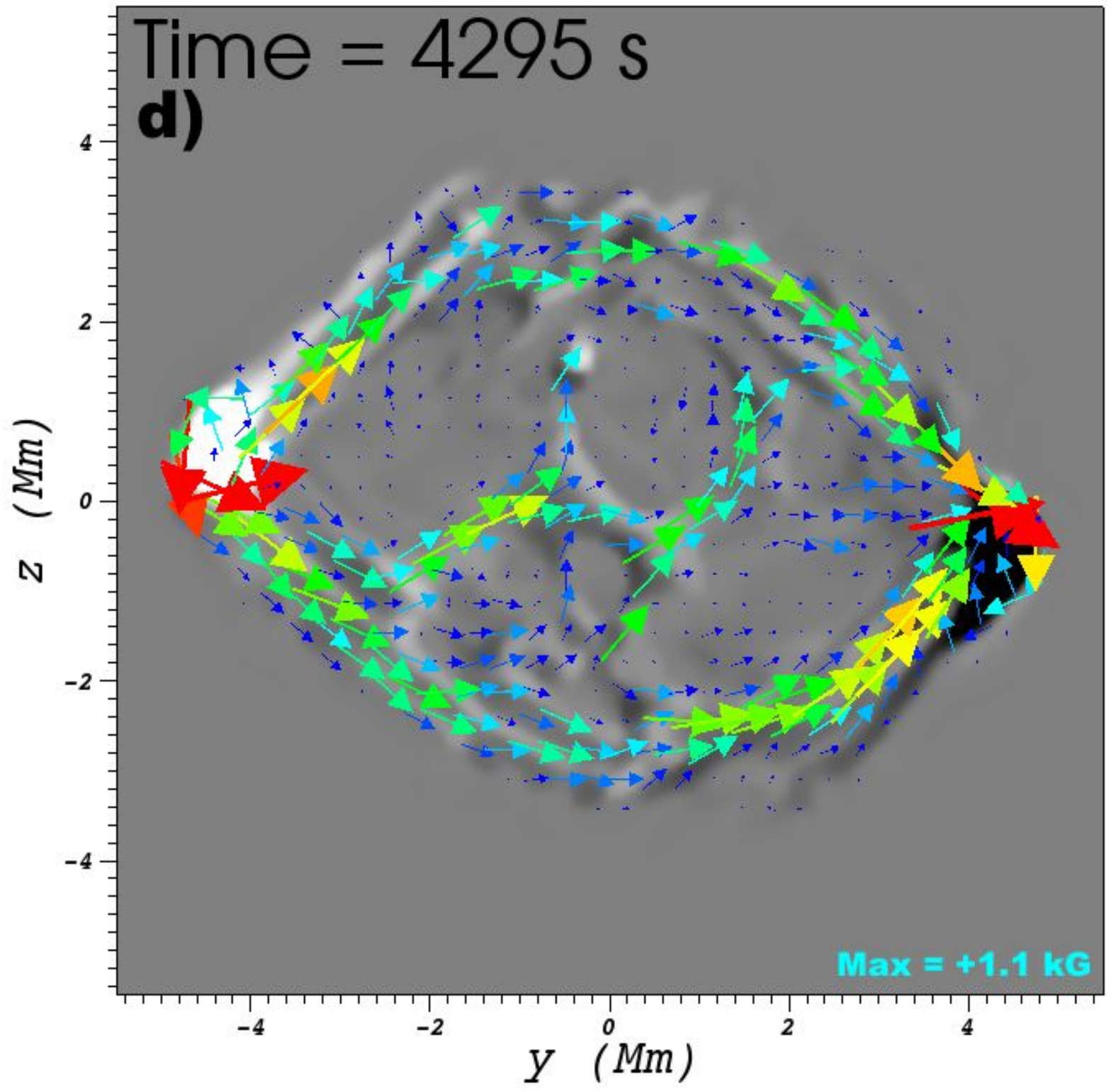}
\centering\includegraphics[scale=0.5, trim=0.0cm 0.0cm 0.0cm 0.0cm,clip=true]{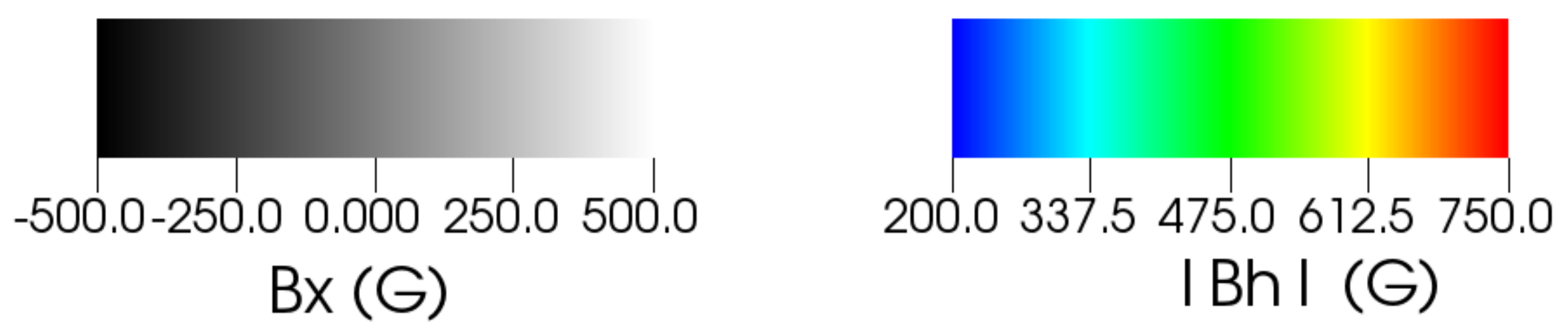}
\caption{Vertical field, $B_x$, on photosphere during the $\zeta=0.5$ simulation, saturated at $\pm500\;\mathrm{G}$. The middle state of the simulation (panel b) is shown with the flux weighted centers of mass (red circles connected with a line), while the final state of the simulation (panel d) has the horizontal field $\vecB_h$ as colored vectors overplotted on the greyscale.}
\label{fig:bx0}
\end{figure*}

The flux rope breaks through the photosphere at $t\approx2955\;\mathrm{s}$, producing two opposite polarity regions whose centroids would be connected by a line oriented approximately along the $z$-direction, as seen in Figure \ref{fig:bx0}a. The flux rope's axial field, oriented approximately in the $y$-direction throughout the emergence, is encircled by an azimuthal field which, at the apex of the loop, is oriented partially along the $z$-direction. It is this field that emerges first, producing the polarities seen in Figure \ref{fig:bx0}a. 
The striations that formed along the flux rope result in the formation of the second, smaller, set of polarities seen in Fig. \ref{fig:bx0}a when the flux rope hits the photosphere. The large and small polarities combine, rotate slightly and spread out as the legs of the flux rope break through the photosphere, forming two polarities with $B_x\sim\pm 500\;\mathrm{G}$ at around $t=3295\;\mathrm{s}$ (Figure \ref{fig:bx0}b). The flux weighted centers of mass of each polarities are depicted with the red circles in panel b. They are defined by
\beg{fwc}
\textbf{L}^\pm = \frac{\int_{S_{ph}}{dy\;dz\; \textbf{s} B_x^\pm(y,z)}}{\int_{S_{ph}}{dy\;dz B_x^\pm(y,z)}} 
\done 
with the distance between them being
\beg{com}
|\textbf{L}_{com}| = |\textbf{L}^+ -\textbf{L}^-|.
\done
Here $B_x^{\pm}$ is the positive/negative magnetic field normal to the photospheric surface $S_{ph}$ and {\color{black}$\textbf{s}=y\unit{y}+z\unit{z}$} is the radial coordinate along the photosphere. $\textbf{L}^\pm$ was evaluated by integrating over the entire active region, defined as locations at $x=0$ where $|B_x|>B_x^{thresh}$. We used $B_x^{thresh}=0.1B_x^{max}$, but the results presented below were qualitatively unchanged for $B_x^{thresh}=20\;\mathrm{G}$. \par
\begin{figure*}
\centering\includegraphics[scale=0.25, trim=0.0cm 0.0cm 0.0cm 0.0cm,clip=true]{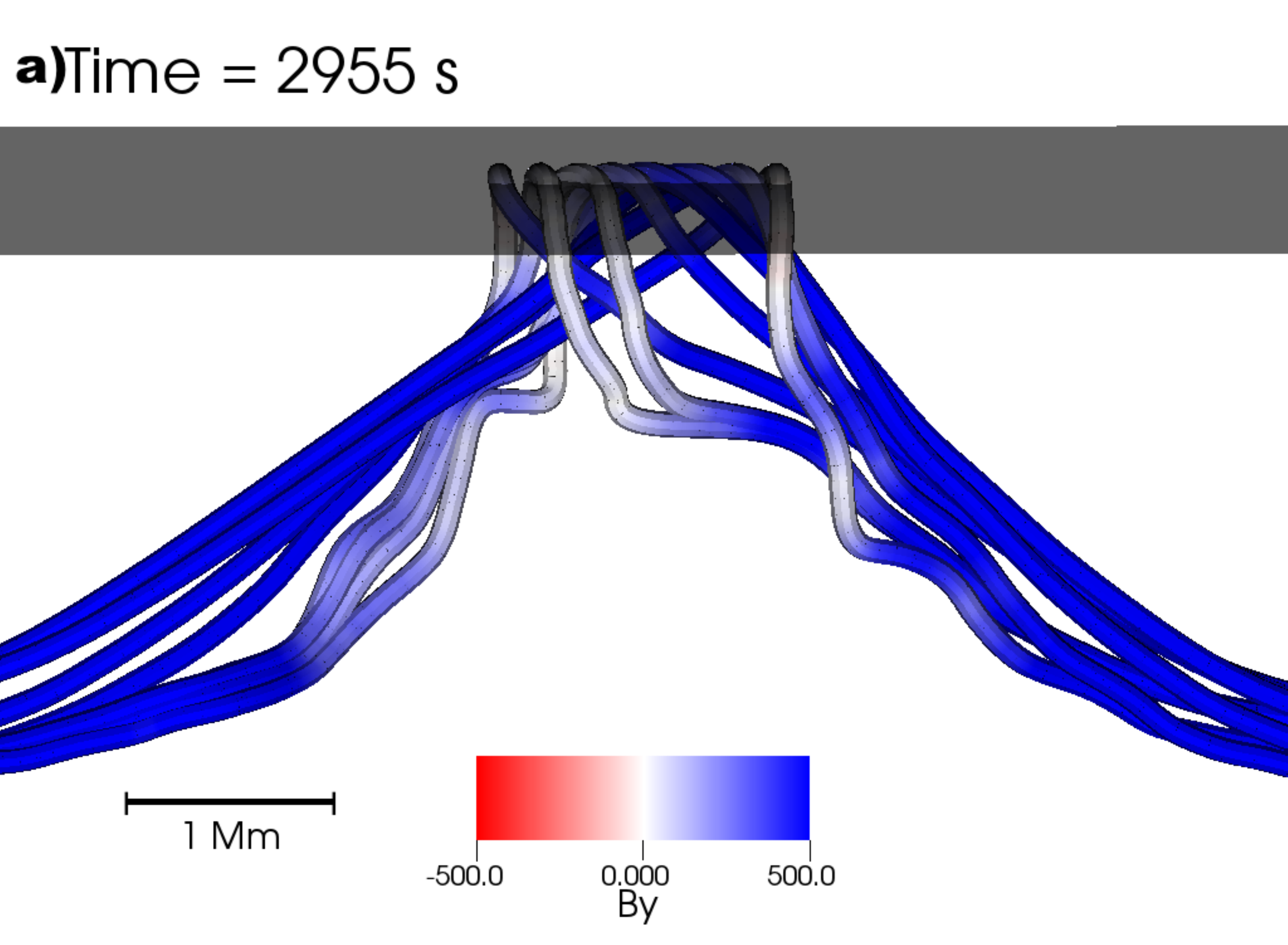}
\centering\includegraphics[scale=0.25, trim=0.0cm 0.0cm 0.0cm 0.0cm,clip=true]{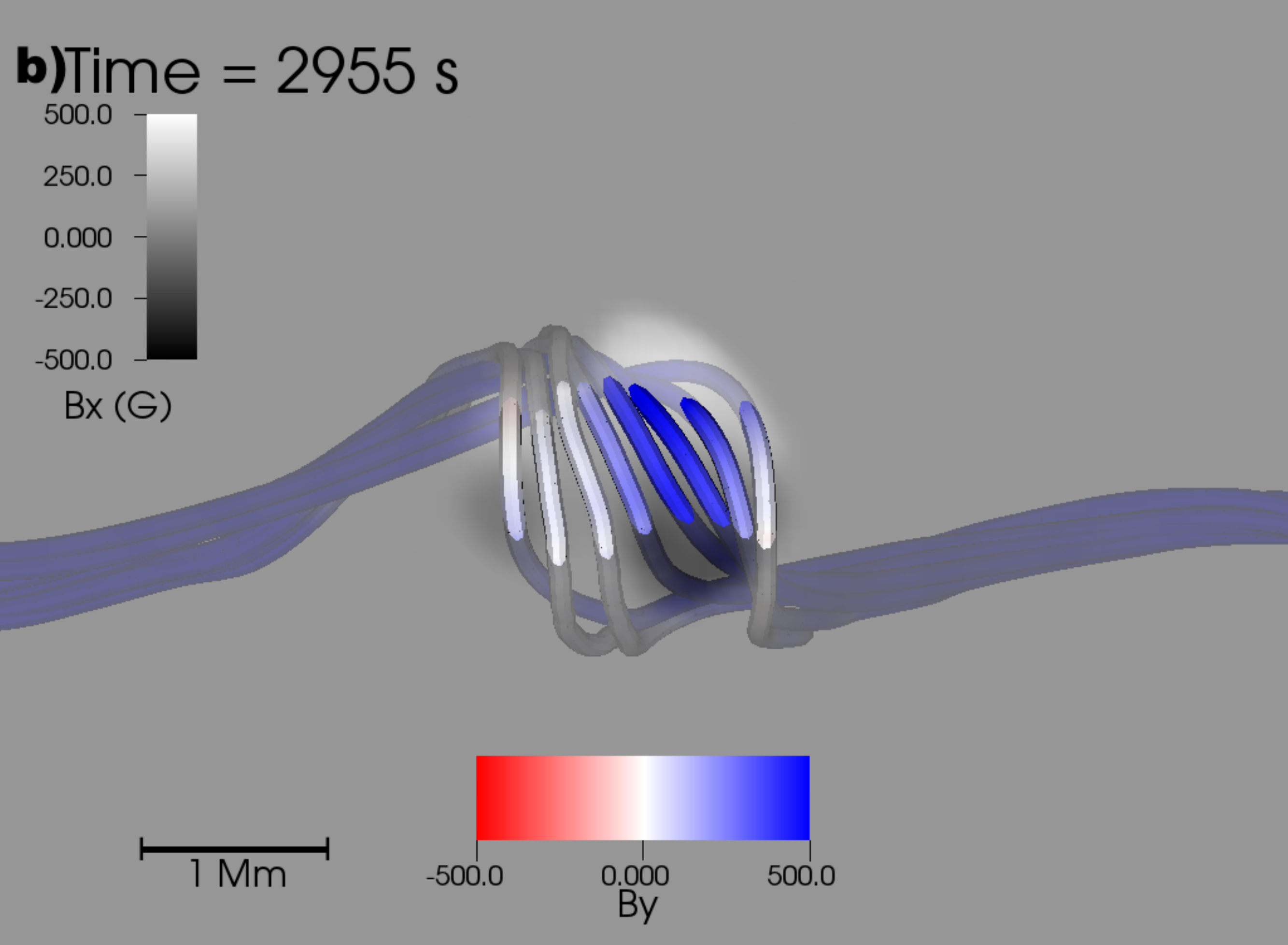}
\centering\includegraphics[scale=0.25, trim=0.0cm 0.0cm 0.0cm 0.0cm,clip=true]{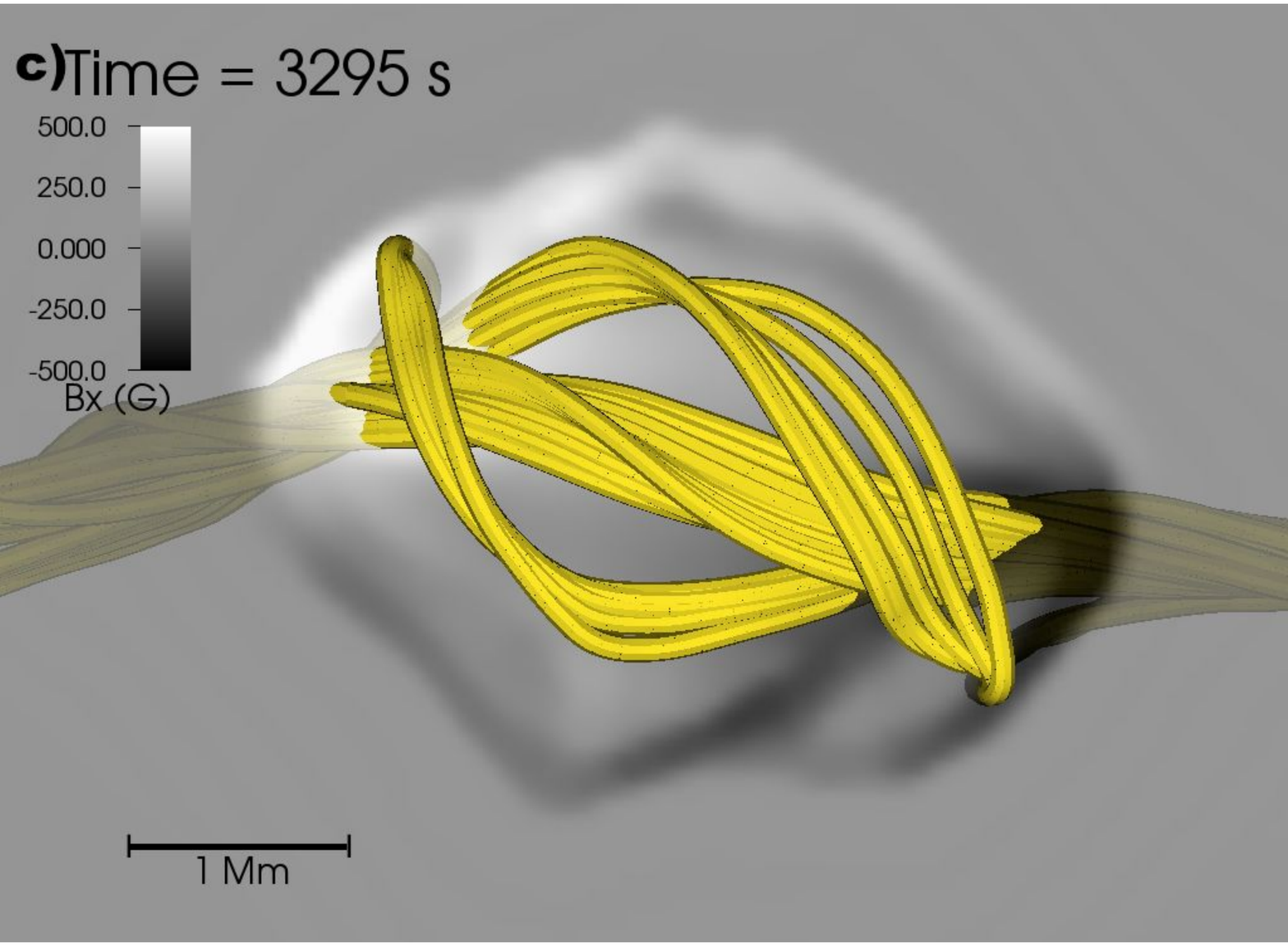}
\centering\includegraphics[scale=0.25, trim=0.0cm 0.0cm 0.0cm 0.0cm,clip=true]{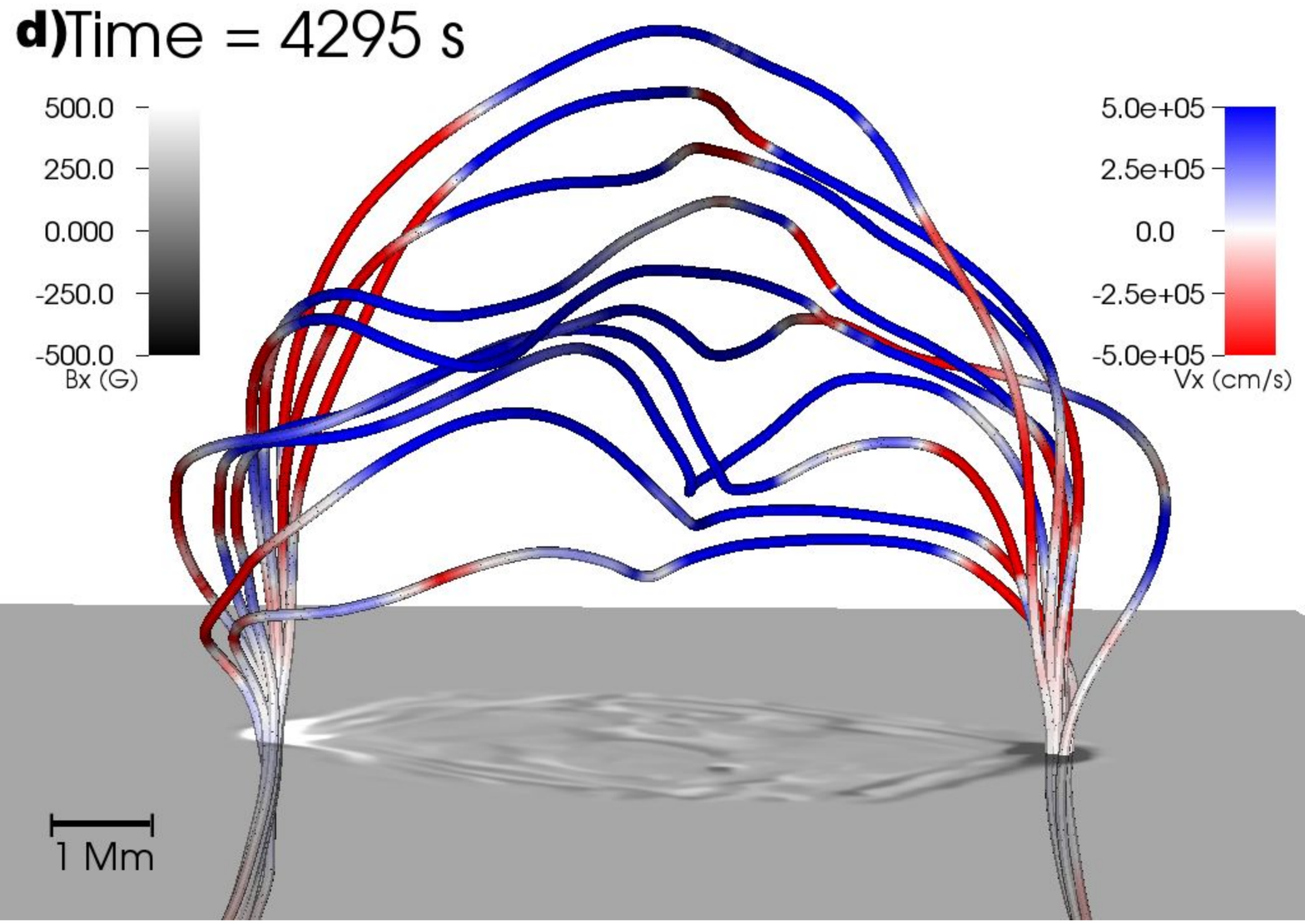}
\caption{Field lines, overplotted on photospheric magnetograms, at various stages of the $\zeta=0.5$ simulation. In a) and b) field lines, colored by $B_y$, are seen from the side and above, respectively. In d) field lines are shown at the final state of the simulation, colored by $v_x$.}
\label{fig:fieldlines0}
\end{figure*}

In Figures \ref{fig:fieldlines0}a,b we plot several field lines traced from the photosphere, and view them from the side and from above. These field lines are colored by the component of their field along the original axial ($y$) direction. The field that emerges into the corona is weakly twisted, and a sheared arcade is evident between the two polarities later in the emergence (Figure \ref{fig:fieldlines0}c). As the legs of the flux rope emerge, the two main polarities continue to separate, and weak flux towards the edges of the flux rope becomes scattered between the two main polarities (Figure \ref{fig:bx0}c). The vectors overplotted in Figure \ref{fig:bx0}d show the horizontal field at $x=0$. The horizontal field is twisted in the two main polarities, and is much stronger than in the region between the two polarities (Figure \ref{fig:bx0}d). As can be seen from Figure \ref{fig:fieldlines0}d, the resulting configuration is approximately an inverted U-loop in the corona, with plasma flowing down the legs of the emerged flux rope, and with the apex of each field line comprising the inverted U-loop rising higher into the corona.
\subsection{$\zeta=1$}\label{sec:1}

\subsubsection{Subsurface Evolution}\label{sec:sub1}
\begin{figure*}
\centering\includegraphics[scale=0.29, trim=0.0cm 1.0cm 4.0cm 0.0cm,clip=true]{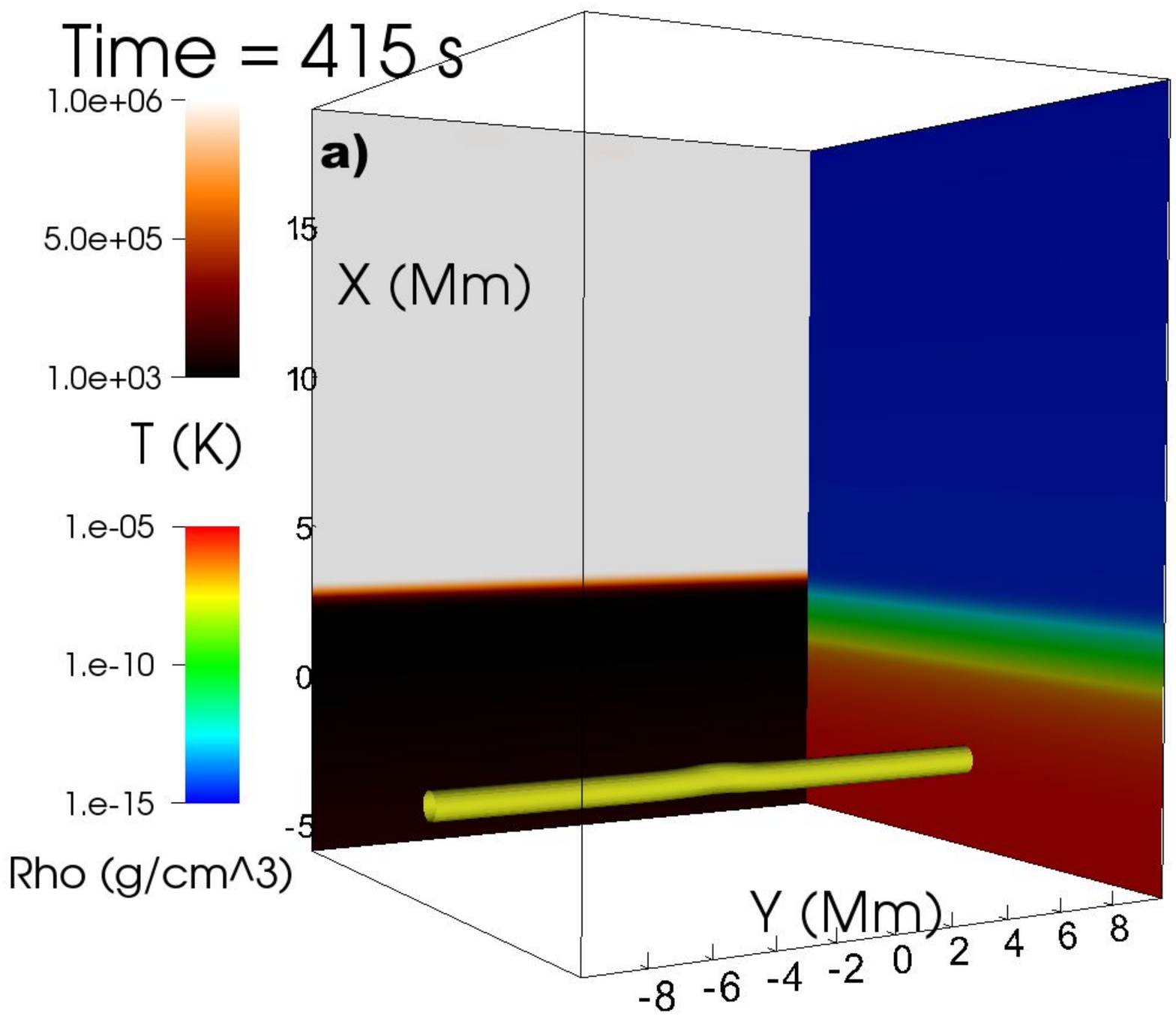}
\centering\includegraphics[scale=0.32, trim=4.0cm 1.0cm 5.0cm 3.0cm,clip=true]{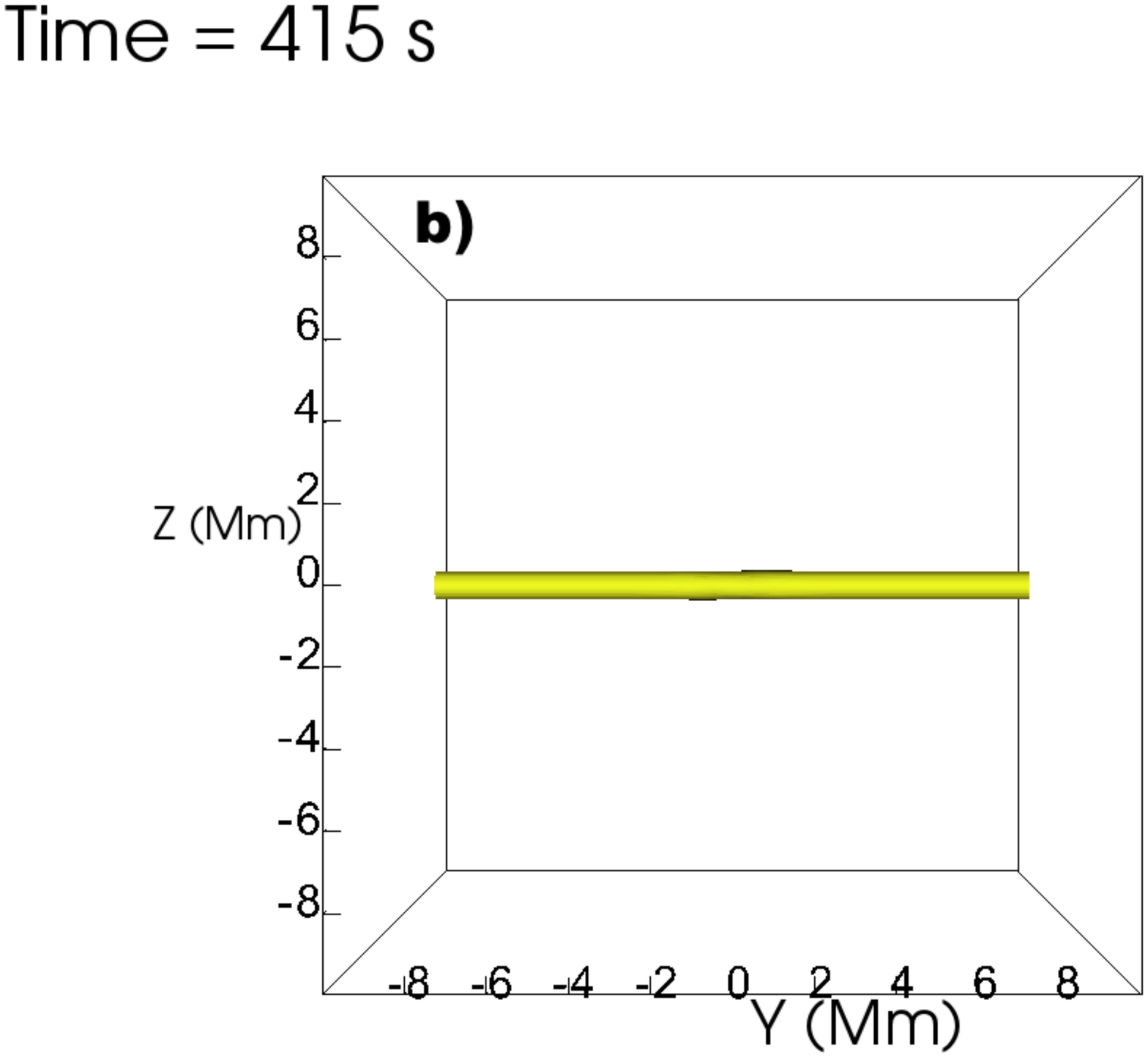}
\centering\includegraphics[scale=0.29, trim=0.0cm 1.0cm 4.0cm 0.0cm,clip=true]{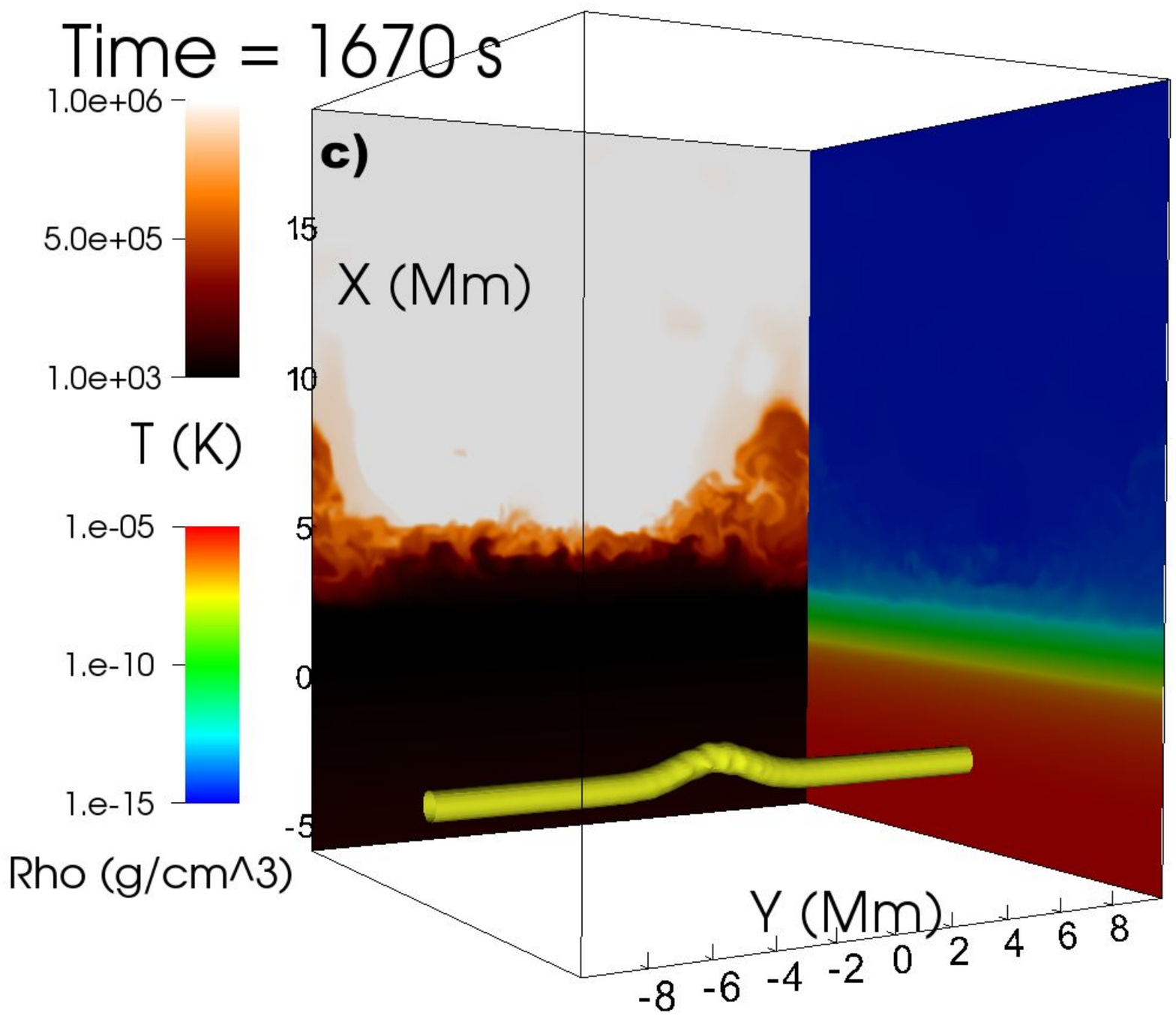}
\centering\includegraphics[scale=0.32, trim=4.0cm 1.0cm 5.0cm 3.0cm,clip=true]{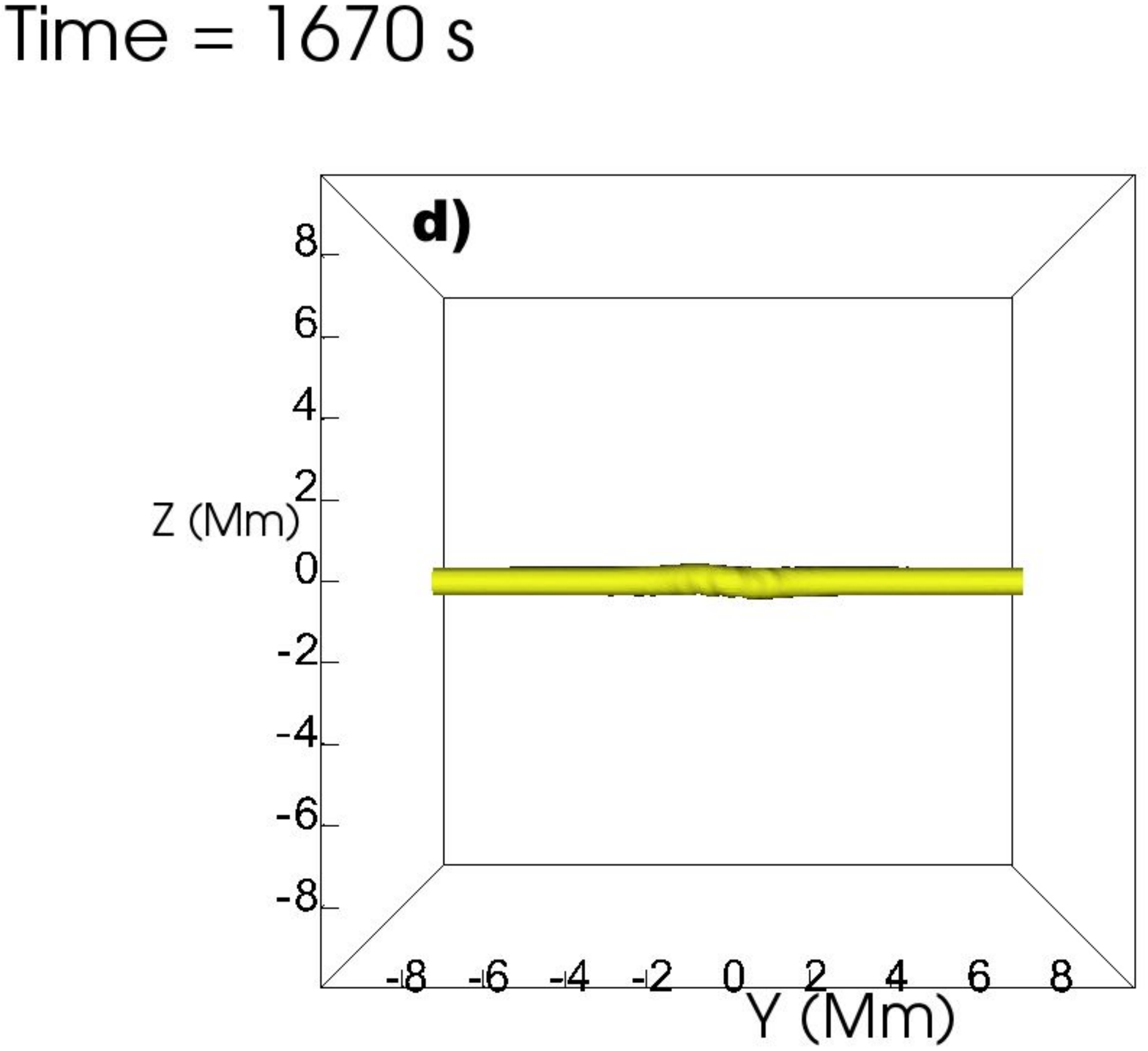}
\centering\includegraphics[scale=0.29, trim=0.0cm 1.0cm 4.0cm 0.0cm,clip=true]{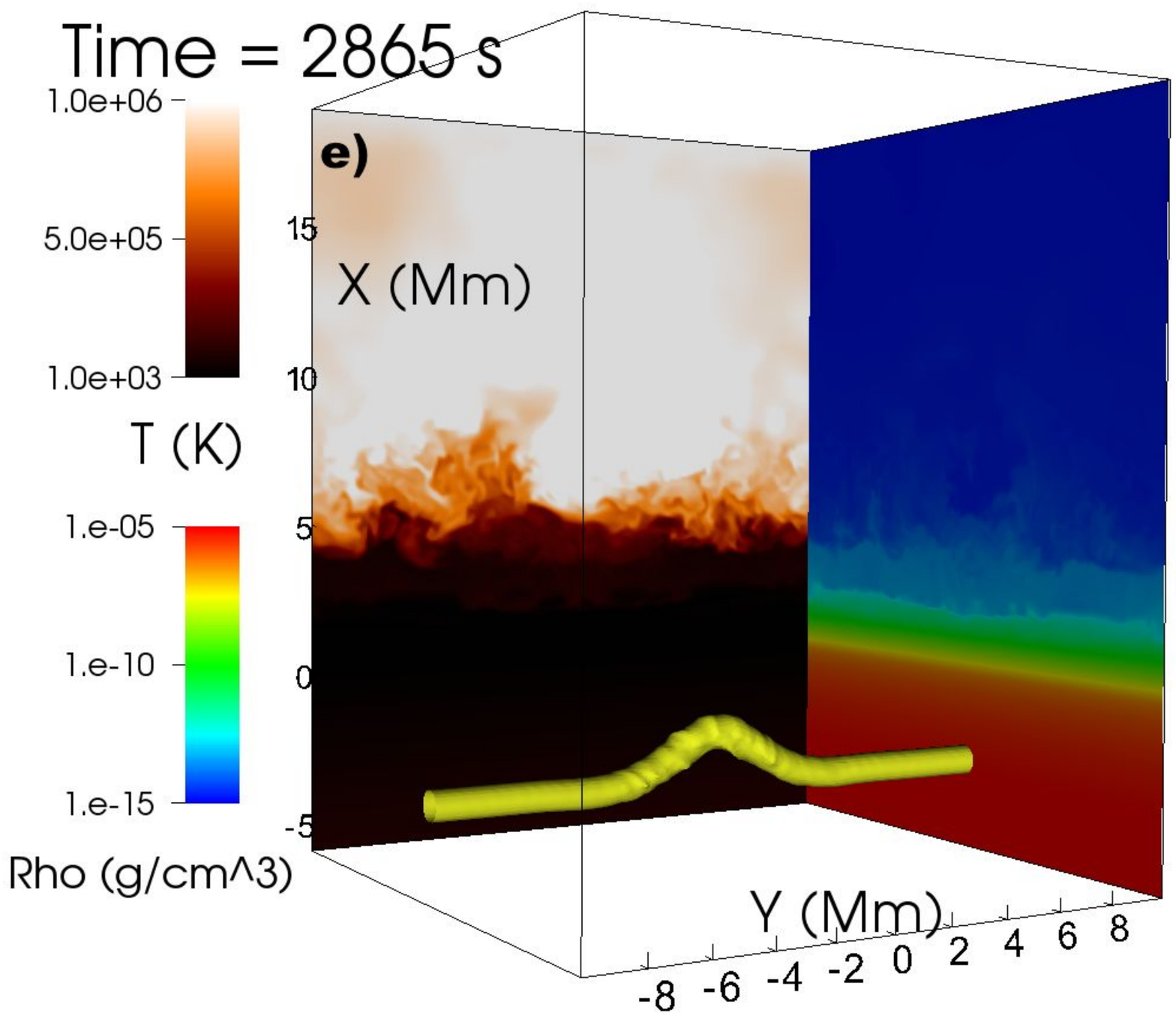}
\centering\includegraphics[scale=0.32, trim=4.0cm 1.0cm 5.0cm 3.0cm,clip=true]{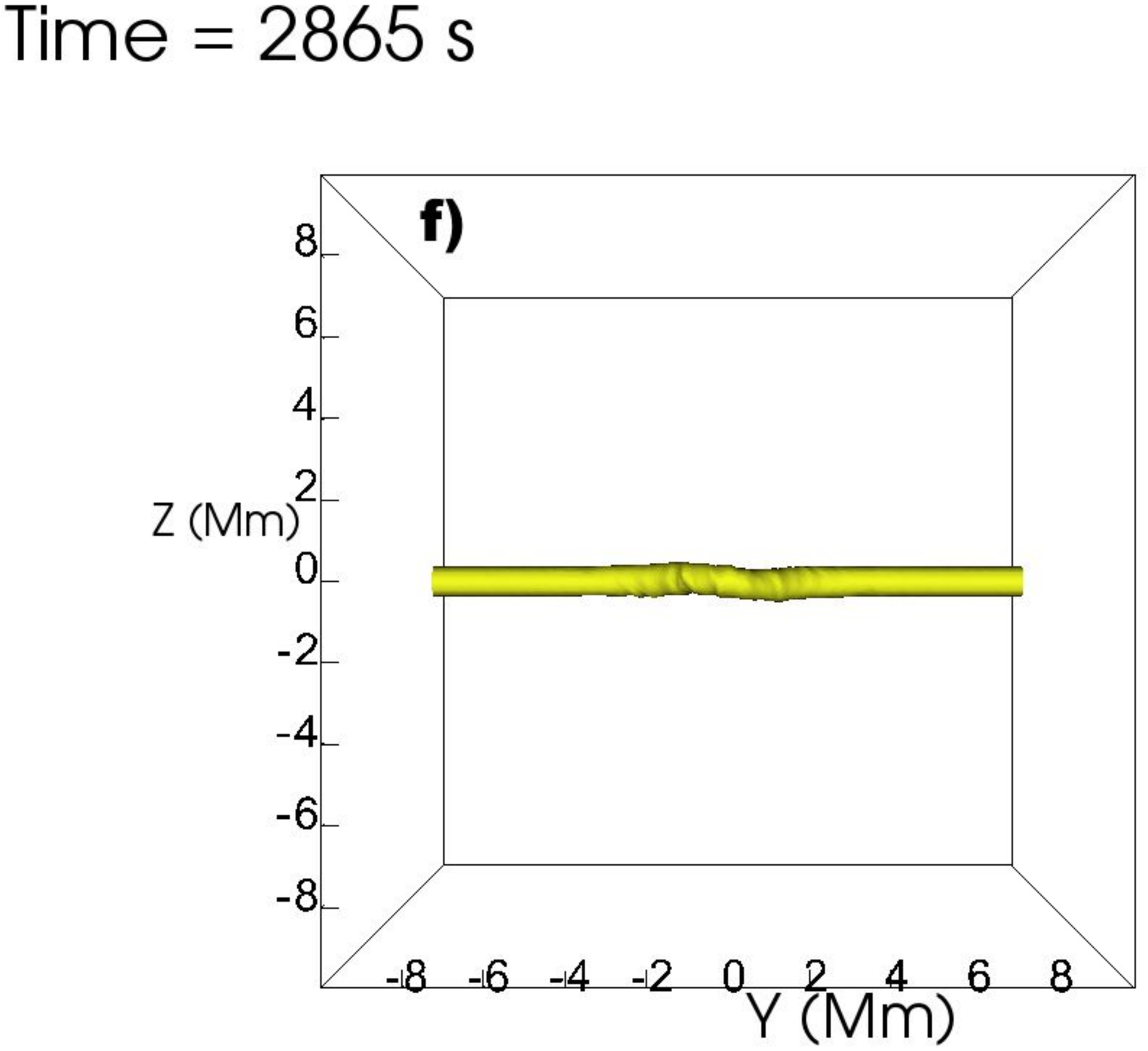}
\centering\includegraphics[scale=0.29, trim=0.0cm 1.0cm 4.0cm 0.0cm,clip=true]{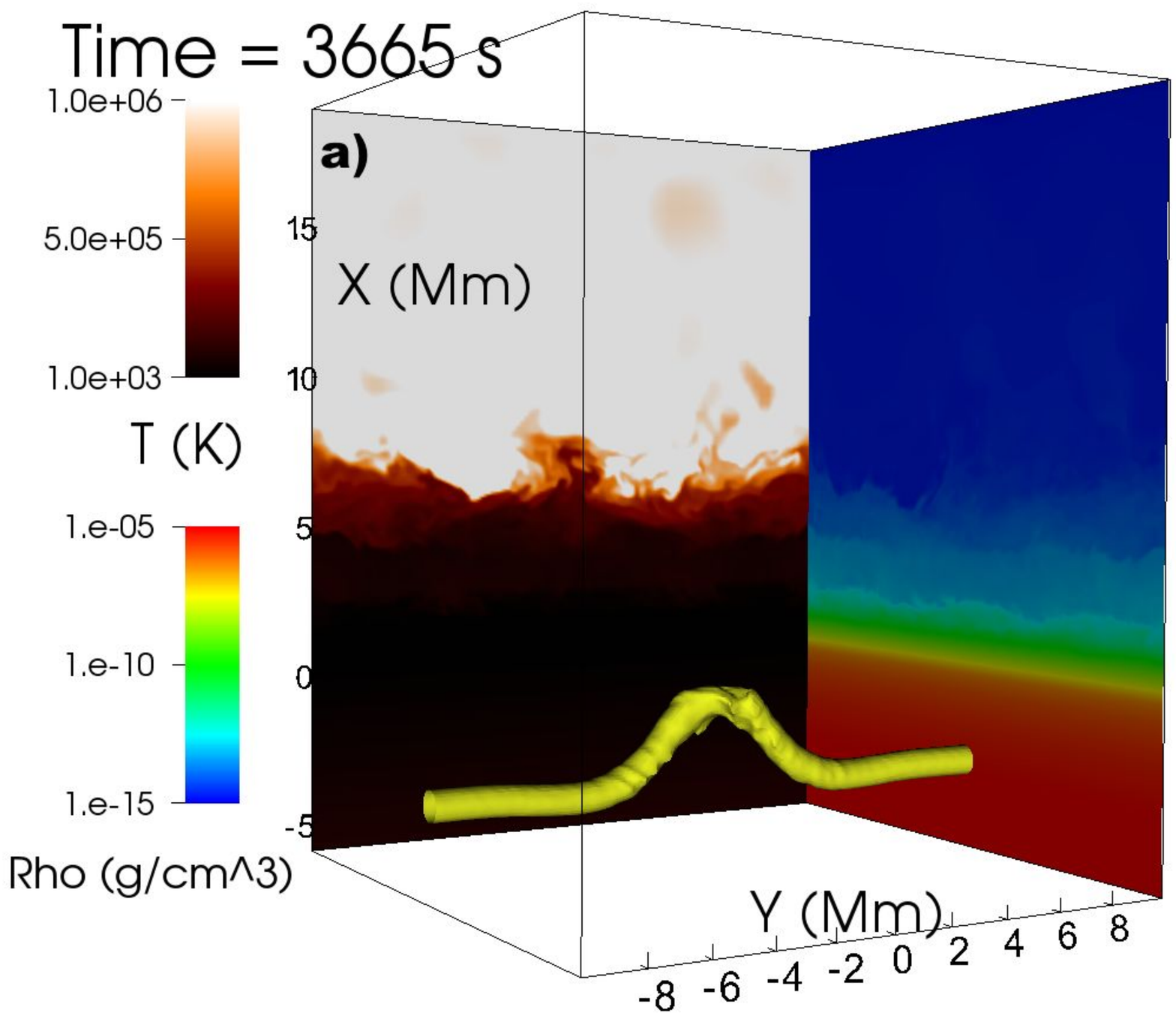}
\centering\includegraphics[scale=0.32, trim=4.0cm 1.0cm 5.0cm 3.0cm,clip=true]{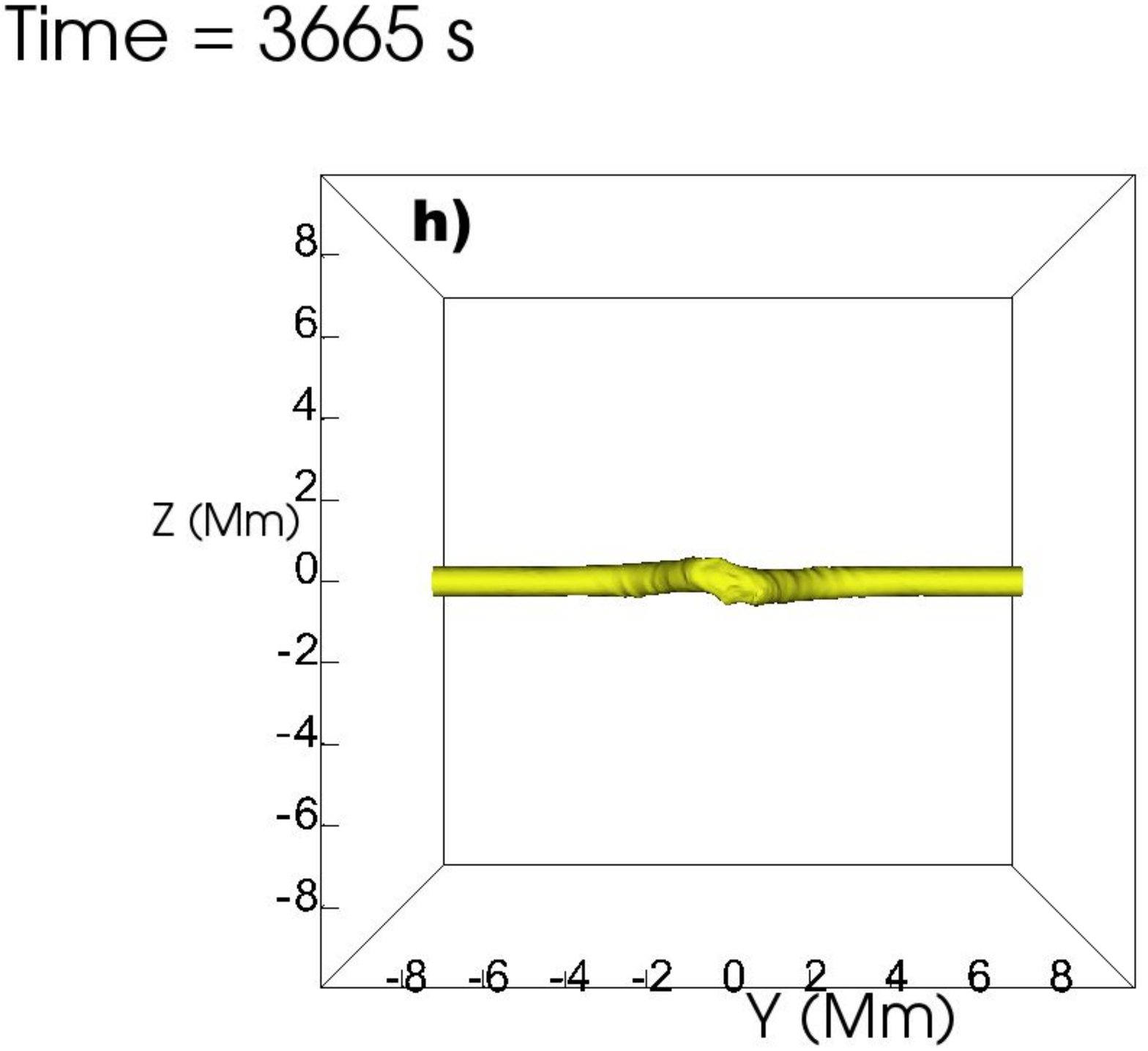}
\caption{Same as Figure \ref{fig:isosurfaces05} for $\zeta=1$}
\label{fig:isosurfaces1}
\end{figure*}

Next, we present the emergence of an initially marginally kink-stable $\zeta=1$ flux rope. The subsurface evolution of this flux rope, shown in Figure \ref{fig:isosurfaces1}, is qualitatively similar to that of the $\zeta=0.5$ case. An $\Omega$-loop rises through the convection zone, with striations forming first along the rising loop and then developing along the flux rope legs. The $\Omega$-loop starts to rotate gently around $t\approx1670\;s$, and is slightly offset from the $z=0$ axis when it hits the photosphere, suggesting that this flux rope may be undergoing the kink instability.

\subsubsection{Photospheric Evolution}\label{sec:phot1}
\begin{figure*}
\centering\includegraphics[scale=0.32, trim=1.0cm 1.0cm 5.5cm 1.0cm,clip=true]{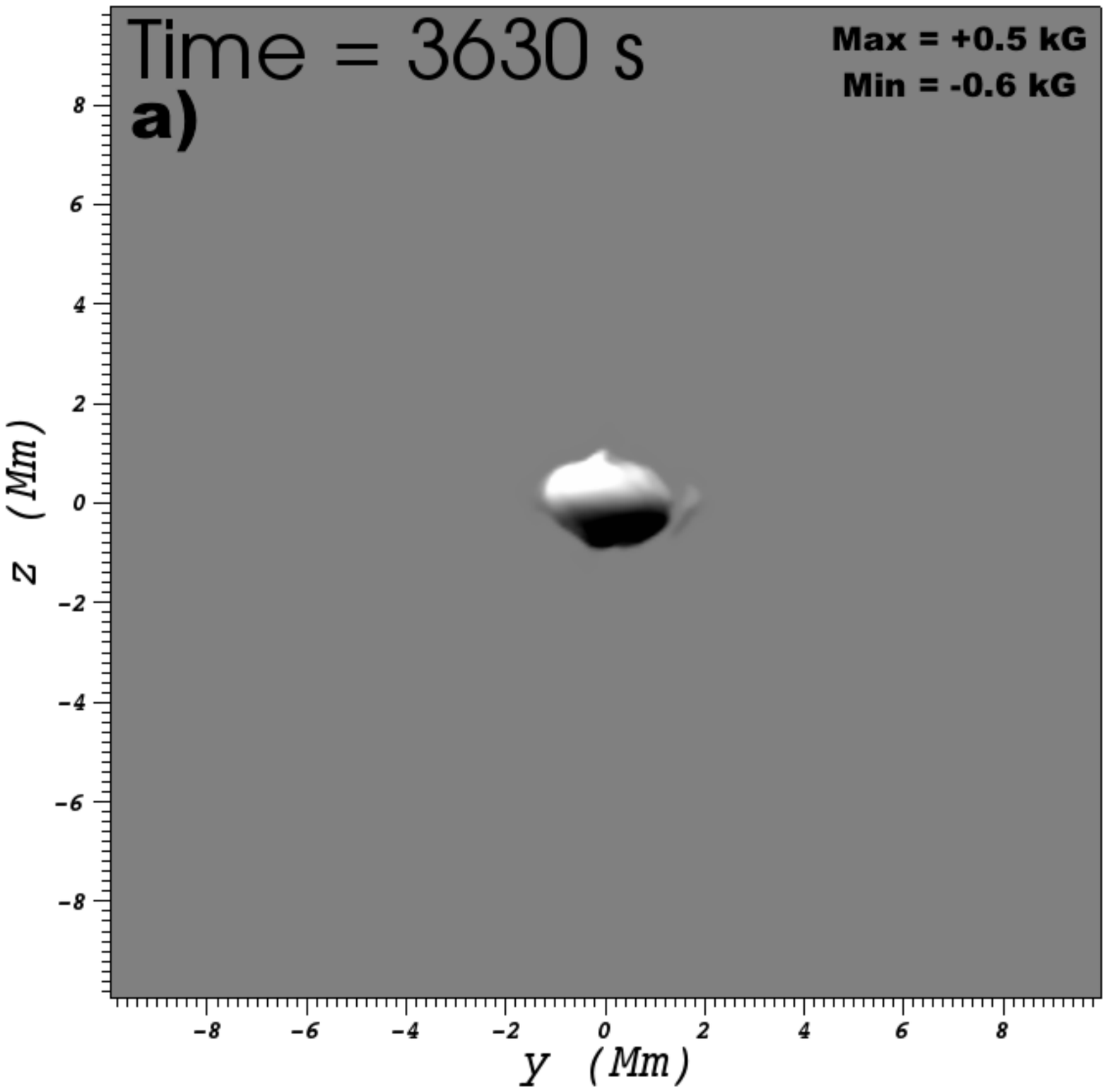}
\centering\includegraphics[scale=0.32, trim=3.0cm 1.0cm 5.5cm 1.0cm,clip=true]{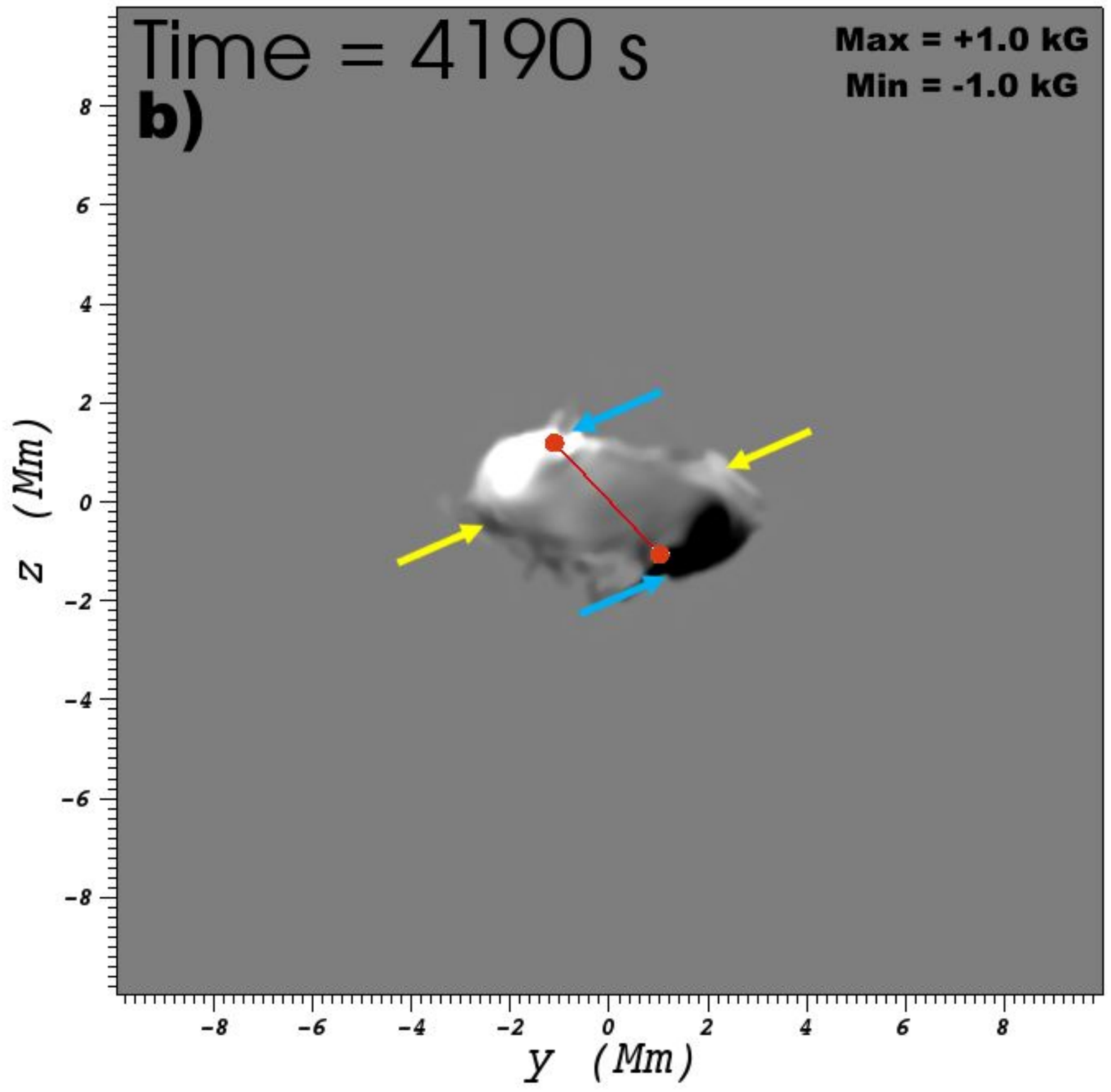}
\centering\includegraphics[scale=0.32, trim=1.0cm 1.0cm 5.5cm 1.0cm,clip=true]{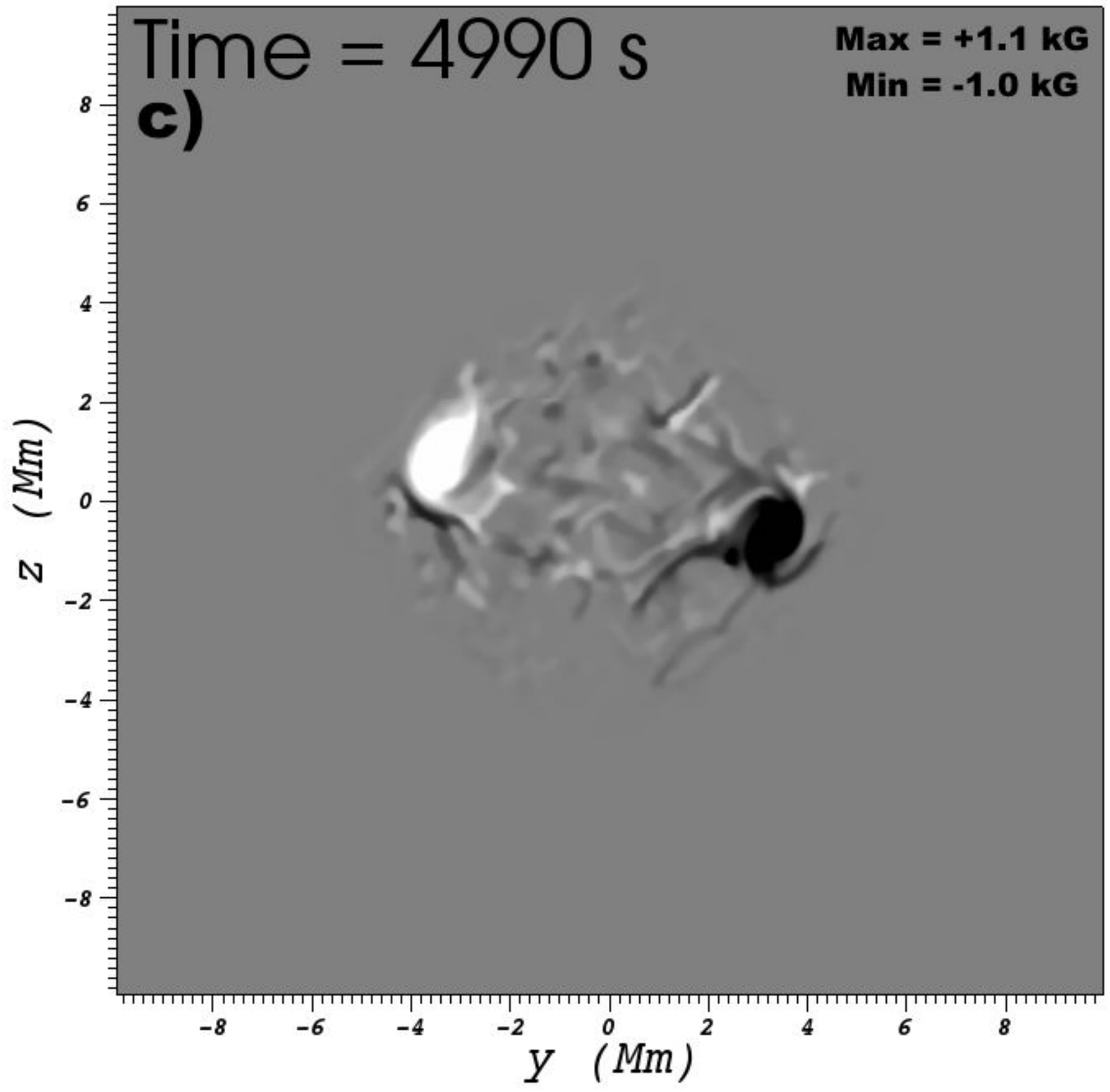}
\centering\includegraphics[scale=0.32, trim=3.0cm 1.0cm 5.5cm 1.0cm,clip=true]{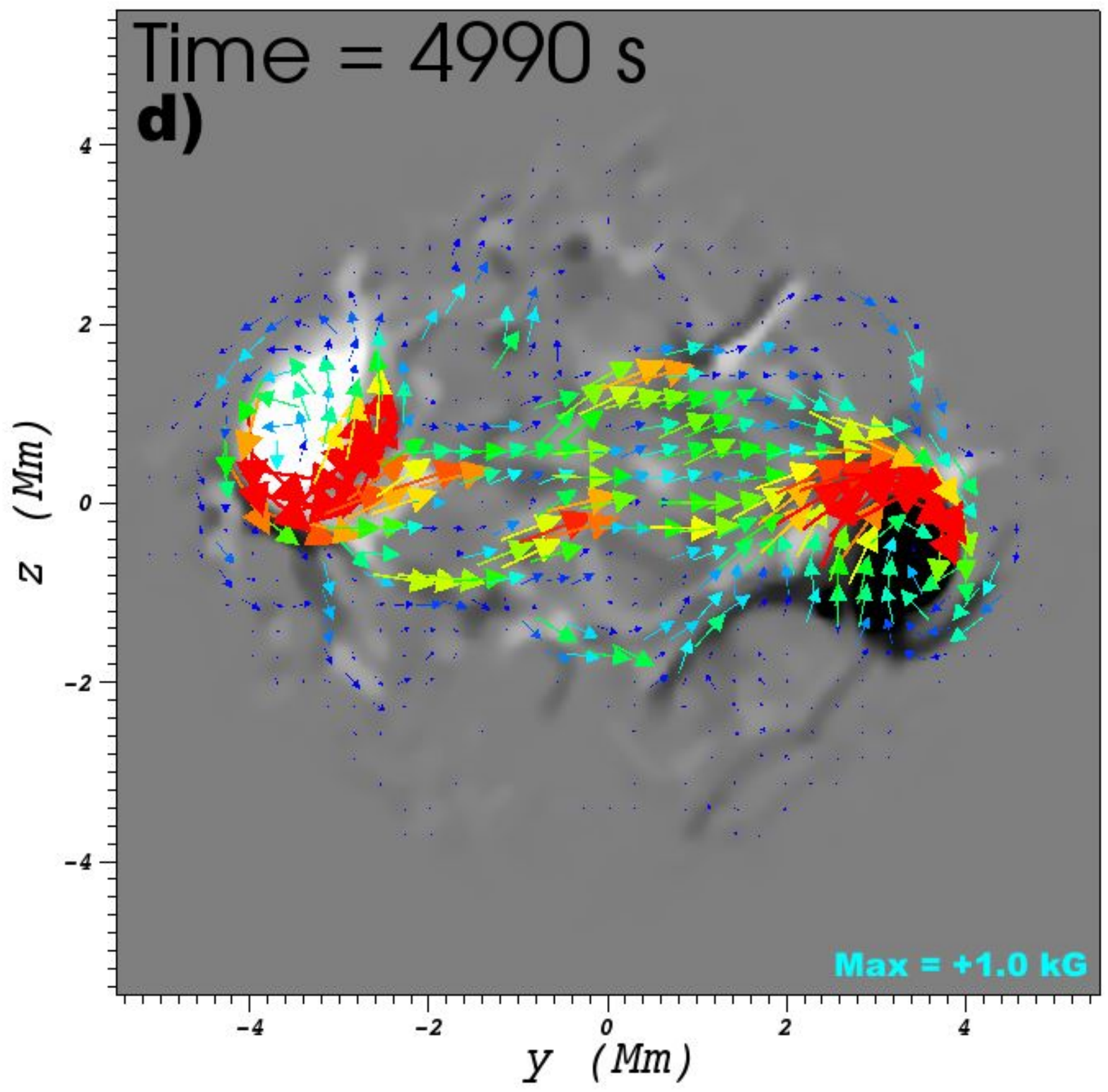}
\centering\includegraphics[scale=0.5, trim=0.0cm 0.0cm 0.0cm 0.0cm,clip=true]{colorbar.pdf}
\caption{Same as Figure \ref{fig:bx0}, for the $\zeta=1$ case. Primary/secondary polarities are denoted by blue/yellow arrows in panel b).}
\label{fig:bx1}
\end{figure*}

\begin{figure*}
\centering\includegraphics[scale=0.25, trim=0.0cm 0.0cm 0.0cm 0.0cm,clip=true]{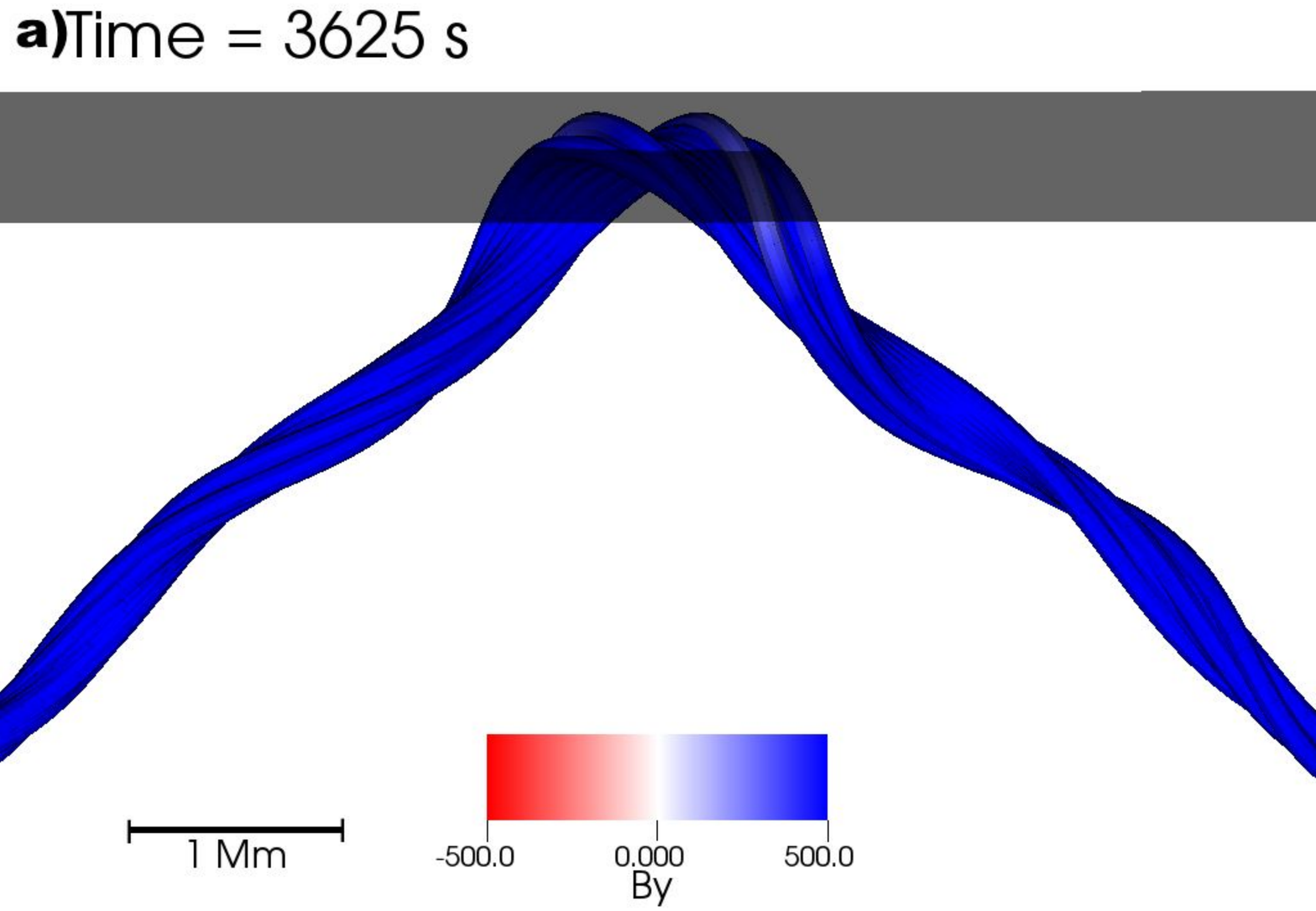}
\centering\includegraphics[scale=0.25, trim=0.0cm 0.0cm 0.0cm 0.0cm,clip=true]{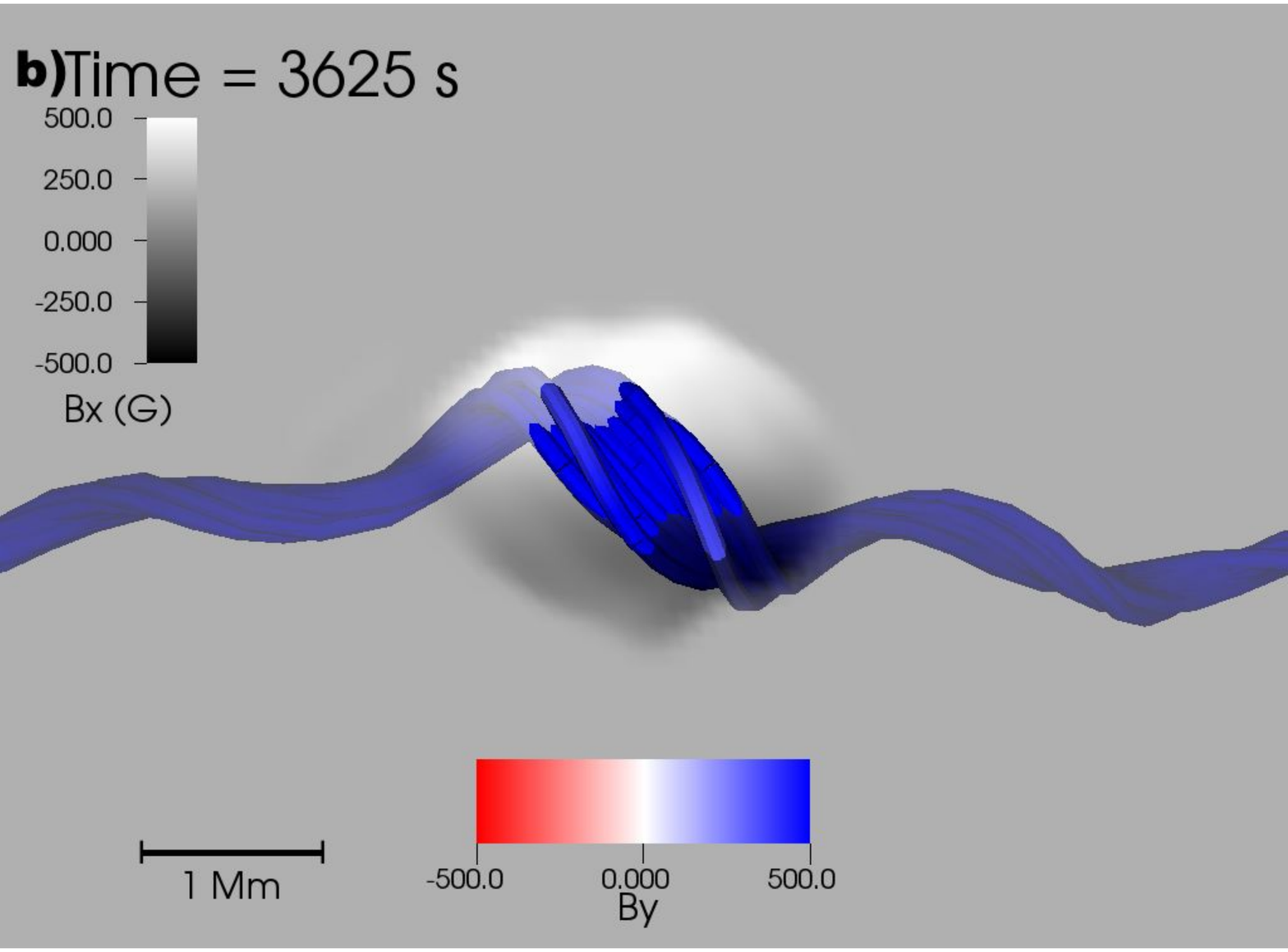}
\centering\includegraphics[scale=0.25, trim=0.0cm 0.0cm 0.0cm 0.0cm,clip=true]{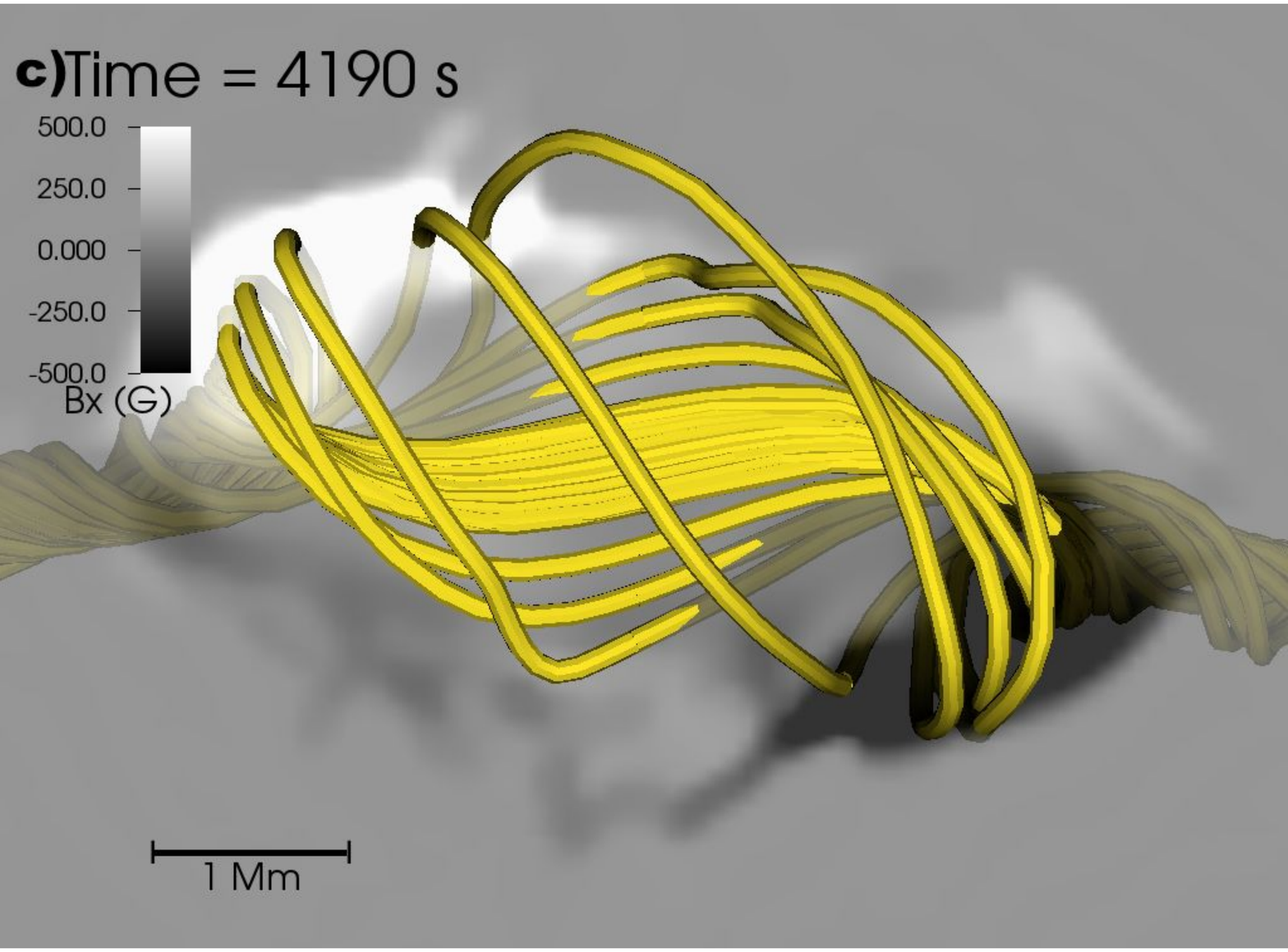}
\centering\includegraphics[scale=0.25, trim=0.0cm 0.0cm 0.0cm 0.0cm,clip=true]{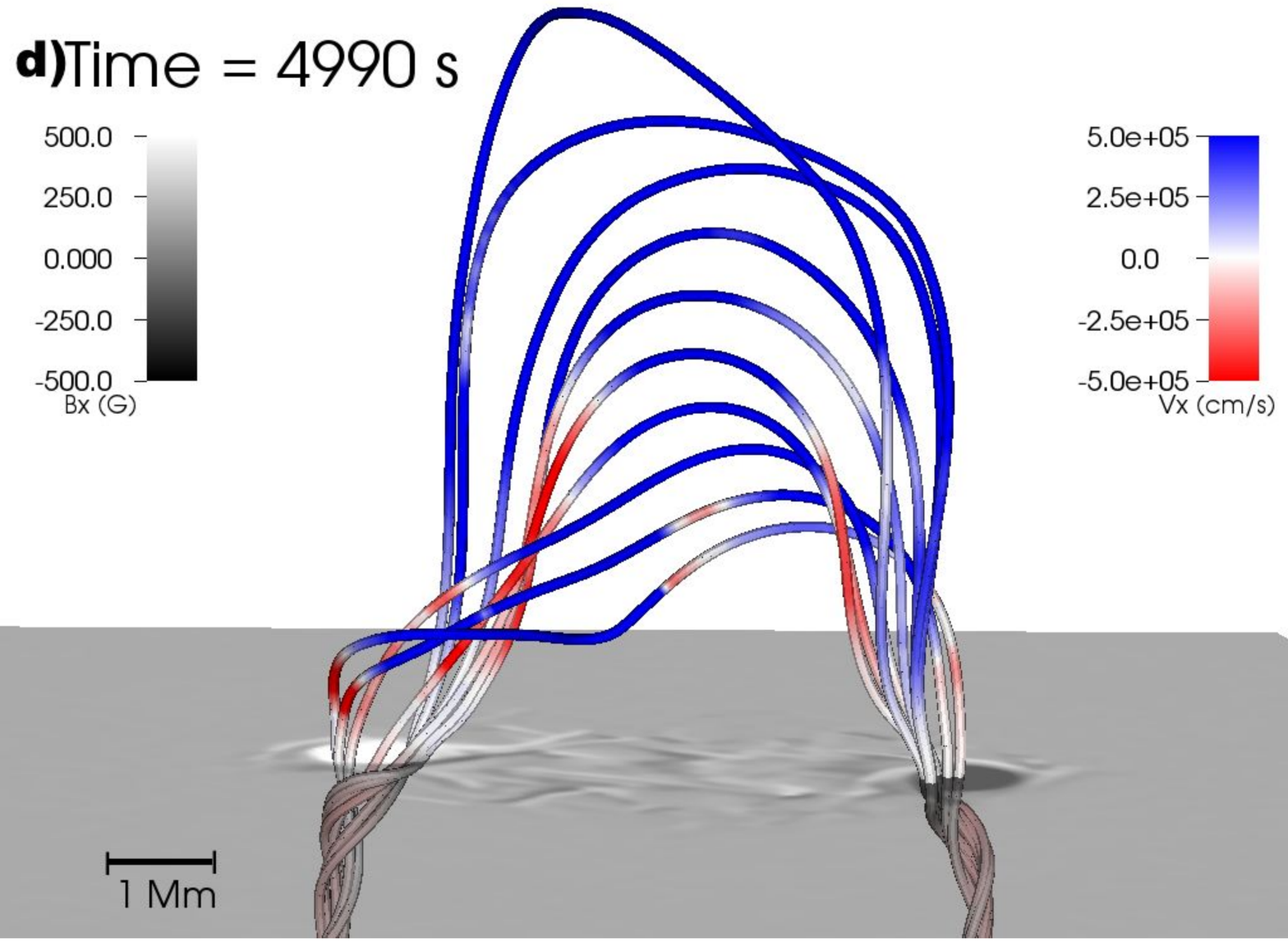}
\caption{Field lines, overplotted on photospheric magnetograms, at various stages of the $\zeta=1$ simulation. In a) and b) field lines, colored by $B_y$, are seen from the side and above, respectively. In d) field lines are shown at the final state of the simulation, colored by $v_x$.}
\label{fig:fieldlines1}
\end{figure*}

The flux rope breaks through the photosphere at $t\approx3625\;\mathrm{s}$ forming two well defined polarities (Figure \ref{fig:bx1}a), whose centroids are again connected by a line oriented approximately along the $z$-direction due to the twist field emerging first. Figures \ref{fig:fieldlines1}a,b show a set of field lines soon after the initial emergence from the side and from above, respectively, again colored by the axial field, $B_y$. They show the two polarities being formed by the apex of the emerging flux rope. This flux rope emerges much more coherently than the previous flux rope because the larger value of $\zeta$ helps prevent the break up of the flux during its rise through the stratified atmosphere.  These polarities gently rotate and separate, with magnetic field strengths approaching $B_x\sim\pm1\;\mathrm{kG}$ around $t=4190\;\mathrm{s}$ (Figure \ref{fig:bx1}b). Interestingly, small opposite polarity regions (yellow arrows) are visible near the primary polarities (blue arrows). At the coronal level, this rotation and separation results in the appearance of a flux rope comprised of sigmoidal field lines along with overlying quasi-potential loops (Figure \ref{fig:fieldlines1}c). Dispersed flux develops near and between the two main polarities as they continue to separate. The horizontal field is primarily twisting around the two main polarities, and is extremely weak between them at $t=4990\;\mathrm{s}$ (Figures \ref{fig:bx1}c,d). The end result is, like the $\zeta=0.5$ case, an inverted U-loop coronal arcade (Figure \ref{fig:fieldlines1}d), with plasma draining along the loop while the loop itself expands upward, carrying plasma with it. \par

\subsection{$\zeta=1.5$}\label{sec:1.5}
\subsubsection{Subsurface Evolution}\label{sec:sub1.5}
\begin{figure*}
\centering\includegraphics[scale=0.3, trim=0.0cm 1.0cm 4.0cm 0.0cm,clip=true]{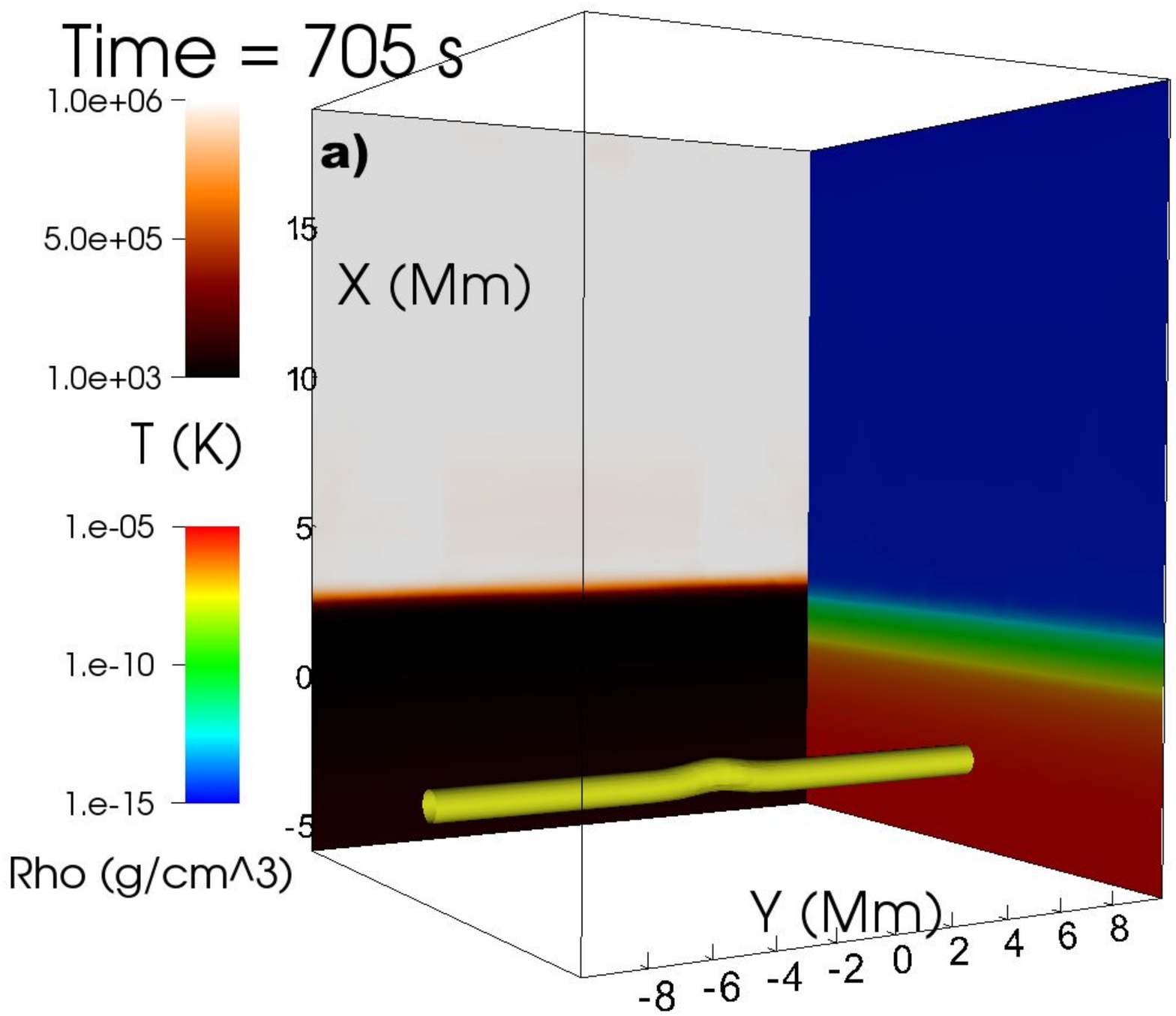}
\centering\includegraphics[scale=0.32, trim=4.0cm 1.0cm 5.0cm 3.0cm,clip=true]{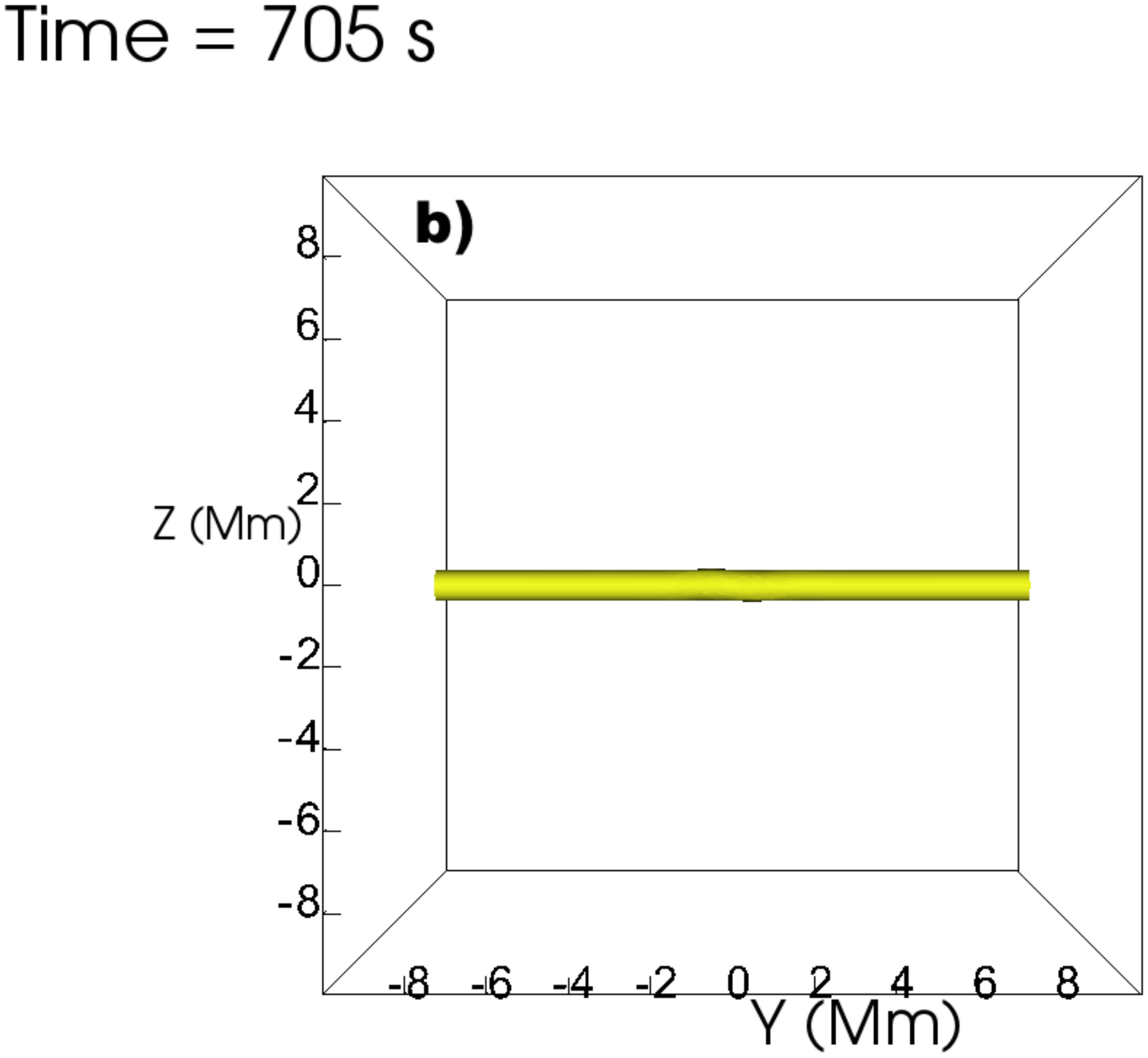}
\centering\includegraphics[scale=0.3, trim=0.0cm 1.0cm 4.0cm 0.0cm,clip=true]{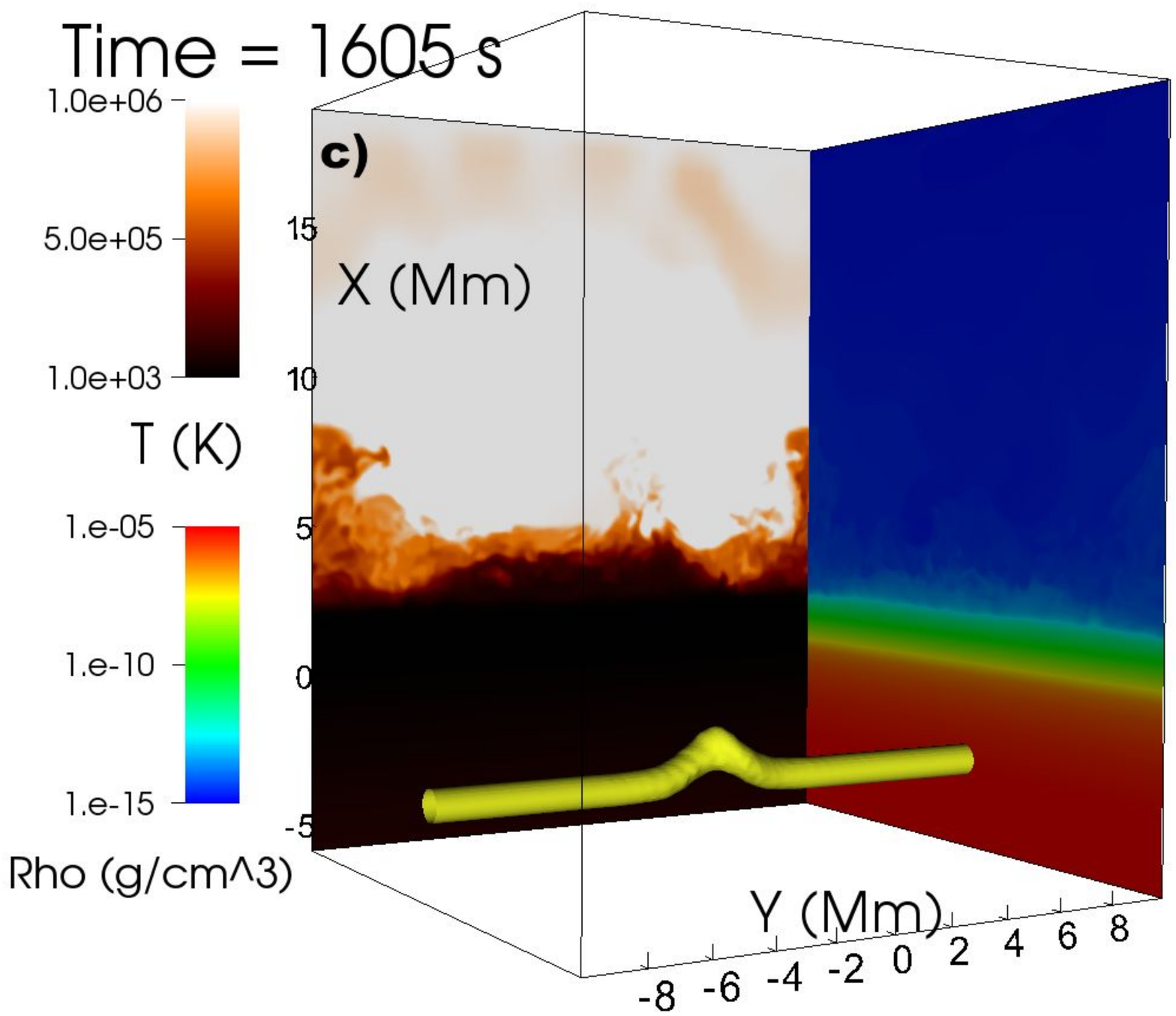}
\centering\includegraphics[scale=0.32, trim=4.0cm 1.0cm 5.0cm 3.0cm,clip=true]{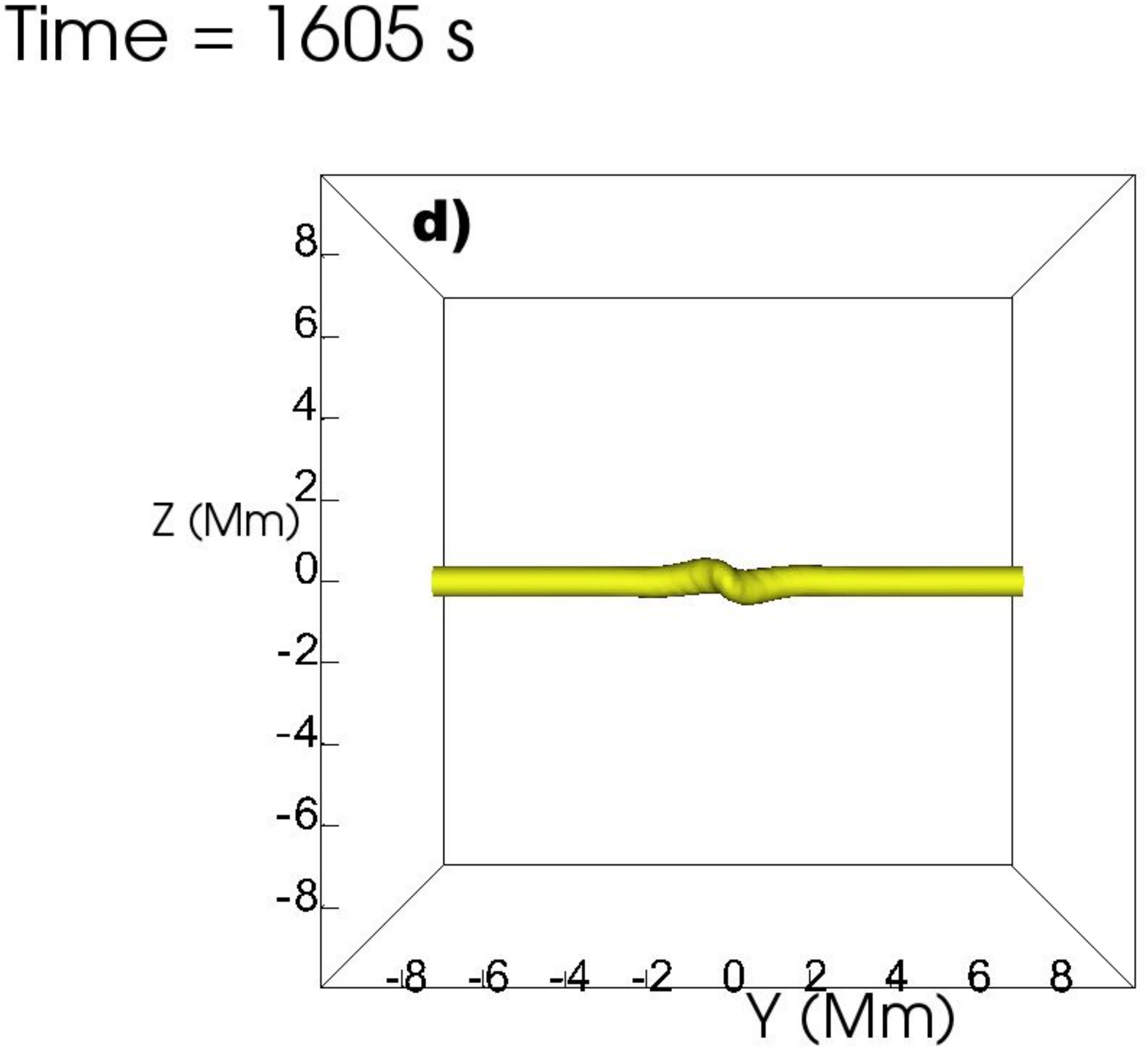}
\centering\includegraphics[scale=0.3, trim=0.0cm 1.0cm 4.0cm 0.0cm,clip=true]{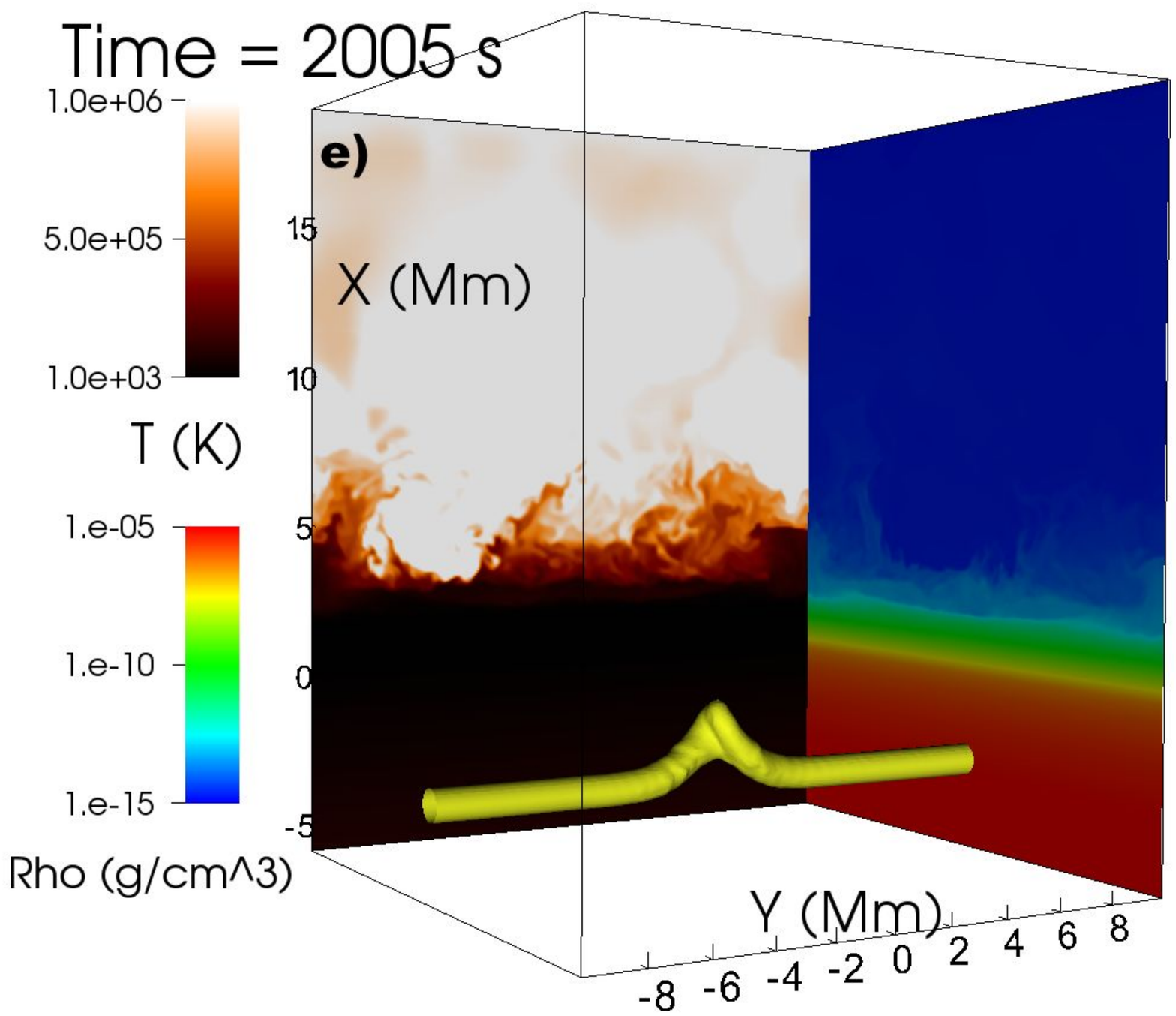}
\centering\includegraphics[scale=0.32, trim=4.0cm 1.0cm 5.0cm 3.0cm,clip=true]{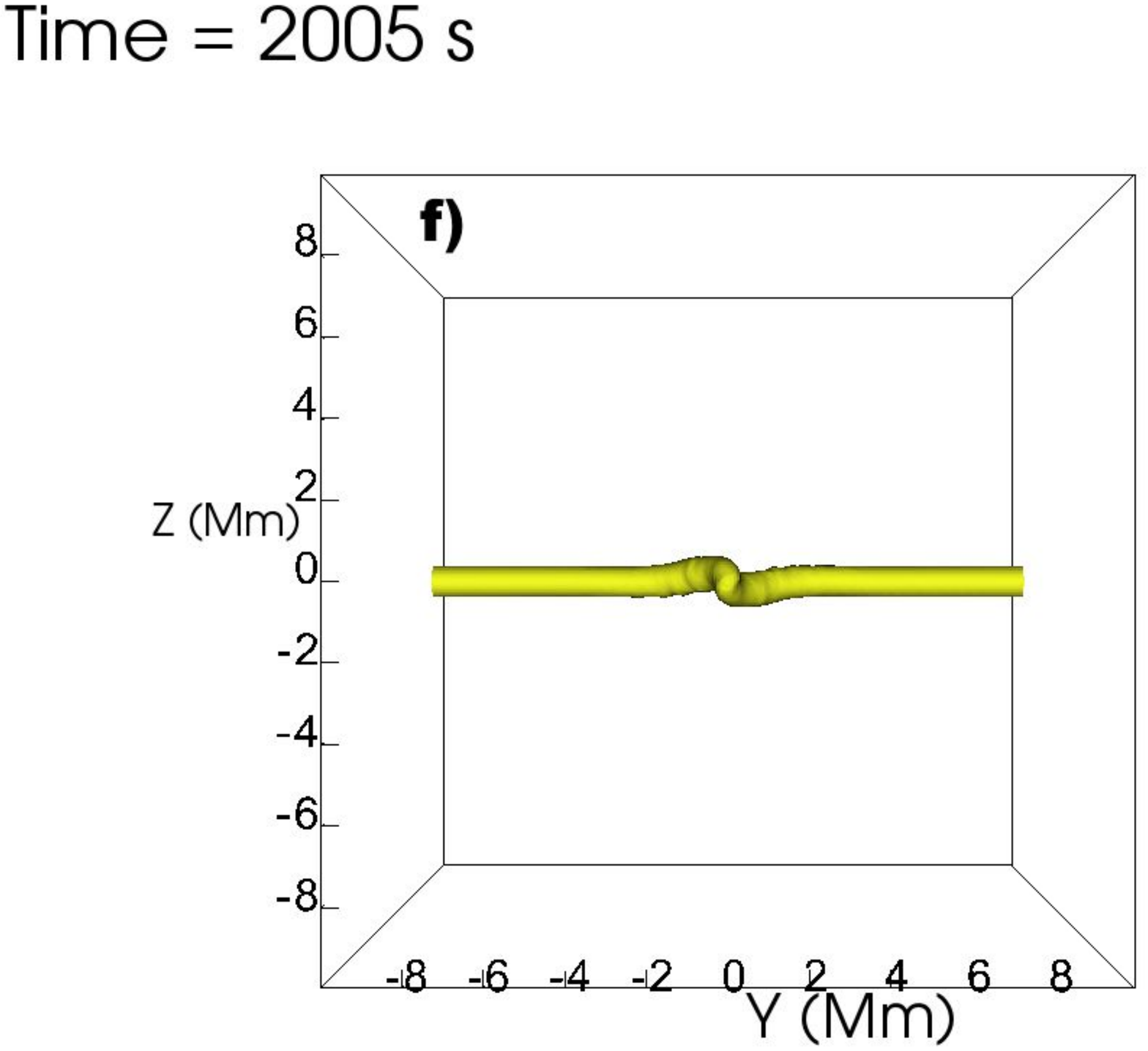}
\centering\includegraphics[scale=0.3, trim=0.0cm 1.0cm 4.0cm 0.0cm,clip=true]{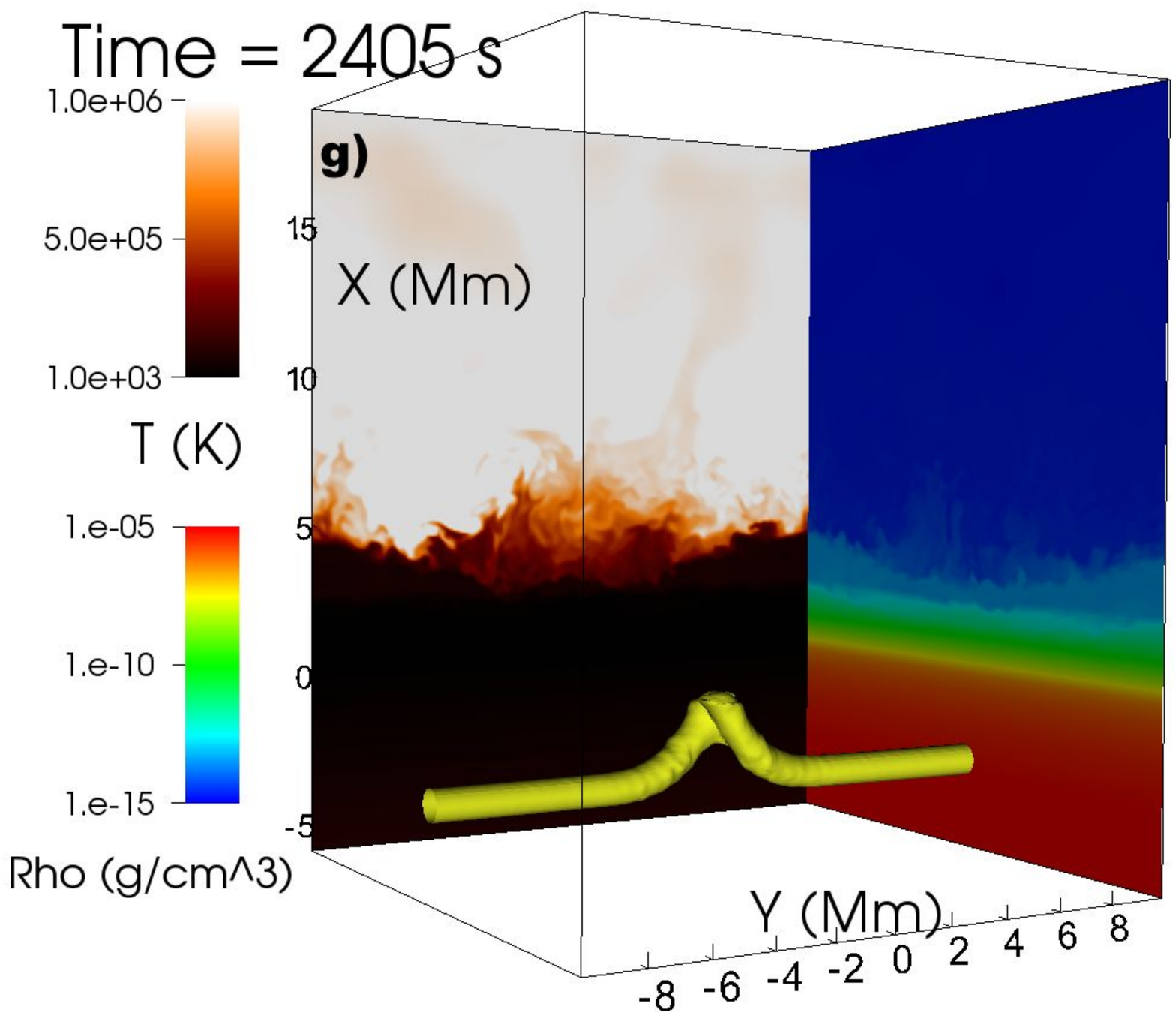}
\centering\includegraphics[scale=0.32, trim=4.0cm 1.0cm 5.0cm 3.0cm,clip=true]{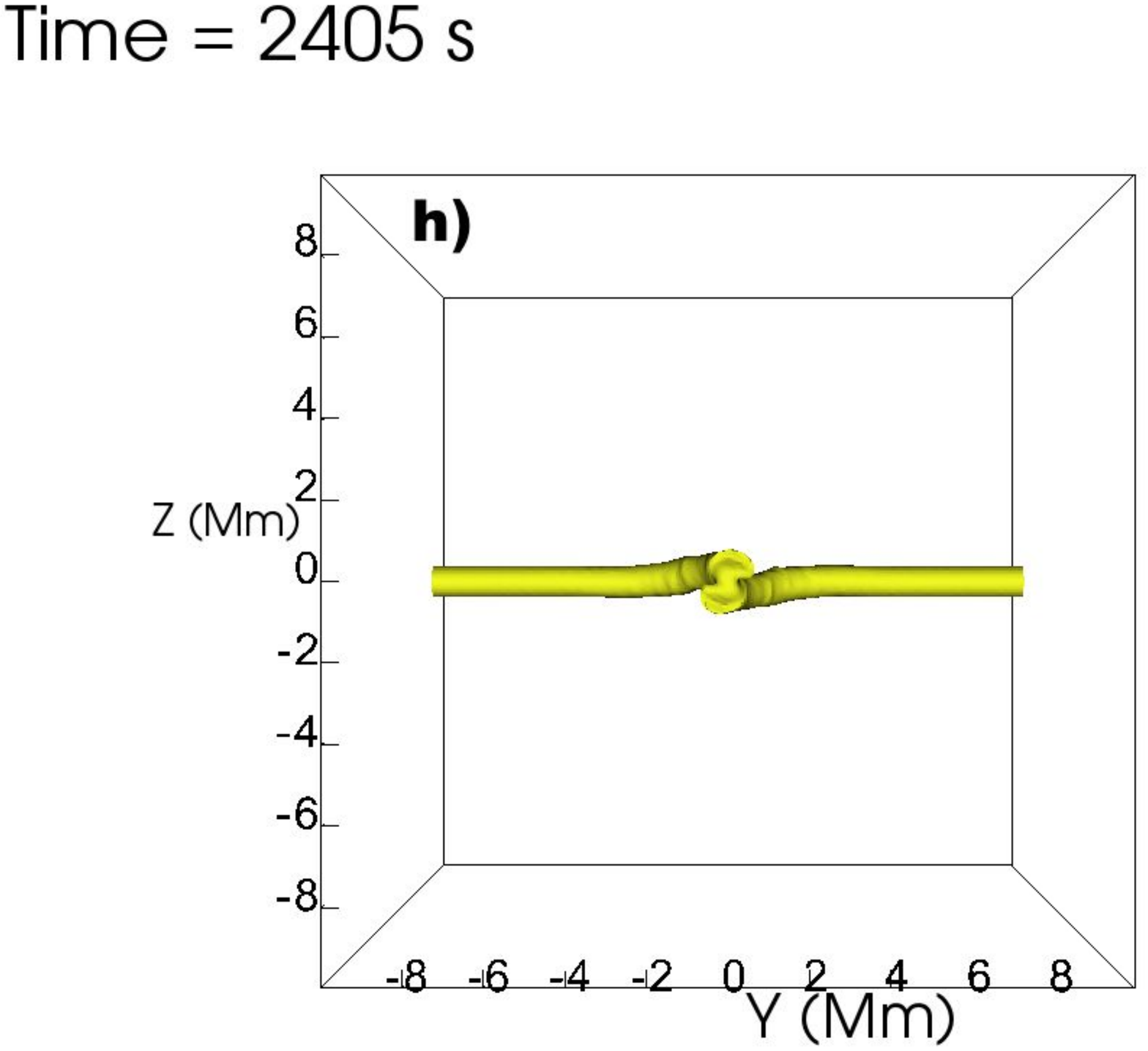}
\caption{Same as Figure \ref{fig:isosurfaces05} for $\zeta=1.5$}
\label{fig:isosurfaces15}
\end{figure*}

We now present a simulation showing an emerging flux rope whose initial twist exceeds the kink instability threshold. The rise of this flux rope through the convection zone is shown in Figure \ref{fig:isosurfaces15}. The initial $\Omega$-loop formed by the density perturbation quickly begins to coherently rotate out of its initial plane, converting twist into writhe. This rotation is much more well-defined than in the previous two cases. Around $t\approx 2000\;s$, the apex of the $\Omega$-loop becomes oriented along the $z$-direction, meaning that it has rotated much further than the two lower twist cases. Striations again form along the rising $\Omega$-loop and extend down the legs of the rising flux rope, although they are not as clearly visible in this simulation as in the previous two.

\subsubsection{Photospheric Evolution}\label{sec:phot1.5}
\begin{figure*}
\centering\includegraphics[scale=0.32, trim=1.0cm 1.0cm 5.5cm 1.0cm,clip=true]{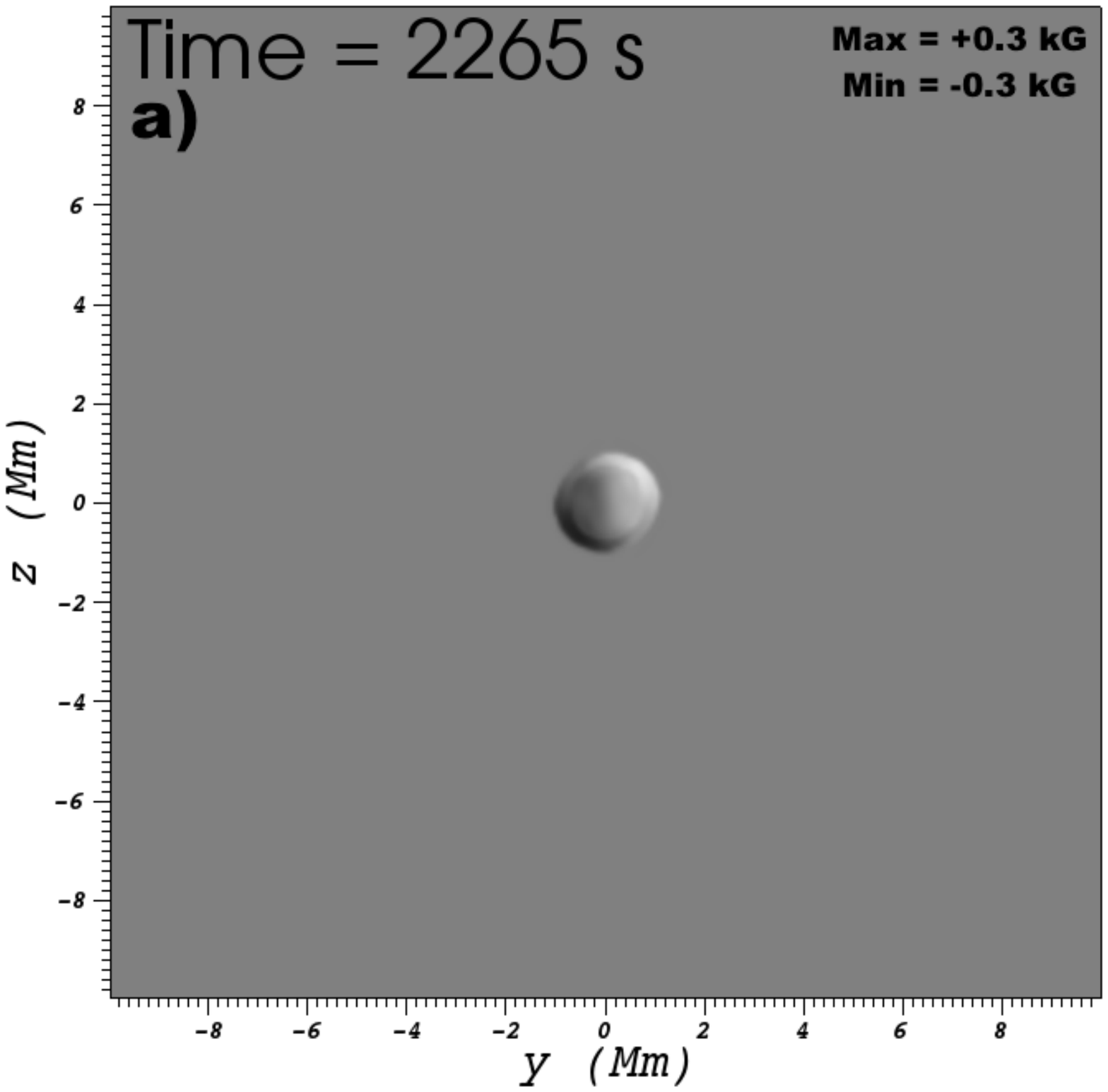}
\centering\includegraphics[scale=0.32, trim=3.0cm 1.0cm 5.5cm 1.0cm,clip=true]{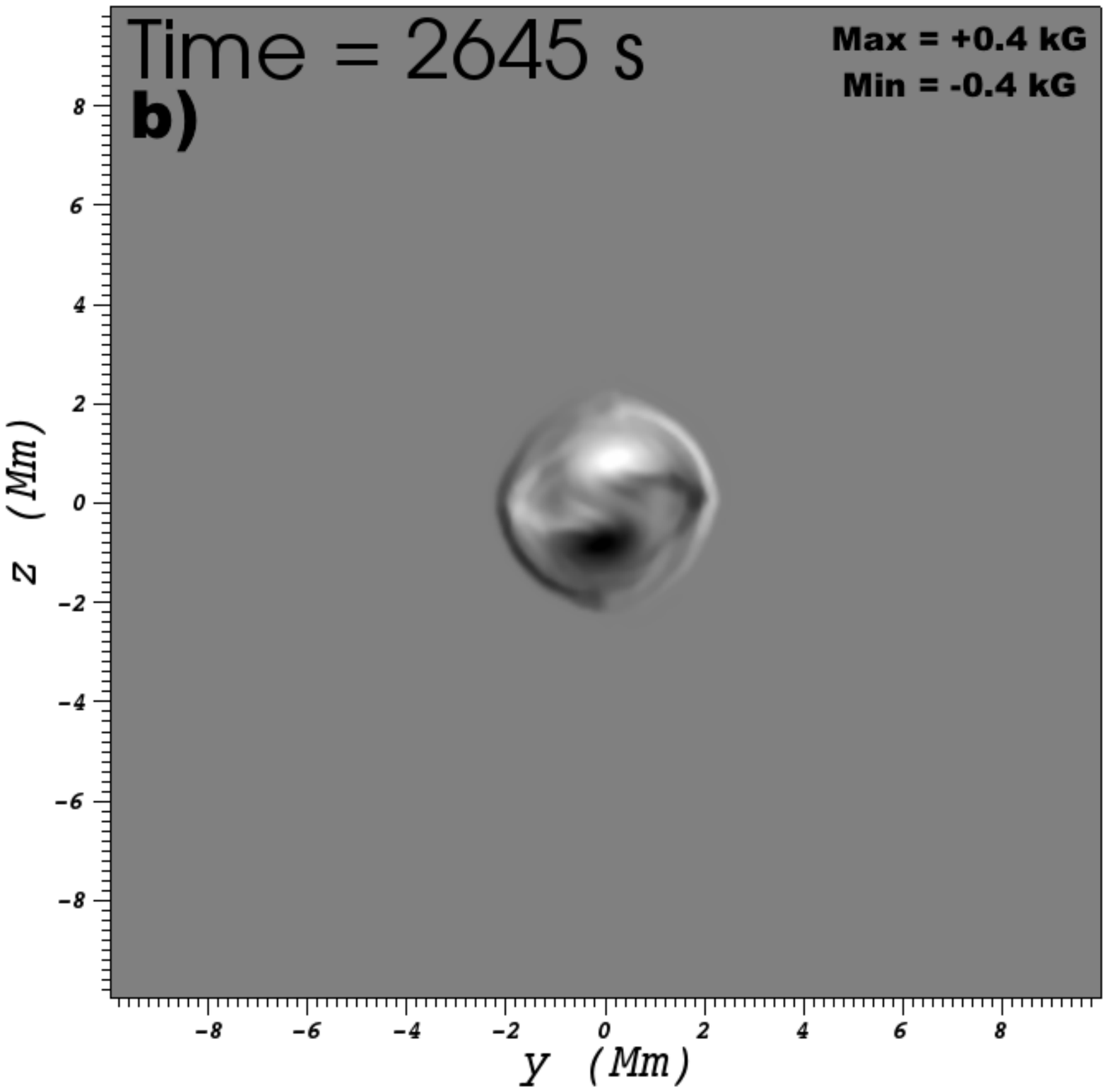}
\centering\includegraphics[scale=0.32, trim=1.0cm 1.0cm 5.5cm 1.0cm,clip=true]{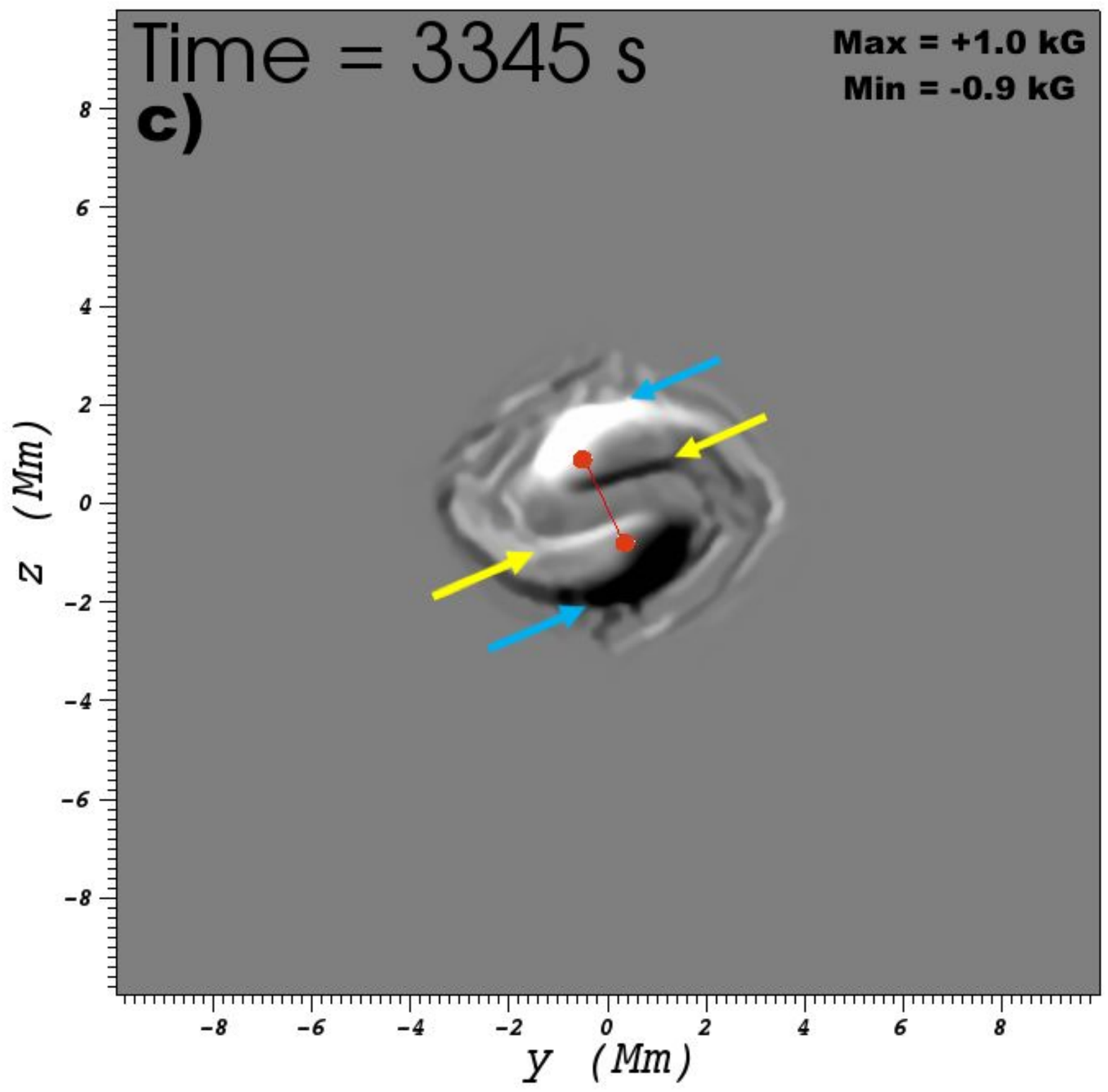}
\centering\includegraphics[scale=0.32, trim=3.0cm 1.0cm 5.5cm 1.0cm,clip=true]{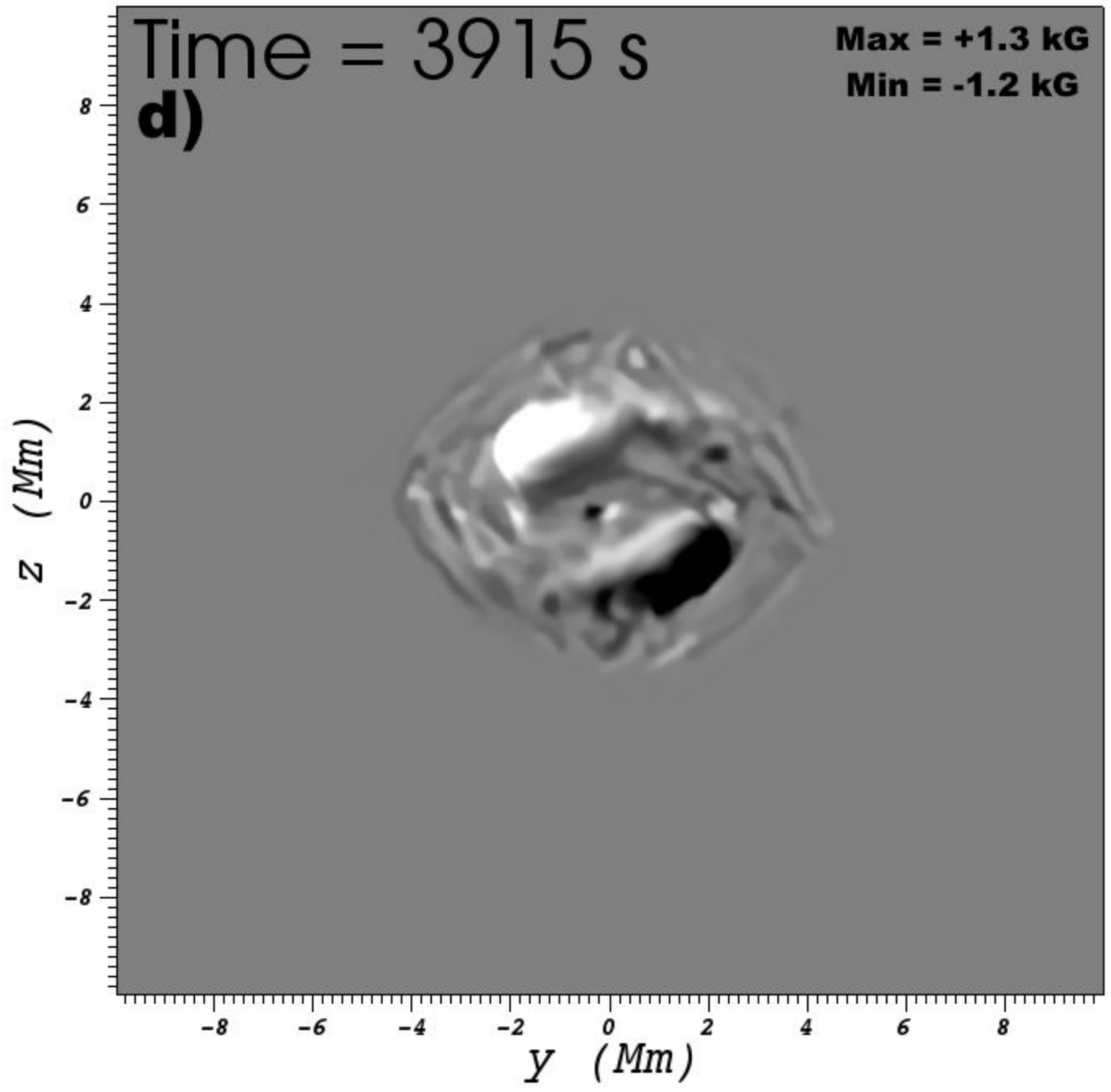}
\centering\includegraphics[scale=0.32, trim=1.0cm 1.0cm 5.5cm 1.0cm,clip=true]{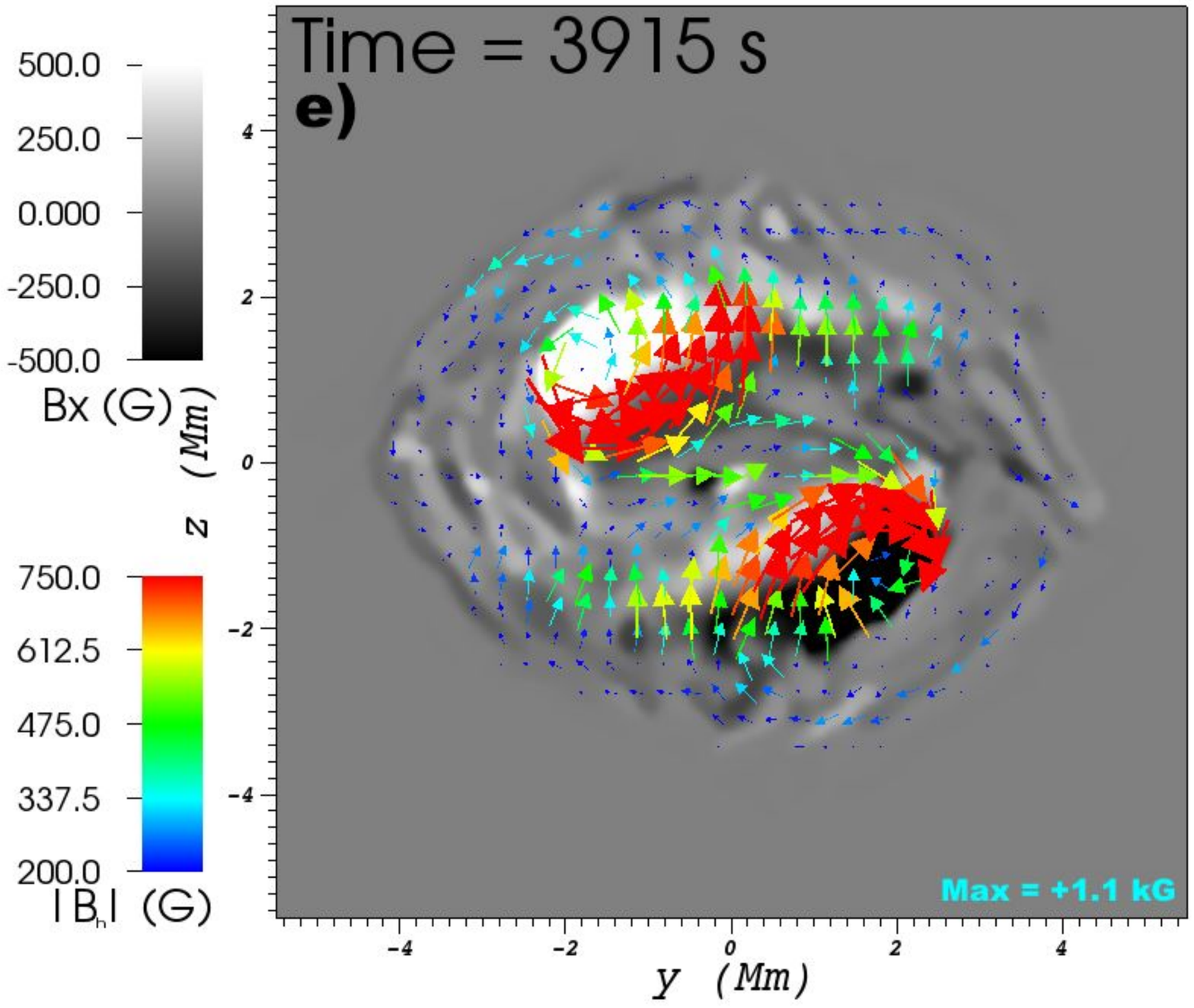}
\caption{Same as Figure \ref{fig:bx0}, for the $\zeta=1.5$ case. Primary/secondary polarities are denoted by blue/yellow arrows in panel c).}
\label{fig:bx1.5}
\end{figure*}

\begin{figure*}
\centering\includegraphics[scale=0.25, trim=0.0cm 0.0cm 0.0cm 0.0cm,clip=true]{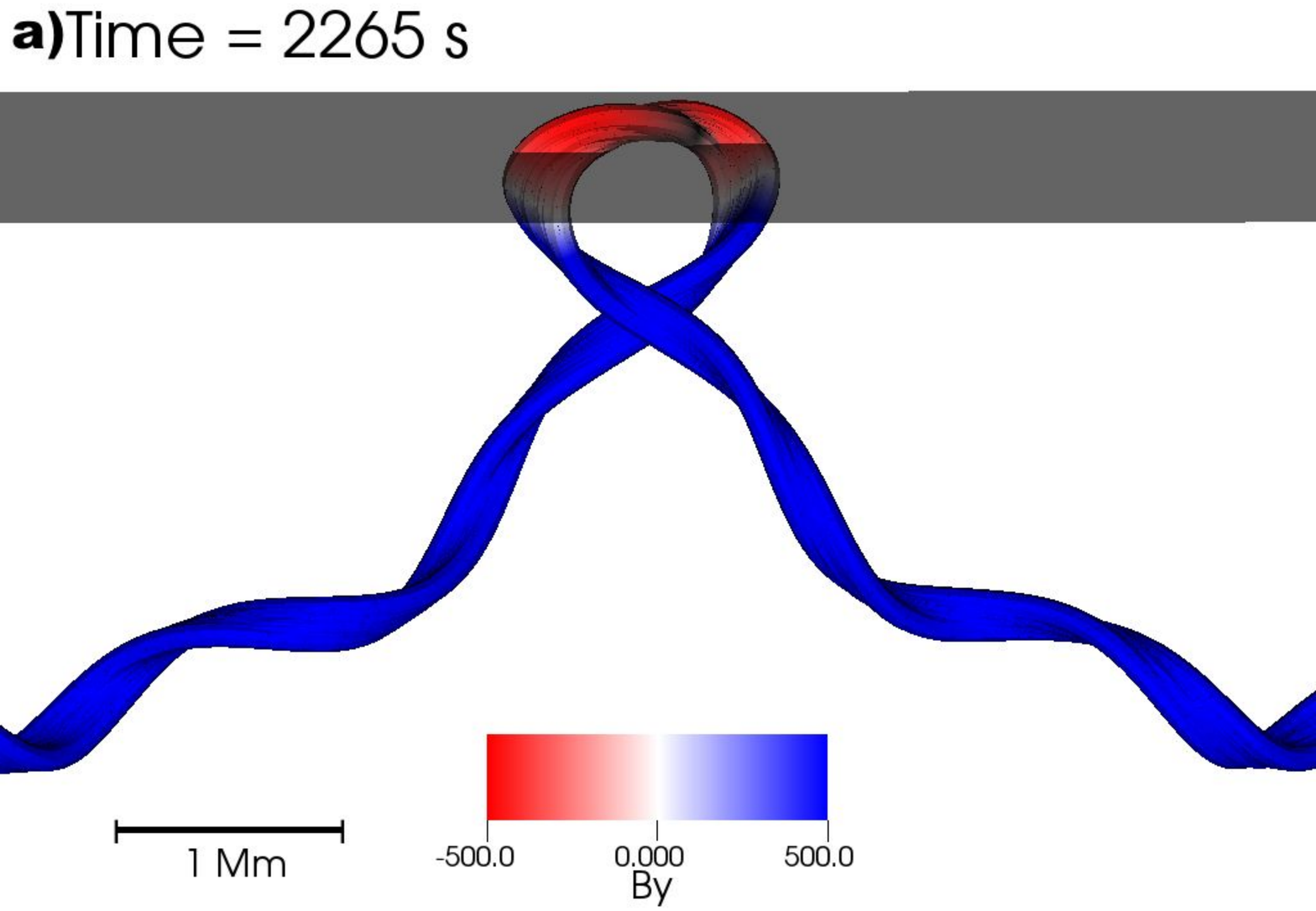}
\centering\includegraphics[scale=0.25, trim=0.0cm 0.0cm 0.0cm 0.0cm,clip=true]{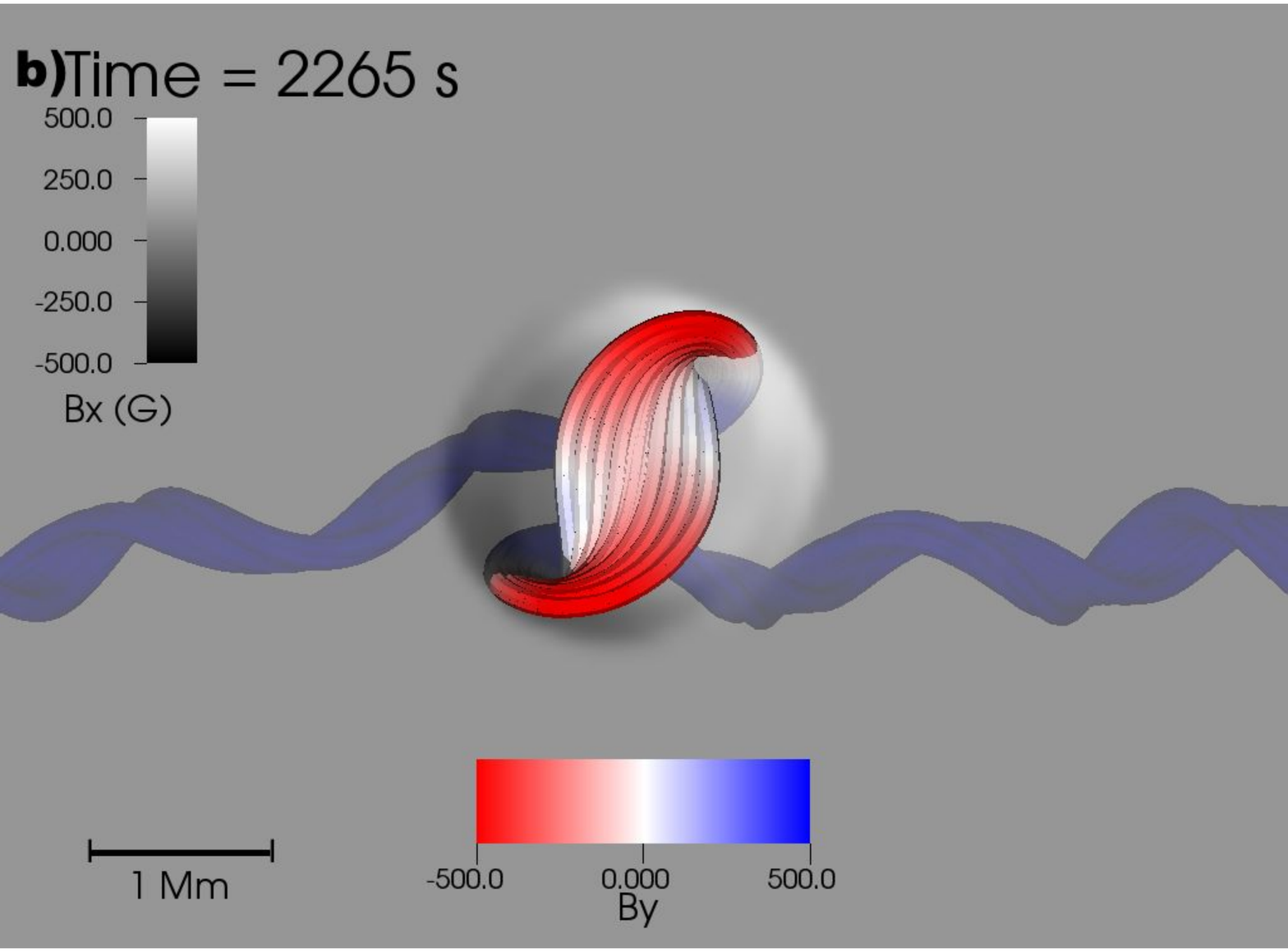}
\centering\includegraphics[scale=0.25, trim=0.0cm 0.0cm 0.0cm 0.0cm,clip=true]{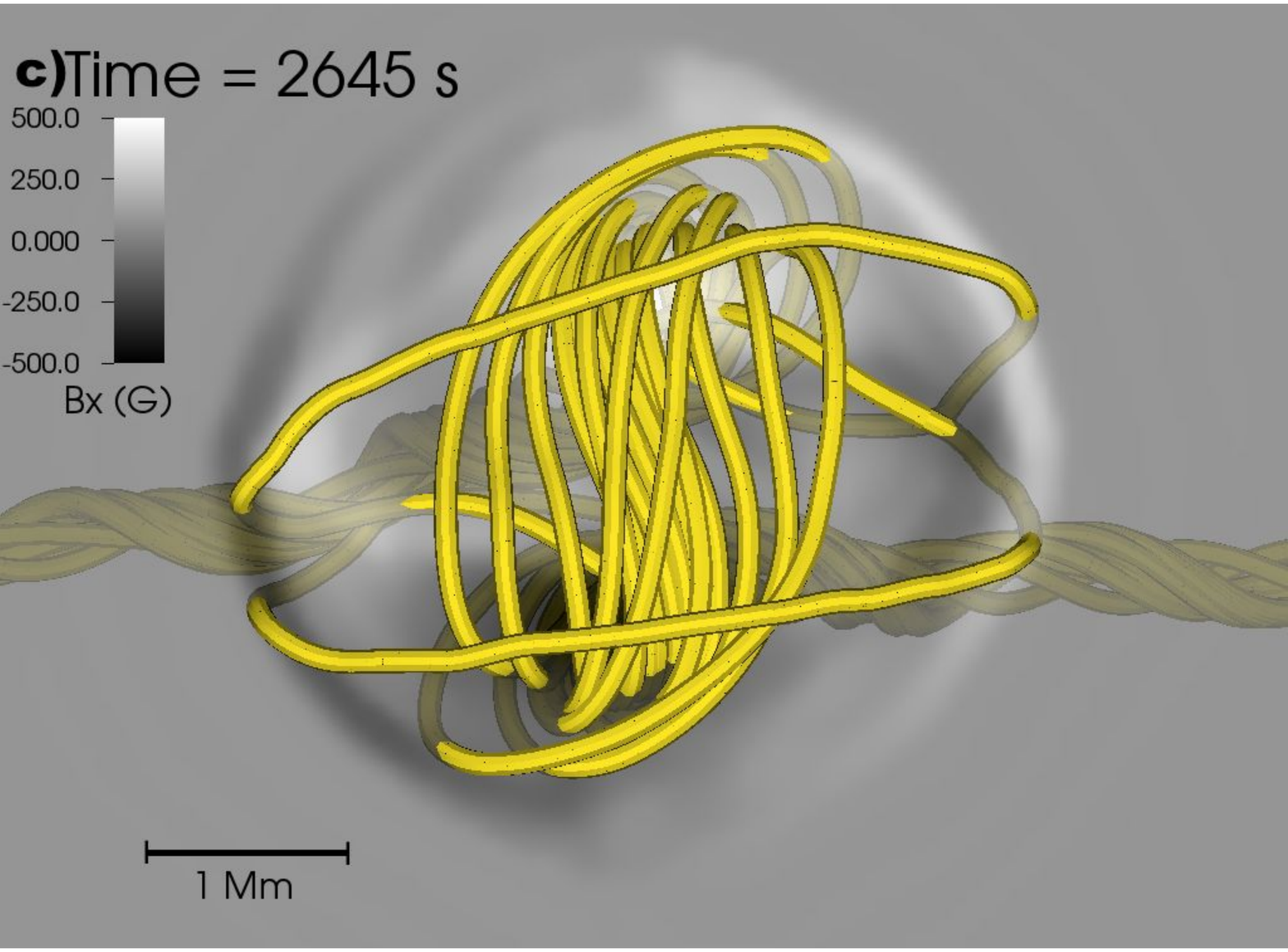}
\centering\includegraphics[scale=0.25, trim=0.0cm 0.0cm 0.0cm 0.0cm,clip=true]{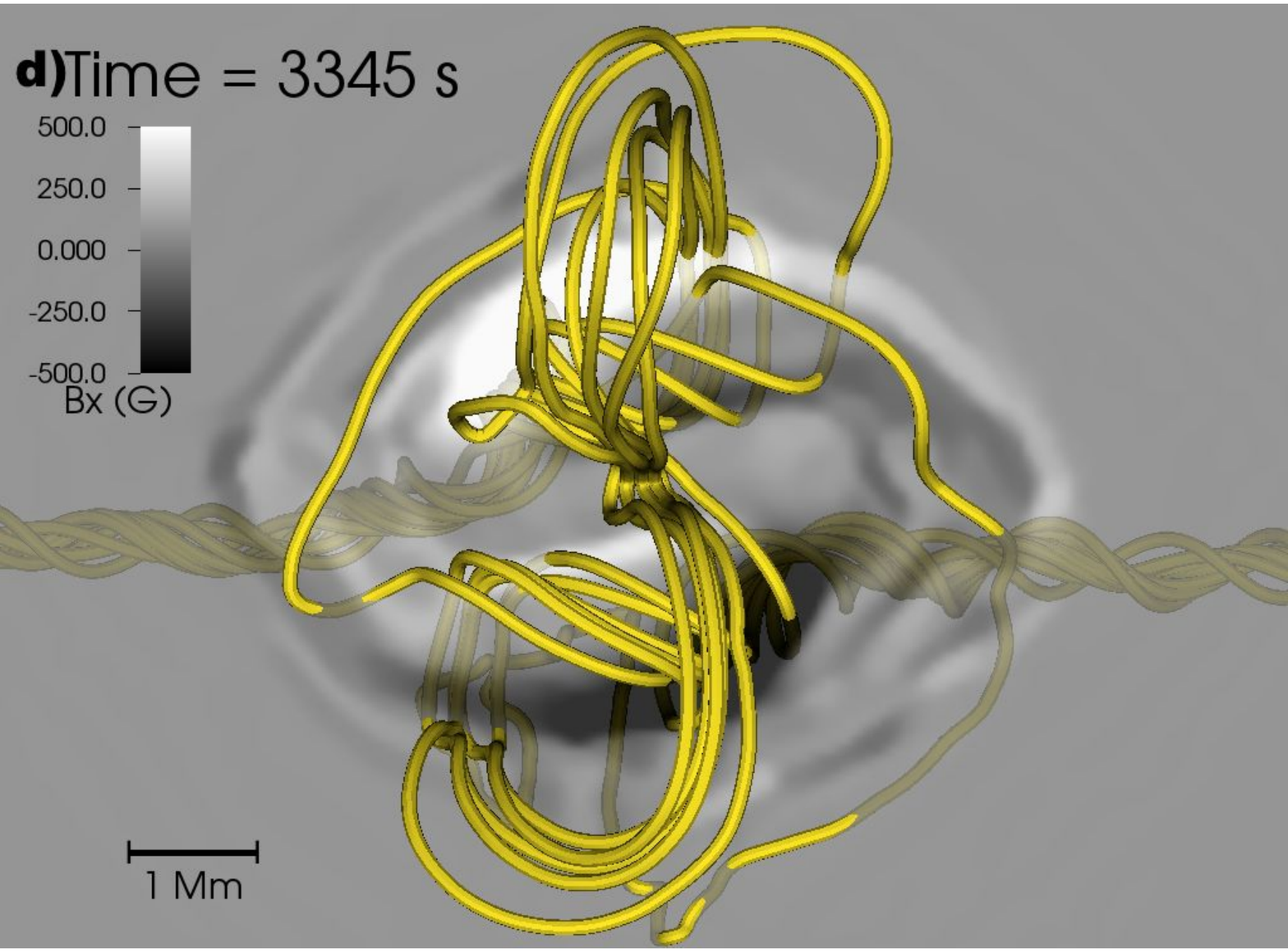}
\centering\includegraphics[scale=0.25, trim=0.0cm 0.0cm 0.0cm 0.0cm,clip=true]{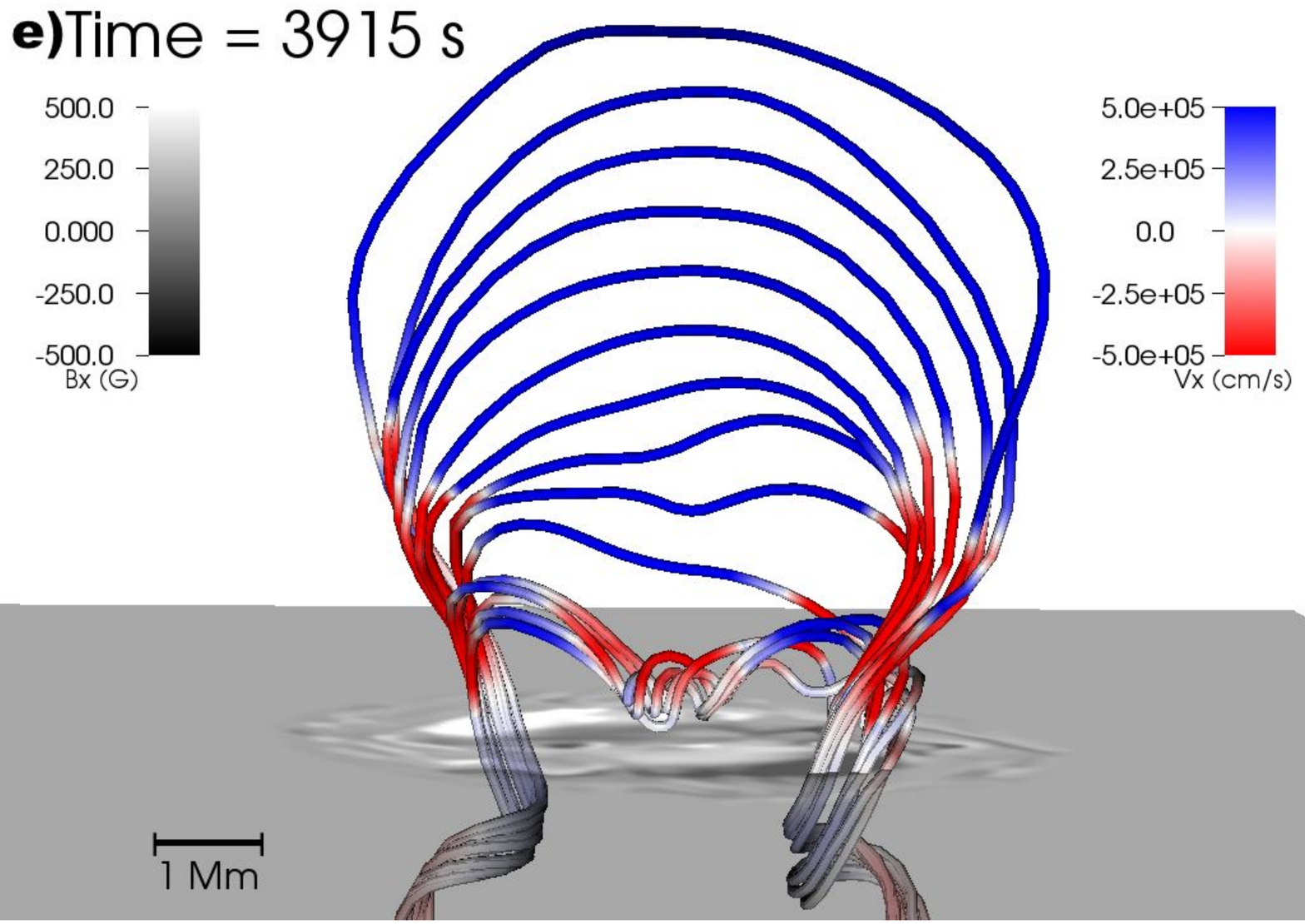}
\caption{Field lines, overplotted on photospheric magnetograms, at various stages of the $\zeta=1.5$ simulation. In a) and b) field lines, colored by $B_y$, are seen from the side and above, respectively. In e) field lines are shown at the final state of the simulation, colored by $v_x$.}
\label{fig:fieldlines1.5}
\end{figure*}

The flux rope begins to emerge through the photosphere at $t\approx2265\;\mathrm{s}$ as a pair of elongated, crescent-shaped polarities (Figure \ref{fig:bx1.5}a). Figures \ref{fig:fieldlines1.5}a,b show the field lines of the flux rope at this time from the side and from above, the former showing the $\ell$ shape formed as a result of the kink. The regions of red $B_y$ indicate locations where the axis of the flux rope has completely reversed from its initial direction as a result of its deformation during the kink. As the emergence proceeds, two pairs of opposite polarities form at the photosphere, and eventually develop into two primary polarities and two secondary, elongated polarities (denoted by the blue and yellow arrows, respectively, in Figure \ref{fig:bx1.5}c). 
Low-lying sheared arcades connecting the adjacent primary and secondary polarities can be seen in Figure \ref{fig:fieldlines1.5}c, along with sigmoidal field lines connecting the two primary polarities. Overlying quasi-potential fields are also apparent. The offset of these two pairs of polarities from each other in the $z$-direction is a result of the two legs of the flux rope becoming offset from each other due to the kink. As a result, the sigmoidal field lines in Figure \ref{fig:fieldlines1.5}c are oriented primarily along the $z$-direction, rather than the $y$-direction, as was observed in the previous two cases. The field lines at this location, comprised of the outer part of the apex of the emerging loop, are concave-down, and plasma is able to drain down the loops, allowing this field to emerge. The secondary pair of polarities arise due to the oblique angle of incidence between the flux rope's legs and the photosphere. Highly twisted field lines in each leg thread the photosphere twice, creating two adjacent polarities (Figures \ref{fig:fieldlines1.5}c,d). This results in the formation of sheared field lines connecting nearby primary and secondary polarities.  This primary and secondary polarity structure will be seen again in higher twist simulations and will be explained in more detail in \S \ref{sec:role_twist}. In the next stage, shown in Figure \ref{fig:fieldlines1.5}d, field lines originating closer to the inside of each leg begin to emerge. Due to the kink, these field lines in each leg are brought into contact and are able to reconnect, creating concave-up field lines at the chromospheric level. As a result of their concave-up shape, the field lines in this stage of emergence are unable to lift the heavy plasma in order to emerge from the chromosphere into the corona. This creates a dip in these field lines, seen in Figure \ref{fig:fieldlines1.5}d. \par 
The photospheric signatures seen in this simulation can be compared to those reported by \citet{Takasao15}, who emerge a flux rope with the same value of $\zeta$, but with a longer initial density perturbation, $\lambda=2.55\;\mathrm{Mm}$, versus $\lambda=1.2\;\mathrm{Mm}$ used here. In particular, by comparing Figure \ref{fig:fieldlines1.5}d here with Figure 7 of \citet{Takasao15}, one can see that the secondary polarities that form in our simulation as a result of the oblique angle of incidence of the flux rope's legs onto the photosphere are also present in the simulation of \citet{Takasao15}. These features are visible both early in their emergence ($t=320\tau$) and later ($t=340\tau$) as a diffuse component of the normal magnetic field separating the two primary polarities. Although \citet{Takasao15} do not discuss these features, we believe they are likely also formed as a result of the oblique angle of incidence of the legs of the flux rope onto the photosphere. \par 
On the other hand, the simulation reported by \citet{Takasao15} displays, at $t=330\tau$ and $t=340\tau$, two long, thin polarities that are pressed up against each other between the two primary polarities. Our photosphere does not, however, show any evidence of such long, thin polarities that are pressed up against each other. \citet{Takasao15} find that these thin polarities form as a result of plasma downflows submerging previously emerged flux. We looked for evidence of such downflows during the middle stages of our emergence but did not see any. 


\par  
During the latter phases of the emergence (Figures \ref{fig:bx1.5}c,d), the primary polarities elongate, reminiscent of the behavior of the so-called 'magnetic tongues' \citep[e.g.,][]{Luoni11,Poisson15,Poisson16}, which originate from the azimuthal magnetic field component of the emerging flux rope. The horizontal magnetic field becomes strongly sheared along the polarity inversion lines separating the primary and secondary polarities (Figure \ref{fig:bx1.5}d,e), since this field is created by short field lines encircling each leg of the flux rope. The final configuration of the field after $25\;\mathrm{min}$ of emergence is a set of overlying loops that spread out laterally and up, as well as a lower lying flux rope onto which is collecting plasma in its concave up dip, as can be seen by the $v_x$ shading in Figure \ref{fig:fieldlines1.5}e. An interesting feature of this emergence simulation is its remarkable symmetry. Transforming the photospheric normal magnetic field by
\beg{transformation}
B_x(y,z)\rightarrow-B_x(-y,-z),
\done 
i.e, rotating it by $180^\circ$ and inverting its sign, leaves the photospheric distribution nearly unchanged for the duration of the simulation. This is not the case in the previous two simulations, where the emerging flux ropes lost their symmetry as they were broken up during their rise through the convection zone. The $\zeta=1.5$ flux rope, on the other hand, stayed coherent throughout its rise, evidently due to the strong azimuthal field supporting the flux rope against perturbations. Since the kink instability is inherently symmetric, the flux rope and, as a result, the photospheric distribution, maintained its invariance for the most part under this transformation. Slight deviations from perfect symmetry may be due to the onset of reconnection inside the flux rope.
\subsection{$\zeta=2$}\label{sec:2}

\subsubsection{Subsurface Evolution}\label{sec:sub2}
\begin{figure*}
\centering\includegraphics[scale=0.3, trim=0.0cm 1.0cm 4.0cm 0.0cm,clip=true]{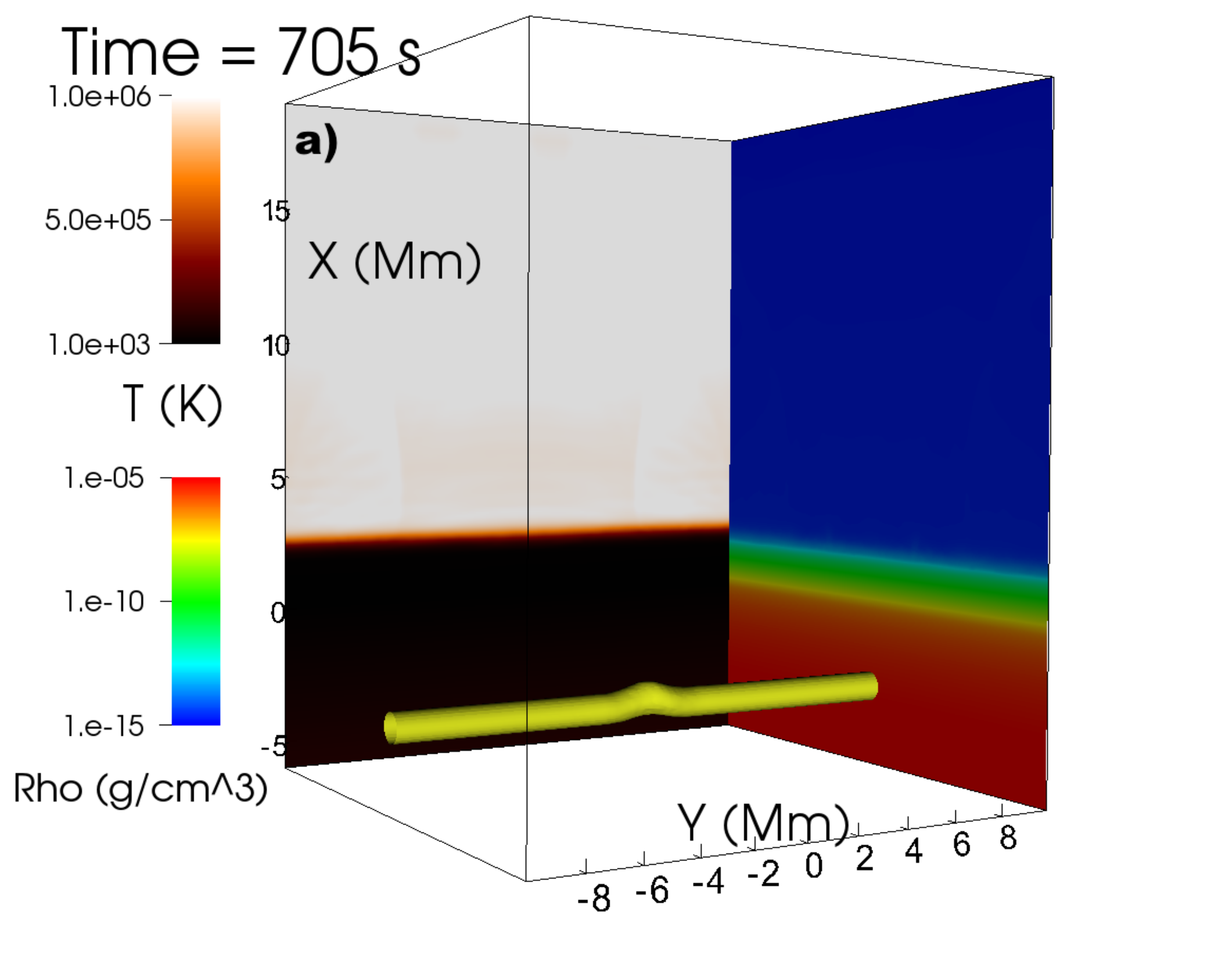}
\centering\includegraphics[scale=0.32, trim=4.0cm 1.0cm 5.0cm 3.0cm,clip=true]{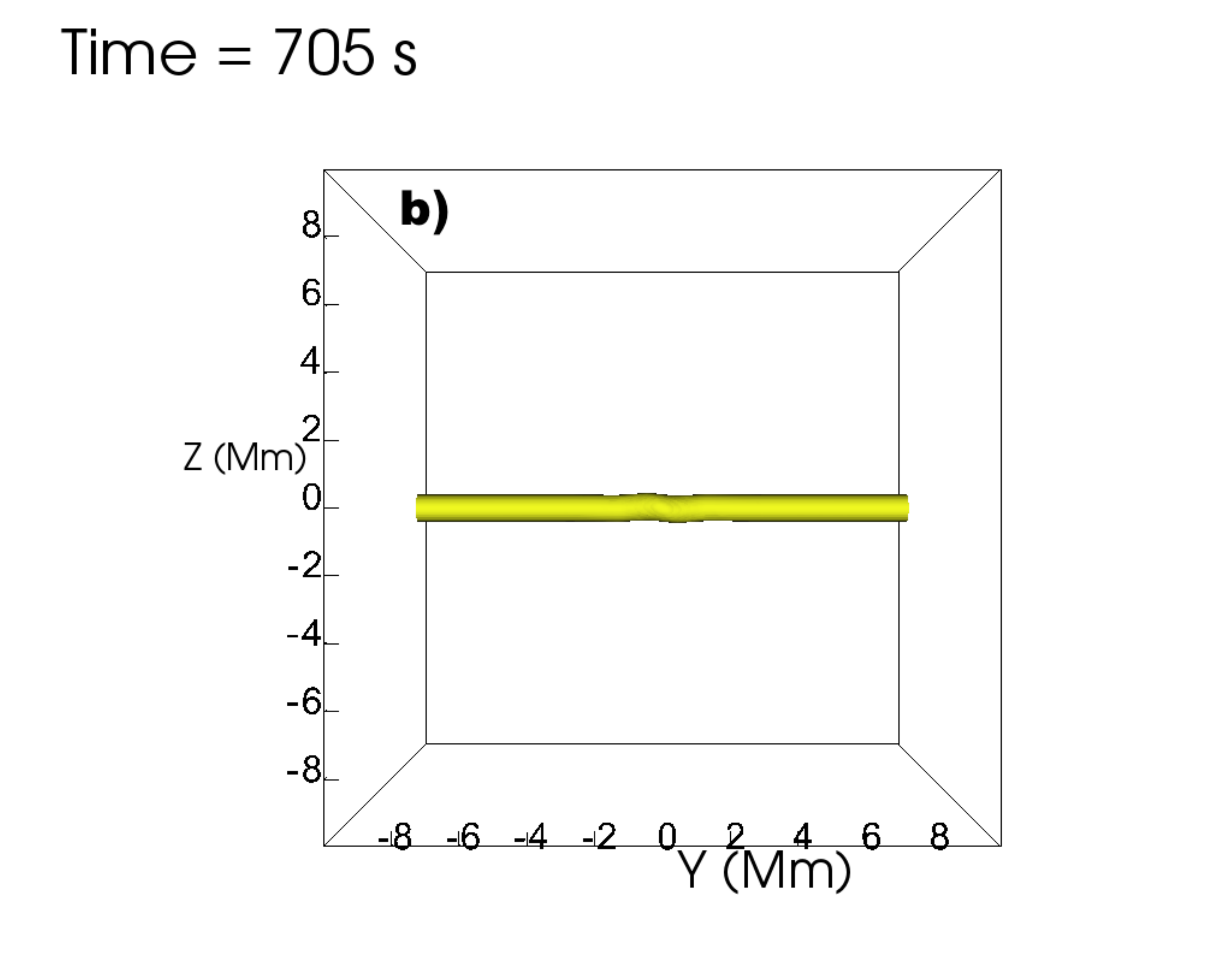}
\centering\includegraphics[scale=0.3, trim=0.0cm 1.0cm 4.0cm 0.0cm,clip=true]{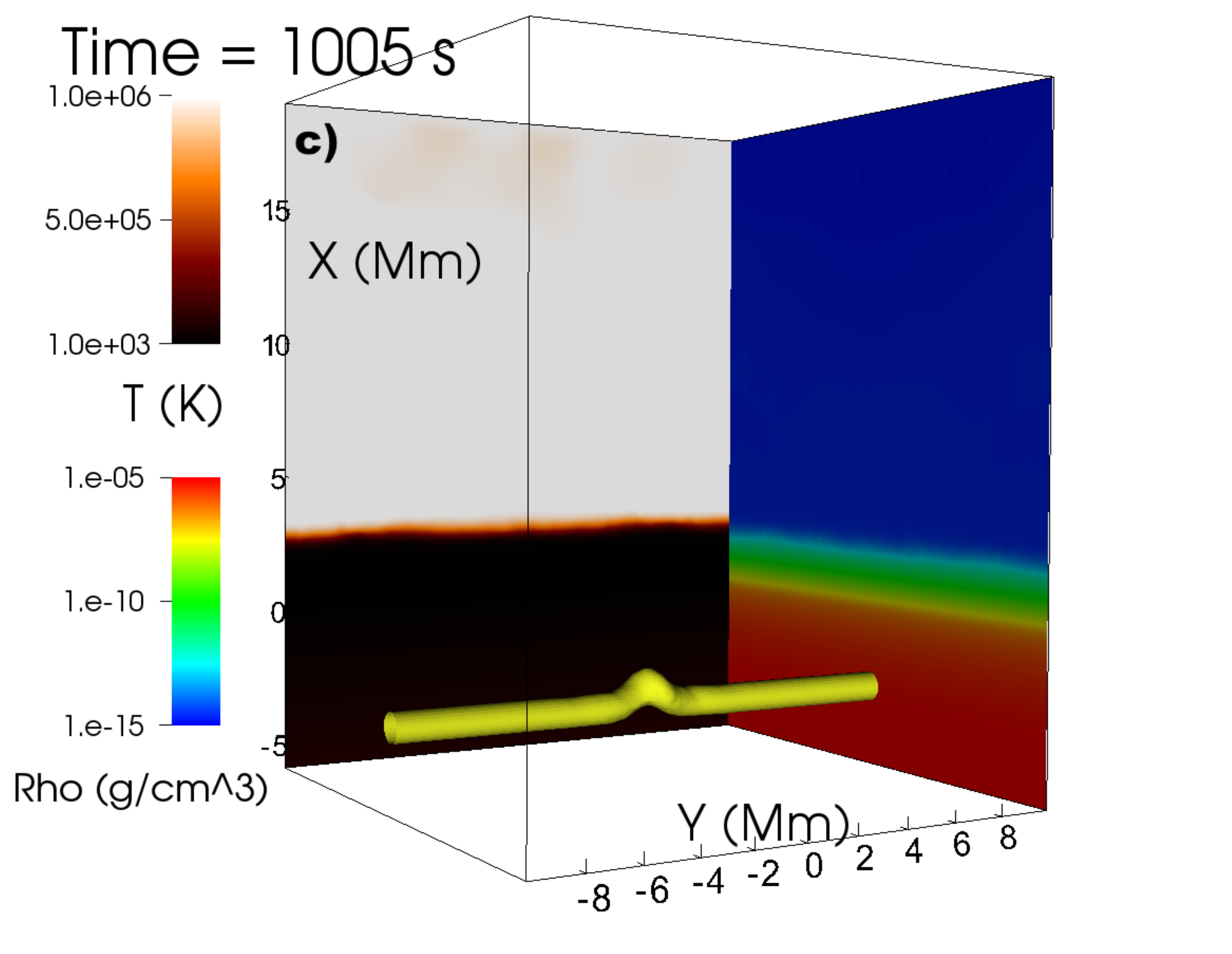}
\centering\includegraphics[scale=0.32, trim=4.0cm 1.0cm 5.0cm 3.0cm,clip=true]{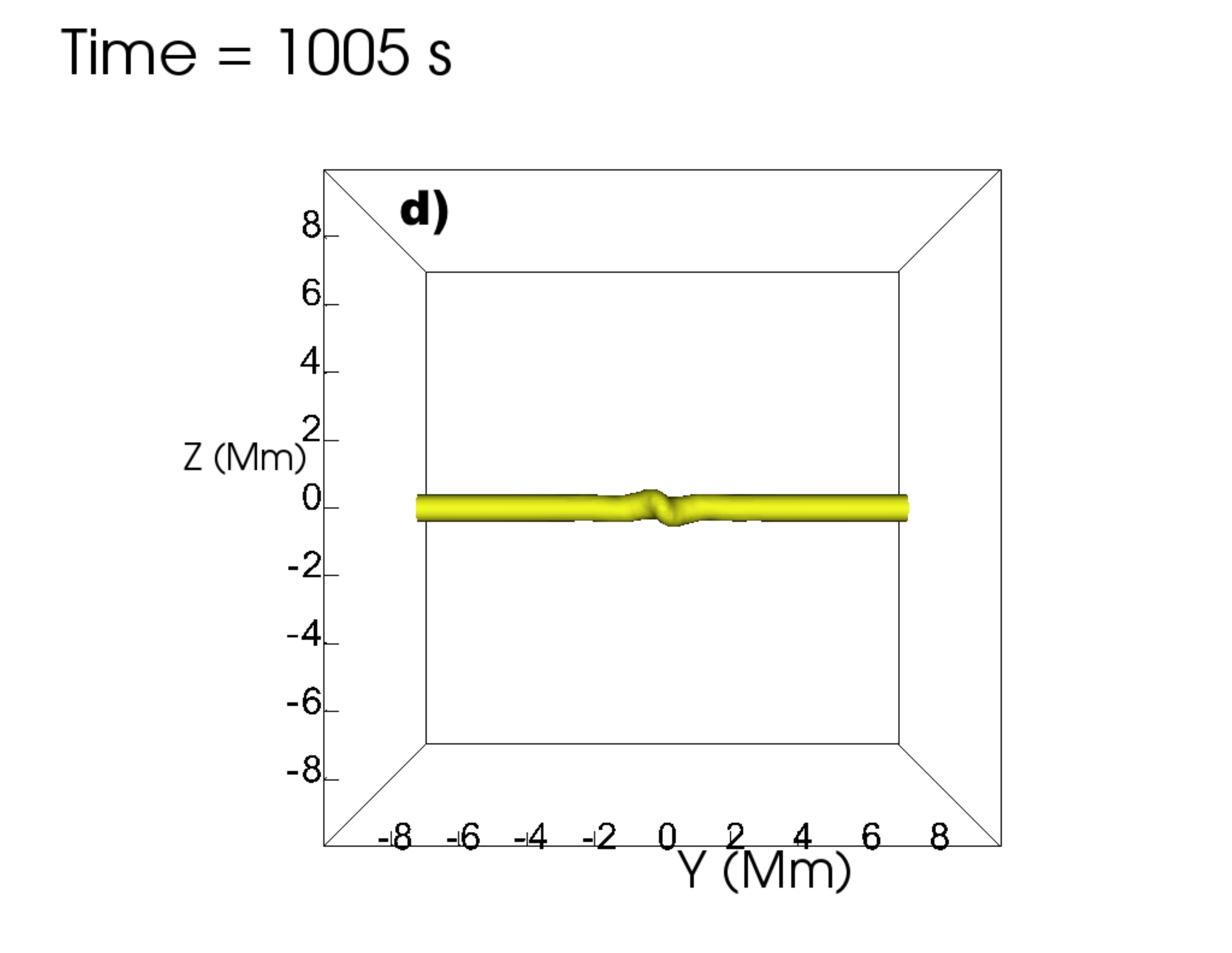}
\centering\includegraphics[scale=0.3, trim=0.0cm 1.0cm 4.0cm 0.0cm,clip=true]{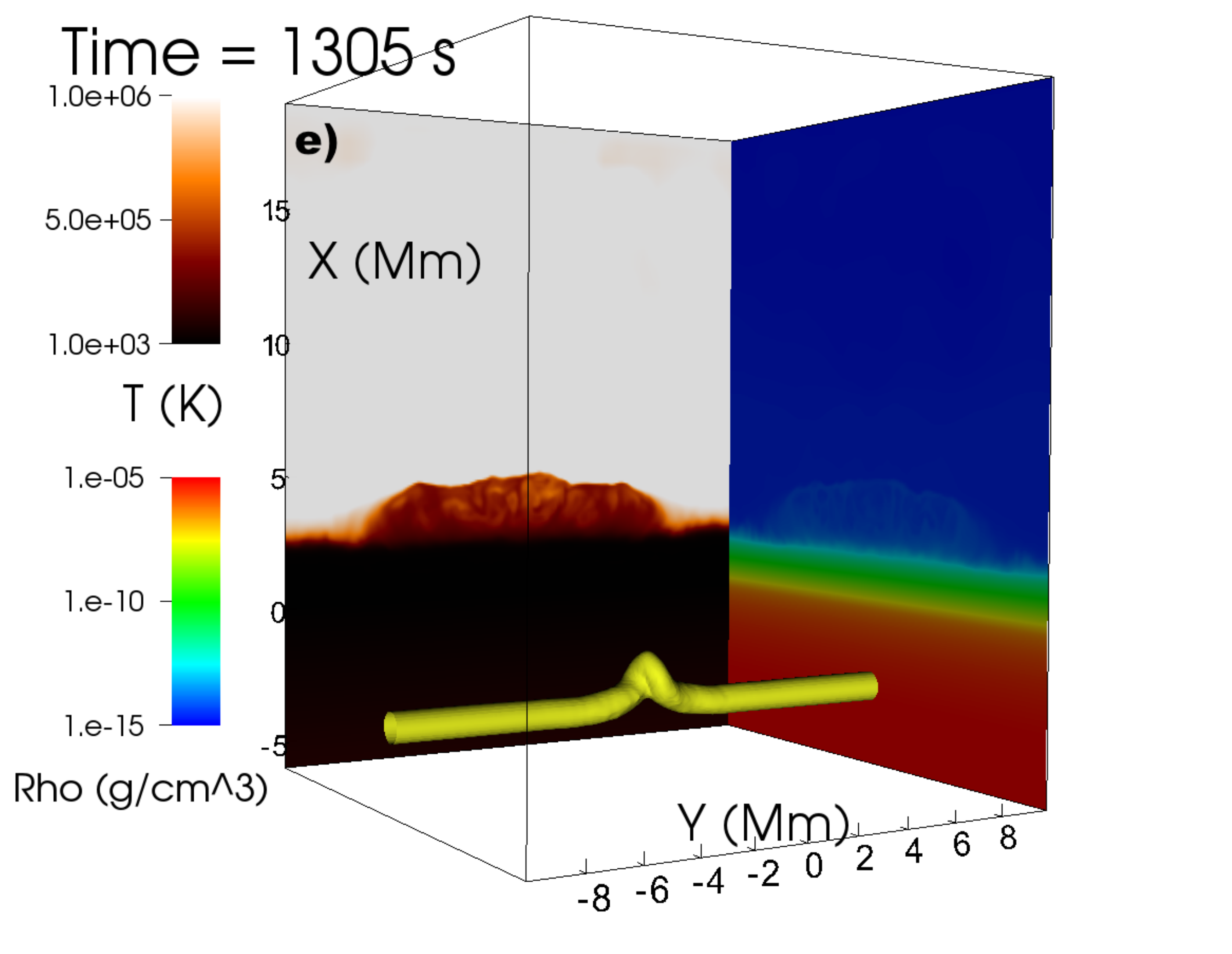}
\centering\includegraphics[scale=0.32, trim=4.0cm 1.0cm 5.0cm 3.0cm,clip=true]{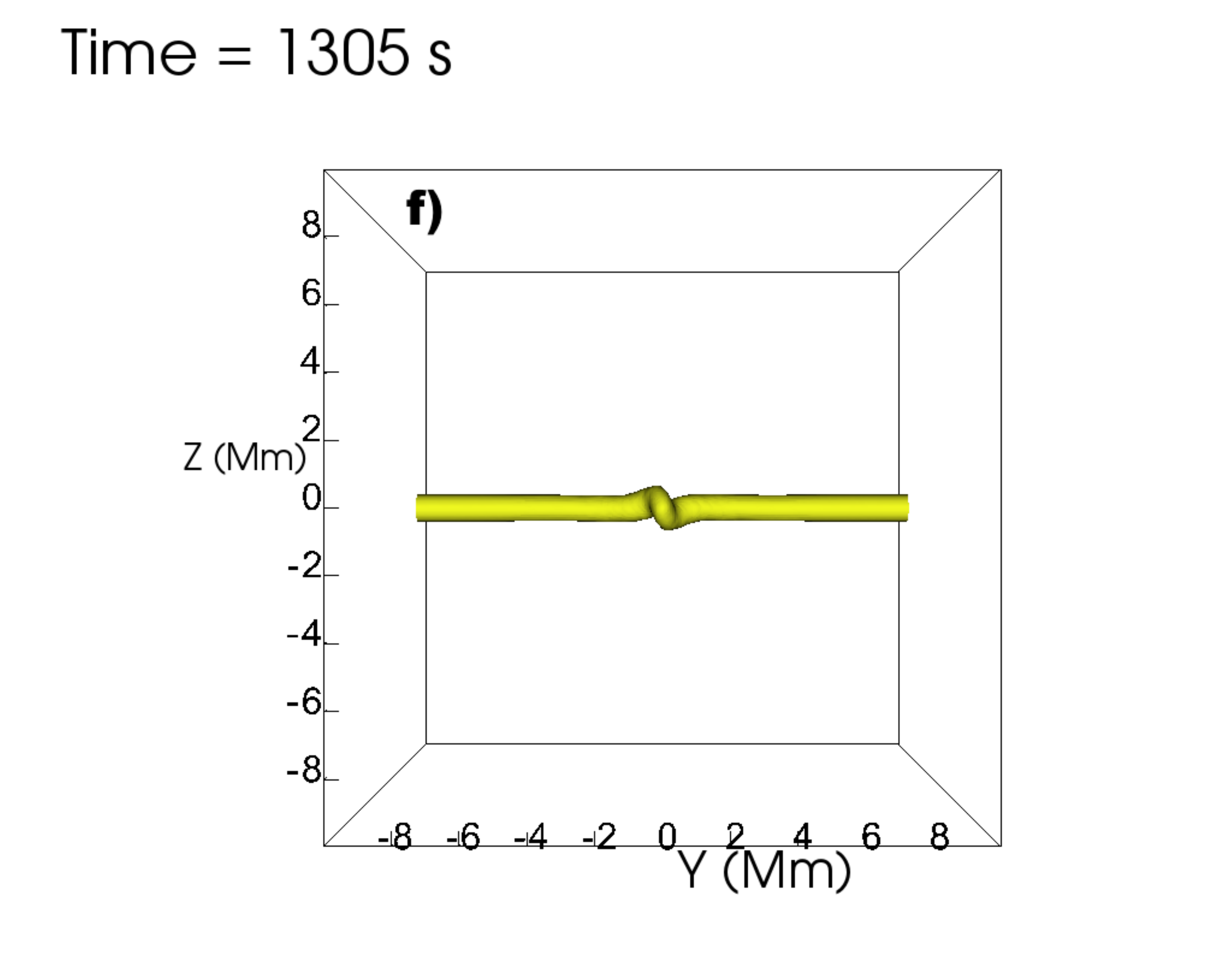}
\centering\includegraphics[scale=0.3, trim=0.0cm 1.0cm 4.0cm 0.0cm,clip=true]{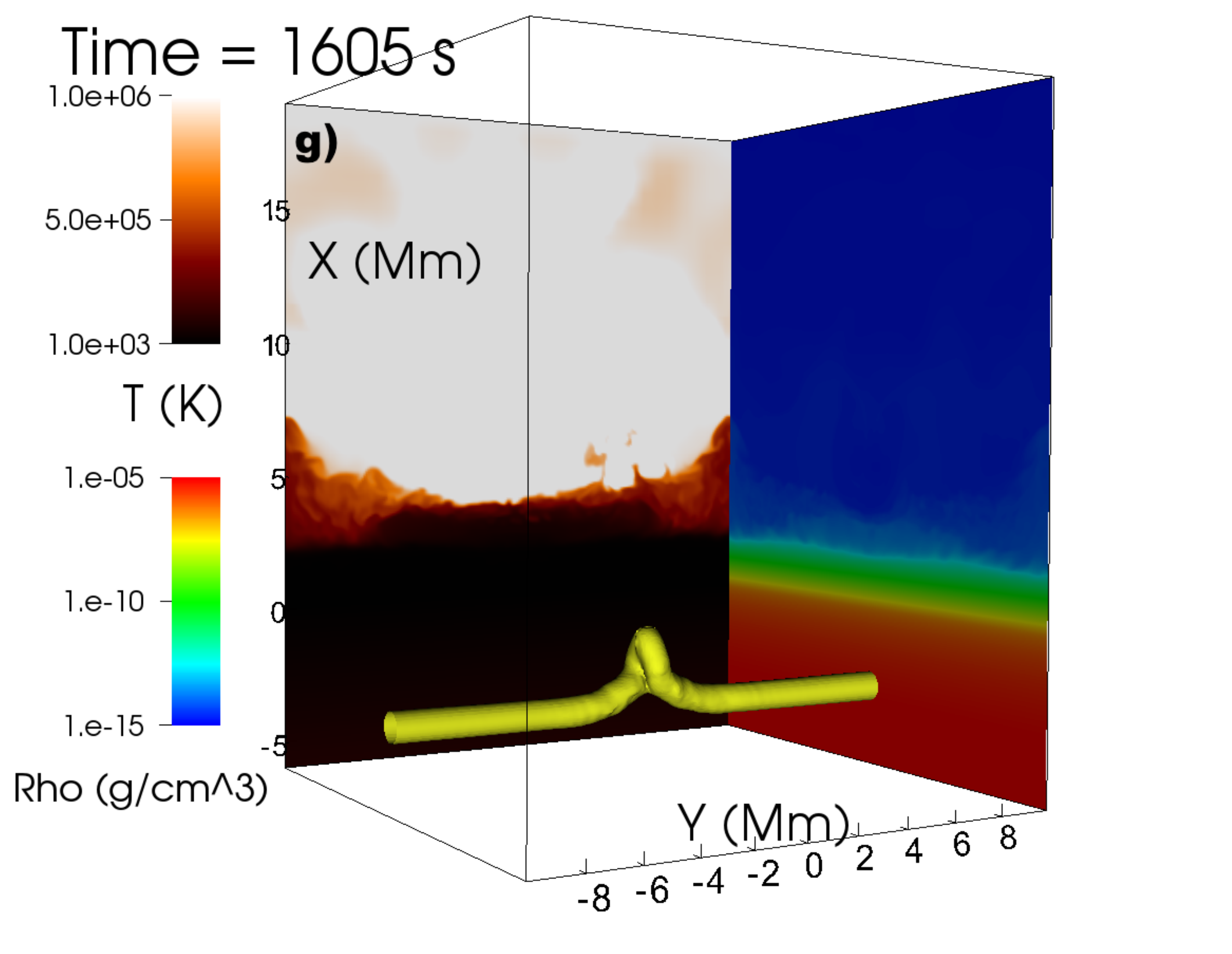}
\centering\includegraphics[scale=0.32, trim=4.0cm 1.0cm 5.0cm 3.0cm,clip=true]{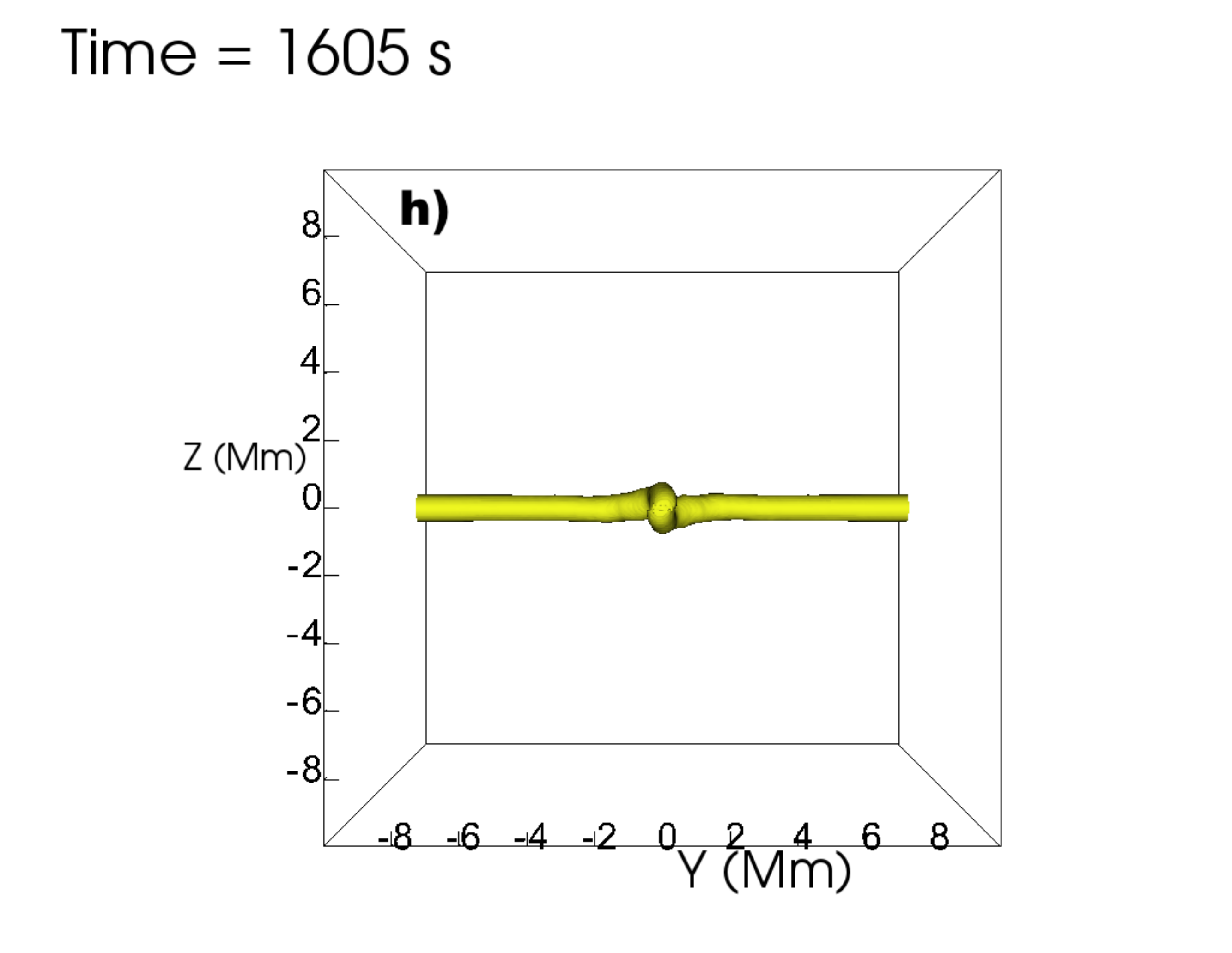}
\caption{Same as Figure \ref{fig:isosurfaces05} for $\zeta=2$}
\label{fig:isosurfaces2}
\end{figure*}

We now present the emergence of a flux rope with $\zeta=2$. In Figure \ref{fig:isosurfaces2}, we show snapshots of its rise through the convection zone at several times. Although the initial density perturbation generates a rising $\Omega$-loop, this behavior is extremely short lived. Already by $t\approx1000\;s$, the $\Omega$-loop has started to rotate out of its initial plane. The $\Omega$-loop becomes aligned to the $z$-axis by $t\approx1605\;s$, and striations only become visible very late in the flux rope's rise through the convection zone.

\subsubsection{Photospheric Evolution}\label{sec:phot2}
\begin{figure*}
\centering\includegraphics[scale=0.32, trim=1.0cm 1.0cm 5.5cm 1.0cm,clip=true]{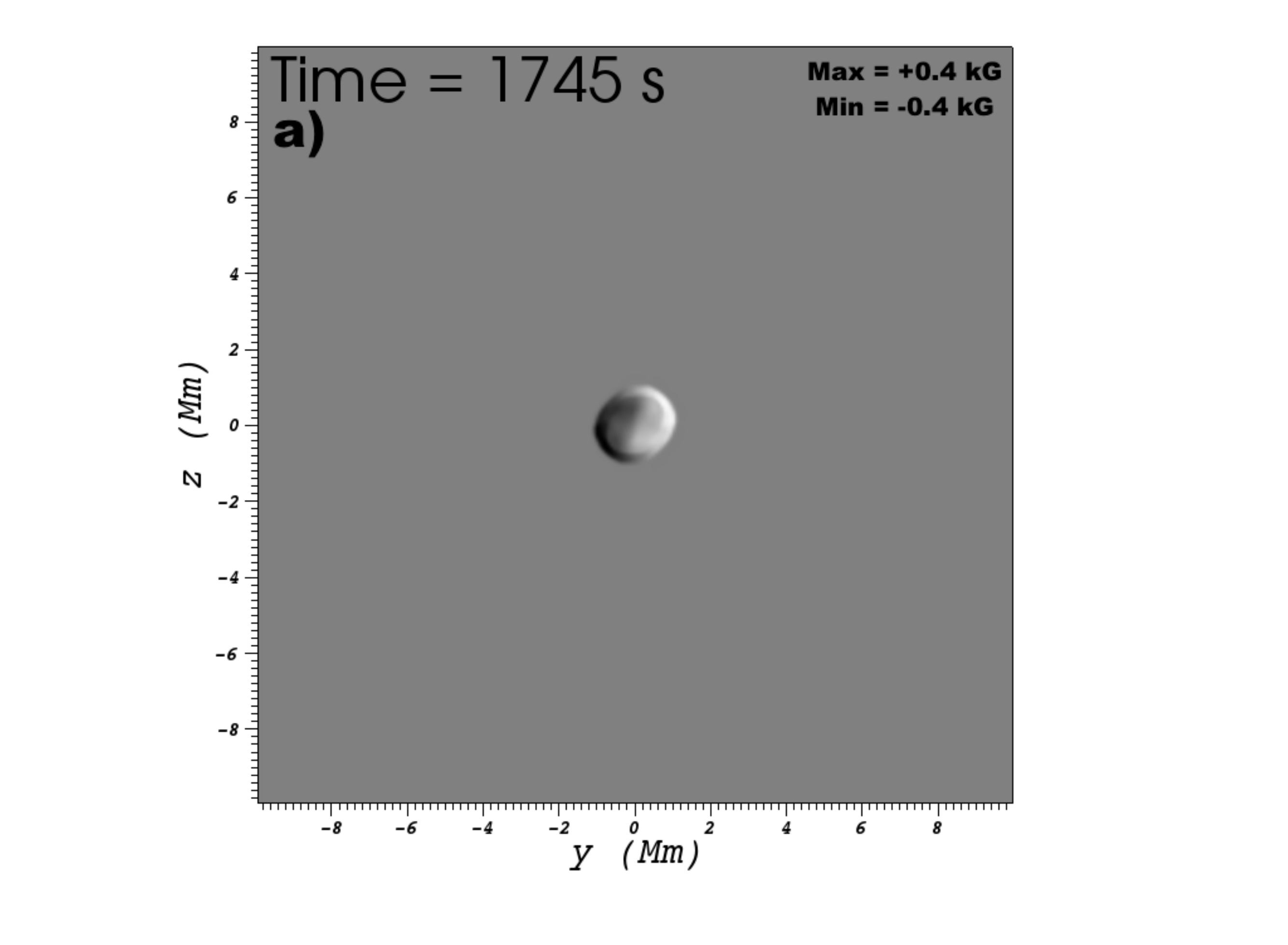}
\centering\includegraphics[scale=0.32, trim=3.0cm 1.0cm 5.5cm 1.0cm,clip=true]{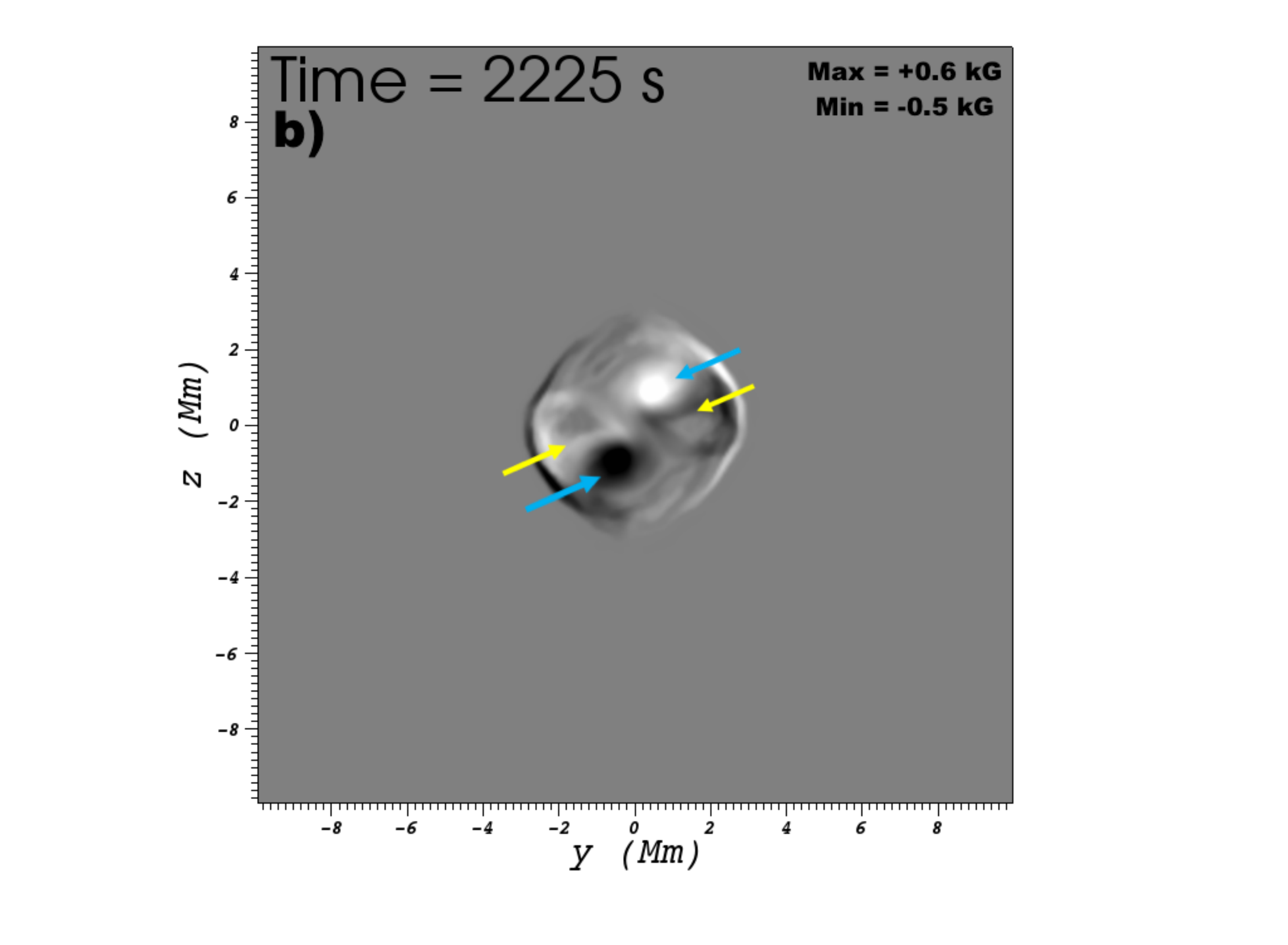}
\centering\includegraphics[scale=0.32, trim=1.0cm 1.0cm 5.5cm 1.0cm,clip=true]{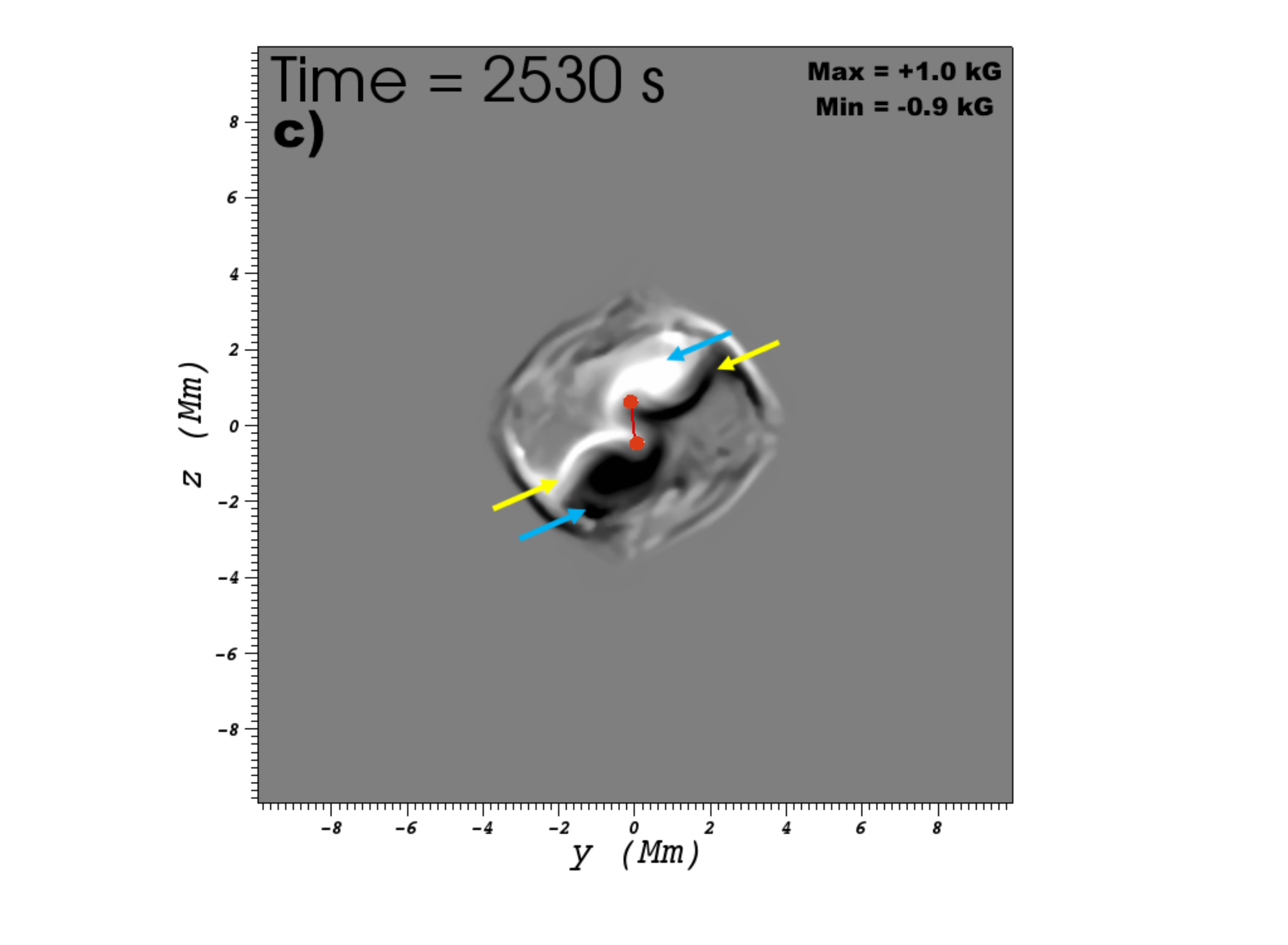}
\centering\includegraphics[scale=0.32, trim=3.0cm 1.0cm 5.5cm 1.0cm,clip=true]{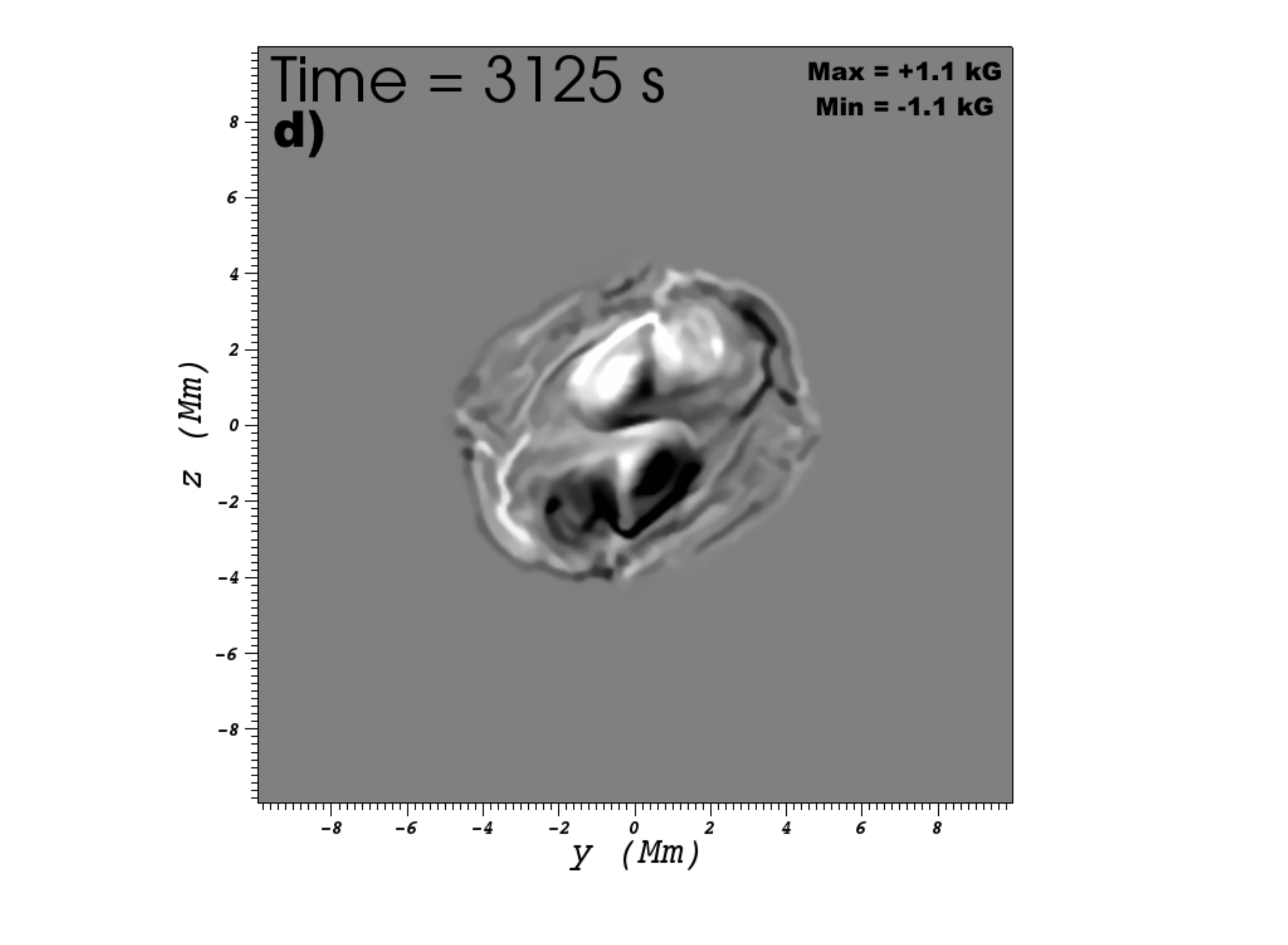}
\centering\includegraphics[scale=0.32, trim=1.0cm 1.0cm 5.5cm 1.0cm,clip=true]{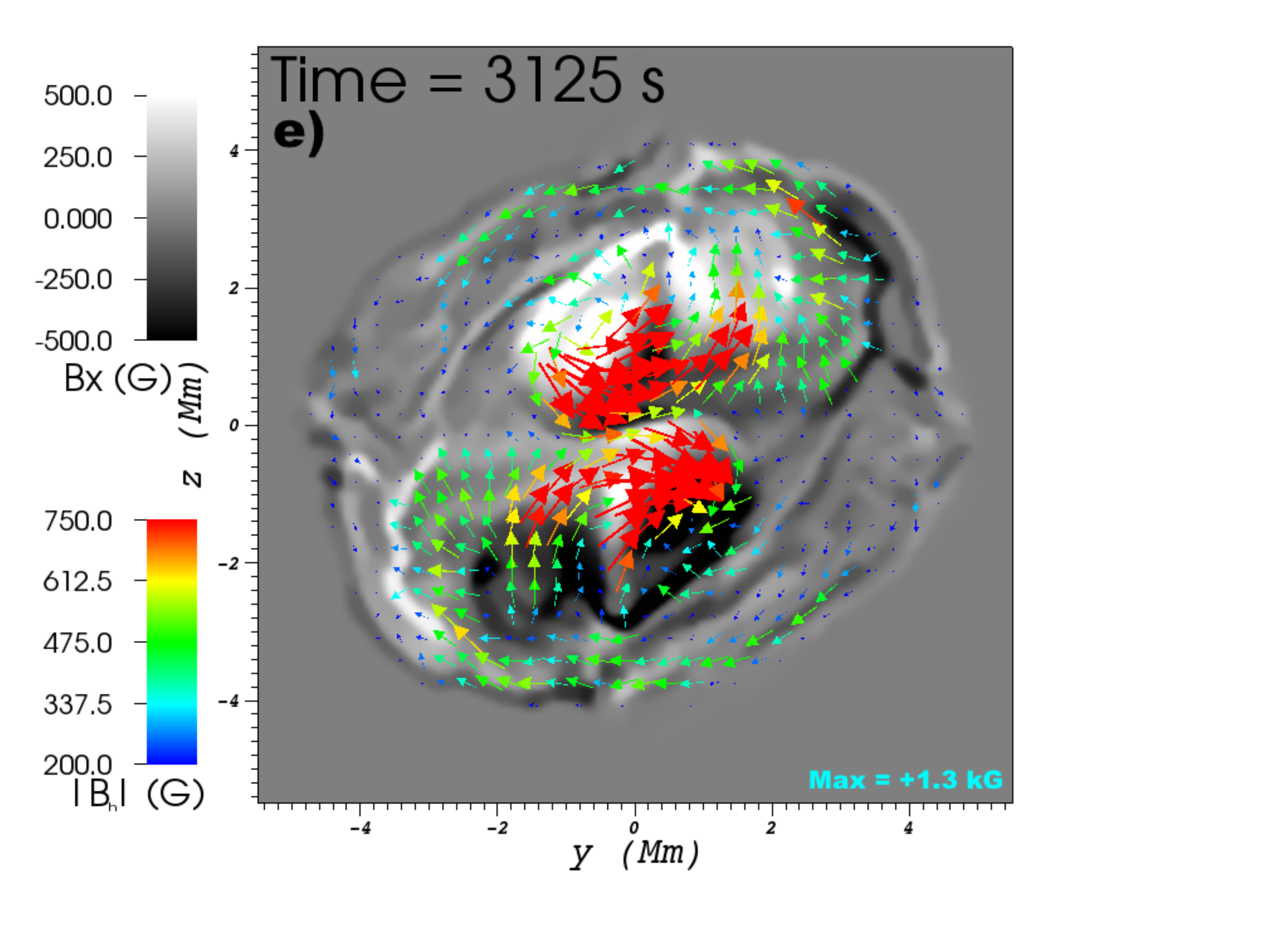}
\caption{Same as Figure \ref{fig:bx0}, for the $\zeta=2$ case. Primary/secondary polarities are denoted by blue/yellow arrows in panels b) and c).}
\label{fig:bx2}
\end{figure*}

\begin{figure*}
\centering\includegraphics[scale=0.25, trim=0.0cm 0.0cm 0.0cm 0.0cm,clip=true]{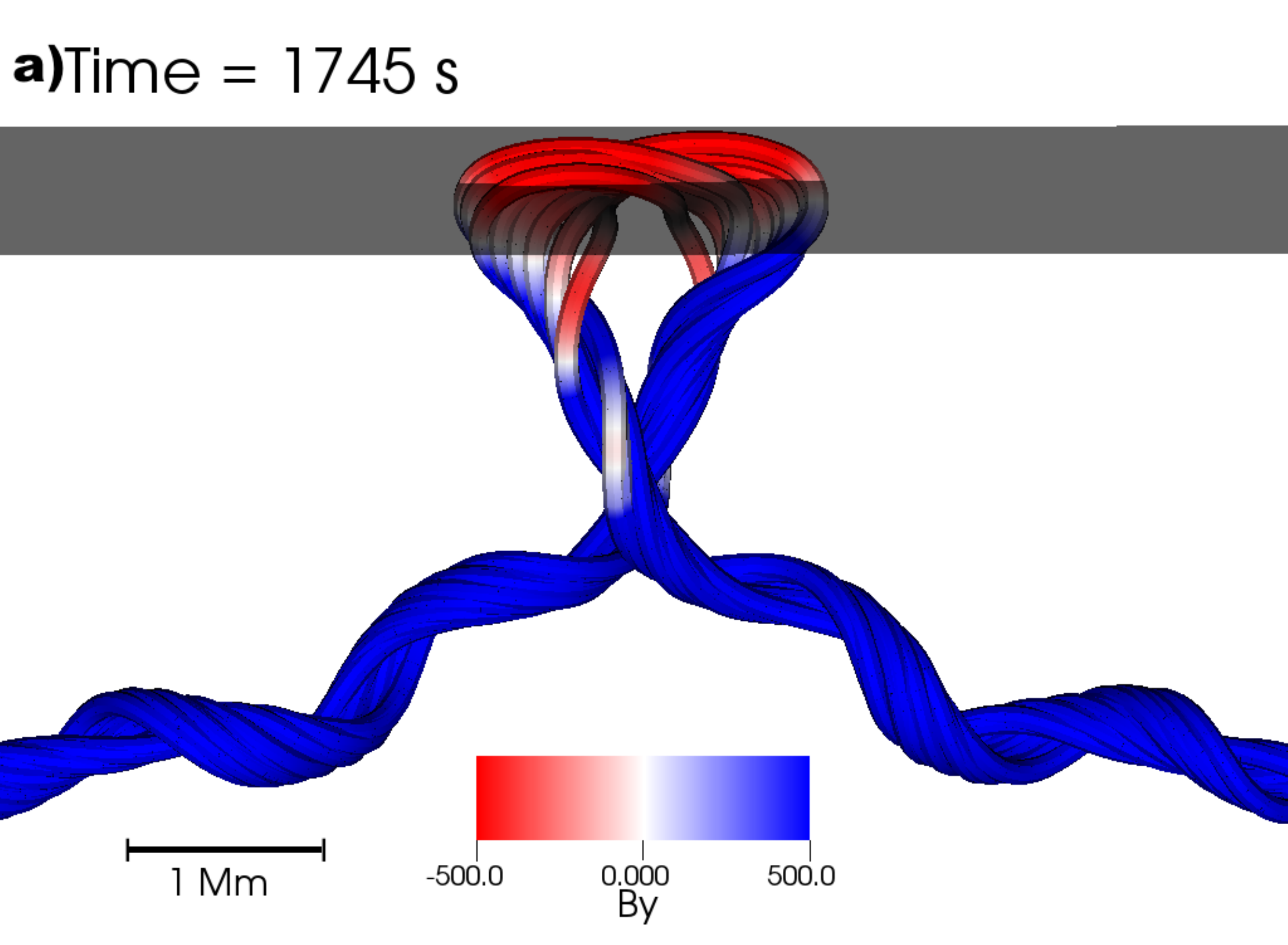}
\centering\includegraphics[scale=0.25, trim=0.0cm 0.0cm 0.0cm 0.0cm,clip=true]{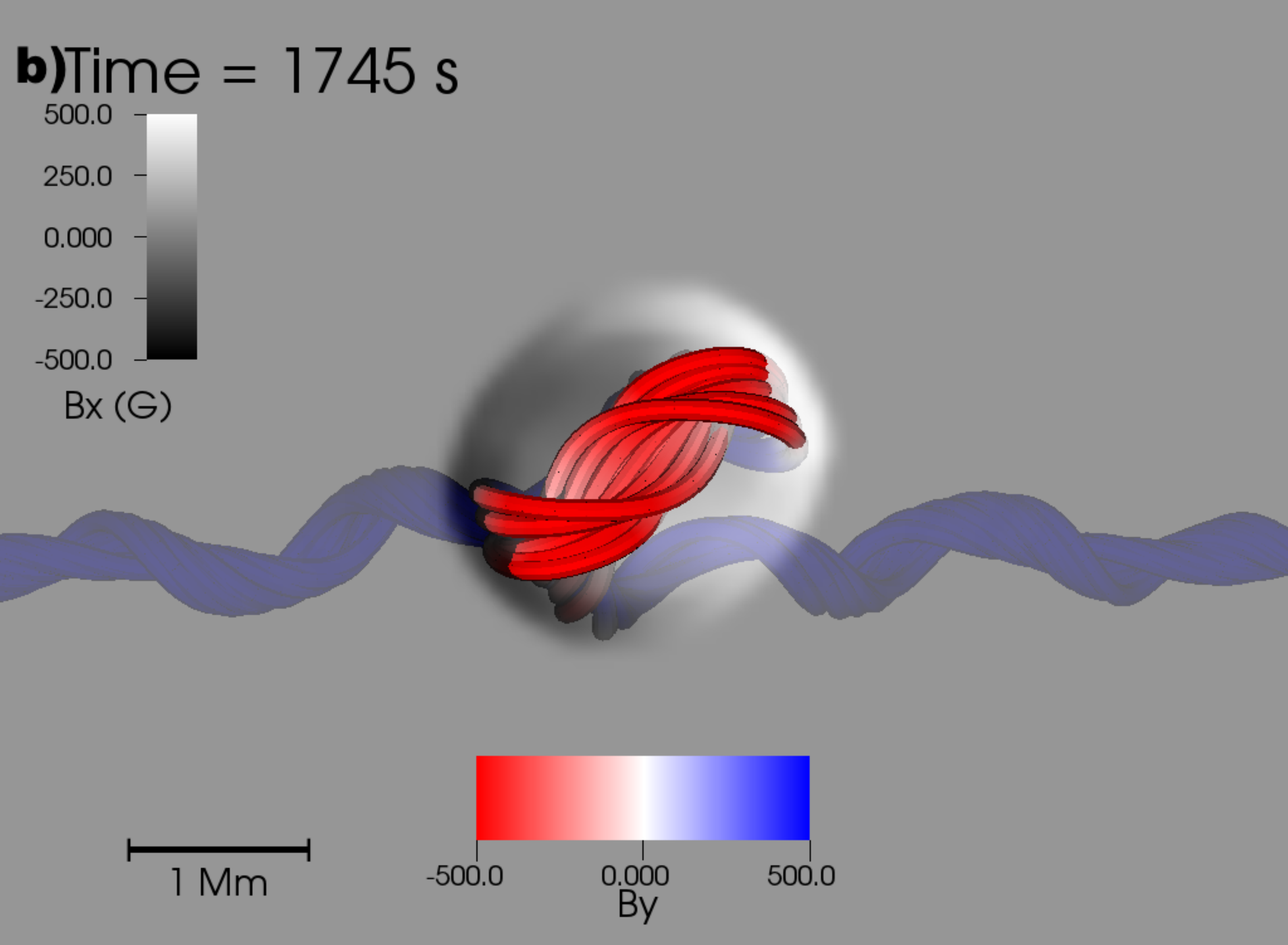}
\centering\includegraphics[scale=0.25, trim=0.0cm 0.0cm 0.0cm 0.0cm,clip=true]{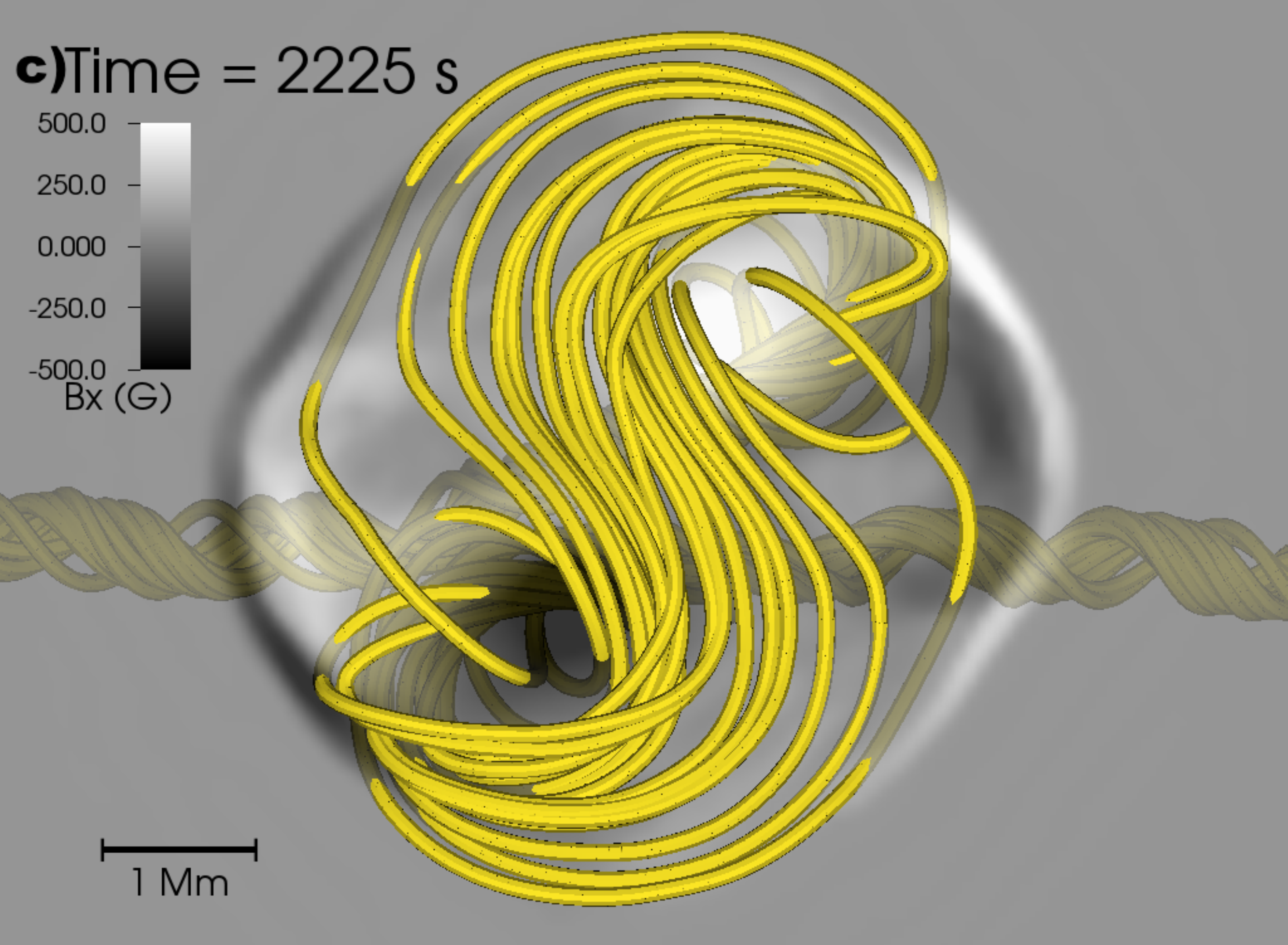}
\centering\includegraphics[scale=0.25, trim=0.0cm 0.0cm 0.0cm 0.0cm,clip=true]{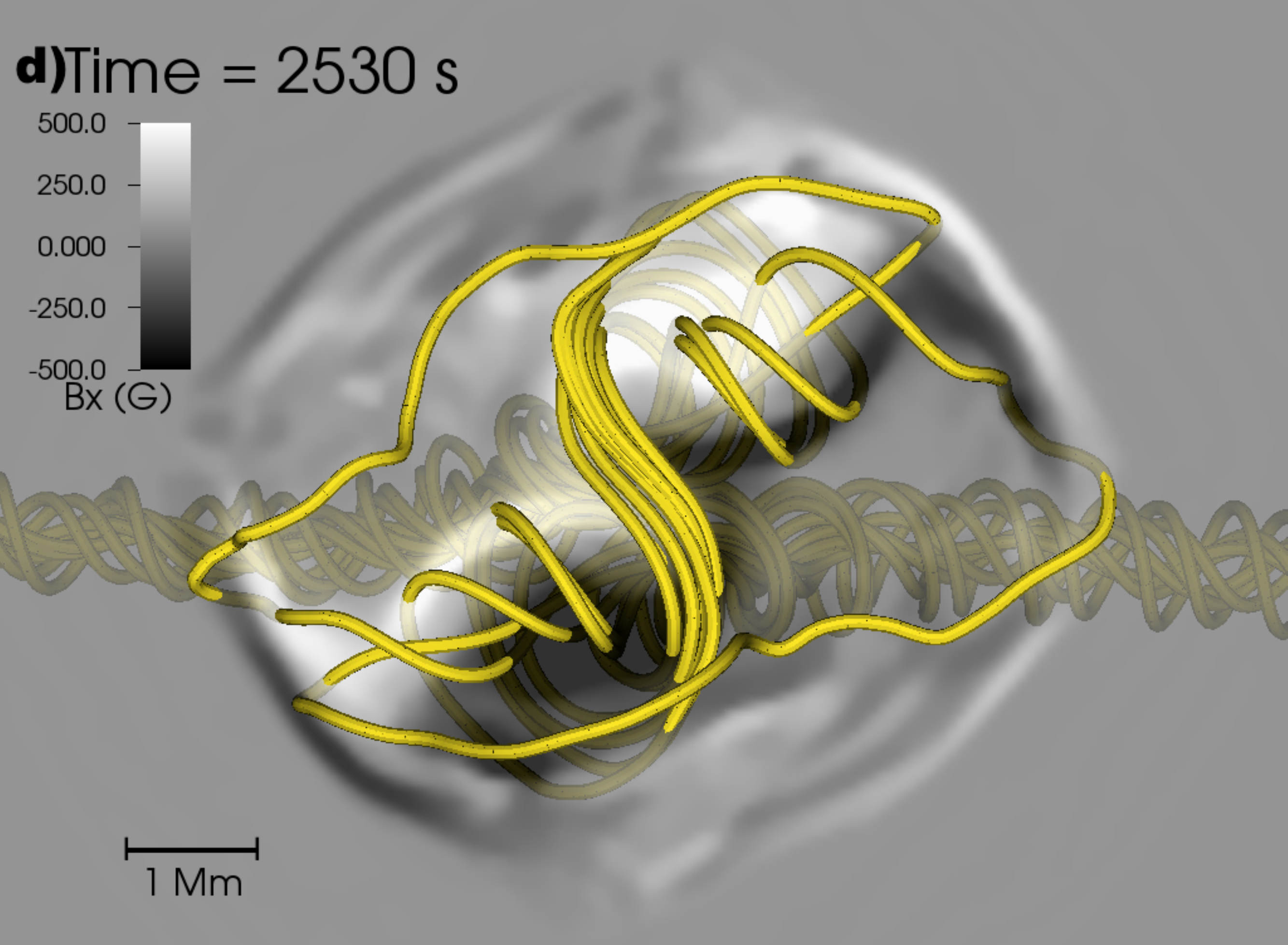}
\centering\includegraphics[scale=0.25, trim=0.0cm 0.0cm 0.0cm 0.0cm,clip=true]{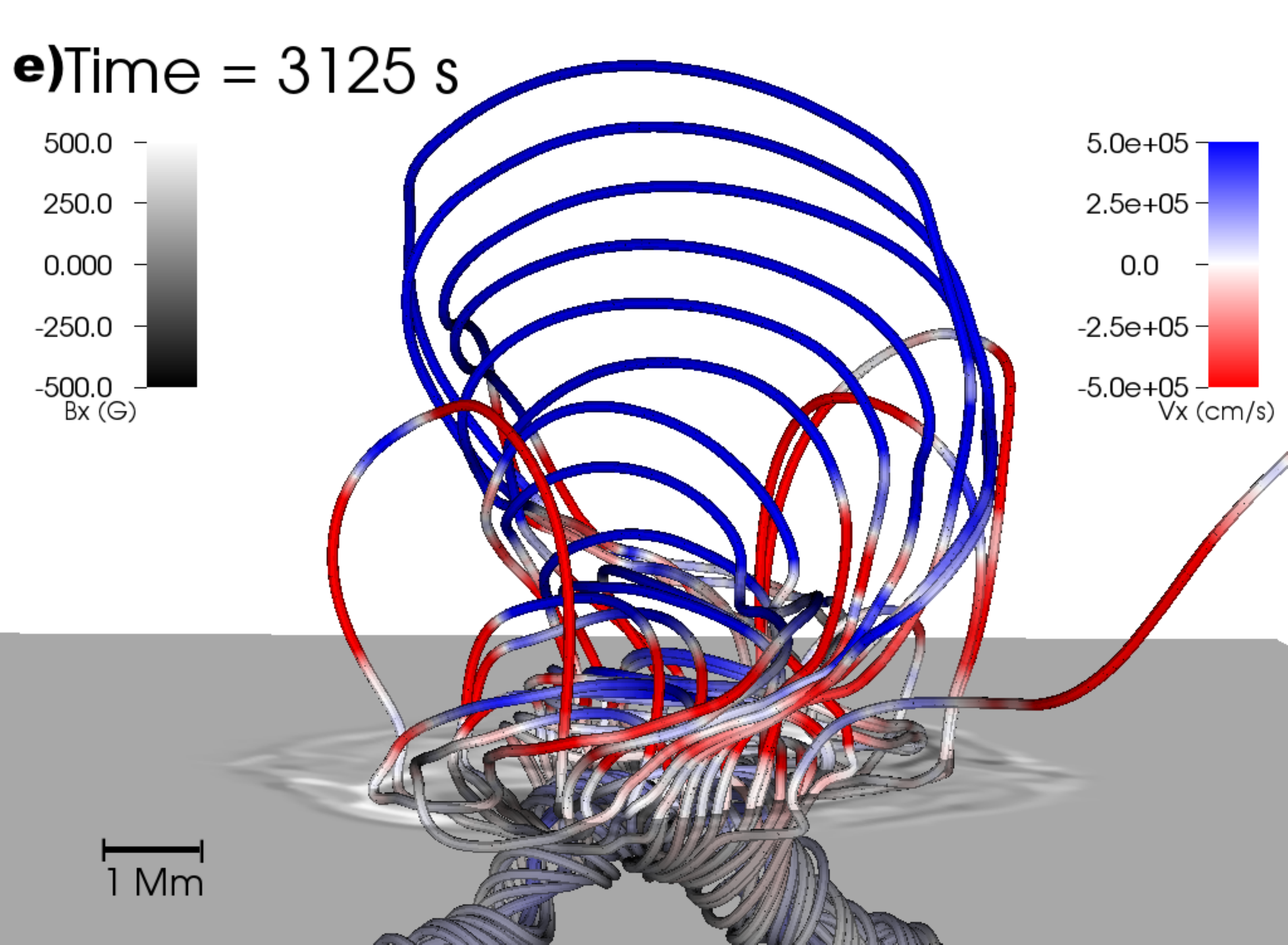}
\centering\includegraphics[scale=0.25, trim=0.0cm 0.0cm 0.0cm 0.0cm,clip=true]{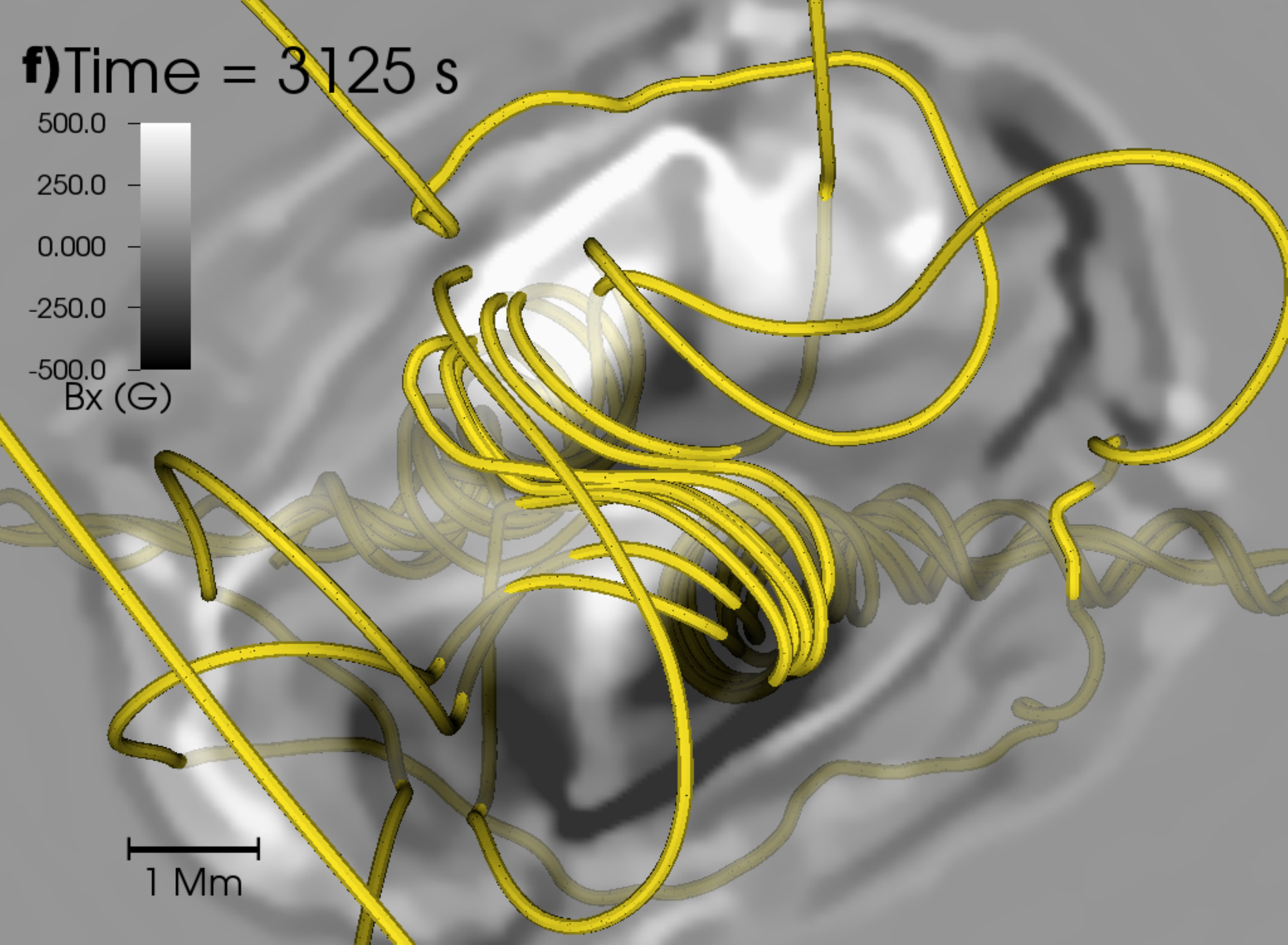}
\caption{Field lines, overplotted on photospheric magnetograms, at various stages of the $\zeta=2$ simulation. In a) and b) field lines, colored by $B_y$, are seen from the side and above, respectively. In e) field lines are shown at the final state of the simulation, colored by $v_x$.}
\label{fig:fieldlines2}
\end{figure*}

The flux rope penetrates the photosphere at $t\approx1745 \;\mathrm{s}$. Like the $\zeta=1.5$ emergence, the two polarities formed by the emerging flux rope develop into elongated, crescent-shaped polarities (Figure \ref{fig:bx2}a). Figures \ref{fig:fieldlines2}a,b show the field lines that emerge first, with the $B_y$ shading demonstrating the $\ell$ shape and almost complete revolution of the flux rope axis by the time it reaches the photosphere. These first two polarities elongate and make a circular shell around two more sets of polarities that subsequently emerge around $t=2225\;\mathrm{s}$ (blue/yellow arrows in Figure \ref{fig:bx2}b), forming a four-leaf clover pattern that is qualitatively similar to the pattern observed in the $\zeta=1.5$ evolution (Figure \ref{fig:bx1.5}b). Once again, the primary polarities are connected by sigmoidal field lines oriented along the $z$-direction, and the secondary polarities are connected to the primary polarities by short, sheared field lines encircling each leg of the emerging flux rope (Figure \ref{fig:fieldlines2}c), with the atmospheric portion of the field line appearing sheared and the subsurface part appearing twisted. Both sets of polarities approach each other and elongate (Figure \ref{fig:bx2}c), forming short, quasi-potential loops and longer sheared arcades in the corona (Figure \ref{fig:fieldlines2}d). Each polarity then rotates about itself, into the configuration shown in Figure \ref{fig:bx2}d, where the primary and secondary polarities contain horizontal field that is highly sheared along the polarity inversion lines (Figure \ref{fig:bx2}e). This stage of the photospheric evolution, with two primary polarities separated by two secondary, elongated polarities, is also qualitatively similar to the photospheric evolution seen in the $\zeta=1.5$ case (Figure \ref{fig:bx1.5}c). The coronal field lines at the end of the simulation are much more complicated than in any of the previous simulations. They are extremely compact, and low-lying field lines appear to be interacting quite strongly with overlying field lines (Figure \ref{fig:fieldlines2}e,f). In this case, in spite of the complexity of the photospheric field, the transformation in Equation \ref{transformation} leaves the photospheric distribution almost entirely unchanged for a large portion of the simulation.

\subsection{$\zeta=4$}\label{sec:4}

\subsubsection{Subsurface Evolution}\label{sec:sub4}
\begin{figure*}
\centering\includegraphics[scale=0.29, trim=0.0cm 1.0cm 4.0cm 0.0cm,clip=true]{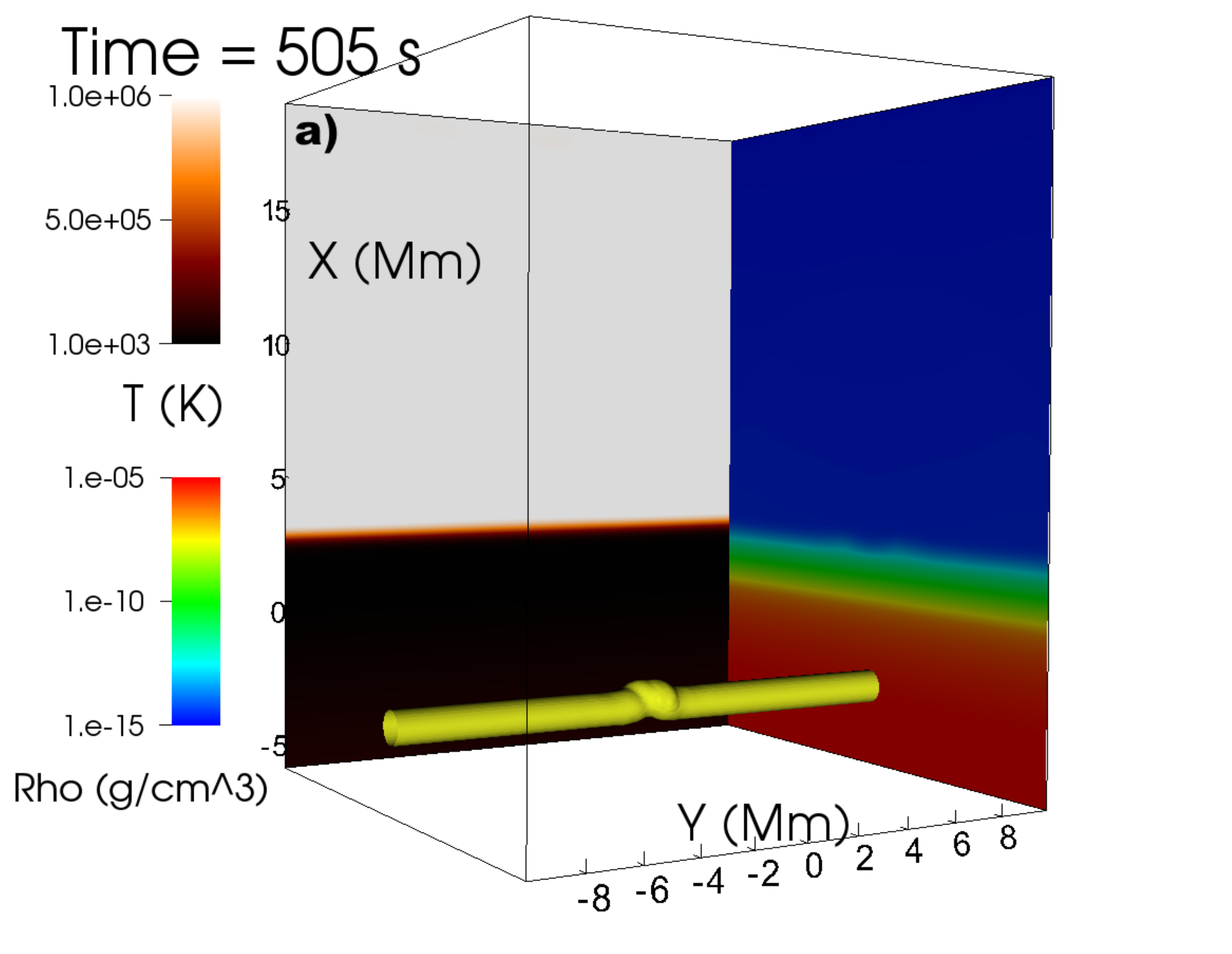}
\centering\includegraphics[scale=0.32, trim=4.0cm 1.0cm 5.0cm 3.0cm,clip=true]{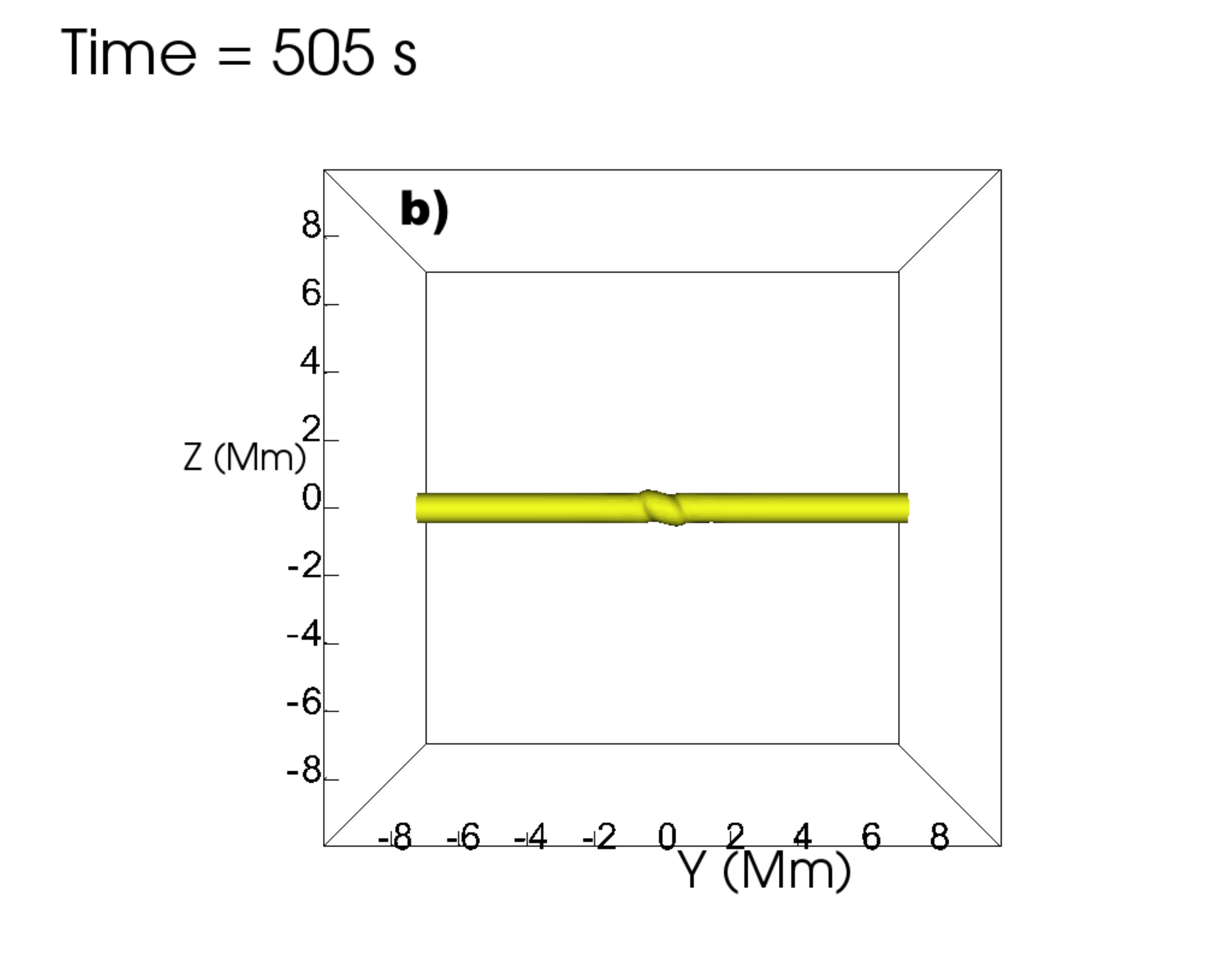}
\centering\includegraphics[scale=0.29, trim=0.0cm 1.0cm 4.0cm 0.0cm,clip=true]{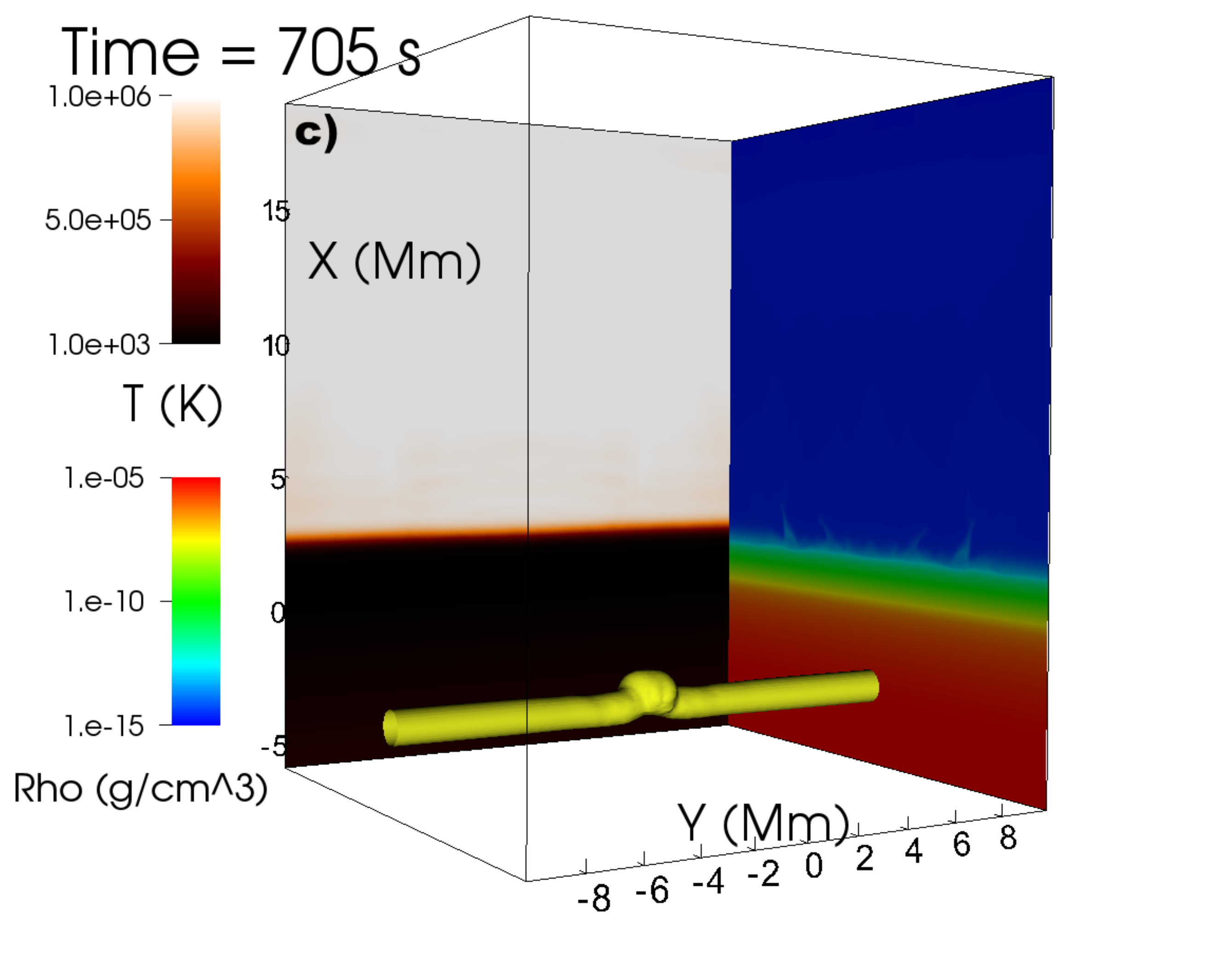}
\centering\includegraphics[scale=0.32, trim=4.0cm 1.0cm 5.0cm 3.0cm,clip=true]{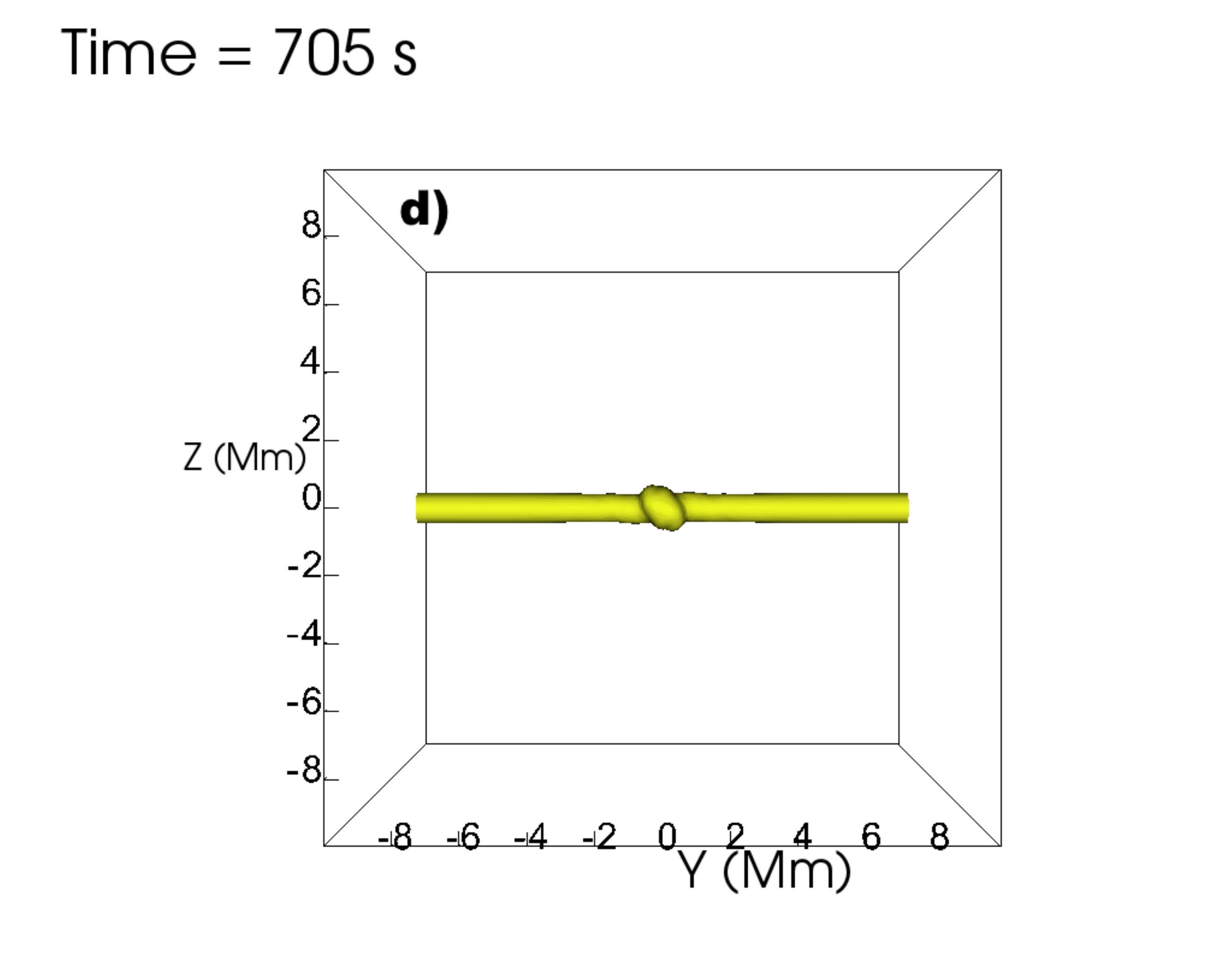}
\centering\includegraphics[scale=0.29, trim=0.0cm 1.0cm 4.0cm 0.0cm,clip=true]{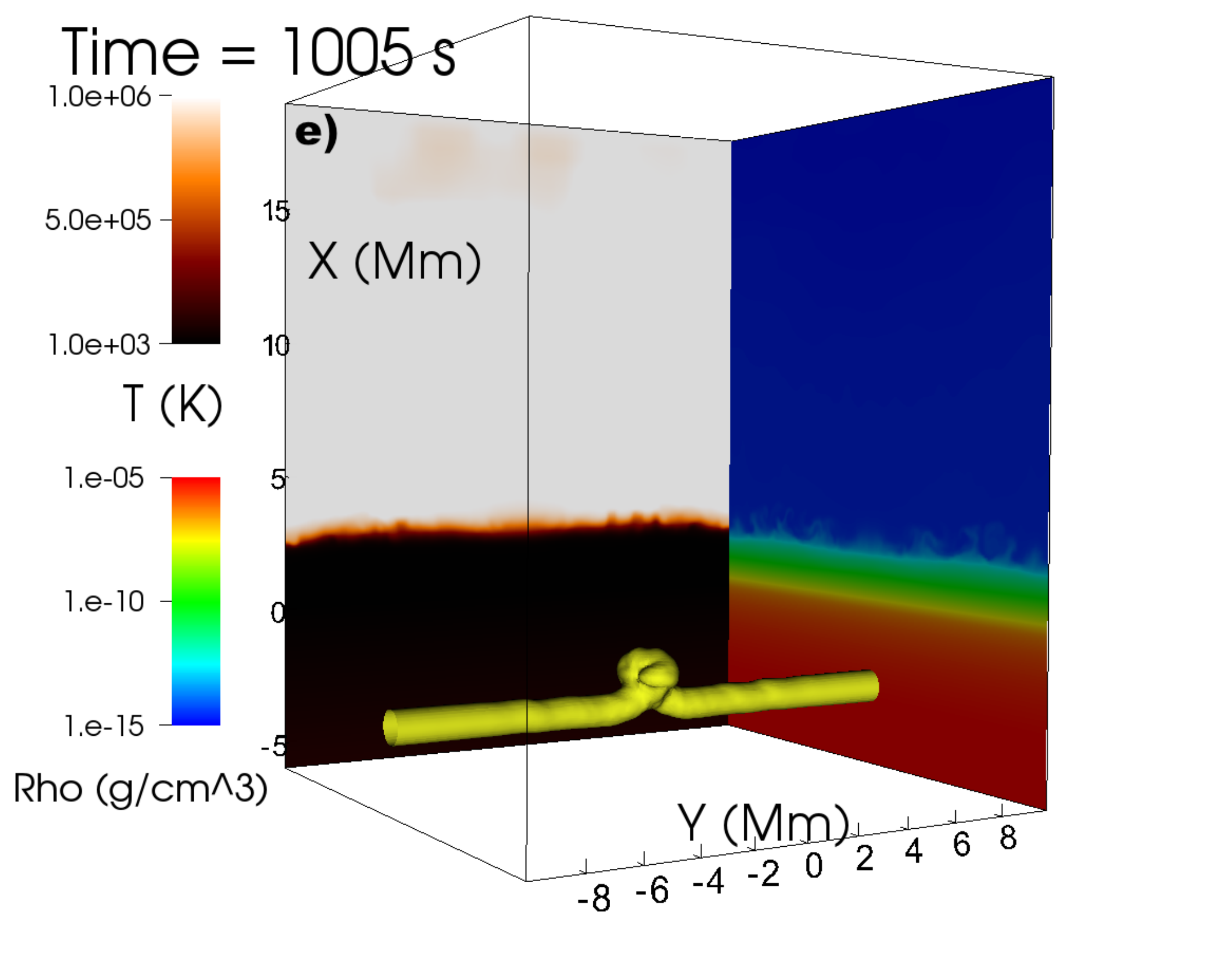}
\centering\includegraphics[scale=0.32, trim=4.0cm 1.0cm 5.0cm 3.0cm,clip=true]{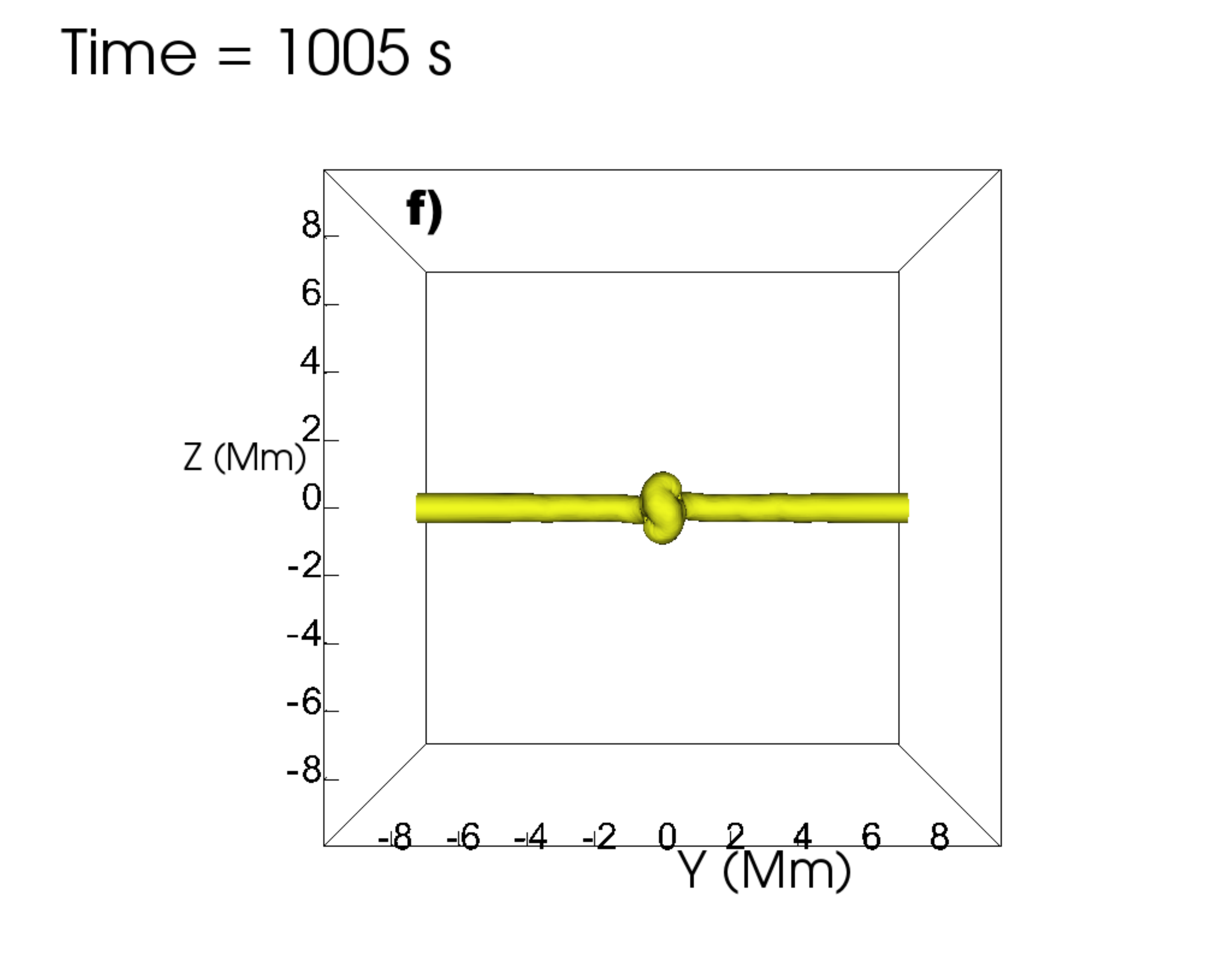}
\centering\includegraphics[scale=0.29, trim=0.0cm 1.0cm 4.0cm 0.0cm,clip=true]{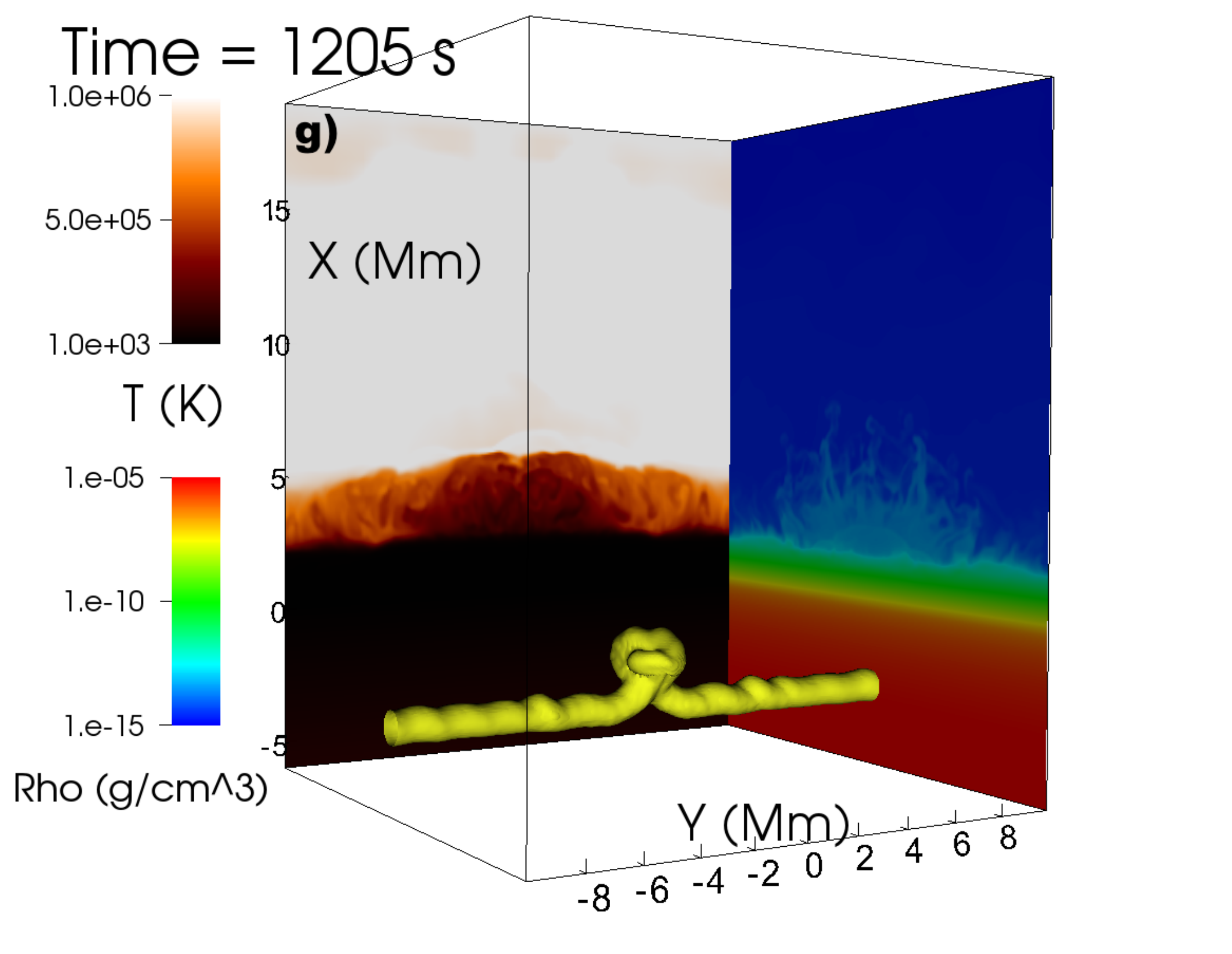}
\centering\includegraphics[scale=0.32, trim=4.0cm 1.0cm 5.0cm 3.0cm,clip=true]{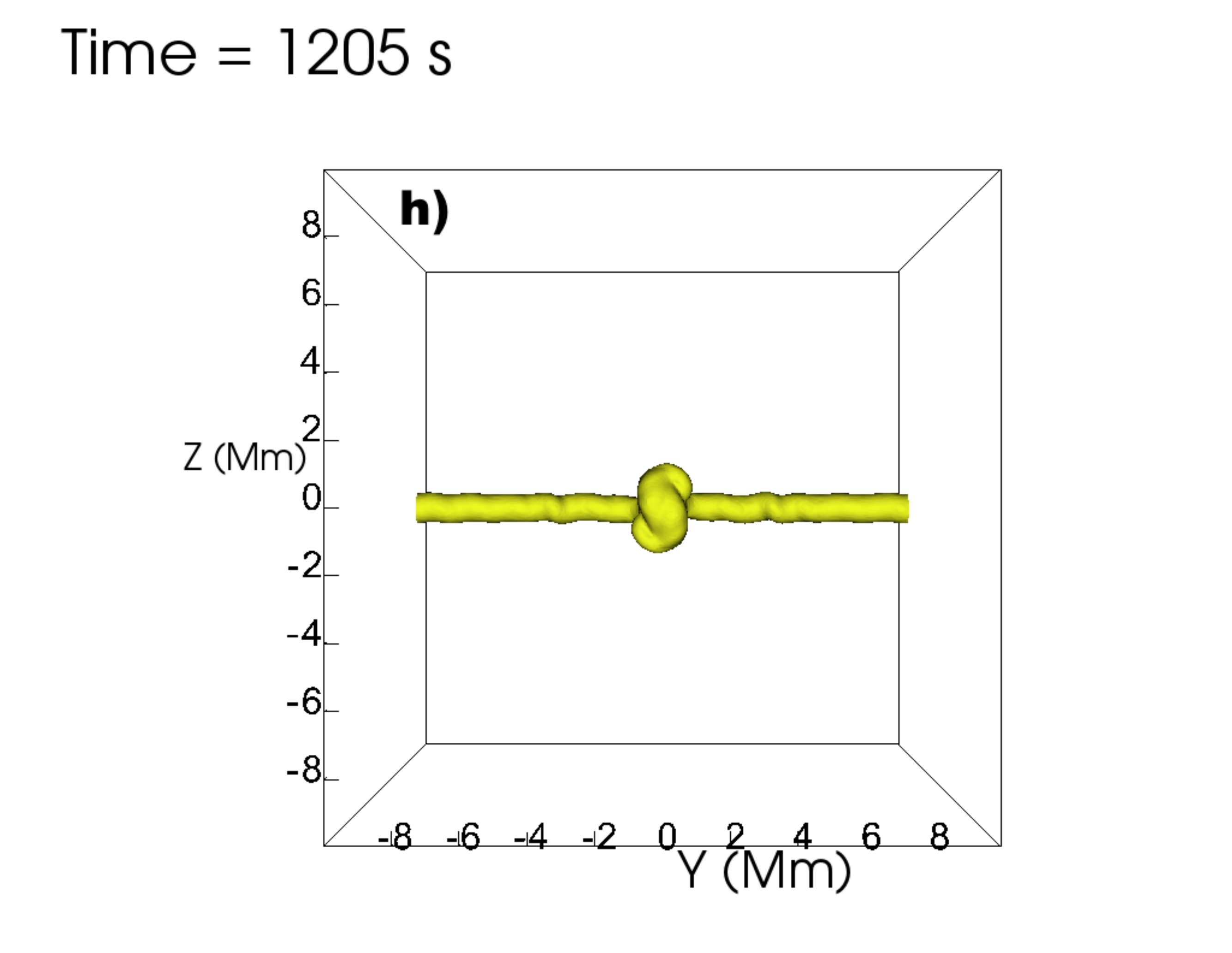}
\caption{Same as Figure \ref{fig:isosurfaces05} for $\zeta=4$}
\label{fig:isosurfaces4}
\end{figure*}

Here we present the emergence of a flux rope with $\zeta=4$. To our knowledge, this is the most highly twisted flux rope that has been emerged in simulations to date. 
The rise of this flux rope through the convection zone is shown in Figure \ref{fig:isosurfaces4}. The rising $\Omega$-loop generated by the mass density deficit is very quickly affected by the onset of the kink instability. As early as $t\approx 505\;s$, the $\Omega$-loop has rotated out of its initial plane. By $t\approx705\;s$, the apex has formed a complicated knot-like structure, and certainly no longer resembles an $\Omega$-loop. As the apex approaches the photosphere, the structure becomes completely aligned with the $z$-axis. Just before the flux rope emerges, at $t\approx1205\;s$, the apex of the rising portion of the flux rope resembles a knot, and appears as an S-shaped structure when looking down on it. The peak of the rising flux rope is highly compact, and very different in structure than the kinks seen in the previous two simulations.

\subsubsection{Photospheric Evolution}\label{sec:phot4}
\begin{figure*}
\centering\includegraphics[scale=0.32, trim=1.0cm 1.0cm 5.5cm 1.0cm,clip=true]{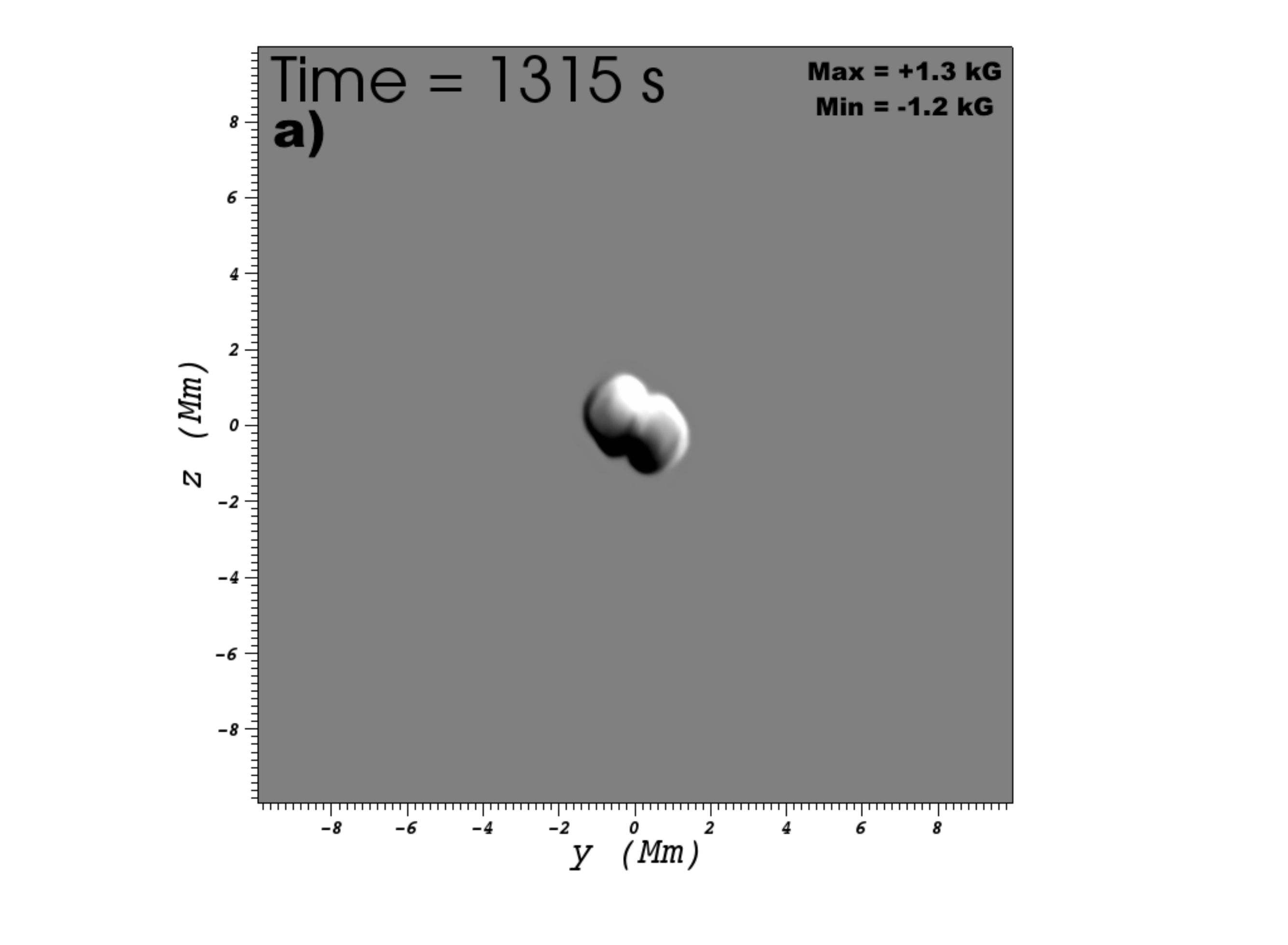}
\centering\includegraphics[scale=0.32, trim=3.0cm 1.0cm 5.5cm 1.0cm,clip=true]{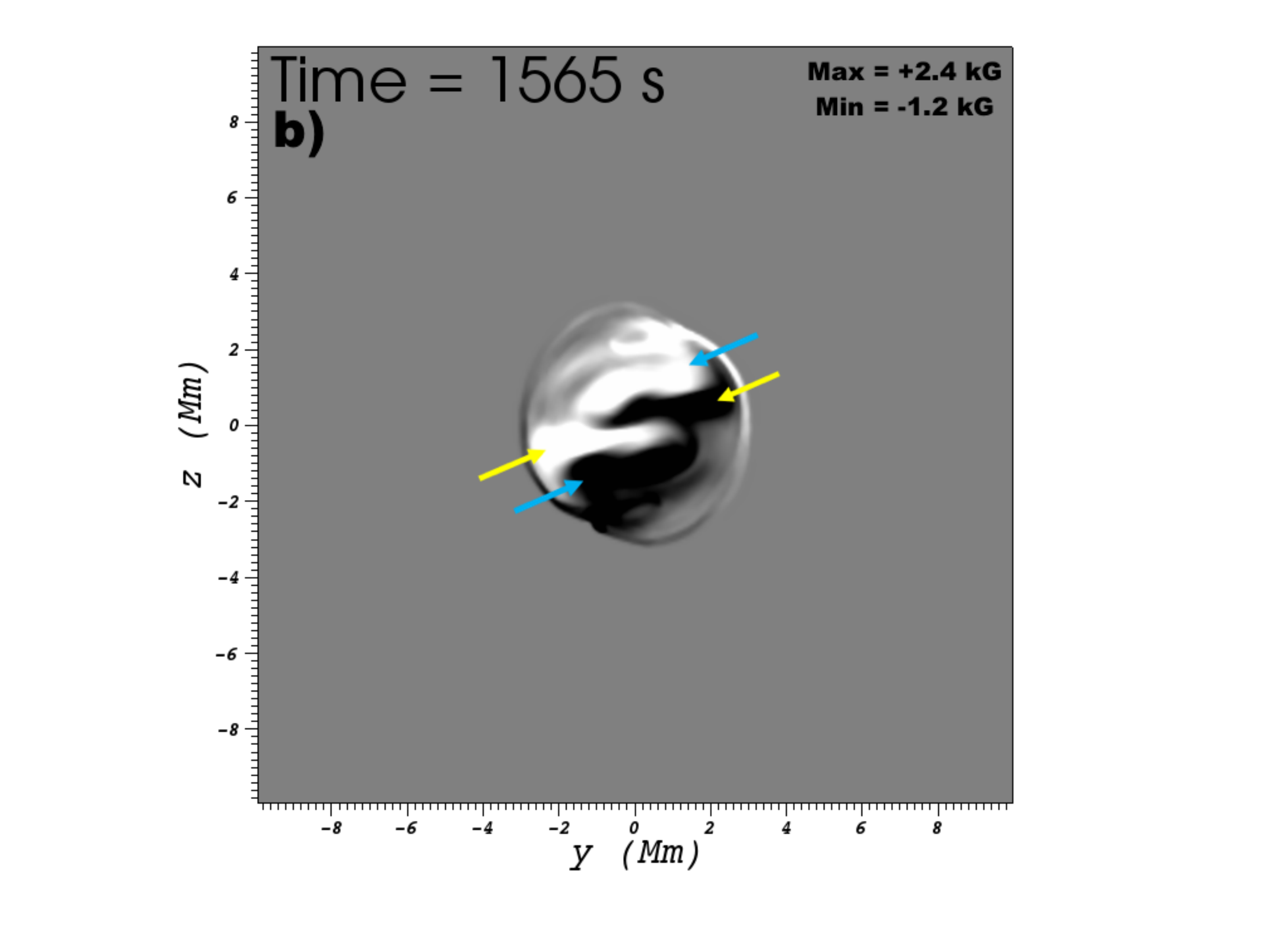}
\centering\includegraphics[scale=0.32, trim=1.0cm 1.0cm 5.5cm 1.0cm,clip=true]{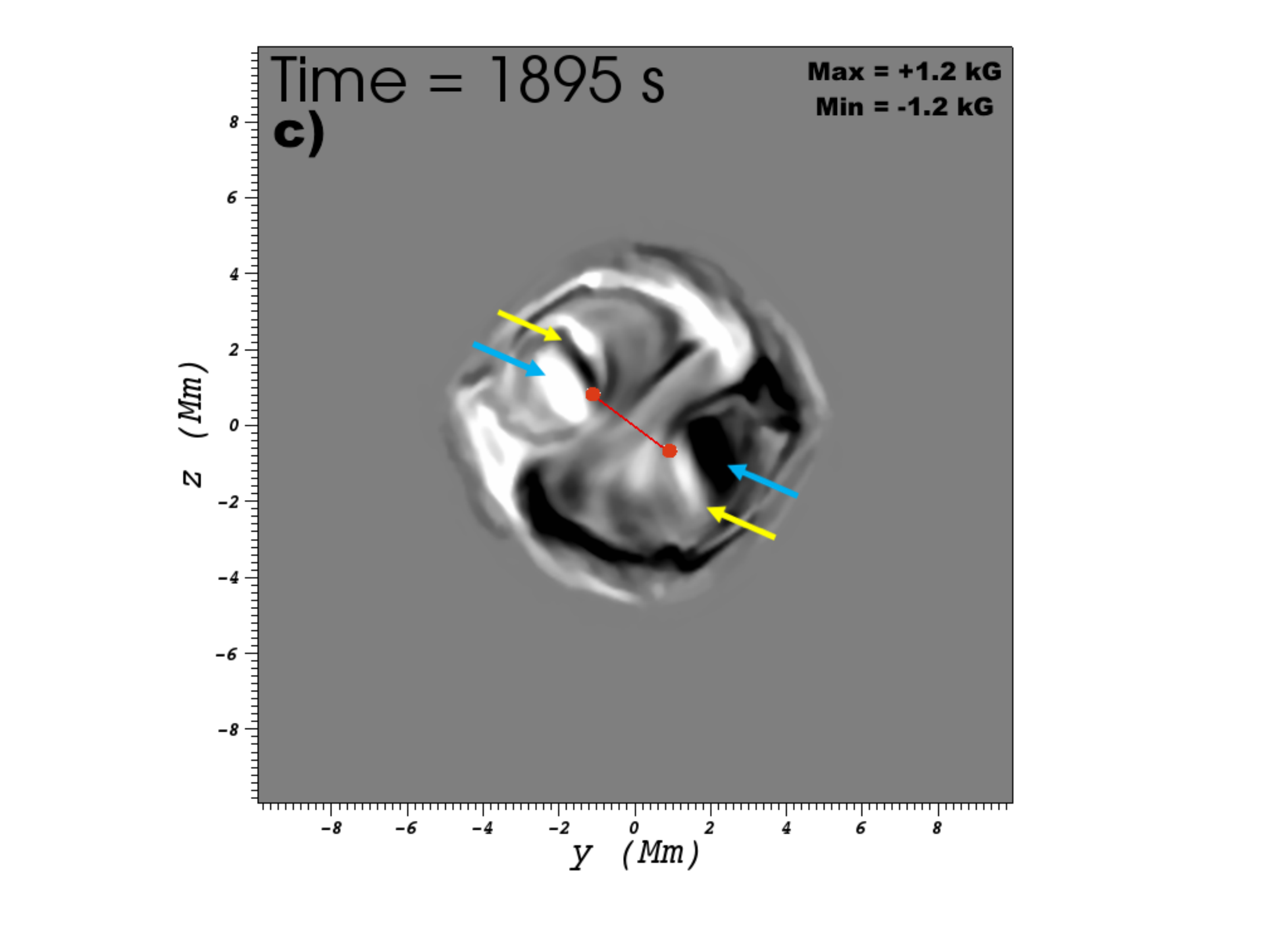}
\centering\includegraphics[scale=0.32, trim=3.0cm 1.0cm 5.5cm 1.0cm,clip=true]{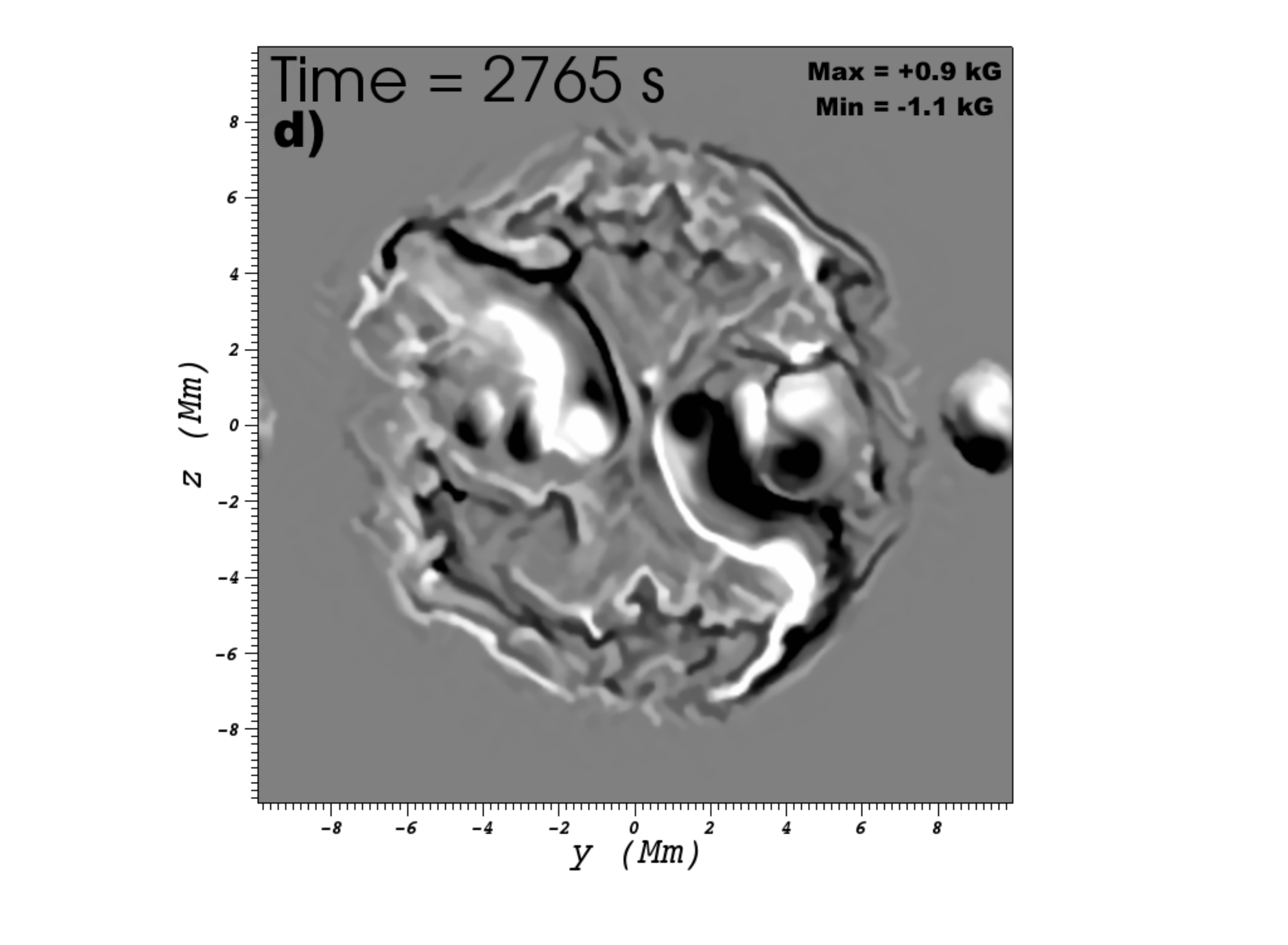}
\centering\includegraphics[scale=0.32, trim=1.0cm 1.0cm 5.5cm 1.0cm,clip=true]{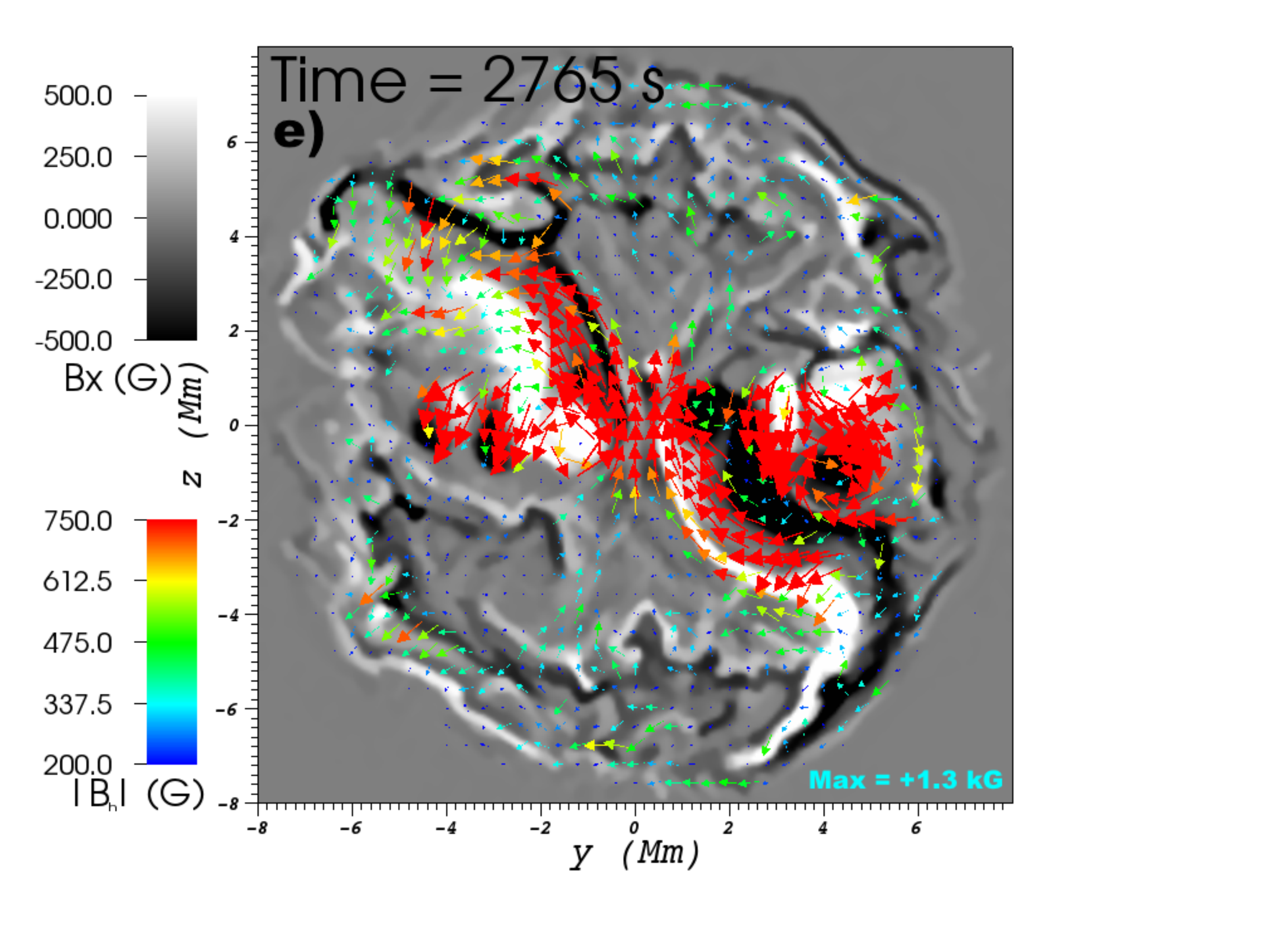}
\caption{Same as Figure \ref{fig:bx0}, for the $\zeta=4$ case. Primary/secondary polarities are denoted by blue/yellow arrows in panels b) and c).}
\label{fig:bx4}
\end{figure*}

\begin{figure*}
\centering\includegraphics[scale=0.25, trim=0.0cm 0.0cm 0.0cm 0.0cm,clip=true]{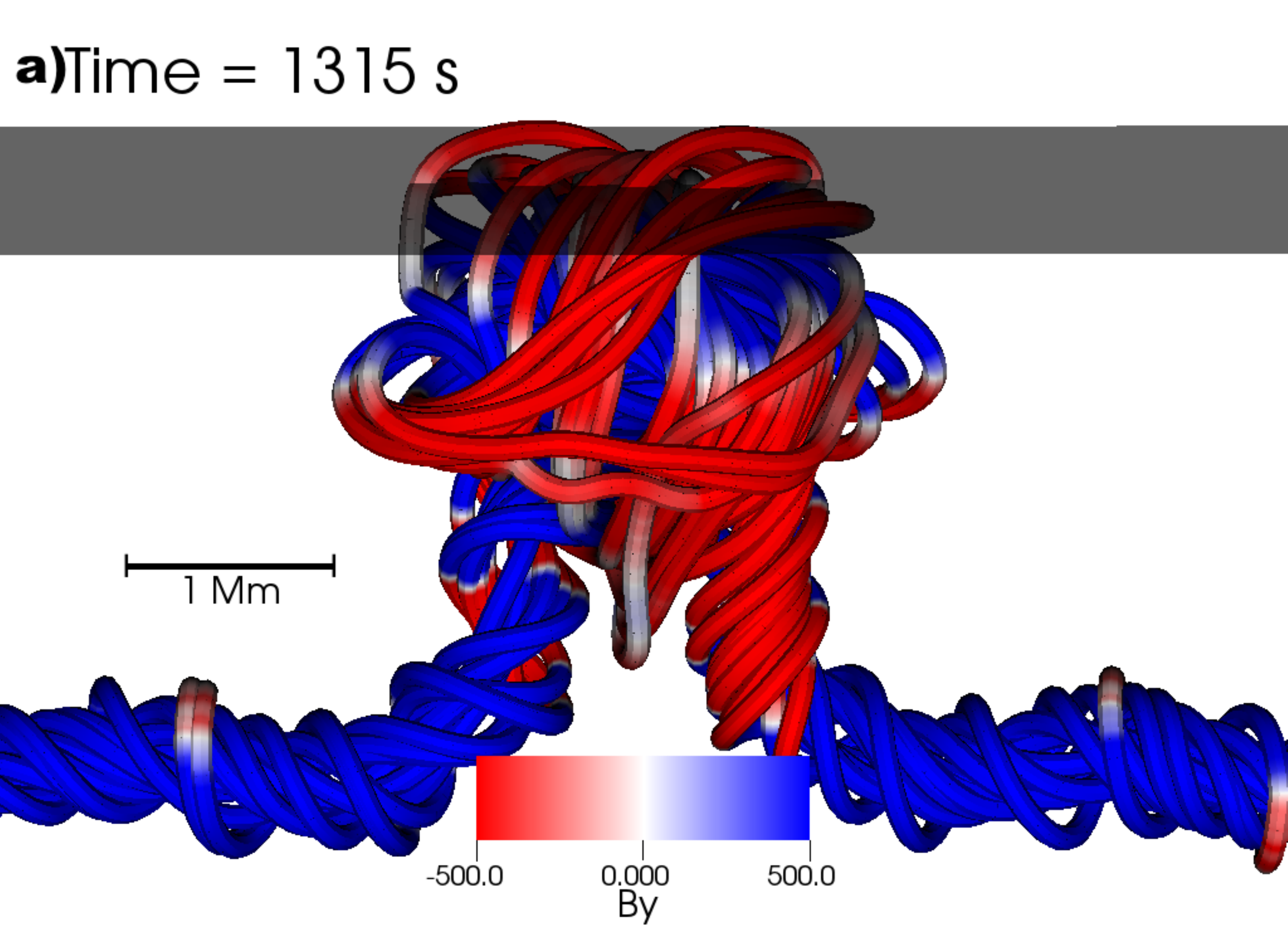}
\centering\includegraphics[scale=0.25, trim=0.0cm 0.0cm 0.0cm 0.0cm,clip=true]{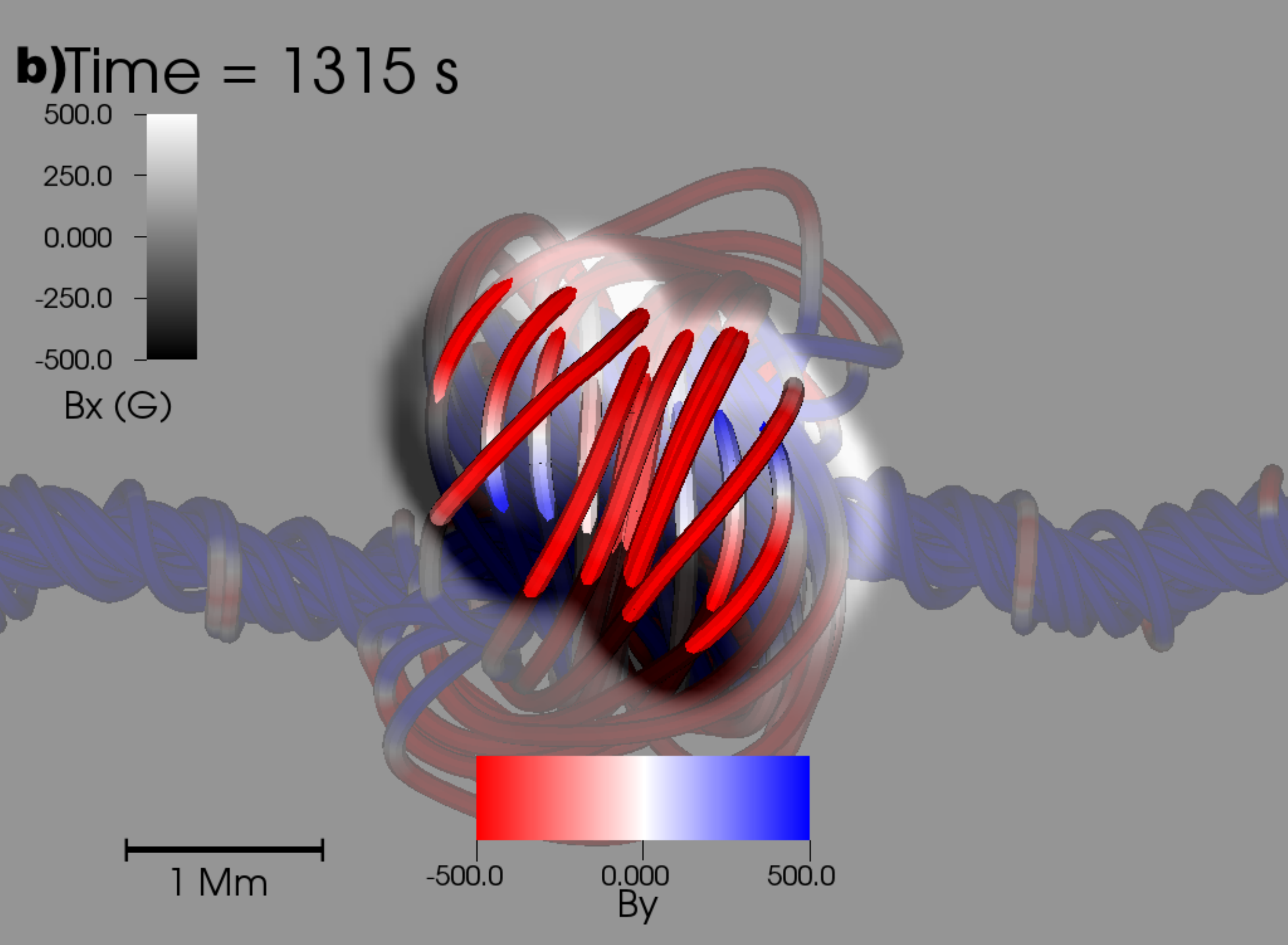}
\centering\includegraphics[scale=0.25, trim=0.0cm 0.0cm 0.0cm 0.0cm,clip=true]{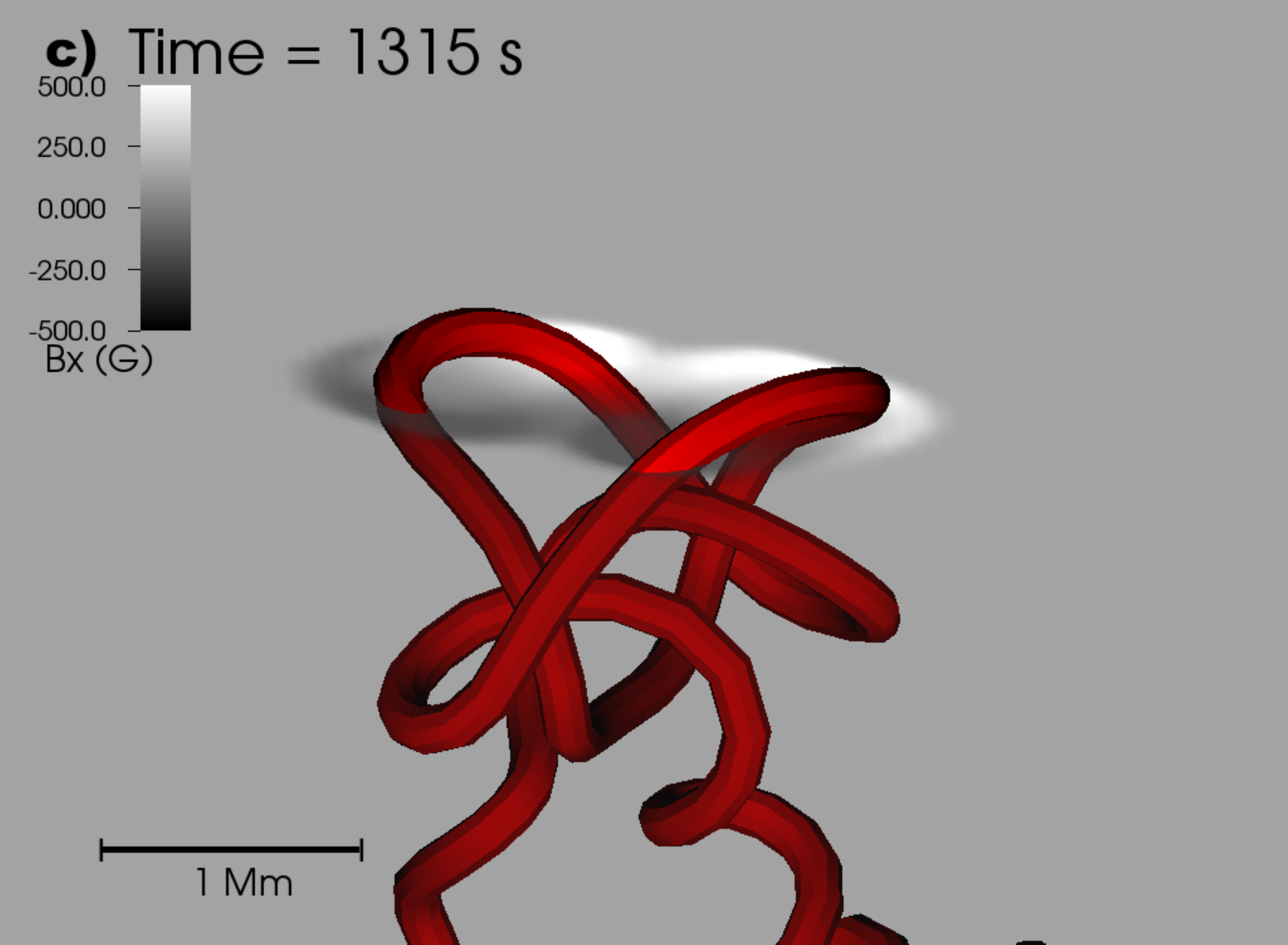}
\centering\includegraphics[scale=0.25, trim=0.0cm 0.0cm 0.0cm 0.0cm,clip=true]{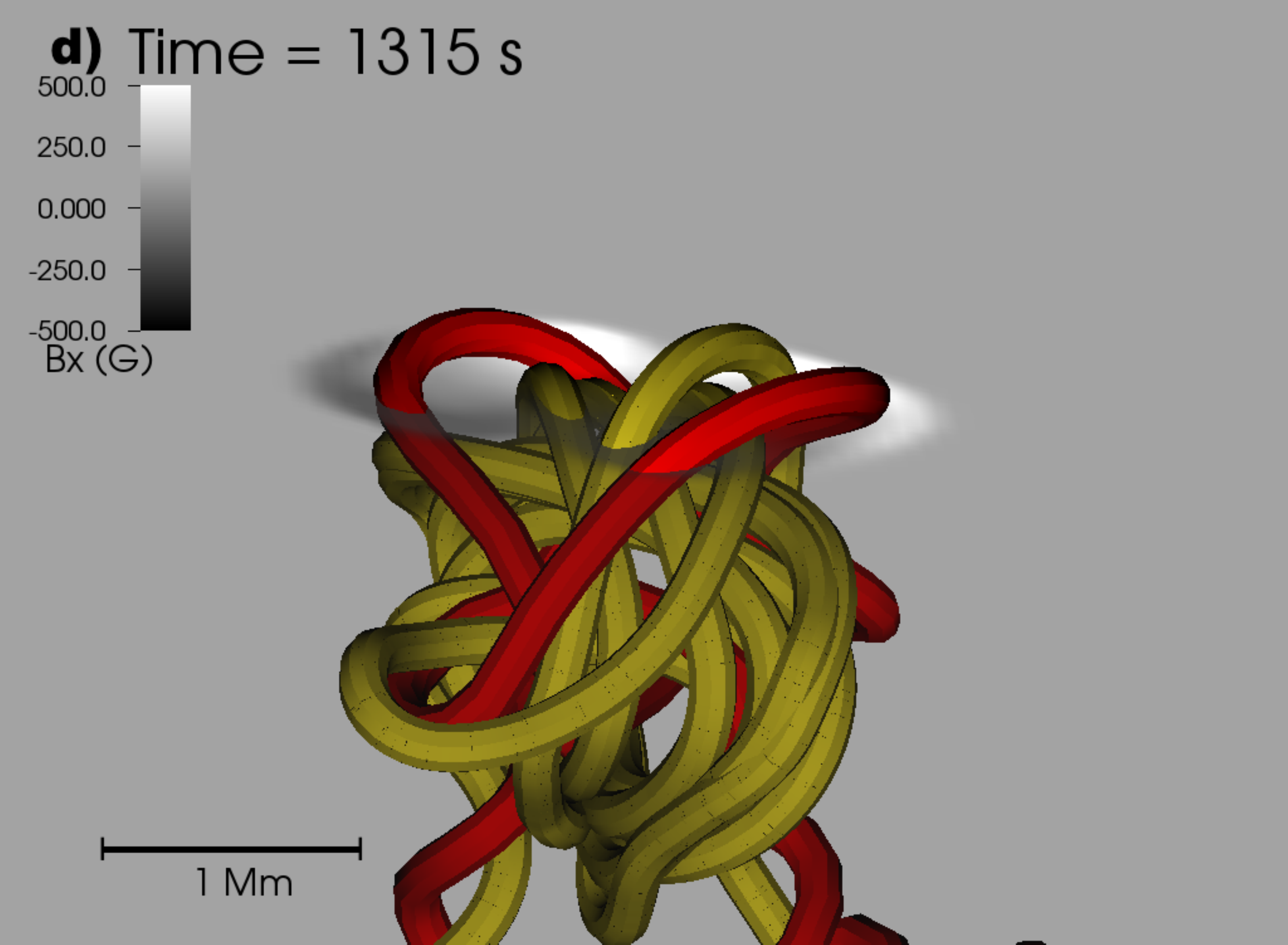}
\centering\includegraphics[scale=0.25, trim=0.0cm 0.0cm 0.0cm 0.0cm,clip=true]{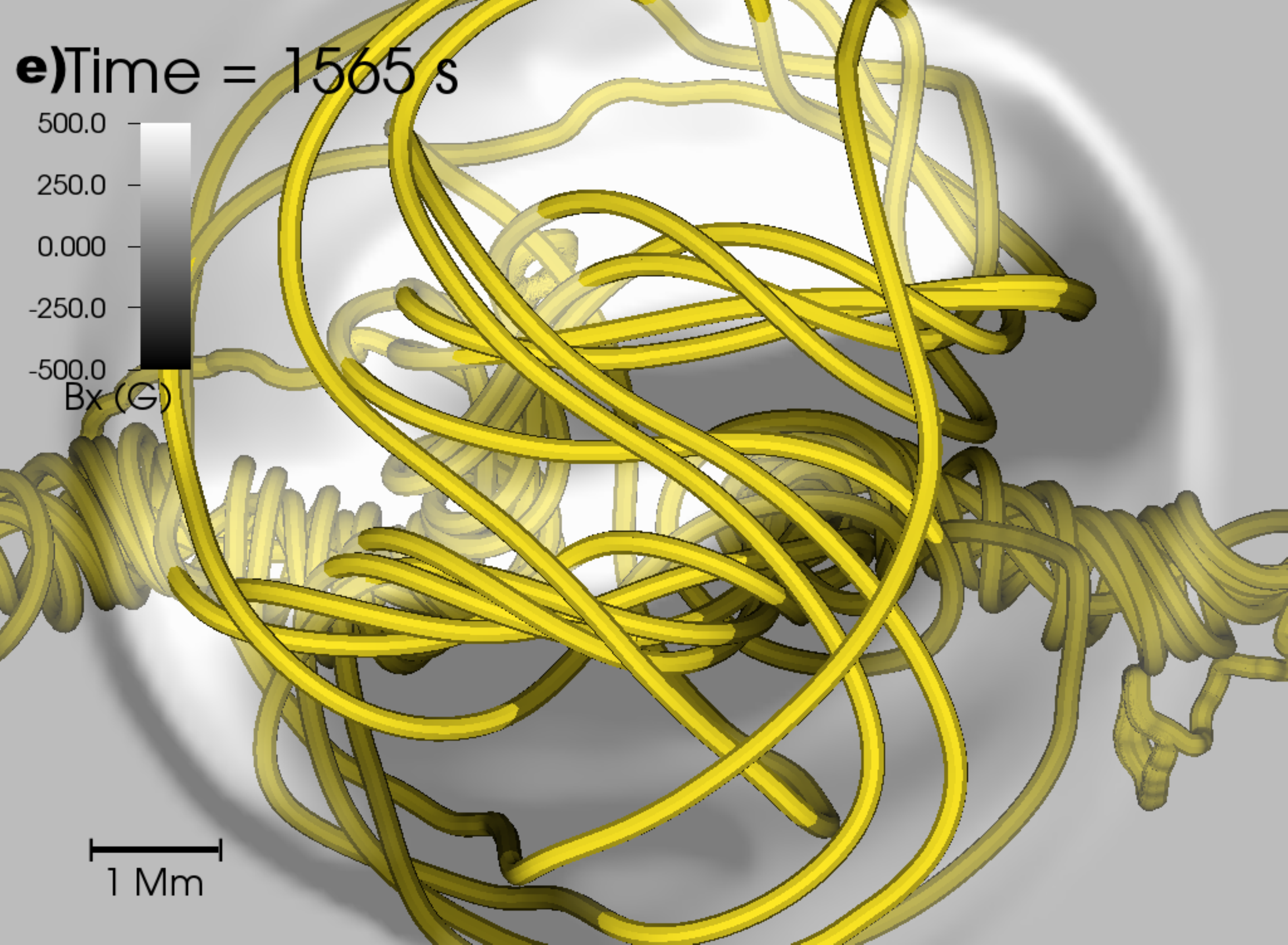}
\centering\includegraphics[scale=0.25, trim=0.0cm 0.0cm 0.0cm 0.0cm,clip=true]{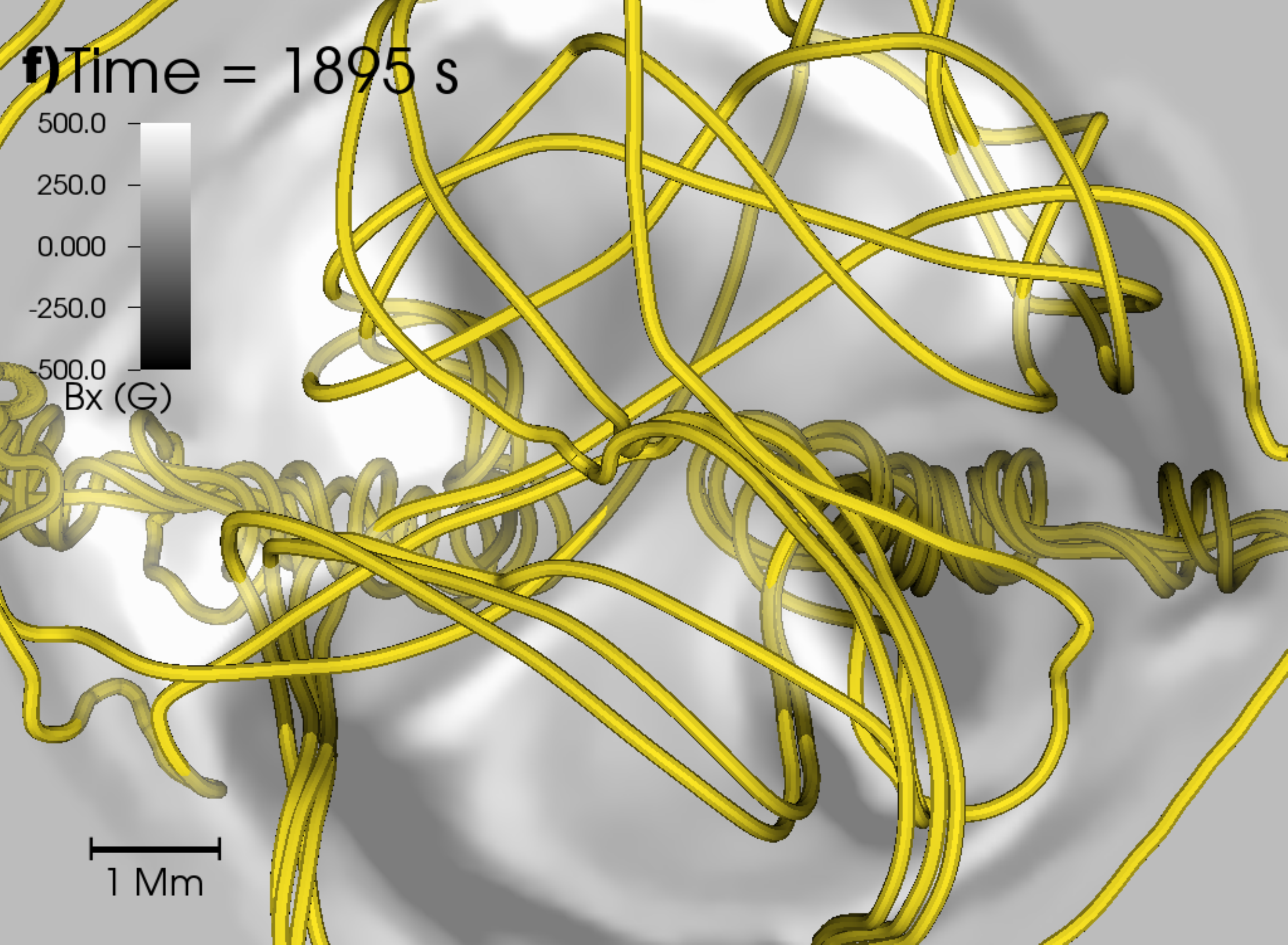}
\centering\includegraphics[scale=0.25, trim=0.0cm 0.0cm 0.0cm 0.0cm,clip=true]{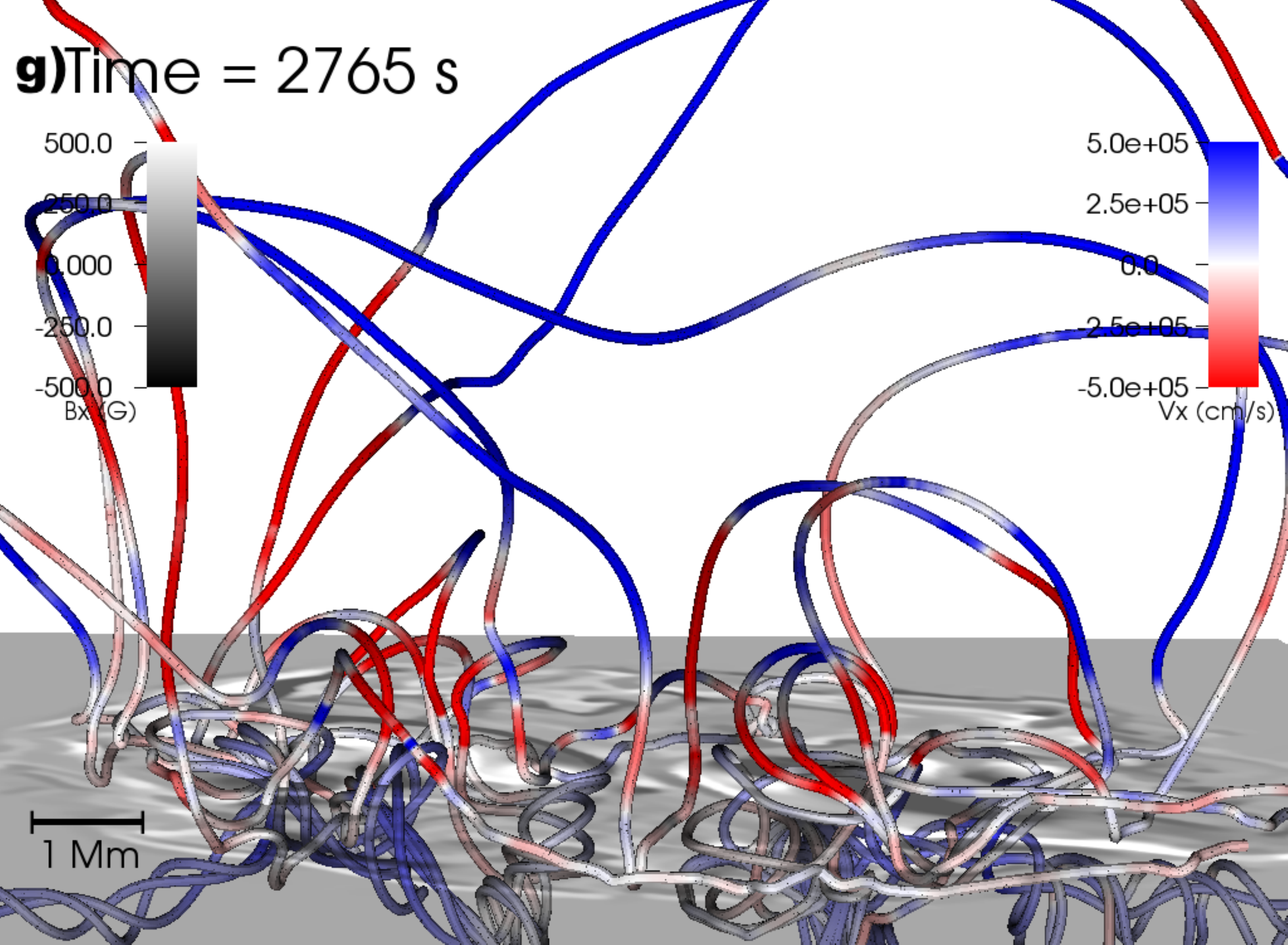}
\centering\includegraphics[scale=0.25, trim=0.0cm 0.0cm 0.0cm 0.0cm,clip=true]{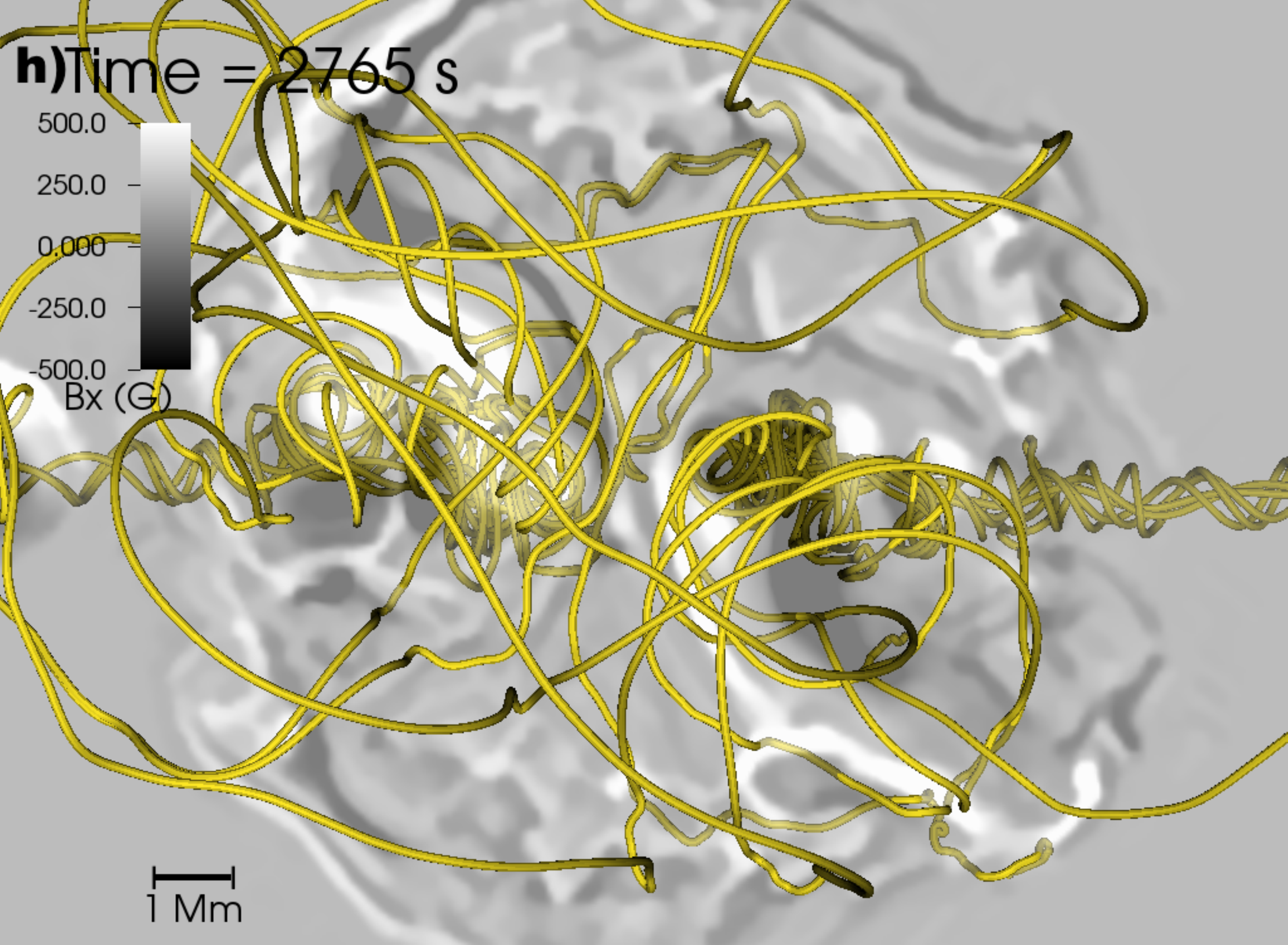}
\caption{Field lines, overplotted on photospheric magnetograms, at various stages of the $\zeta=4$ simulation. In a) and b) field lines, colored by $B_y$, are seen from the side and above, respectively. In c), a single knotted field line is shown, with several other field lines shown wrapped up in the knotted field line in d). In g) field lines are shown at the final state of the simulation, colored by $v_x$.}
\label{fig:fieldlines4}
\end{figure*}

This flux rope reaches the photosphere at $t\approx1315\;\mathrm{s}$, where it emerges in the shape of a figure-8 (Figure \ref{fig:bx4}a). Figures \ref{fig:fieldlines4}a,b show the field lines at and below the photosphere at this instant. There is clearly an extremely compact, highly tangled structure emerging into the corona. There is also some evidence that a double-overhand knot has formed from a succession of reconnection events during the rise of the flux rope, which may be a general result for an extreme kink \citep{Linton98}. This is shown in Figure \ref{fig:fieldlines4}c, where a single red field line is shown looping around on itself in a way that cannot be untangled without cutting the ends.  In Figure \ref{fig:fieldlines4}d, this field line is shown wrapped around a single yellow field line that wraps itself into an extremely compact bundle.
The next stage of the evolution, seen at $t=1565\;\mathrm{s}$ in Figures \ref{fig:bx4}b and \ref{fig:fieldlines4}e shows many of the same features as the $\zeta=1.5$ and $\zeta=2$ simulations: primary polarities connected to each other via sigmoidal coronal field lines and connected to elongated secondary polarities by short, sheared field lines encircling the two offset legs of the emerging flux ropes. In this case, however, the field lines are much more complicated, since the compactness of the emerging flux rope results in long, wandering field lines in the corona. Nevertheless, even the third stage of the evolution, shown in Figures \ref{fig:bx4}c and \ref{fig:fieldlines4}f, at $t=1895\;\mathrm{s}$, shows two primary polarities and two elongated, secondary polarities (indicated by the blue and yellow arrows, respectively, in Figure \ref{fig:bx4}c). This time, however, there is also another set of two elongated polarities encircling the central set. Numerous polarity inversion lines exist at the photosphere at this stage and persist until the end of the simulation, with strongly sheared horizontal fields between the primary and secondary polarities (Figure \ref{fig:bx4}d,e). The field lines in the corona are extremely tangled, and no coherent structures can be discerned (Figures \ref{fig:fieldlines4}g,h). While some regions on the photosphere remain wholly unchanged by the symmetry transformation Equation \ref{transformation}, some of the dispersed flux in the interior of the active region is not invariant under this transformation, likely due to copious reconnection reorganizing the field.  {\color{black} Additionally, the polarities seen in Figure \ref{fig:bx4}d near $(y,z)=(8.5,0)$ are due to the emergence of secondary buoyant sections along the rising flux rope. These sections are produced by the strong non-linearity of the kink mode in such a highly twisted flux rope, where it produces kinks along the entire axis of the flux rope (these can be seen in Figures \ref{fig:isosurfaces4}g,h).}\par 

\subsection{The Role of Twist in Flux Emergence}\label{sec:role_twist}
For flux ropes with $\zeta\leq1$, the subsurface behavior is determined, to a large extent, by the initial density perturbation. An $\Omega$-loop is formed due to the Gaussian shape of the density perturbation (Eq. \ref{dens_perturb}), and this $\Omega$-loop rises, mostly undisturbed, through the convection zone. As the flux ropes continue to rise, and the helical field lines emerge into the corona, the legs become slanted toward the photosphere, hitting it at an angle. This behavior turns out to have important implications for the signatures observed at the photosphere. When the flux rope reaches the photosphere, it forms two distinct polarities resembling simple, bipolar active regions which gently rotate and separate as the flux rope emerges, with dispersed, weak field appearing between the two polarities. This evolution is typical of many previous simulations \citep{Manchester04,Murray06, Archontis09,Leake13}. \par 
The behavior of flux ropes with $\zeta>1$, however, is determined less by the initial density perturbation, and more by the properties of the kink instability. In the convection zone, these flux ropes undergo a deformation of their central axis, whereby the apex of the $\Omega$-loop rotates to form an $\ell$-shaped loop, and the legs of the rising structure become offset from the plane formed by the initial $\Omega$-loop. Like the weakly twisted flux ropes, the legs of the highly twisted flux ropes become slanted, and hit the photosphere at an angle. A common feature of the photospheric evolution of the $\zeta\ge1$ flux ropes was their apparent four-fold structure, with two strong, primary polarities separated by two elongated, secondary polarities of alternating sign. Sigmoidal field lines connected the two primary polarities, while the secondary polarities were connected to their adjacent primary polarities with short, highly sheared field lines. The sigmoidal field lines connecting the primary polarities were due to the connection between the two off-center legs of the flux rope, while the field lines connecting the secondary to the primary polarities were due to the field lines encircling each individual leg of the flux rope. An obvious question is: why was this behavior not observed in the lowest twist simulation with $\zeta=0.5$, despite the legs of that flux rope being incident on the photosphere at an oblique angle as well? \par
\begin{figure*}
\centering\includegraphics[scale=0.27, trim=0.0cm 0.0cm 0.0cm 0.0cm,clip=true]{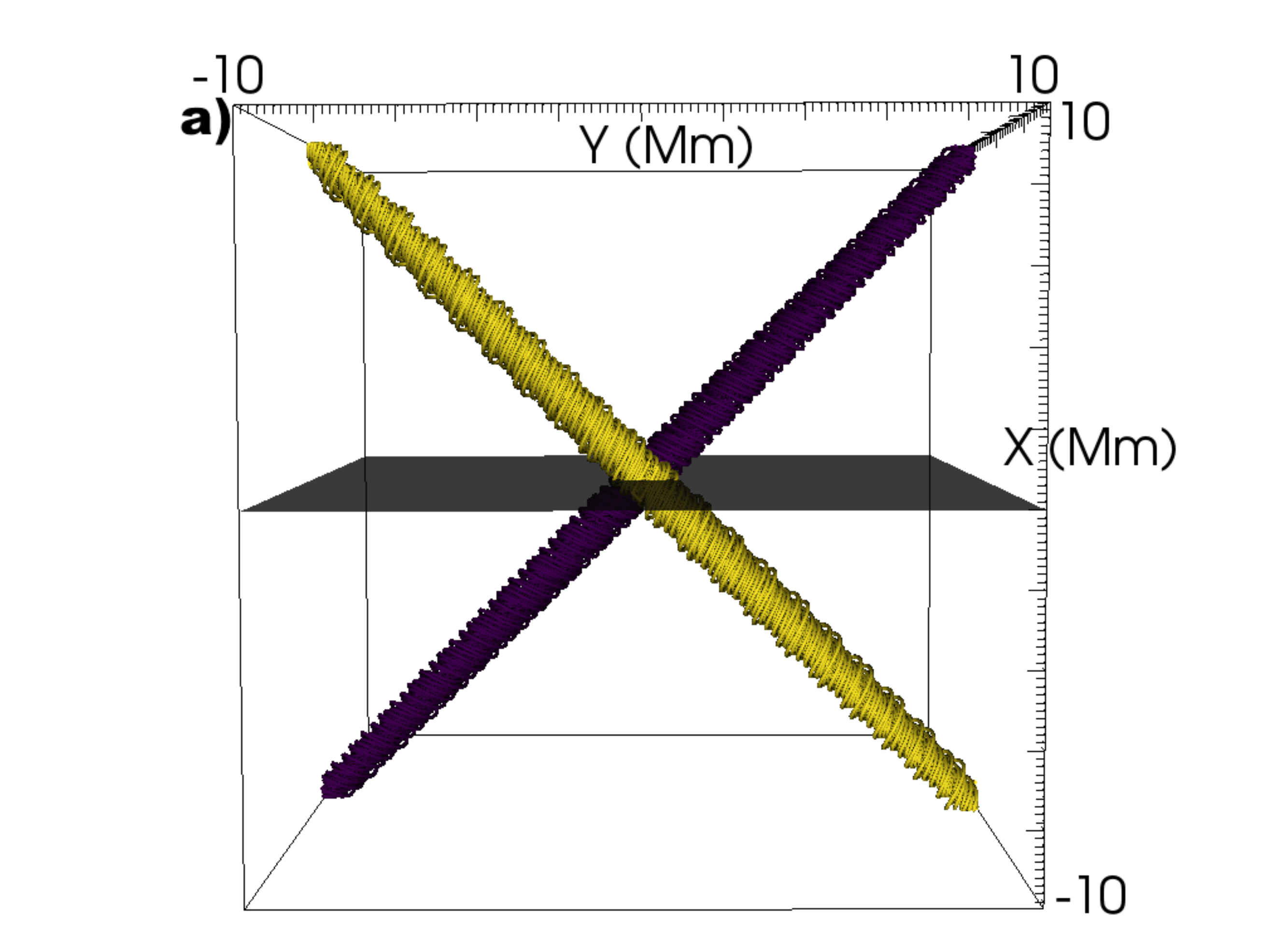}
\centering\includegraphics[scale=0.27, trim=0.0cm 0.0cm 0.0cm 0.0cm,clip=true]{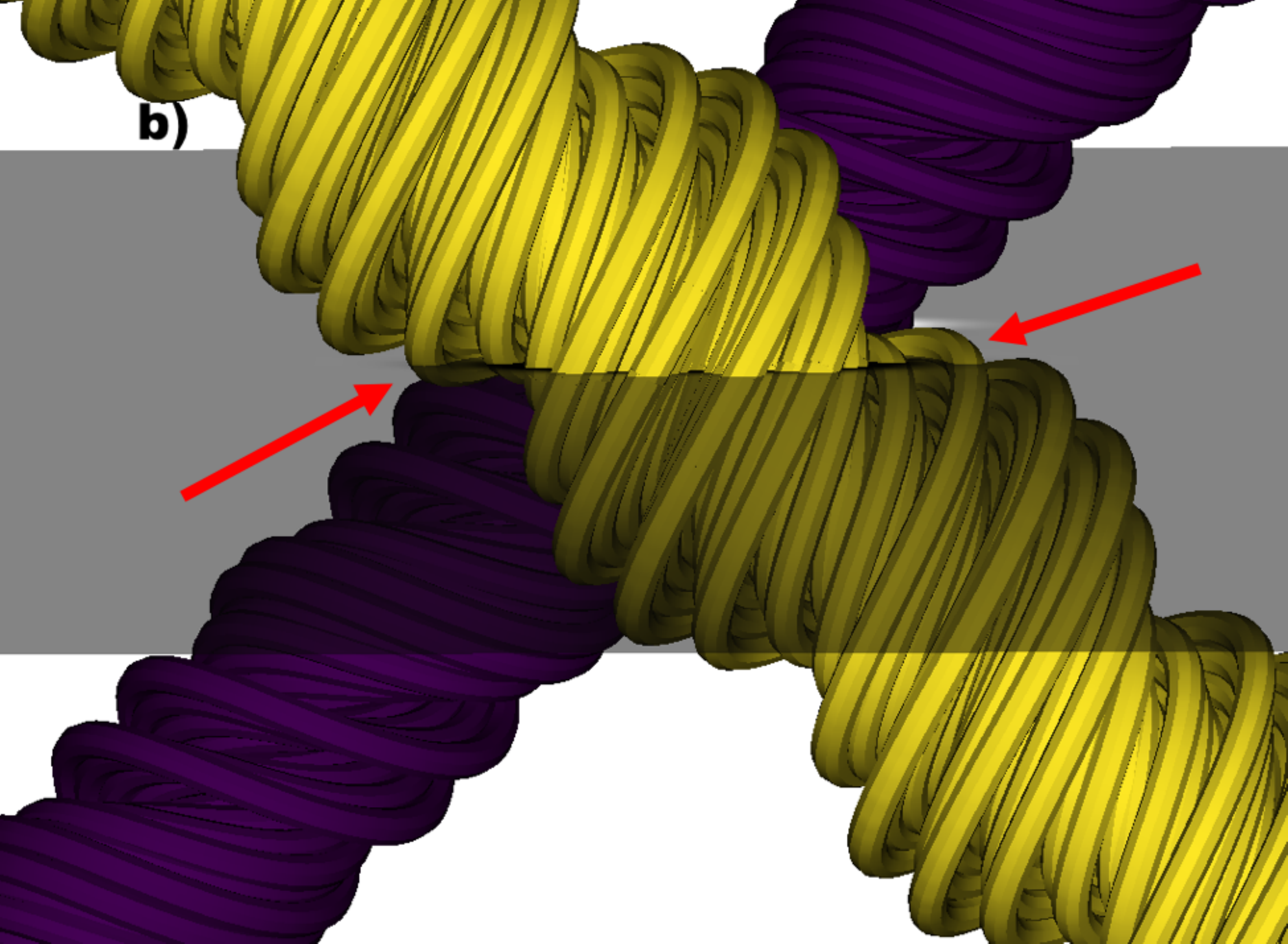}
\centering\includegraphics[scale=0.3, trim=1.0cm 1.7cm 6.0cm 0.7cm,clip=true]{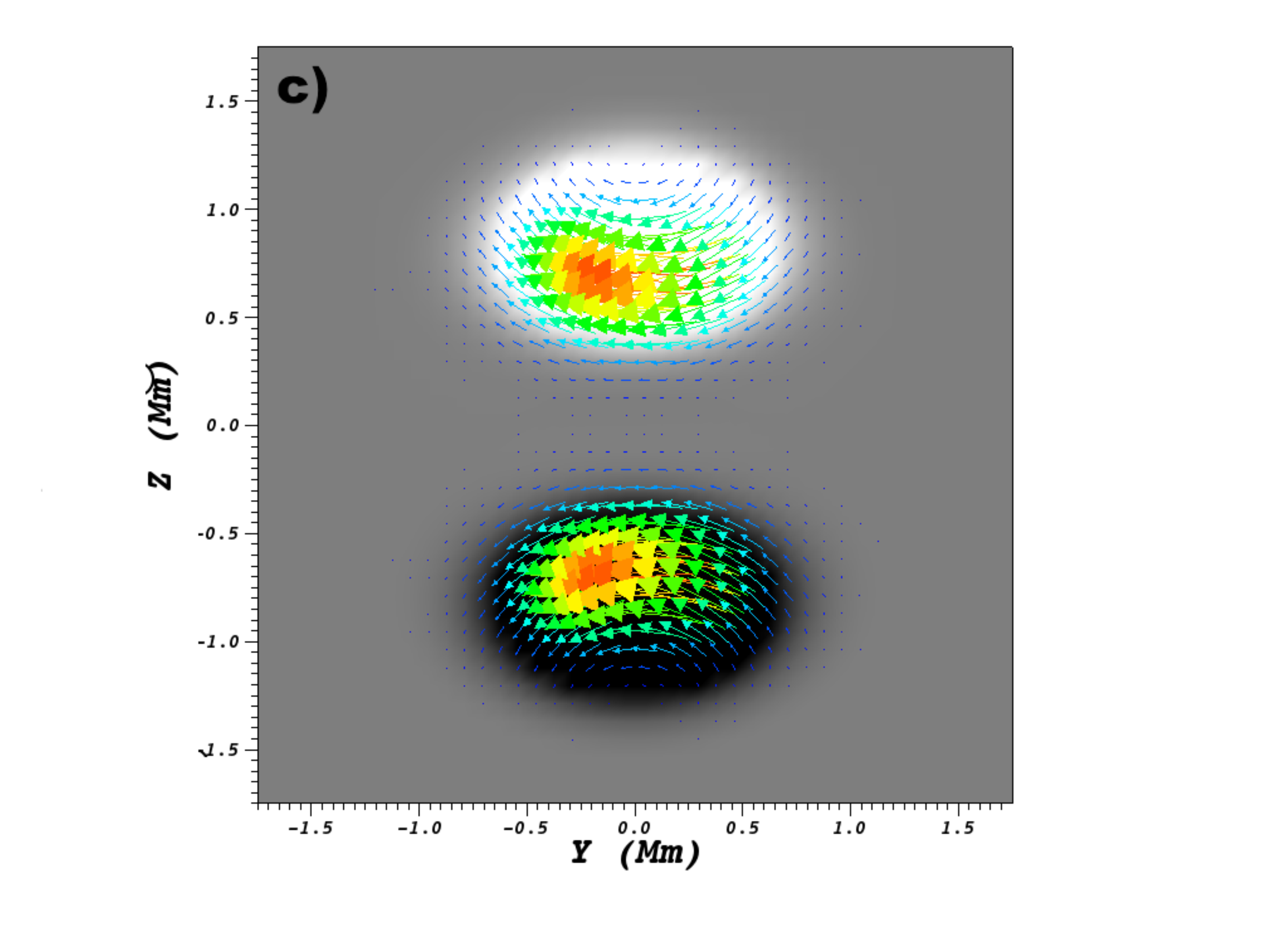}
\centering\includegraphics[scale=0.3, trim=1.0cm 1.7cm 6.0cm 0.7cm,clip=true]{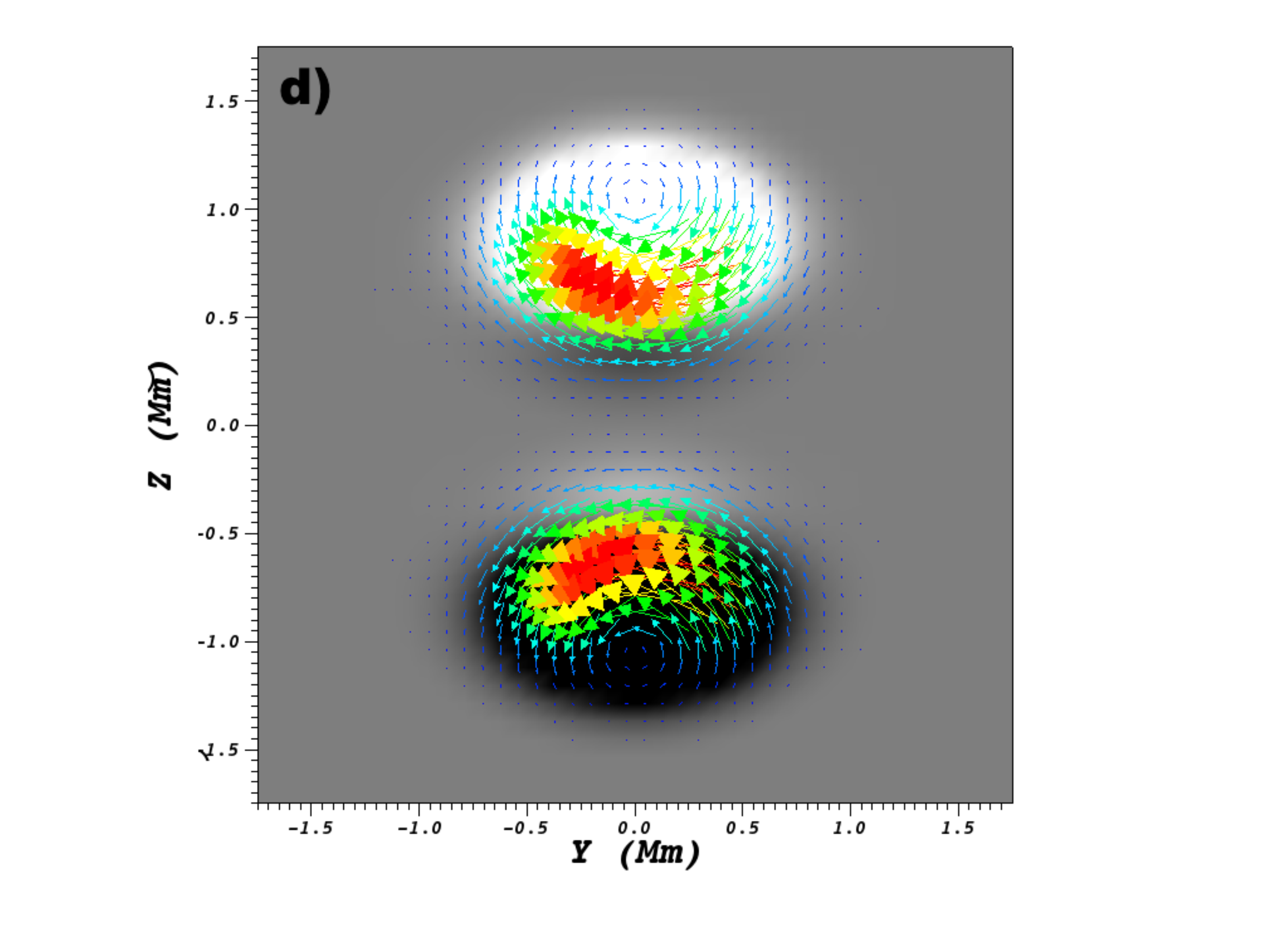}
\centering\includegraphics[scale=0.3, trim=1.0cm 1.7cm 6.0cm 0.7cm,clip=true]{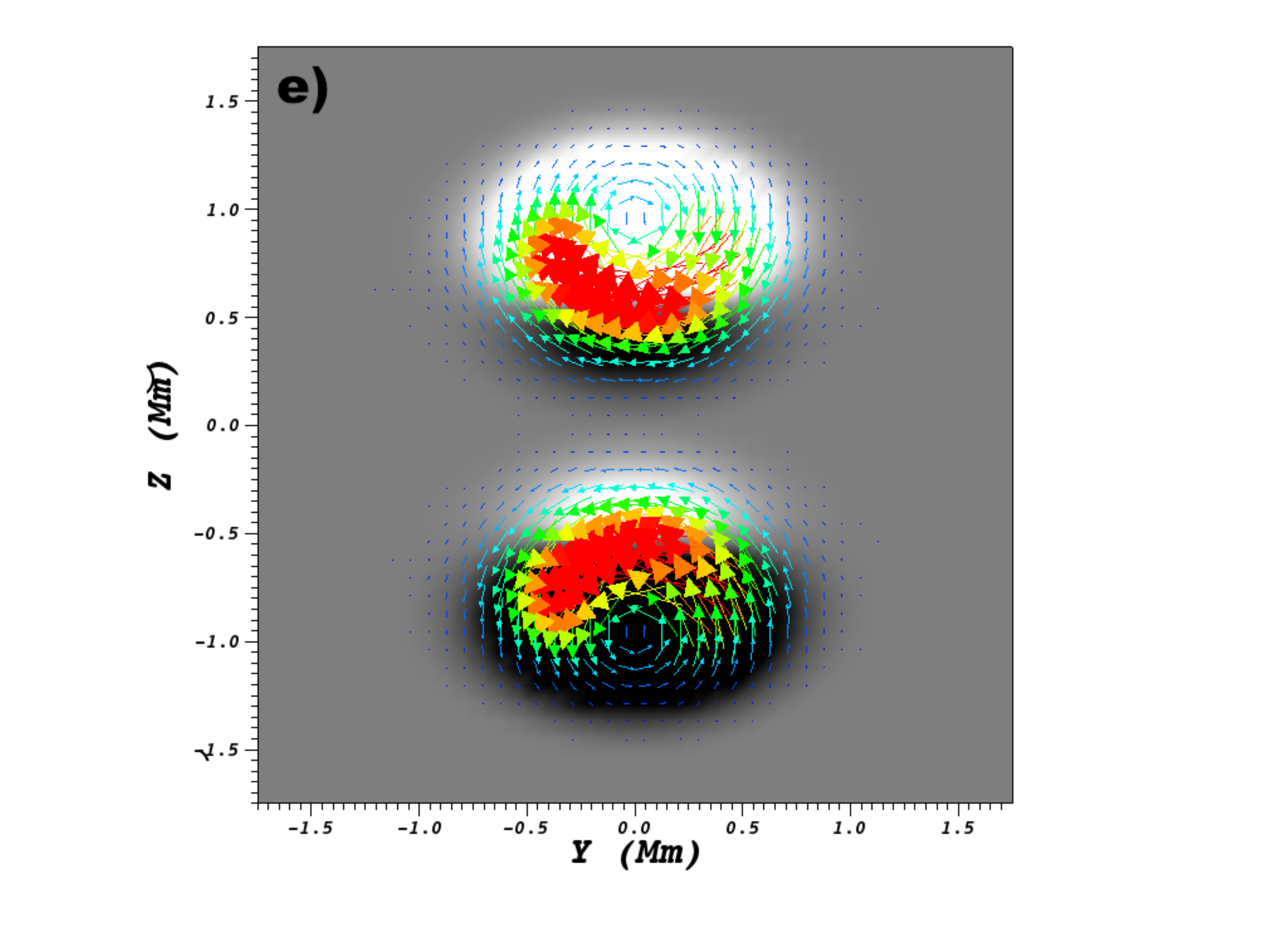}
\centering\includegraphics[scale=0.3, trim=1.0cm 1.7cm 6.0cm 0.7cm,clip=true]{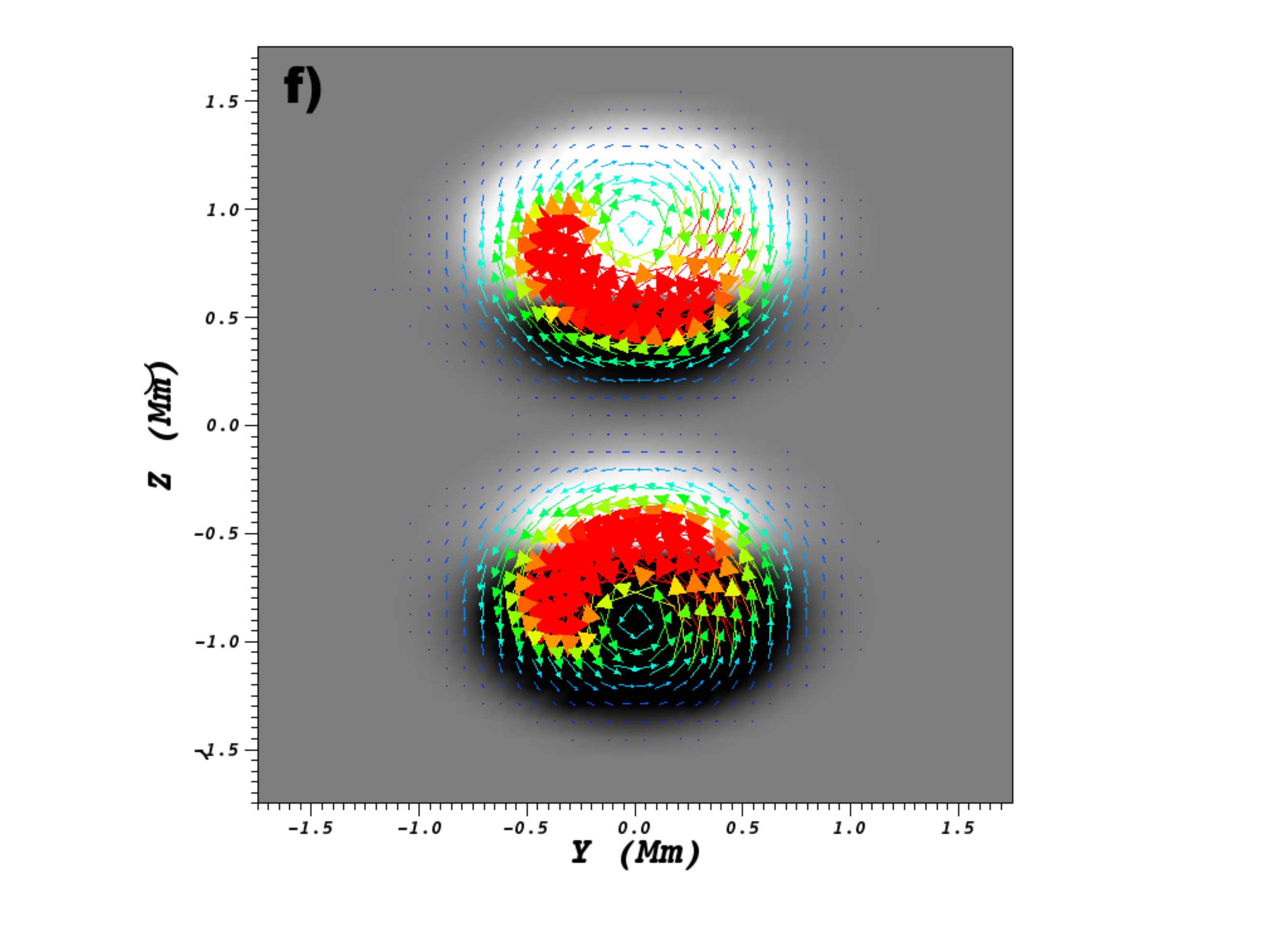}
\centering\includegraphics[scale=0.3, trim=0.0cm 1.7cm 6.0cm 0.7cm,clip=true]{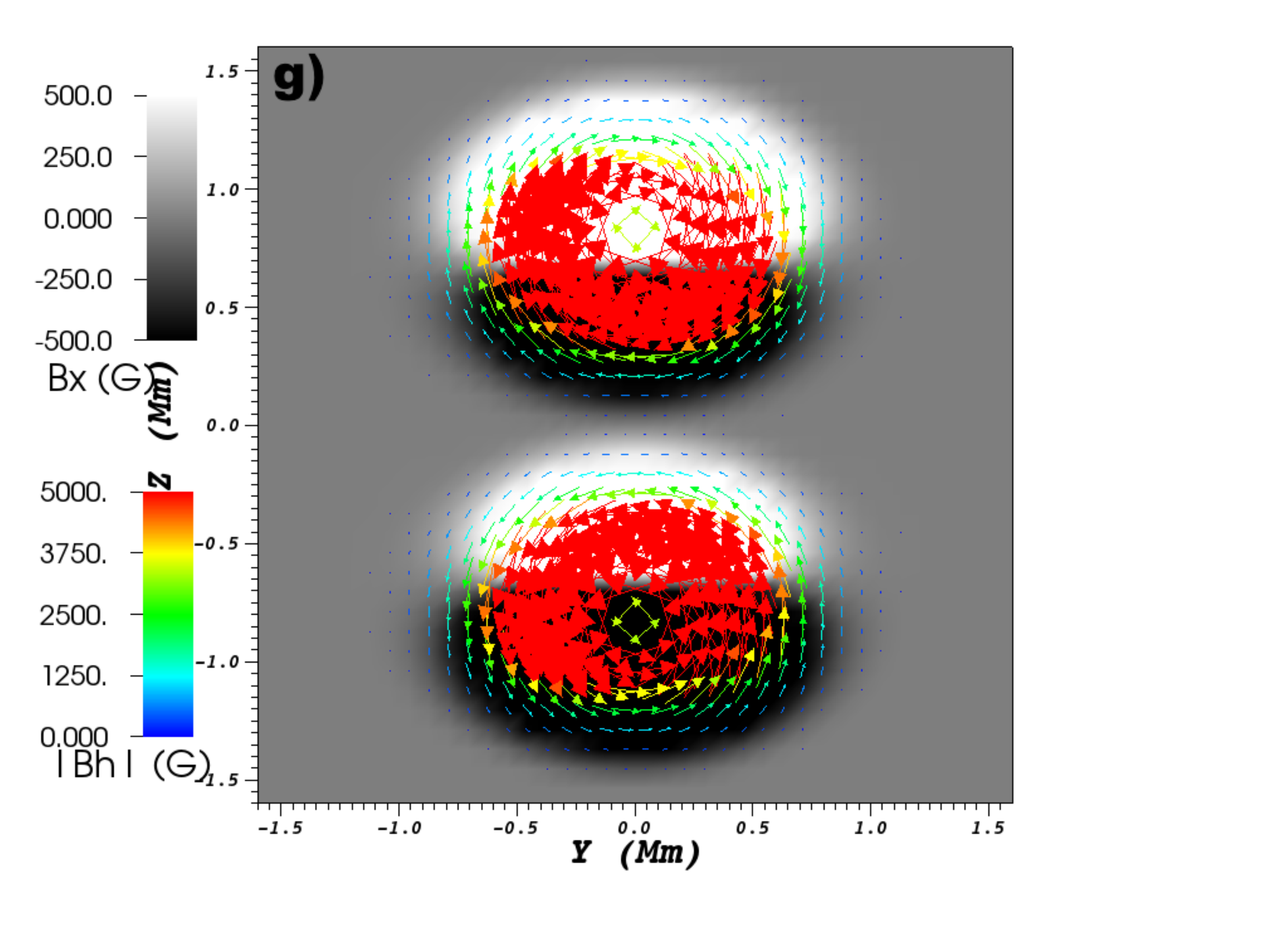}
\caption{a) Two flux ropes, with a peak field strength of $B_0=6.5\;\mathrm{kG}$, oriented at $\pm45^\circ$ to the vertical. b) Zoom in of a), with red arrows denoting example field lines which cross the photosphere twice. $B_x$ (greyscale) and $B_h$ (colored vectors) due to two flux ropes oriented at a $45^\circ$ angle to the vertical with c) $\zeta=0.5$, d) $\zeta=1$, e) $\zeta=1.5$, f) $\zeta=2$, g) $\zeta=4$. The color table applies to panels c-g.}
\label{fig:twoFRs}
\end{figure*}

 
A simple explanation is that these signatures come from the photosphere cutting through a twisted flux rope at an oblique angle with respect to its axis. Figure \ref{fig:twoFRs} illustrates the concept. Two twisted flux ropes, with magnetic field structures defined by Equation \ref{fluxrope} cut through the photospheric plane as shown in Figure \ref{fig:twoFRs}, panels a and b. The axial magnetic field of the two flux ropes is specified to be oppositely directed, representing the two legs of an emerging $\Omega$-loop flux rope, and the flux ropes are incident on the photosphere at an oblique angle, taken to be $45^\circ$ in this example. The direction of field lines through the photospheric plane depends on both their twist and angle of incidence, relative to the normal to the photosphere. For sufficiently strongly twisted field lines, or for large angles of incidence, field lines will go through the photospheric plane not once but twice (or even more), in opposite directions on either side of the flux rope axis, producing oppositely signed vertical magnetic field. This can be seen in the zoom-in in panel b, where the red arrows point to example field lines that go through the photospheric plane twice. Figures \ref{fig:twoFRs}c-g show the photospheric vertical magnetic field due to the two flux ropes shown in Figures \ref{fig:twoFRs}a and b for different values of the twist. As the twist of each flux rope increases, a second polarity appears in the vertical magnetic field distribution of each flux rope. 
\par 
Quantitatively, this reversal in the photospheric vertical field, $B_x$, is due to the fact that both the axial and azimuthal components of the flux rope have a vertical component when projected onto the photosphere (for a non-zero angle of incidence of the flux rope axis). On one side of the axis of each flux rope, the projection of the azimuthal field into the vertical direction will be in the same direction as the projection of the axial field into the vertical direction and will add constructively to increase $B_x$. On the other side of the axis of each flux rope, the azimuthal field projection will be in the opposite direction of the axial field projection and will add destructively, reducing $B_x$. If the azimuthal field projection is strong enough or the angle of incidence is sufficiently oblique, the azimuthal field projection will dominate the axial field projection on one side of each flux rope's axis, and will produce a photospheric signature with different signs of $B_x$ on either side of each flux rope's axis (assuming, of course, that the flux rope radius extends far enough). As a result, the photospheric distribution of $B_x$ will consist of two oppositely signed polarities for each flux rope leg, for a total of two pairs of polarities. This is precisely what is observed in the high twist simulations. Each flux rope leg, in addition to having a strong azimuthal field, rises through the photosphere at an oblique angle, resulting in a dominant, primary polarity, and a smaller, secondary polarity. In the lowest twist flux rope, the twist and angle of incidence together were insufficient to create the four-fold structure observed for the high twist simulations. \par 
A lower limit for the magnitude of the flux rope twist required for seeing the four-fold structure can be calculated 
mathematically by projecting a flux rope oriented in some arbitrary direction onto the photospheric plane. This is done as follows. A constant twist flux rope, such as the one given in Equation \ref{fluxrope}, oriented along some arbitrary axis $\hat{\xi}$ has a magnetic field given by
\beg{Barb}
\vecB(\rho) =
\begin{cases}
B_\xi(\rho)\hat{\xi} + B_\psi(\rho) \hat{\psi} & \text{if } \rho\leq R\\
0 & \text{if } \rho>R
\end{cases}
\done 
for the cutoff radius $R$ with axial and azimuthal components 
\beg{componentsearb}
\begin{split}
B_\xi(\rho) &= B_0e^{-\rho^2/a^2}, \\
B_\psi(\rho) &= \zeta \frac{\rho}{a}B_\xi(\rho),
\end{split}
\done
where $\rho$ is the radial coordinate, $\xi$ is the direction along the flux rope and $\psi$ is the polar coordinate around the flux rope perimeter. If the flux rope intersects the photosphere at an angle $\theta$ with respect to the vertical $x$-axis, i.e.,
\beg{intersect}
\begin{split}
\unit{x}\cdot\hat{\xi} &= \cos{\theta},\\
\unit{x}\cdot\hat{\psi} &= \pm \sin{\theta},
\end{split}
\done
the axial and azimuthal fields of the flux rope produce the normal photospheric field $B_x(y,z)$ given by 
\beg{projection}
\begin{split}
B_x(y,z) &= \vecB(\rho)\cdot\unit{x} \\
         &= B_\xi(\rho)\cos{\theta} \pm B_\psi(\rho)\sin{\theta} \\
         & = B_\xi(\rho)\Big(\cos{\theta}\pm\zeta\frac{\rho\sin{\theta}}{a}\Big).
\end{split}
\done
Here the $\pm$ sign represents the different direction at which the azimuthal field $B_\psi$ intersects the $x$-axis on either side of the flux rope axis. On one side of the flux rope axis, $\unit{x}\cdot\hat{\psi}>0$, the projection of the azimuthal field $B_\psi$ adds to the projection of the axial field $B_\xi$, so $B_x$ will be positive/negative if $B_\xi$ is positive/negative. On the other side of the flux rope axis, $\unit{x}\cdot\hat{\psi}<0$, the azimuthal field $B_\psi$ projects in the opposite $x$-direction than the axial field $B_\xi$, and in this case, the projection of the magnetic field will change sign, from positive/negative to negative/positive, if
\beg{rhochangesign}
\rho > \frac{a}{\zeta} \cot{|\theta|}
\done 
The smallest twist for which $B_x$ will change sign at the photosphere at $\rho=a$ is
\beg{criterion}
\zeta = \cot{|\theta|}.
\done 
For the flux ropes shown in Figure \ref{fig:twoFRs}, $\theta=\pm45^\circ$, which implies that the normal component of the photospheric field will change sign at $\rho=a$ for an inclined flux rope with $\zeta=1$. Note, however, that the flux rope extends out beyond $\rho=a$, up to $\rho=R$, with $R=6a$. Thus a reversal of $B_x$ will still occur outside $\rho=a$, as can be seen in Figure \ref{fig:twoFRs}d. Even the lowest twist case shown in Figure \ref{fig:twoFRs}c displays a weak reversal at $\rho=2a$, though it is not visible with the color scale in the Figure.
In these cases, the in-plane horizontal vector field is primarily parallel to the polarity inversion line separating the secondary polarity from the adjacent primary polarity, exactly as observed in the high twist simulations presented above. Interestingly, it is aligned parallel to the polarity inversion line, but it is mostly concentrated off the polarity inversion line, except for $\zeta=2$. As $\zeta$ increases, the azimuthal field increasingly overwhelms the axial field on one side of the flux rope axis, creating the reversals in $B_x$ evident in Figure \ref{fig:twoFRs}d-f. \par 
\par
Thus, the primary and secondary polarities are a direct consequence of flux ropes with large values of $\zeta$ being incident on the photosphere at an oblique angle. The mathematical arguments presented above explain why these features of the photospheric field do not appear for the lowest twist flux rope and why the photospheric signatures in the higher twist simulations display at least four, rather than two, strong polarities; two for each leg of the flux rope. \par 
Of course, the above arguments do not take into account distortions of the flux ropes that occur during their rise through the convection zone. As the flux ropes rise through the convection zone and untwist themselves in the corona, their profile deviates from the simple form given in Equation \ref{Barb}, resulting in photospheric features that are more complicated than the simple picture presented in Figure \ref{fig:twoFRs}, but the signatures observed at the photosphere are well represented by simply having two oppositely directed flux ropes oriented at angle to the photosphere.\par 
\par 

\subsection{Simulation Comparison}\label{sec:compare}
\begin{figure*}
\centering\includegraphics[scale=0.45, trim=0.0cm 0.0cm 0.0cm 0.0cm,clip=true]{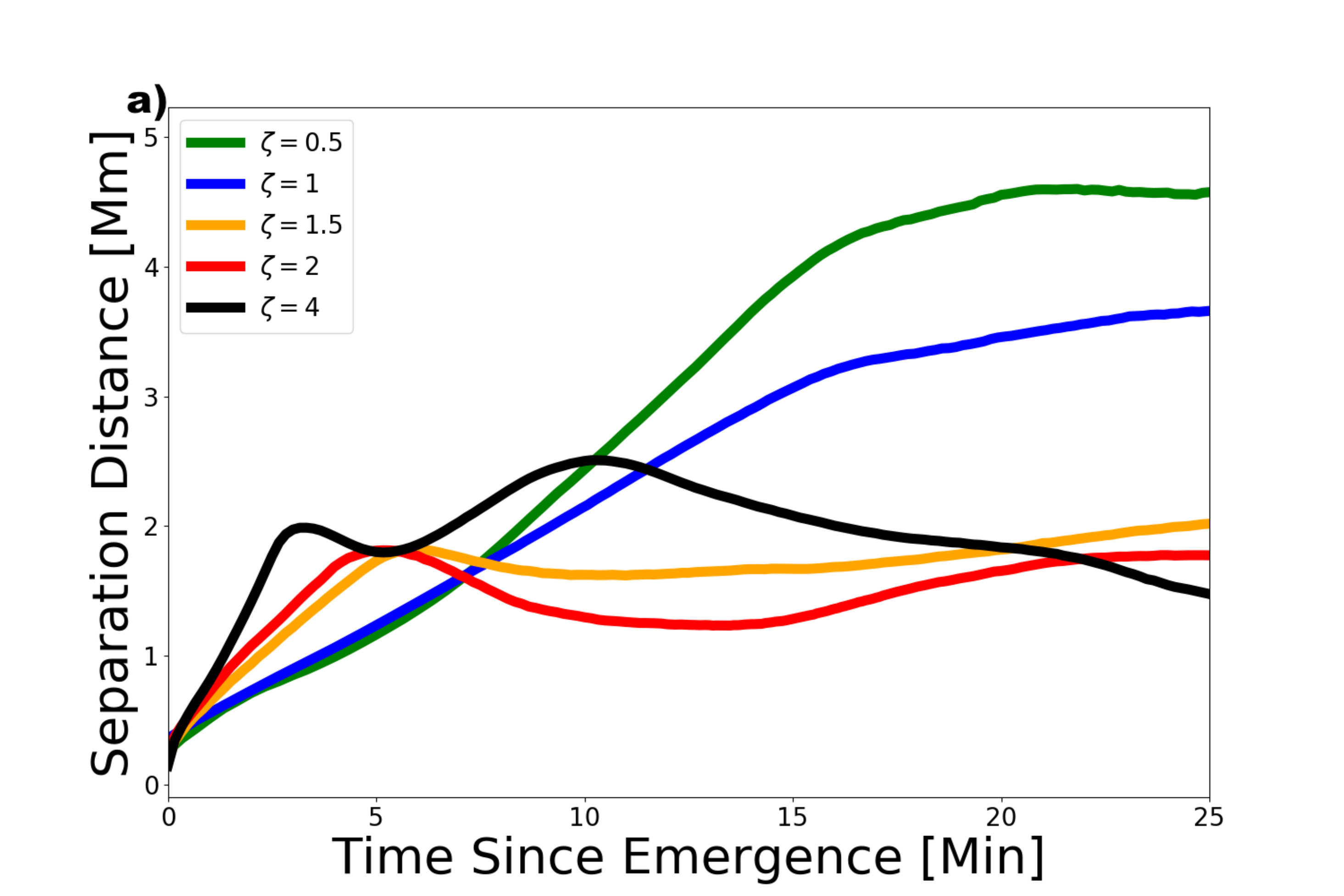}
\centering\includegraphics[scale=0.45, trim=0.0cm 0.0cm 0.0cm 0.0cm,clip=true]{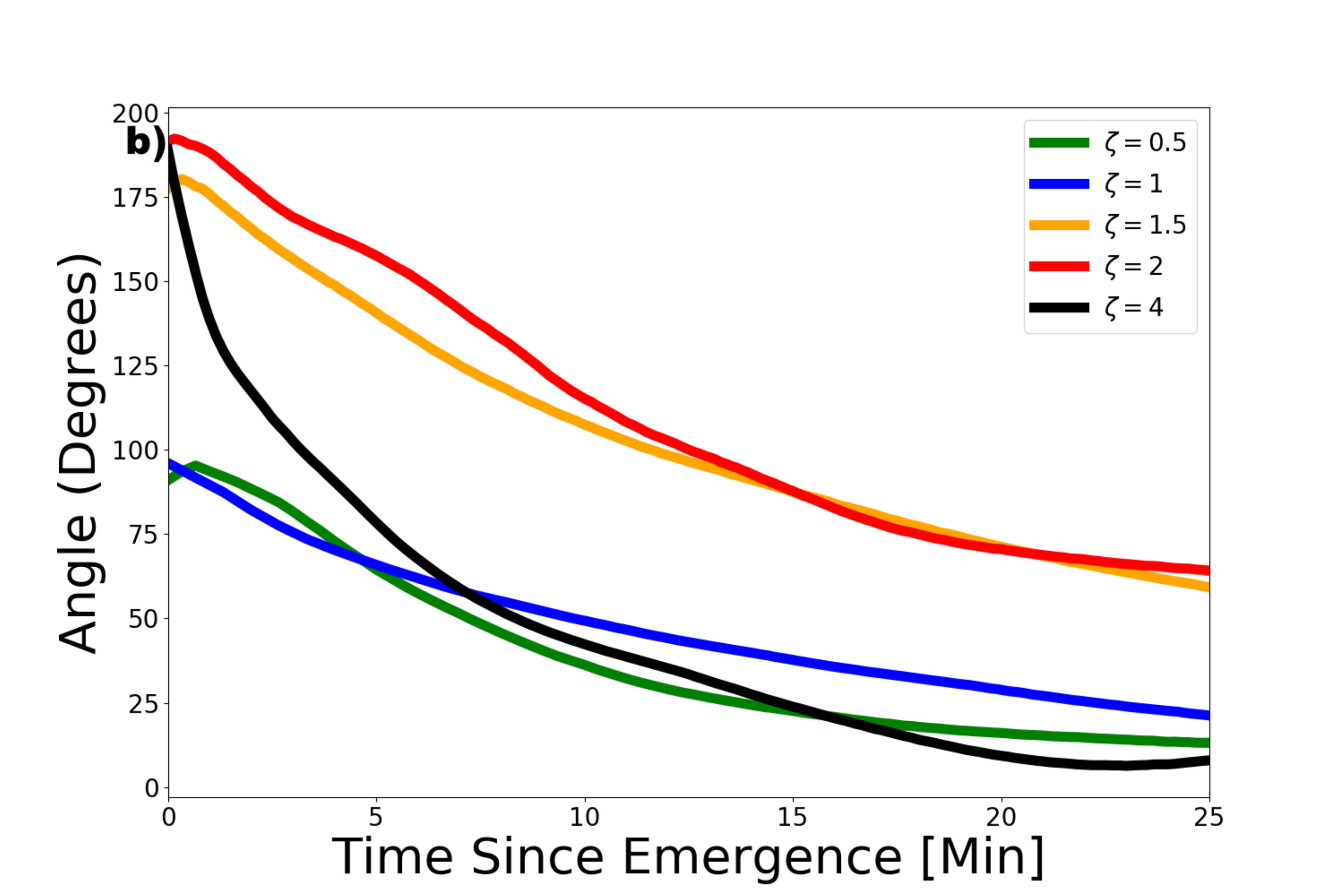}
\caption{Top: The distance between the flux weighted centers of mass, $|\textbf{L}_{com}|$. Bottom: The angle between the line connecting the flux weighted centers of mass and the $y$-axis, $\theta_{com}$ vs.\ time for each simulation. $t=0$ is defined as the instant a photospheric pixel first exceeds $B_x = 30\;\mathrm{G}$.}
\label{fig:seprot}
\end{figure*}

\begin{figure*}
\centering\includegraphics[scale=0.45, trim=0.0cm 0.0cm 0.0cm 0.0cm,clip=true]{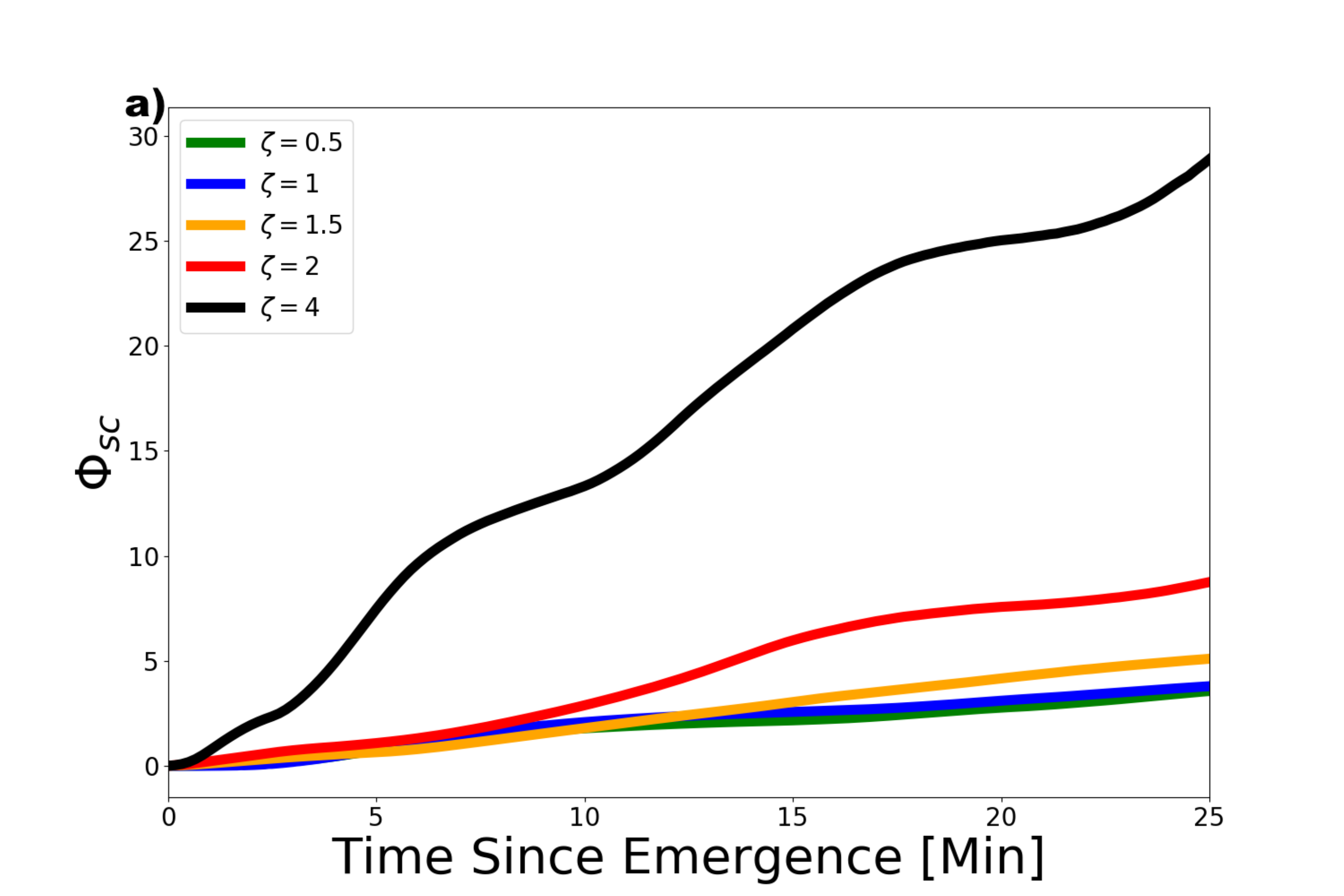}
\centering\includegraphics[scale=0.45, trim=0.0cm 0.0cm 0.0cm 0.0cm,clip=true]{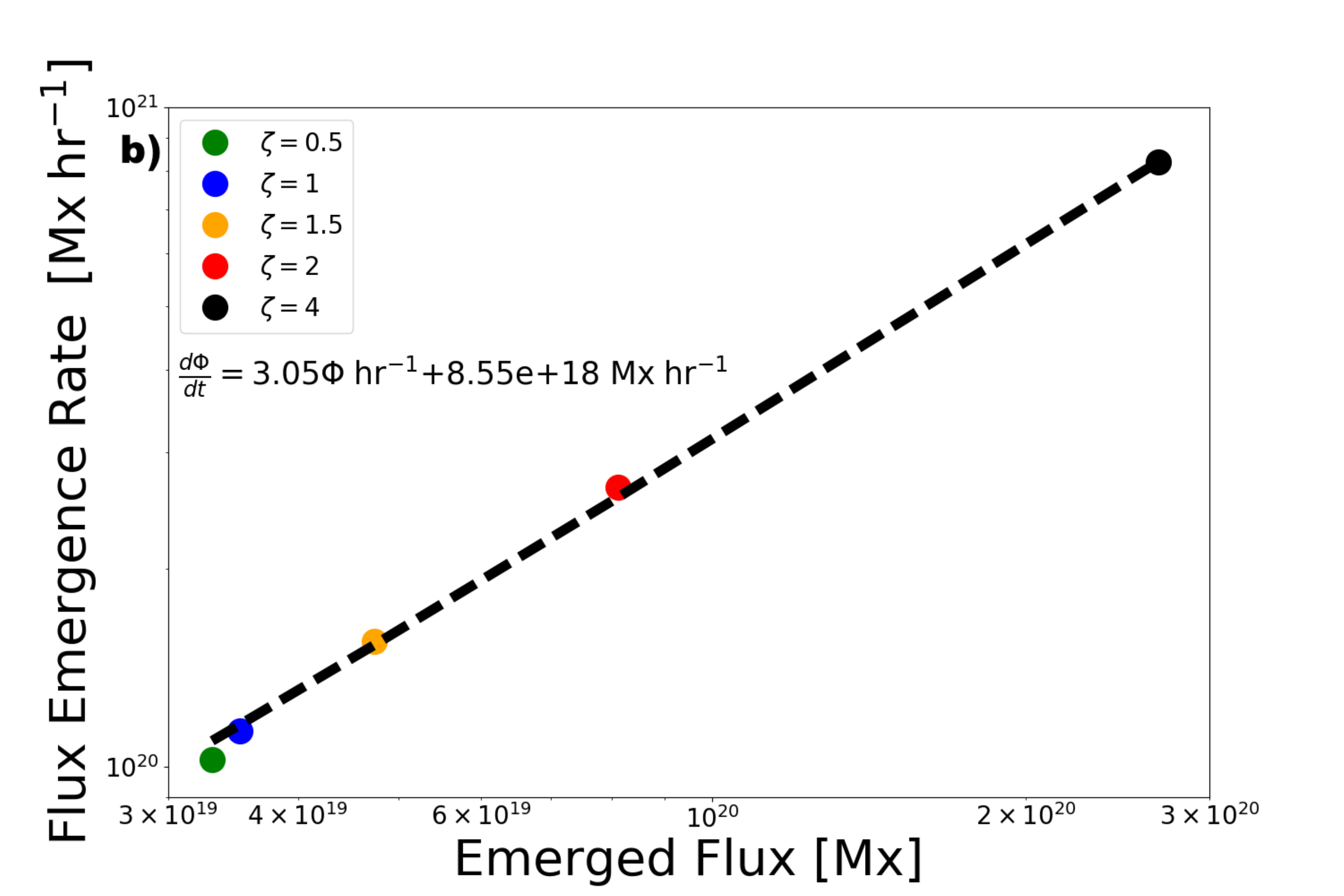}
\caption{Top: Unsigned scaled photospheric flux $\Phi_{sc}$ vs. time for each simulation. Bottom: Flux emergence rate vs.\ emerged flux, plotted following \citet{Norton17}}
\label{fig:flux}
\end{figure*}

\begin{figure*}
\centering\includegraphics[scale=0.35, trim=0.0cm 0.0cm 0.0cm 0.0cm,clip=true]{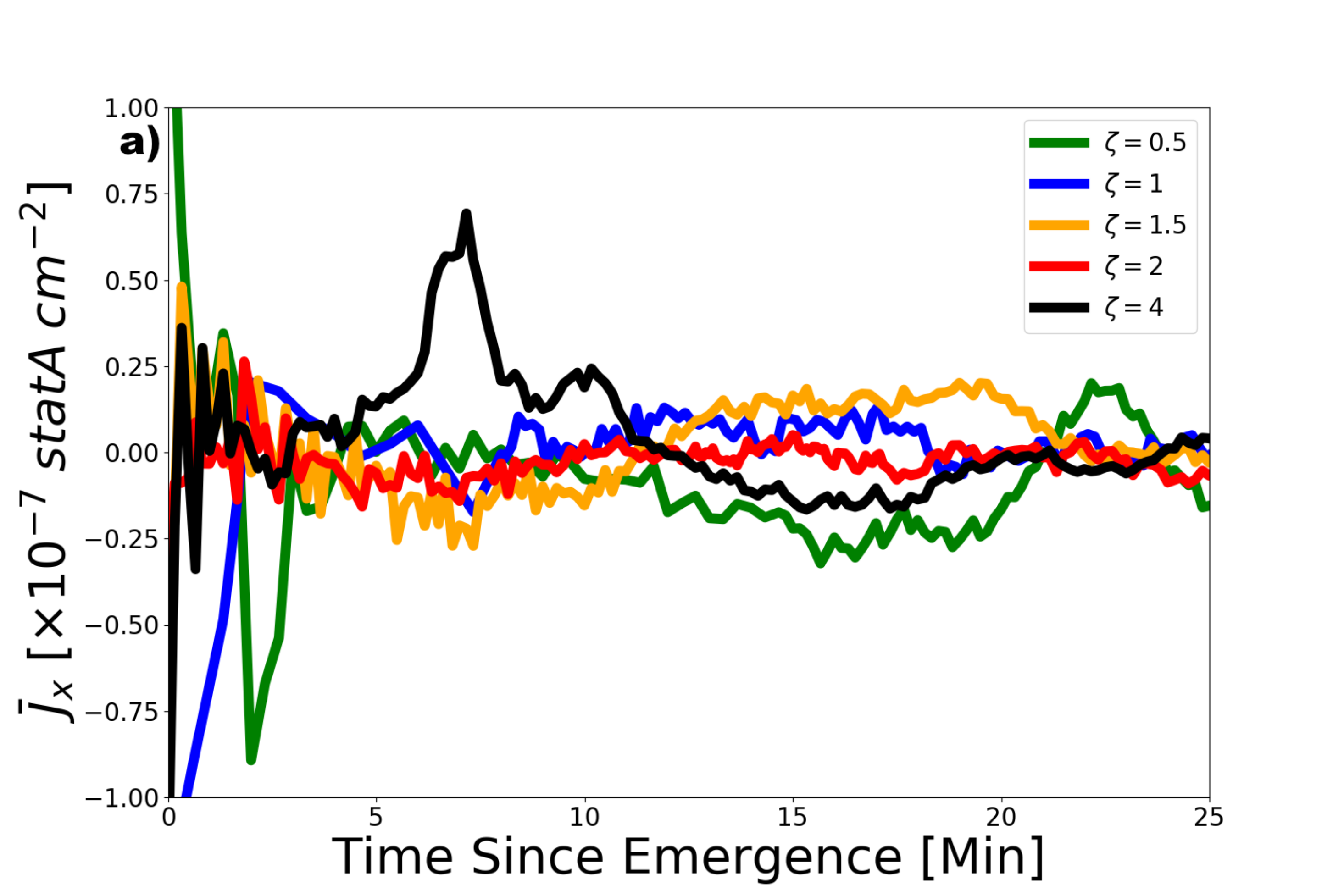}
\centering\includegraphics[scale=0.35, trim=0.0cm 0.0cm 0.0cm 0.0cm,clip=true]{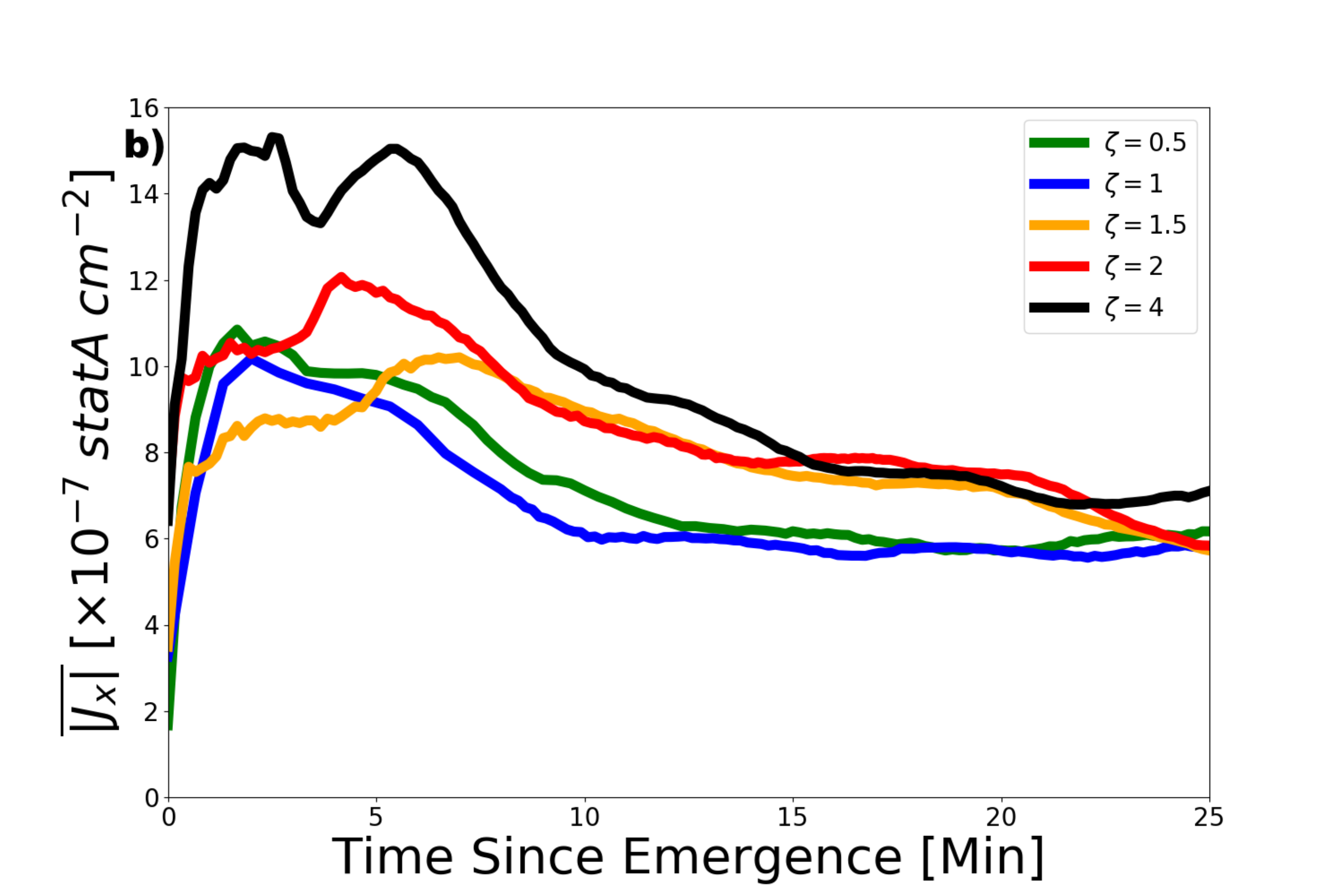}
\centering\includegraphics[scale=0.35, trim=0.0cm 0.0cm 0.0cm 0.0cm,clip=true]{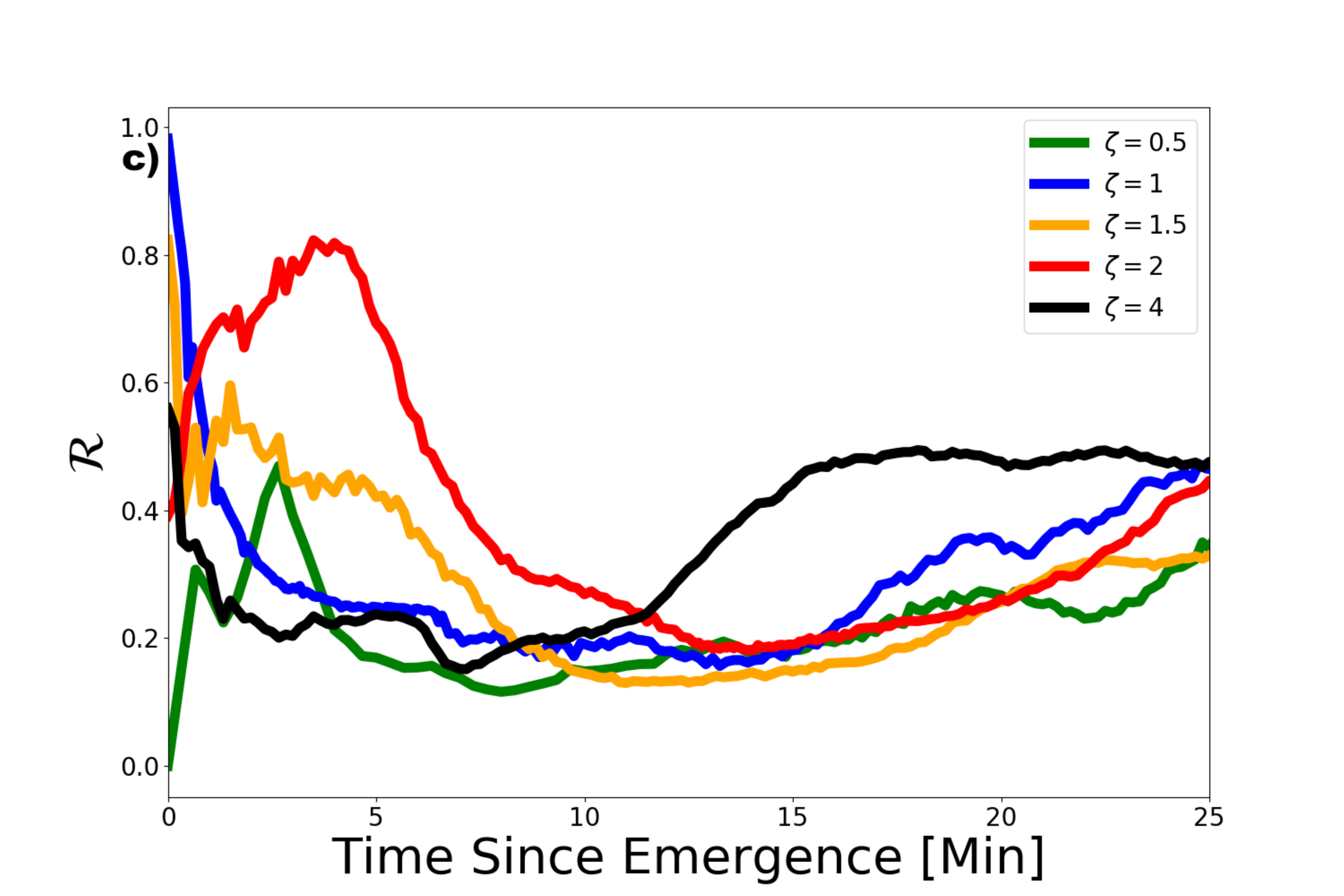}
\caption{Average signed vertical current density $\bar{J}_x$ (top), average unsigned current density $\overline{|J_x|}$ (middle) and current neutralization ratio $\mathcal{R}$ (bottom) as functions of time for each simulation.}
\label{fig:currents}
\end{figure*}

\begin{figure*}
\centering\includegraphics[scale=0.3, trim=0.0cm 0.0cm 0.0cm 0.0cm,clip=true]{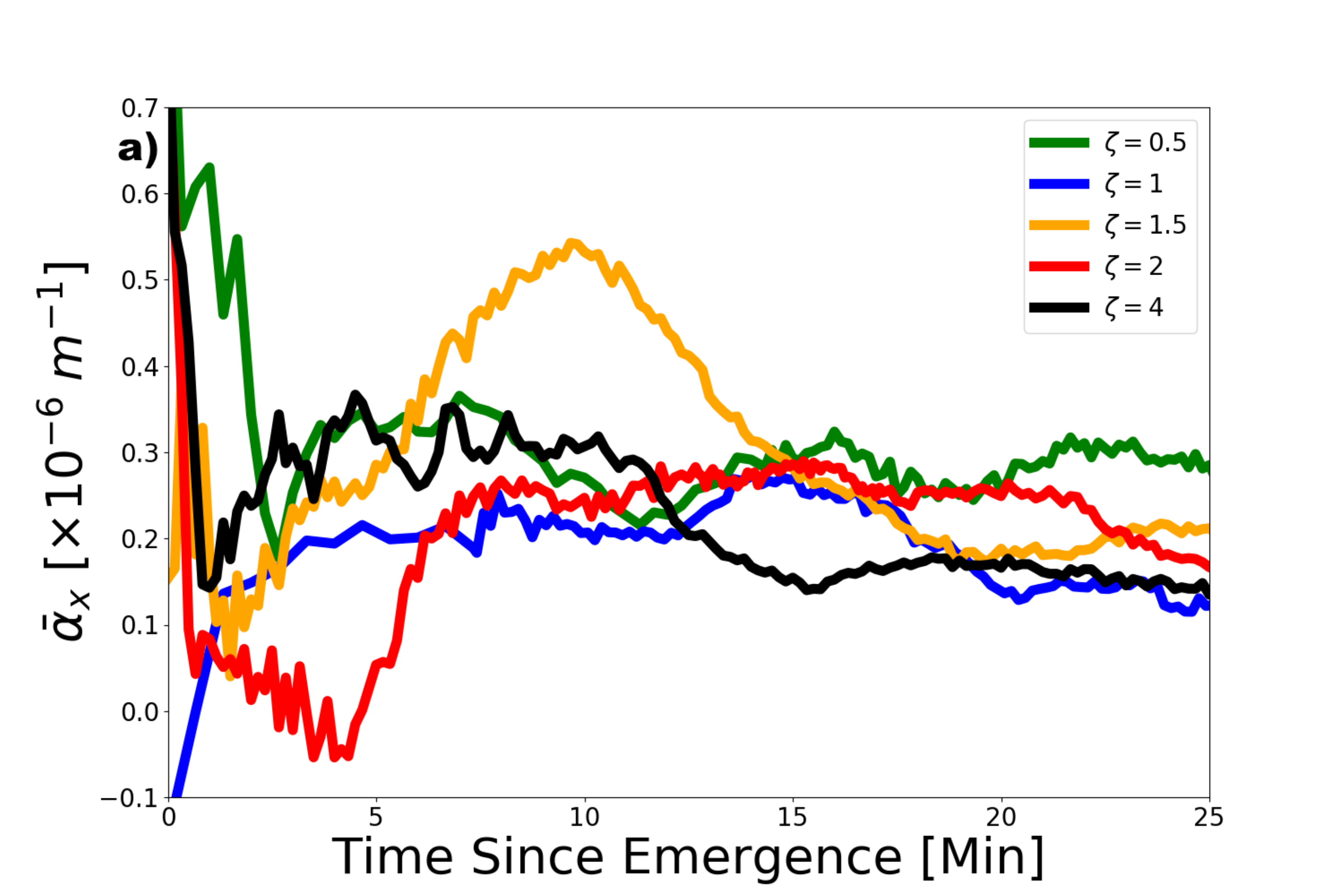}
\centering\includegraphics[scale=0.3, trim=0.0cm 0.0cm 0.0cm 0.0cm,clip=true]{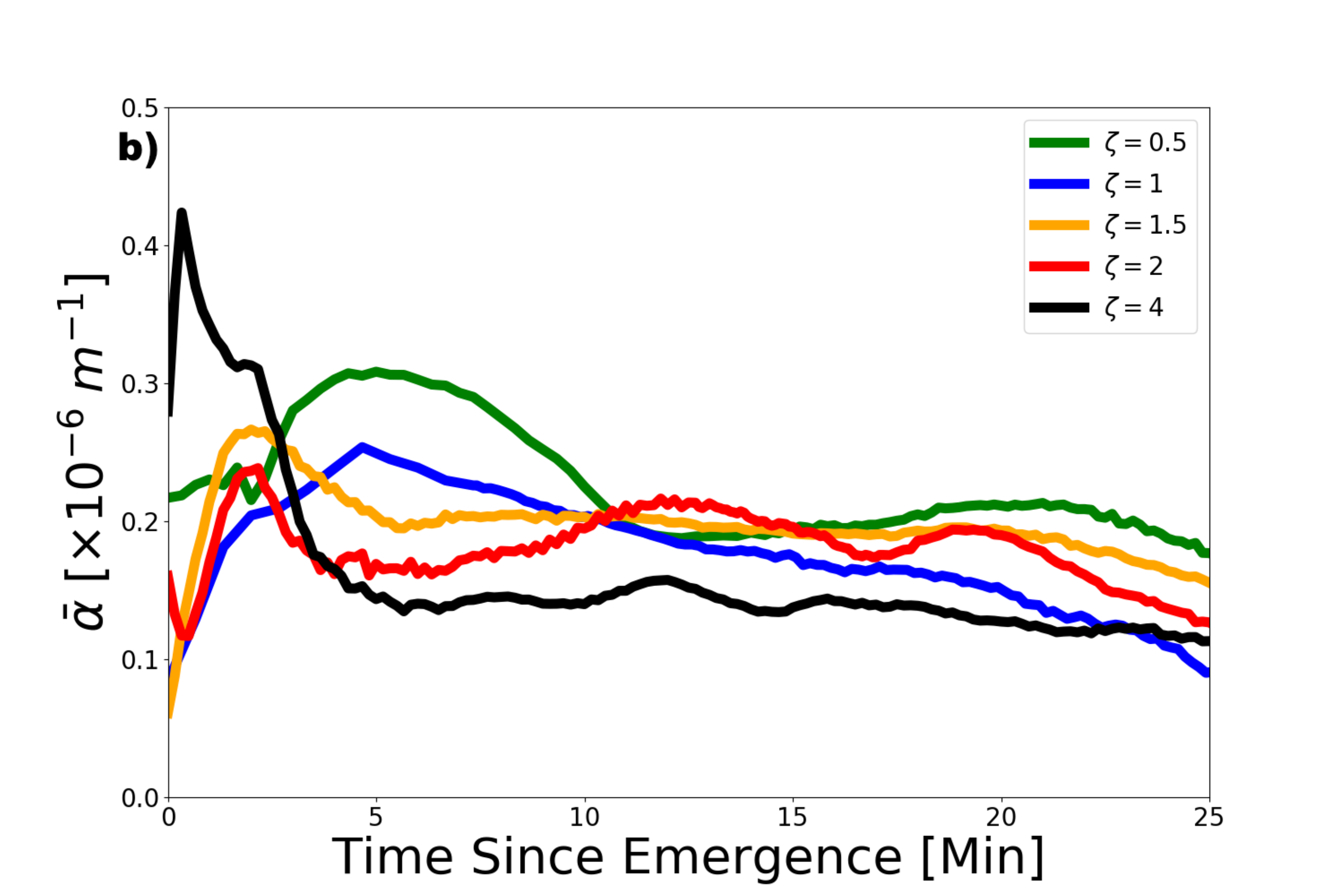}
\centering\includegraphics[scale=0.3, trim=0.0cm 0.0cm 0.0cm 0.0cm,clip=true]{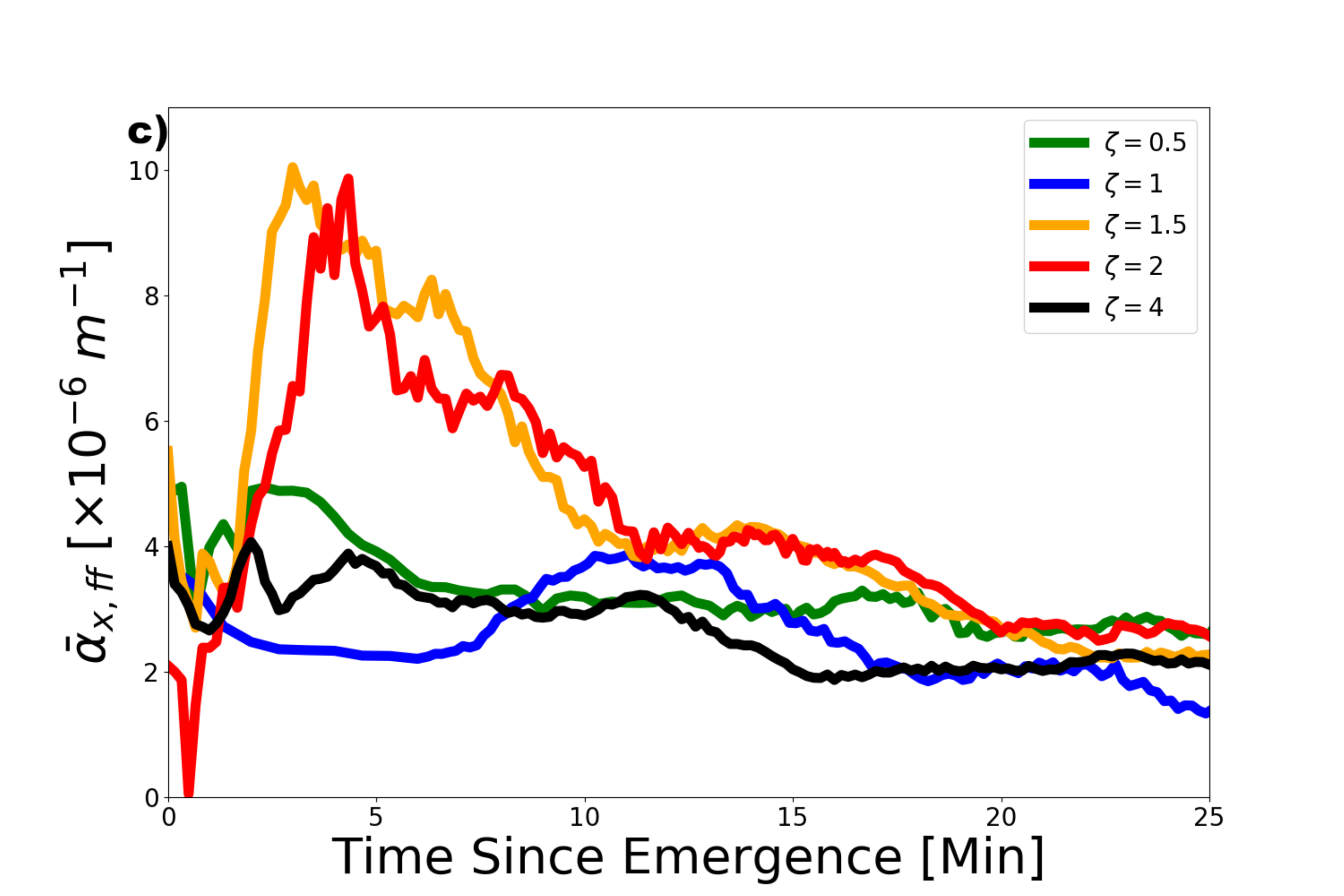}
\centering\includegraphics[scale=0.3, trim=0.0cm 0.0cm 0.0cm 0.0cm,clip=true]{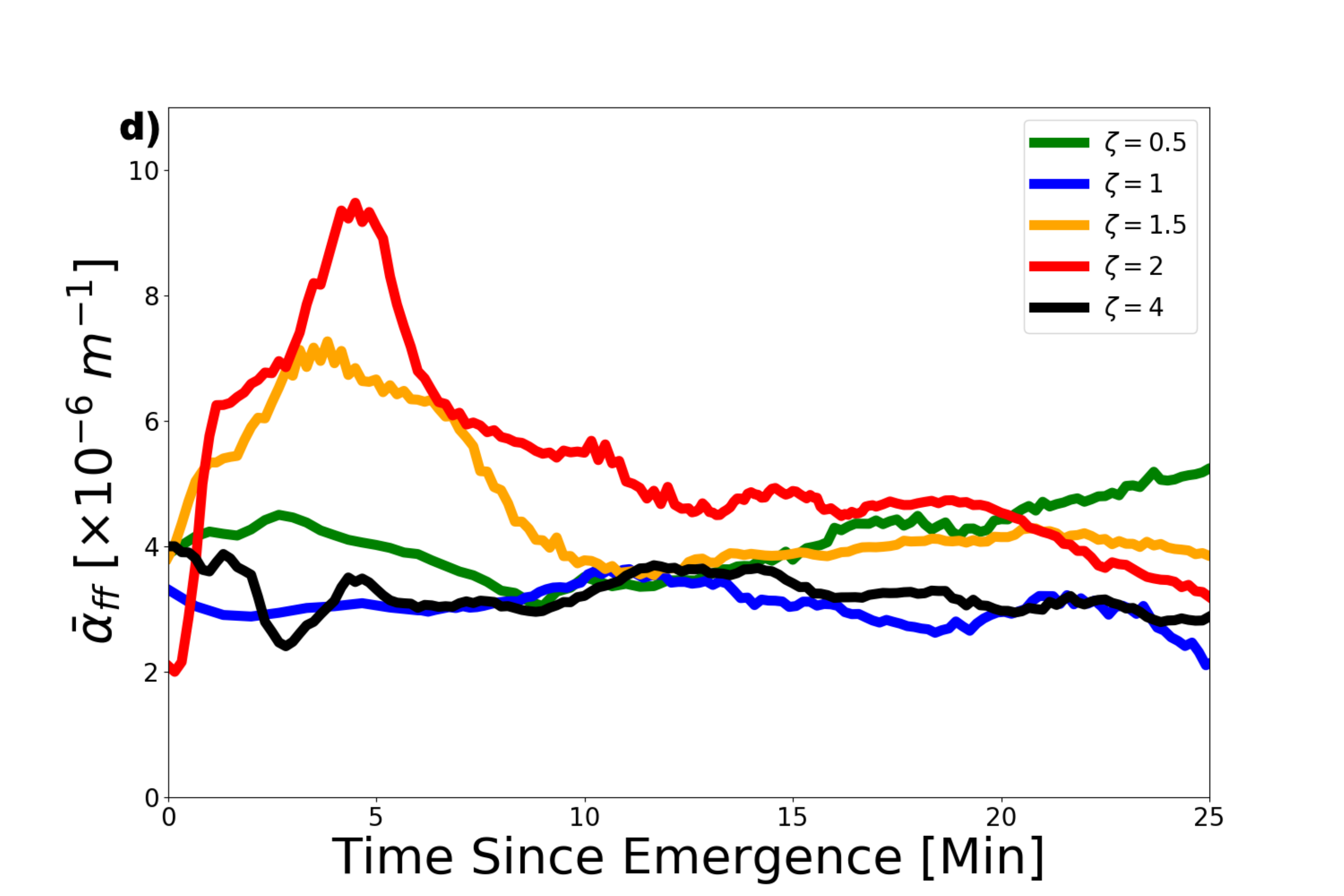}
\caption{a) $\bar{\alpha_x}$ b) $\bar{\alpha}$ c) $\bar{\alpha}_{x,ff}$, d) $\bar{\alpha}_{ff}$, as functions of time for each simulation.}
\label{fig:alpha}
\end{figure*}

Figures \ref{fig:seprot}-\ref{fig:alpha} show the evolution of various photospheric parameters for each of the simulations as a function of time since emergence, defined as the time at which a single photospheric pixel first exceeds {\color{black} $B_x^{thresh}=30\;\mathrm{G}$.  For each of the quantities described below, we integrate over the entire photospheric plane, excluding pixels where $|B_x|<B_x^{thresh}$. Hence, even regions far from the primary polarities, such as the secondary polarities in Figure \ref{fig:bx4}d located at $(y,z)=(8.5,0)$, are included in the calculation.}\par 
We write the unsigned flux at the photosphere as
\beg{totalflux}
\Phi_{ph} = \int_{S_{ph}}{dy\;dz |B_x(y,z)|},
\done
and plot $\Phi_{ph}$ scaled by the initial axial flux in the flux rope
\beg{flux}
\Phi_{sc} = \frac{\Phi_{ph}}{\Phi_0}.
\done
We calculate the average vertical current density, signed and unsigned, taken from the $x$-component of Equation \ref{defJ}, $\bar{J}_x$. We calculate the average vertical twist parameter
\beg{alpha_x}
\bar{\alpha}_x \;=\; \frac{\int_{S_{ph}}{dy\; dz\;\alpha_x(y,z)}}{\int_{S_{ph}}{dy\;dz}},
\done
where
\beg{defalphax}
\alpha_x = \frac{(\curl{\vecB})_x}{B_x}
\done 
is a commonly used proxy for the twist at the photospheric level \citep{Pevtsov94}. For completeness, we also calculate the average twist parameter,
\beg{defalpha}
\alpha = \frac{(\nabla\times\vecB)\cdot\vecB}{B^2},
\done
despite the fact that is not measurable in observations. We calculate the distance between the flux weighted centers of mass $L_{com}=|\textbf{L}_{com}|$ (cf. Equation \ref{com}) and
the angle between the flux weighted centers of mass and the initial flux rope axis
{\color{black}
\beg{angle}
\theta_{com} = \arctan\left({\frac{\unit{z}\cdot\textbf{L}_{com}}{\unit{y}\cdot\textbf{L}_{com}}}\right).
\done
}
Finally,we calculate the current neutralization ratio \citep{Torok14,Vemareddy15,Liu17}
\beg{nr}
\mathcal{R} = \frac{I_{ret}}{I_{dir}}
\done
where $I_{ret}$ and $I_{dir}$ are the return and direct currents, respectively, and are defined as
\beg{currentdef}
\begin{split} 
I_{ret} & = \int_{S_{ph}}{dy\;dz\;\kappa_-(y,z)\;J_x(y,z)} \\
I_{dir} & = \int_{S_{ph}}{dy\;dz\;\kappa_+(y,z)\;J_x(y,z)}
\end{split}
\done
with
\beg{xidef}
\begin{split}
\kappa_-(y,z) &=
\begin{cases}
1 & \text{if } J_x(y,z)B_x(y,z)<0\\
0 & \text{otherwise } 
\end{cases}
\\
\kappa_+(y,z) & =
\begin{cases}
1 & \text{if } J_x(y,z)B_x(y,z)>0\\
0 & \text{otherwise}.
\end{cases}
\end{split}
\done 
$\mathcal{R}$ has been suggested as an important parameter in determining flaring behavior \citep{Dalmasse15,Kontogiannis17,Liu17}. We also investigate the energetics of each simulation by defining the magnetic energy in the entire simulation domain at the start of each simulation as 
\beg{Wm0}
W_m^0 = \frac{1}{8\pi}\int{B^2(t=0)\;dV},
\done 
and the magnetic, potential and free energies in the coronal volume ($x>0$), respectively, as follows:
\beg{Wm}
W_m =  \frac{1}{8\pi}\int_{x>0}{B^2\;dV},
\done 
\beg{Wmp}
W_{m,p} = \frac{1}{8\pi}\int_{x>0}{B_p^2\;dV},
\done 
\beg{free}
\mathcal{F} = W_m - W_{m,p}.
\done 
We calculate the scaled parameters
\beg{scaled}
\begin{split}
\Tilde{W}_m & = \frac{W_m}{W_m^0} \\ 
\Tilde{W}_{m,p} & = \frac{W_{m,p}}{W_m^0} \\
\Tilde{\mathcal{F}} &=\frac{\mathcal{F}}{W_m^0}.
\end{split}
\done 
In the above expressions, $\vecB_p$ is the potential magnetic field in the corona, which satisfies the equation
\beg{defBp}
\nabla\times \vecB_p = 0.
\done 
\par

The separation distance between the flux weighted centers of mass is one of the defining characteristics known to differentiate \dss from other sunspot group types \citep{Kunzel65}. The evolution of the separation distance between the flux weighted centers of mass in our simulations, shown with the red dots in Figures \ref{fig:bx0}b, \ref{fig:bx1}b, \ref{fig:bx1.5}c, \ref{fig:bx2}c, and \ref{fig:bx4}c, displays clearly different behavior between the initially weakly and highly twisted flux ropes, as seen in the top panel of Figure \ref{fig:seprot}. The more weakly twisted flux ropes undergo significant separation, in one case more than double that of the highly twisted flux ropes. In the weakly twisted cases, there is clearly a general trend by the opposite polarities to move apart for the majority of the emergence. The kink-unstable cases, in contrast, have a peak separation distance of only $2-2.5\;\mathrm{Mm}$, after which the separation does not increase, and may even decrease. Thus, the flux ropes with higher initial twist display, at later times in the emergence process, a much more compact photospheric flux distribution, in agreement with the observed behavior of \dss. Based on this plot, there appears to be a bimodal distribution of separation distances between highly and weakly twisted flux ropes. It will be interesting to check this with observations.\par 

Rotation of the active region is another photospheric observable that is one of the defining characteristics of \dss \citep{Kurokawa87}. The angle $\theta_{com}$ between the line connecting the flux weighted centers of mass and the initial flux rope axis changes quite a bit for each simulation, as shown in the bottom panel of Figure \ref{fig:seprot}. The lower twist flux ropes emerge with $\theta_{com}\sim90^\circ$, corresponding to the fact that the azimuthal flux encircling the axial flux emerges first. The polarities then rotate back around to $\sim20^\circ$, corresponding to the emergence of almost purely axial flux in the legs of the flux rope. The kink-unstable flux ropes, meanwhile, emerge with $\theta_{com}$ greater than around $180^\circ$, before rotating to a value near $70^\circ$ (for $\zeta=1.5$ and $\zeta=2$) and $\sim15^\circ$ for $\zeta=4$. The total rotation undergone by the emerging flux ropes increases with increasing initial twist. If $\theta_{com}<90^\circ$ is assumed to be the angle of emergence that obeys Hale's law, as it should be for a typical weakly twisted emerging flux rope, then the three highest twist flux ropes obviously violate Hale's law, in agreement with the observed behavior of \dss, and the two lowest twist flux ropes nearly violate Hale's law. In both cases, the active region rotates into a configuration that satisfies Hale's law. It should be noted, however, that this calculation is obviously affected by the presence of the secondary polarities created by the projection of the twisted flux onto the photosphere, possibly disguising an ever larger contrast between weakly and strongly twisted flux ropes.\par 

One of the simplest quantities to measure observationally at the photospheric level, the unsigned magnetic flux, plotted as a function of time for each simulation in the top panel of Figure \ref{fig:flux}, is commonly measured to characterize an emerging active region \citep[e.g.,][]{Sun17}. In our simulations, although each flux rope has, initially, the same axial flux, the total flux at the photospheric level is determined by both the axial and azimuthal fluxes. As a result, $\Phi_{sc}$ is largest for $\zeta=4$, and is smallest for $\zeta=0.5$ and $\zeta=1$. Interestingly, between about $4$-$10$ minutes after emergence (emergence is defined here as the time when $B_x$ first exceeds $30\;\mathrm{G}$ on the photosphere), or about $25\%$ of the total simulation time, the $\zeta=1.5$ case has less flux going through the photosphere than $\zeta=1$. In general, however, after equal emergence times, we find more flux at the photospheric level for higher initial flux rope twist.\par

In the bottom panel of Figure \ref{fig:flux}, we plot the flux emergence rate against the emerged flux, following the method of \citet{Norton17}. Here, the emerged flux is defined as 
\beg{emerged_flux}
\Phi_{em} = \Phi_2(t_2) - \Phi_1(t_1)
\done
with $\Phi_2=0.9\Phi_{max}$ and $\Phi_1=0.1\Phi_{max}$, where $\Phi_{max}$ is the peak emerged flux. Thus $\Phi_{em}$ is the amount of flux that has emerged during the interval $\Delta t=t_2-t_1$. The flux emergence rate is defined as
\beg{FER}
\dot{\Phi}_{em} = \frac{\Phi_{em}}{\Delta t}
\done
where $\Delta t=t_2-t_1$ is the interval of time during which the flux $\Phi_{em}$ is emerged. \par 
Plotted this way, the flux emergence rate increases linearly with increasing flux (initial twist). The fluxes observed here ($10^{19-20}\;\mathrm{Mx}$) are within the observed range, as are the emergence rates ($10^{20-21}\;\mathrm{Mx\;hr^{-1}}$). However, the rate of emergence for a given flux is an order of magnitude higher than observed by \citet{Norton17}, though in line with simulated results \citep{Cheung07,Cheung08}. \par 

The magnitude of the vertical current density, often used as a proxy for energetic processes such as magnetic reconnection \citep{Knizhnik18}, is a fundamental quantity that can be determined directly through observations. Spikes in current density have been shown to coincide with X-class flares in \dss \citep{Vemareddy15}. In our simulations, the average unsigned vertical current density $\bar{J}_x$ strongly oscillates during each simulation, as shown in Figure \ref{fig:currents}a. There does not seem to be any dependence of $\bar{J_x}$ on the initial flux rope twist. At certain times during the emergence, the $\zeta=0.5$ or the $\zeta=1.5$ cases have higher values of $\bar{J}_x$ than the other cases, making it difficult to say anything definitive about the emerging structure from these quantities. $\bar{J}_x$ is different, in general, than the magnitude of the full vector current density, Equation \ref{defJ}, but the latter cannot be calculated from photospheric vector magnetograms. In our simulations, however, $\vecJ$ is easily measurable, and we find qualitatively and quantitatively very similar behavior of $\bar{J}_x$ and $\bar{J}$ (not shown here). We also measure the average signed vertical current density $\overline{|J_x|}$ for each simulation, shown in Figure \ref{fig:currents}b. There is some evidence that flux ropes with higher twist have larger average signed current densities than do flux ropes with smaller twist, but this breaks down around $t\approx 24\;\mathrm{min}$ into the emergence.\par 

The current neutralization ratio, easily measurable at the photosphere, has been found to be further from unity for flaring active regions than for non-flaring active regions \citep{Liu17}. $\mathcal{R}=1$ corresponds to a neutralized current at the photosphere, i.e., the direct current is balanced by the return current. Given the propensity of \dss for flaring behavior, it is natural to ask whether this deviation from unity is observed in our simulations. The bottom panel of Figure \ref{fig:currents} shows that the currents in our simulation are quite non-neutralized, and $\mathcal{R}$ is not obviously dependent on the flux rope's initial twist. There seems to be a general trend of rapid decrease followed by gradual increase of the value of $\mathcal{R}$ for all of the flux ropes, but the specific ordering does not seem to depend on $\zeta$. During some phases of emergence, $\zeta=2$ has higher values of $\mathcal{R}$ than other simulations, with $\zeta=0.5$ or $\zeta=4$ having the lowest values, and at other times $\zeta=4$ has the highest value and $\zeta=0.5$ or $\zeta=1.5$ has the lowest value. In all cases, however, currents do not seem to be neutralized at the photospheric level. \par 

$\bar{\alpha}_x$ has traditionally been the parameter used to study active region twist \citep{Pevtsov94,Pevtsov95,Leka96,Holder04} since it can be calculated directly from observations using photospheric vector magnetograms. However, in our simulations, the evolution of $\bar{\alpha}_x$, shown in Figure \ref{fig:alpha}a, does not display any obvious twist-dependent behavior. It is not generally the case that a larger $\bar{\alpha}_x$ corresponds to a larger value of $\zeta$. In fact, towards the end of the simulation, the $\zeta=0.5$ case has the largest value of $\bar{\alpha}_x$. The $\zeta=1$ and $\zeta=4$ flux ropes have the smallest values of $\bar{\alpha}_x$ throughout a significant fraction of the emergence process. Although each flux rope continuously emerges, the value of $\bar{\alpha}_x$ does not continuously increase, instead showing peaks and troughs during the course of each emergence. This would seem to indicate that, $\bar{\alpha}_x$ does not, by itself, reveal anything about the structure of the emerging flux rope. Although $\alpha_x$ is the quantity that can be measured from photospheric magnetogram, the twist parameter is actually $\alpha$ as defined in Equation \ref{defalpha}. This quantity can be measured directly in our simulations, though not in observations. The time evolution of $\bar{\alpha}$ is plotted in Figure \ref{fig:alpha}b. Notably, the plots of $\bar{\alpha}$ do not resemble the plots of $\bar{\alpha}_x$, except that they share a lack of a clear dependence on twist. This indicates that $\bar{\alpha}_x$ may not be a good proxy for $\bar{\alpha}$ at the non-force free photosphere. \par 
From the figures, it appears that little, if anything, can be ascertained from the temporal evolution of either $\bar{\alpha}_x$ or $\bar{\alpha}$, since the photosphere is not force free, and $\vecJ\times\vecB=0$ is expected to be a poor approximation. However, there are certain locations where the photosphere can be taken to be force free. These locations can be determined as follows. The current density and magnetic field are related via
\beg{JxB}
|\vecJ\times\vecB| = |\vecJ||\vecB|\sin\theta,
\done 
\beg{JdotB}
\vecJ\cdot\vecB = |\vecJ||\vecB|\cos\theta,
\done 
with $\theta$ the angle between $\vecJ$ and $\vecB$. Therefore
\beg{forcefreeness}
\Sigma_1 + \Sigma_2 = 1
\done 
identically, where 
\beg{sigma1}
\Sigma_1 = \frac{\left(\vecJ\cdot\vecB\right)^2}{|\vecJ|^2|\vecB|^2}, 
\done
and 
\beg{sigma2}
\Sigma_2 = \frac{|\vecJ\times\vecB|^2}{|\vecJ|^2|\vecB|^2}.
\done 
Thus, locations on the photosphere where $\Sigma_1$ ($\Sigma_2$) is large (small) correspond to locations where the magnetic field is close to force free. Therefore, we define the `force free twist parameters' as
\beg{ffalphas}
\begin{split}
\alpha_{x,ff}(y,z) &=
\begin{cases}
\alpha_x(y,z) & \text{if } \Sigma_1(y,z)>\Sigma_1^{thresh}\\
0 & \text{otherwise } 
\end{cases}
\\
\alpha_{ff}(y,z) &=
\begin{cases}
\alpha(y,z) & \text{if } \Sigma_1(y,z)>\Sigma_1^{thresh}\\
0 & \text{otherwise}.
\end{cases}
\end{split}
\done
In panels c,d of Figure \ref{fig:alpha}, we plot the average unsigned vertical and full force free twist parameters for each simulation, having chosen $\Sigma_1^{thresh}=0.8$. Here, in contrast to panels a) and b), one can see that there is a discernable correlation between the $\bar{\alpha}_x$ and $\bar{\alpha}$. However, there is still no discernable correlation between $\zeta$ and the magnitude of either $\bar{\alpha}_x$ or $\bar{\alpha}$. Especially early in the emergence process, the $\zeta=1.5$ and $\zeta=2$ flux ropes have the largest values, while the $\zeta=0.5$ and $\zeta=1$ flux ropes have the smallest values. The $\zeta=4$ flux rope seems to fall somewhere in the middle, destroying any dependence of these quantities on initial twist. 
{\color{black}  
This results in the average photospheric $\alpha$, for all values of $\zeta$, clustering around a narrow range, particularly later in the emergence. For example, $\bar{\alpha}_x$ clusters within the range 
$10-30\times10^{-8}\;\mathrm{m}^{-1}$ for these simulations. The fact that this is the observed result from flux ropes whose original peak alpha ranged from $330-2700\times10^{-8}\;\mathrm{m}^{-1}$ illustrates how different the photospheric $\alpha$ can be from the initial, peak $\alpha$ of a convection zone flux rope.
}
\par
\begin{figure*}
\centering\includegraphics[scale=0.38, trim=0.0cm 1.0cm 5.5cm 1.0cm,clip=true]{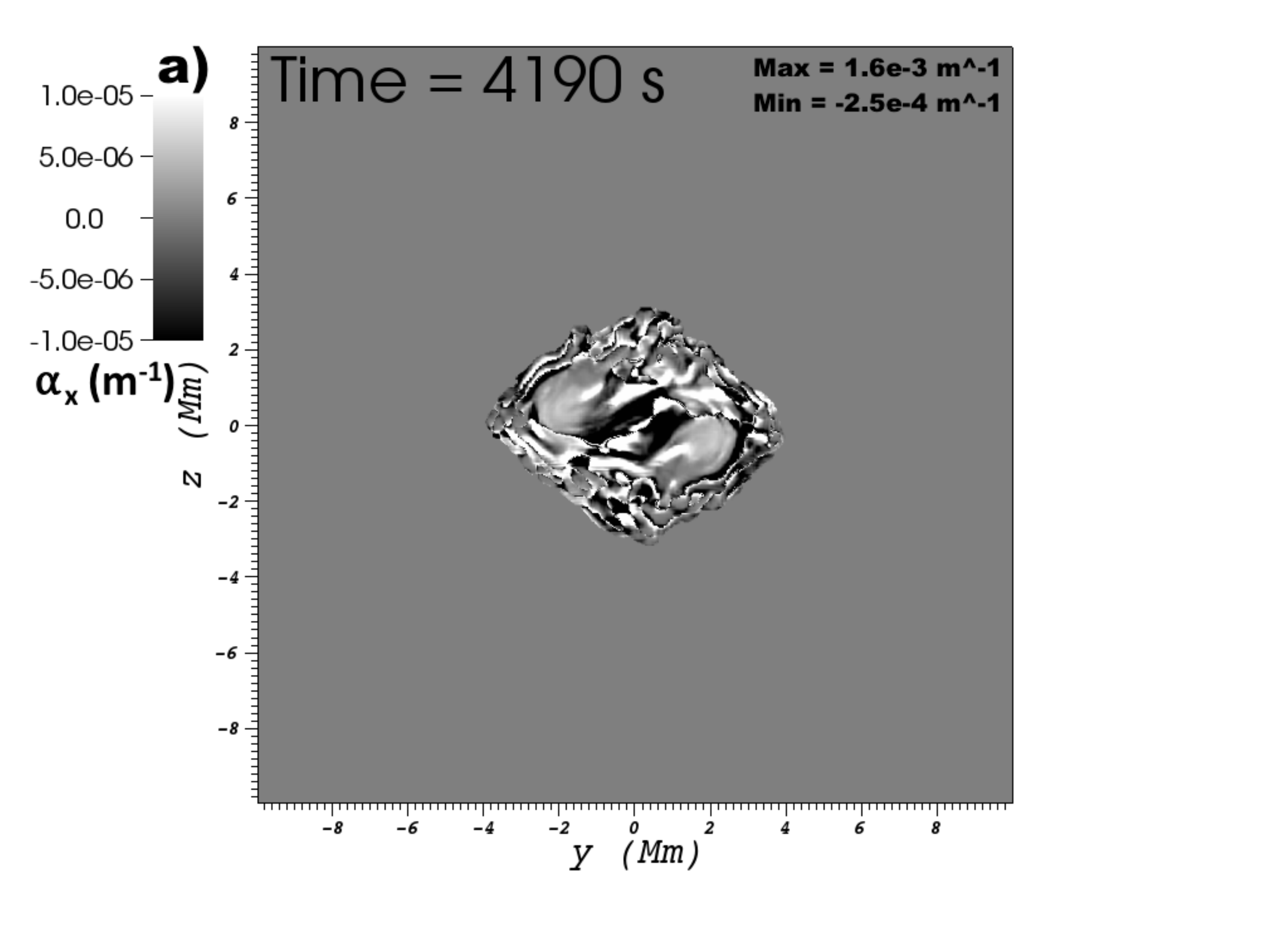}
\centering\includegraphics[scale=0.38, trim=0.0cm 1.0cm 5.0cm 1.0cm,clip=true]{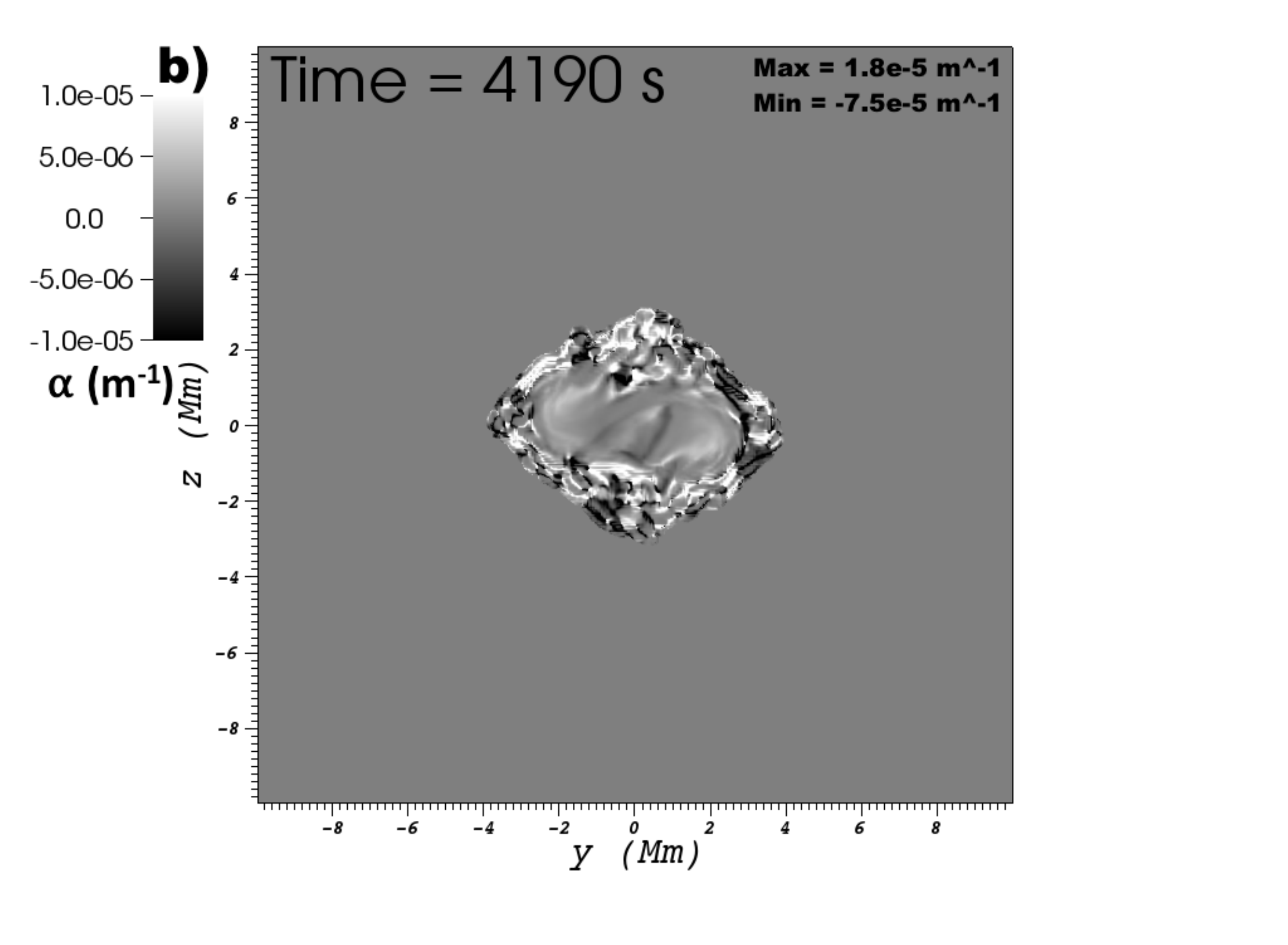}
\centering\includegraphics[scale=0.38, trim=0.0cm 1.0cm 5.5cm 1.0cm,clip=true]{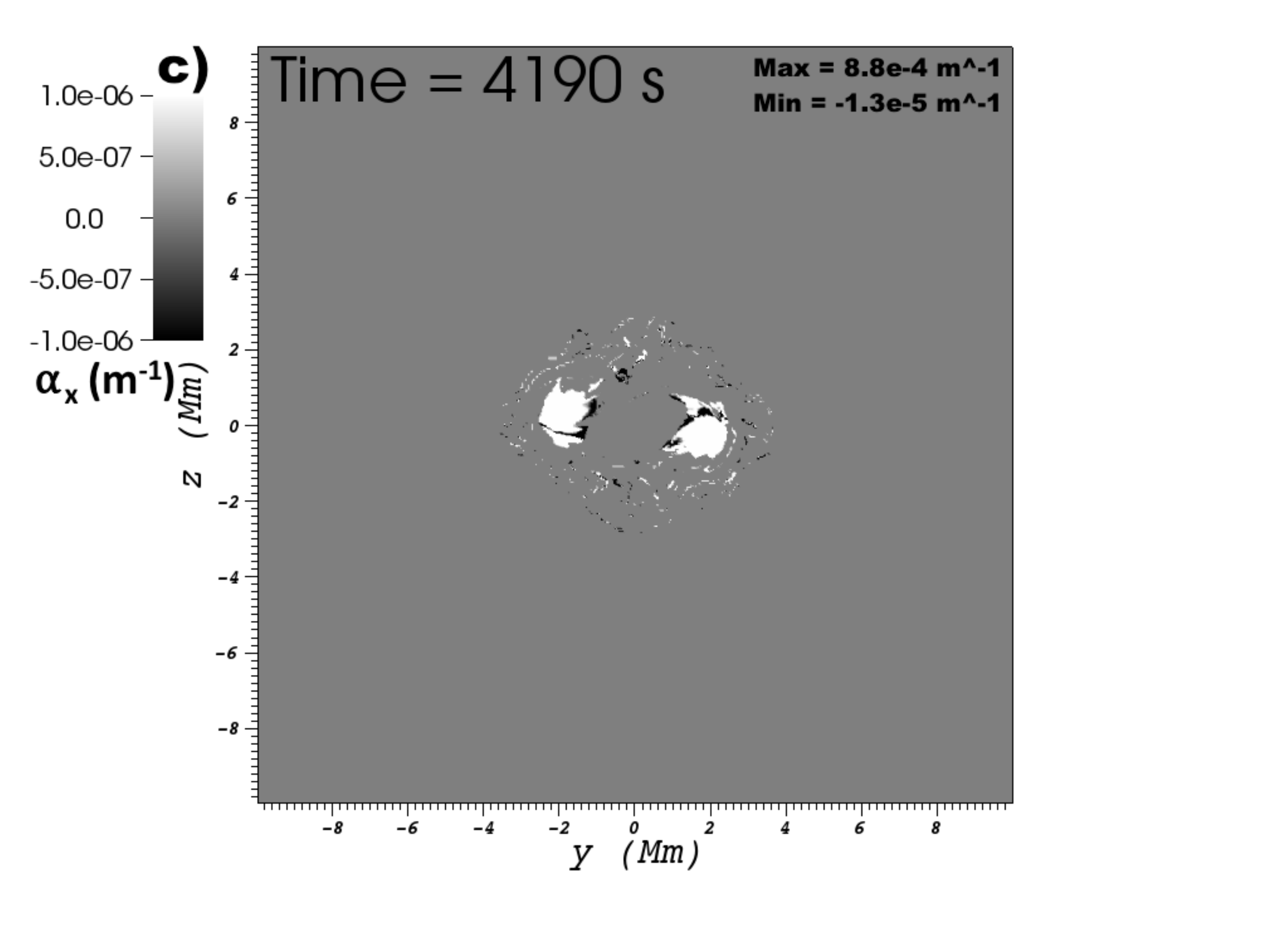}
\centering\includegraphics[scale=0.38, trim=0.0cm 1.0cm 5.0cm 1.0cm,clip=true]{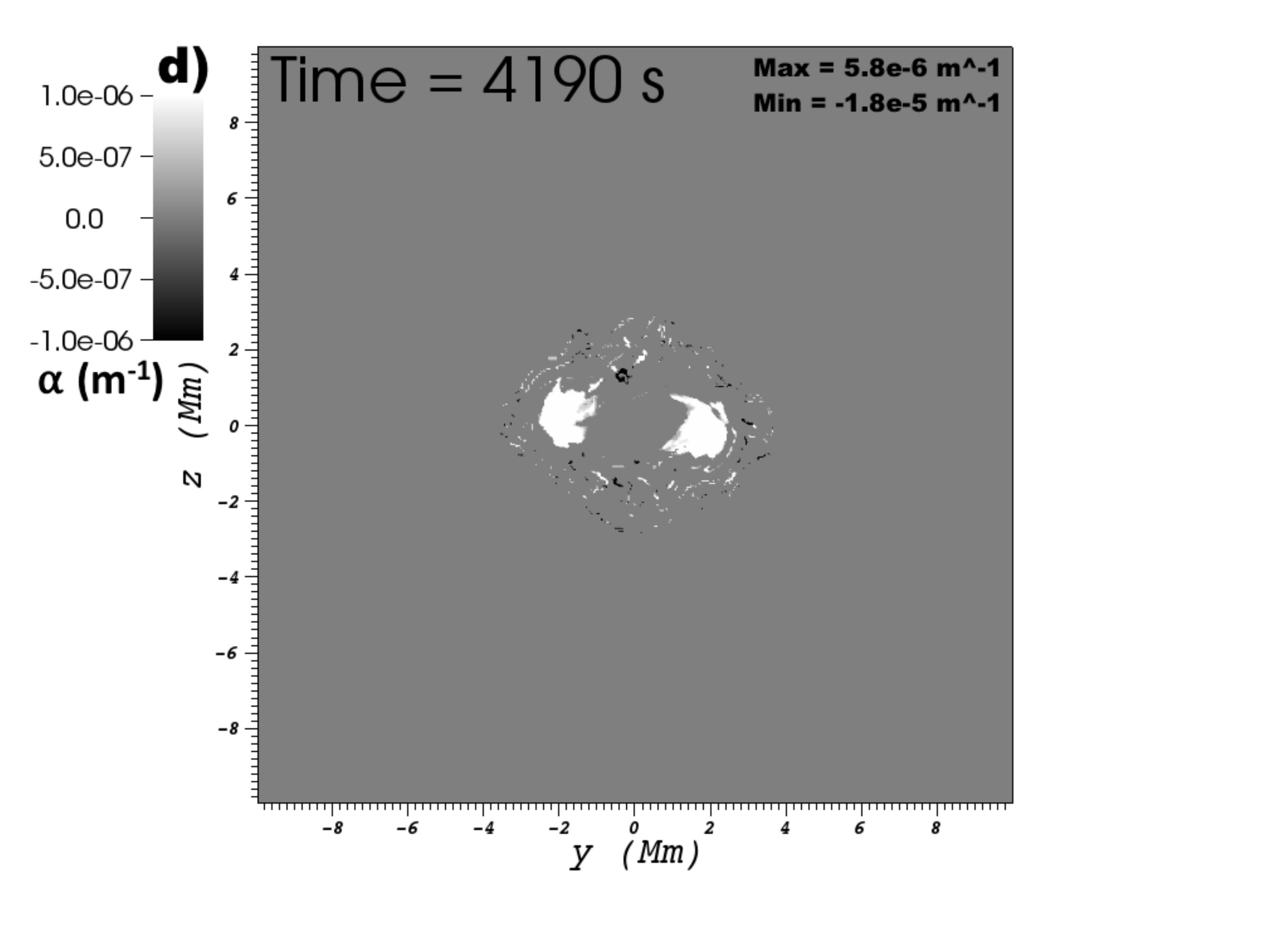}
\caption{a) $\alpha_x$ b) $\alpha$, c) $\alpha_{x,ff}$, and d) $\alpha_{ff}$ for $\zeta=1$.}
\label{fig:alphas1}
\end{figure*}

 
In Figure \ref{fig:alphas1}, we plot $\alpha_x$, $\alpha$, $\alpha_{x,ff}$, and $\alpha_{ff}$ at a representative snapshot of the $\zeta=1$ simulation. Immediately obvious from these figures is the complexity of the signal in $\alpha_x$ and $\alpha$. No single sign of either dominates. Meanwhile, the sign of the photospheric distribution of $\alpha_{x,ff}$ and $\alpha_{ff}$ is overwhelmingly positive, reflecting the fact that $\alpha$ in the initial flux rope was positive. It is evident, from looking at these figures, that a simple average over the twist parameter, as is done in several papers \citep[e.g.][]{Pevtsov94,Pevtsov95,Leka96,Holder04} is bound to be affected by the cancellation of positive/negative values of approximately equal magnitude. Thus, averaging over $\alpha_x$ (or even $\alpha$ if it could be measured), will produce a value of minimal significance. The Figure also address the question of whether $\alpha_x$ is a reasonable proxy for $\alpha$. It is clear that $\alpha_x$ is much noisier than $\alpha$ and dominates near polarity inversion lines outside the main polarities, whereas $\alpha$ is relatively smooth in the interior of the active region (near $y=z=0$). Comparing $\alpha_x$ and $\alpha$ reveals that they do not match well, especially in the interior of the active regions. On the other hand, comparison of $\alpha_{x,ff}$ and $\alpha_{ff}$ reveals significantly more similarity, indicating that $\alpha_{x,ff}$ is a good proxy for $\alpha_{ff}$, supporting the expectation that $\alpha_x$ should be a good proxy for $\alpha$ in force-free fields.
This calls into question the usefulness of measurements of $\alpha_x$ at the photosphere as a proxy for $\alpha$, since identifying which regions are force-free in observations is, at present, a challenge. 
{\color{black} This result also justifies, a posteriori, our choices of $\zeta$ which are significantly larger than the values required to explain observations \citep{Pevtsov95, Longcope99}. Since there appears to be no relationship between the values of either $\alpha_x$ or $\alpha$ observed at the photosphere and the initial flux rope twist, the twist inferred from $\alpha$ using theoretical arguments may not necessarily be representative of its true value. Each of the highly twisted flux ropes initially had a much larger $\alpha$, but the conversion of twist into writhe as a result of the kink instability decreased $\alpha$ to values very similar to those measured for the lower twist flux ropes. Furthermore, the salt-and-pepper nature of the distribution of photospheric $\alpha$ and $\alpha_x$ means that any strong positive values may be canceled by strong negative values, decreasing the overall average. }
\par

\begin{figure}
\centering\includegraphics[scale=0.35, trim=0.0cm 0.0cm 0.0cm 0.0cm,clip=true]{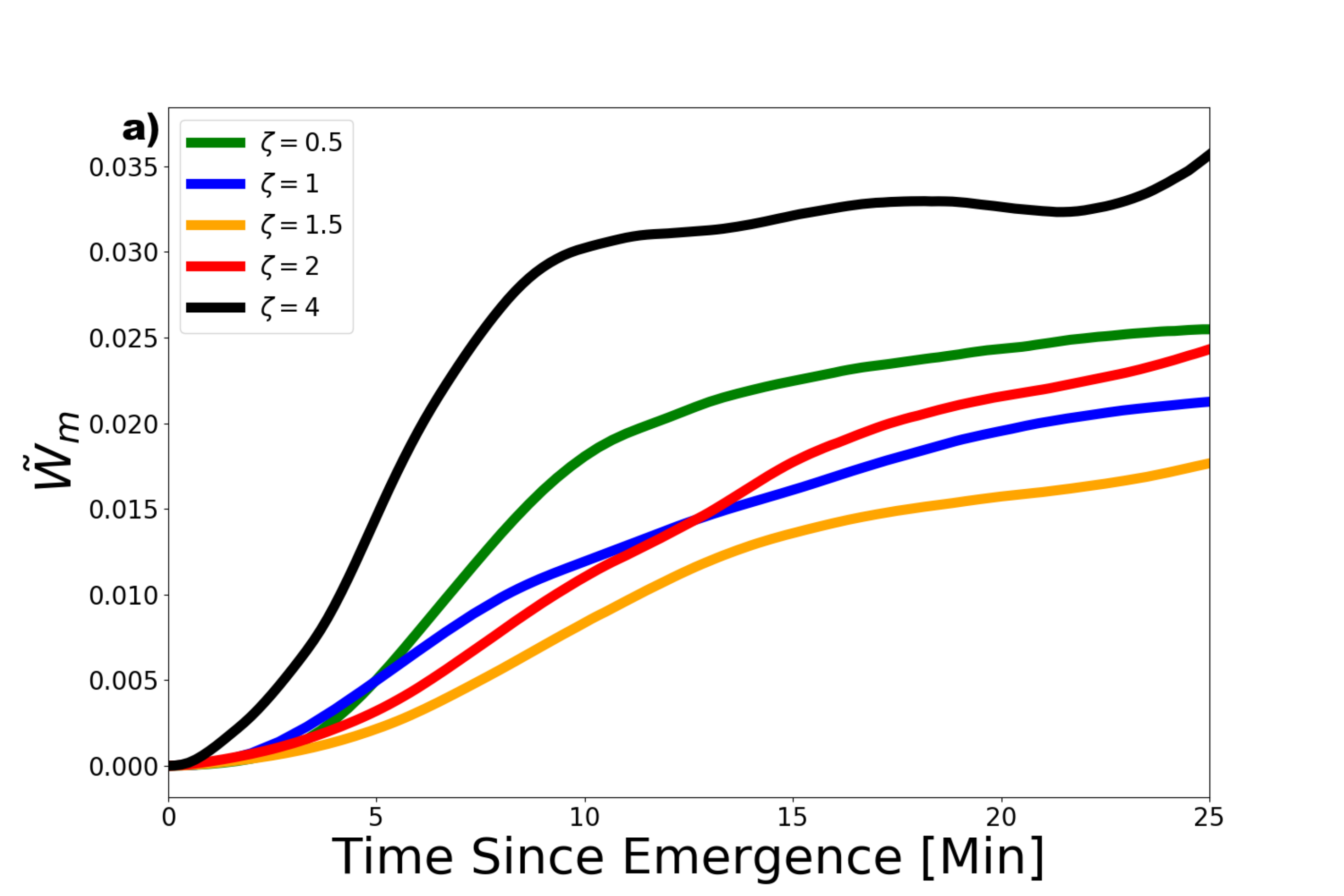}
\centering\includegraphics[scale=0.35, trim=0.0cm 0.0cm 0.0cm 0.0cm,clip=true]{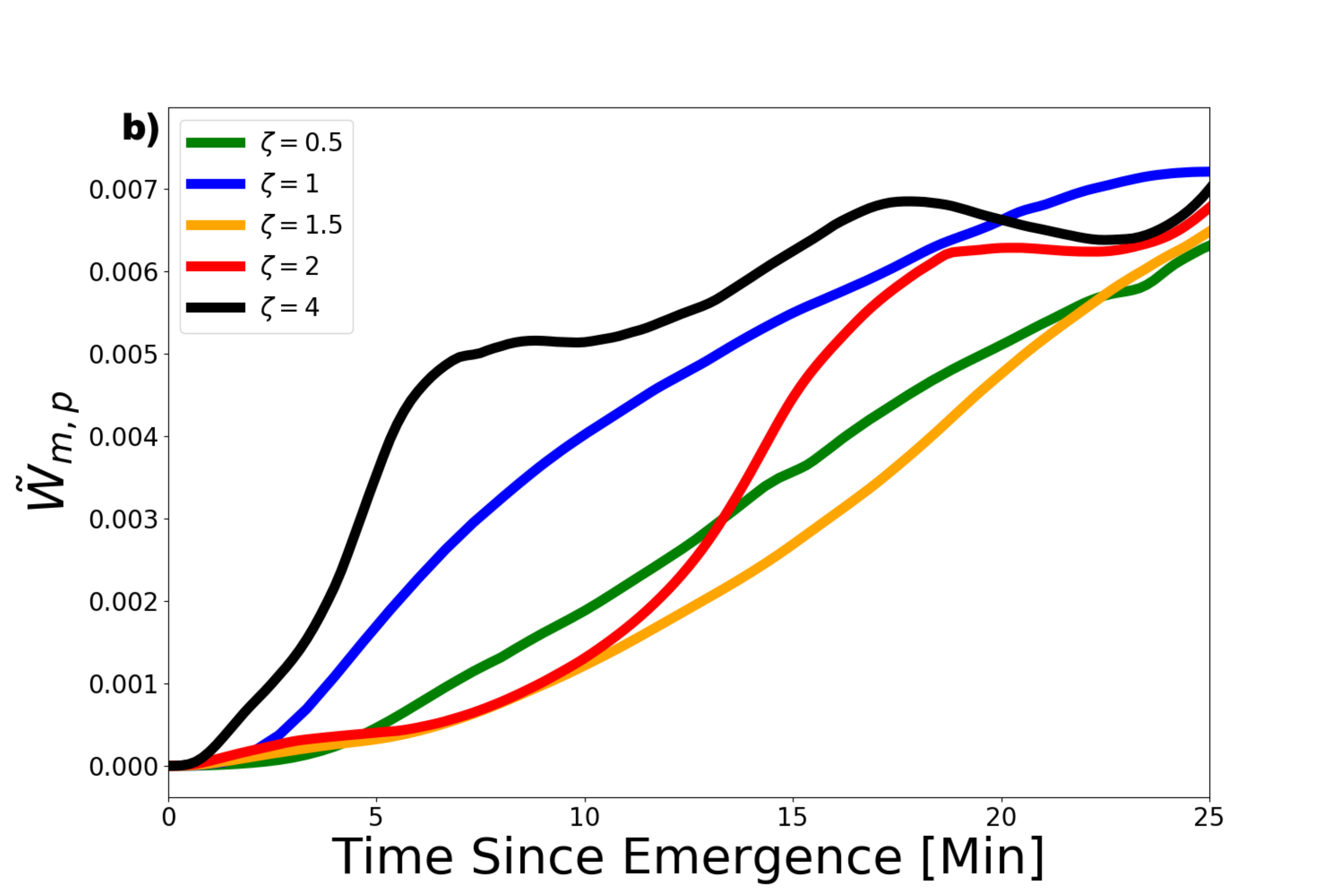}
\centering\includegraphics[scale=0.35, trim=0.0cm 0.0cm 0.0cm 0.0cm,clip=true]{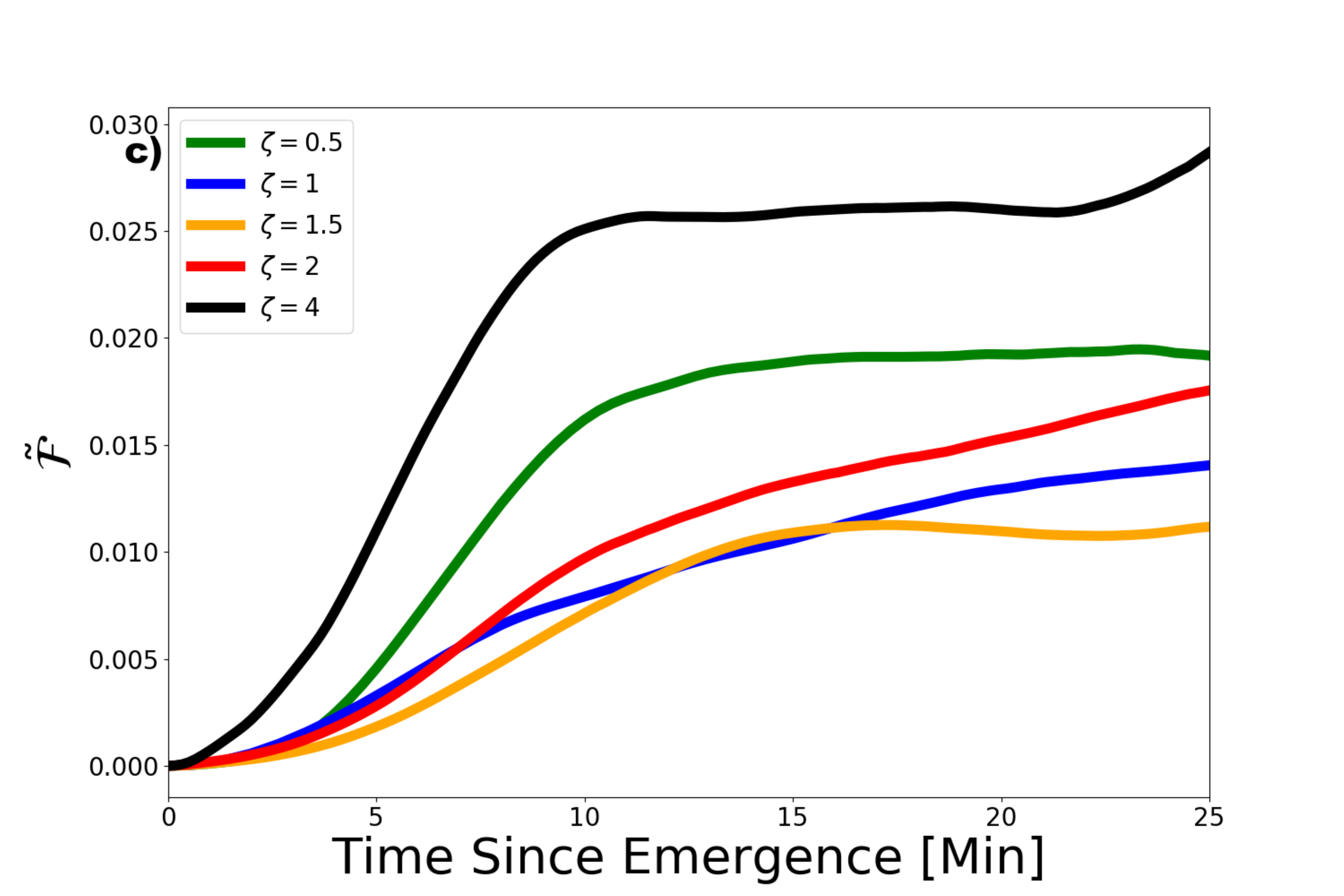}
\caption{Top: Magnetic energy in the corona as a function of time, scaled by the initial magnetic energy in the whole volume. Middle: Potential energy in the corona as a function of time, scaled by the initial magnetic energy in the whole volume. Bottom: Free energy in the corona as a function of time, scaled by the initial magnetic energy in the whole volume.}
\label{fig:energy}
\end{figure}

Most of the quantities described above are observationally measurable from photospheric data alone. However, some of the most important quantities, from the standpoint of energy release, are the magnetic, potential, and free energies in the corona. Although the magnetic and free energies rely on knowledge of the magnetic field in the entire corona, rather than just the photosphere, they are directly related to how much energy can be released in a flaring event. To understand how these quantities scale with initial flux rope twist, we plot the magnetic and potential energies in Figure \ref{fig:energy}. Neither the magnetic nor the potential energies show a tendency to be larger for large initial twist and smaller for small initial twist. While the $\zeta=4$ simulation has the largest magnetic energy for the duration of the simulation, the $\zeta=0.5$ flux rope has the second most, and the $\zeta=1.5$ flux rope has the least throughout the simulation. The $\zeta=1.5$ and $\zeta=2$ flux ropes have somewhat less free energy than the $\zeta=0.5$ flux rope. Both the $\zeta=0.5$ and $\zeta=4$ flux ropes have approximately constant free energy starting around $10\;\mathrm{min}$ after the initial emergence, while the $\zeta=1.5$ flux rope has approximately constant free energy starting around $15\;\mathrm{min}$ after the initial emergence. The $\zeta=1$ and $\zeta=2$ flux ropes continue to increase their free energy through their emergence.   {\color{black} The free energies measured here are quite large: about $80\%$ of the total magnetic energy is in the form of free energy. Meanwhile, the nonlinear force-free extrapolations of \citet{DeRosa09} of a stable AR magnetogram found that approximately $20\%$ of the total magnetic energy in the region was in the form of free magnetic energy. This large difference in free-energy content could arise from the assumptions inherent in the nonlinear force-free extrapolations, which use manifestly non force-free photospheric magnetograms as boundary conditions, or because the AR studied by \citet{DeRosa09} did not contain a \ds and was not flare active during the observation, and even later did not produce anything larger than a C8 flare. It is possible that a similar analysis conducted on a \ds containing AR would find that a much larger fraction of the total magnetic energy was stored as free energy.}\par 
\section{Discussion}\label{sec:implications}
The lack of understanding of the formation of \dss is one of the most important impediments preventing proper prediction of solar energetic events. The emergence of highly twisted, kink unstable flux ropes has been hypothesized to be the source of a large percentage of \dss \citep{Fan99,Linton96,Toriumi17}. In this work, we test this model using numerical simulations of the emergence of both initially kink-stable and -unstable flux ropes from the convection zone, through the photosphere, into the corona. We focused mainly on AR properties that can be measured using line-of-sight and vector magnetogram observations and related them to the subsurface properties of emerging flux ropes.\par 
By calculating the flux weighted centers of mass of each polarity, we demonstrated that initially highly twisted flux ropes remain much more compact, and rotate much more than, their lower twist, kink-stable counterparts. We also showed that while low twist flux ropes obey Hale's law, kink unstable flux ropes appear to violate it. Our results demonstrate that at the photosphere, emerging kinked flux ropes behave in a manner that is both qualitatively and quantitatively similar to the most well-documented behavior of \dss \citep{Kunzel65,Smith68, Kurokawa87,Zirin87,Tanaka91,LF00,LF03}. However, quantitative statements of exactly what is meant by `compactness' and `rotation' are sparse, so it is difficult to compare with observations. \par 
On the other hand, our work also indicates that quantitative measurements of many \ds properties are unlikely to produce results which are vastly different than those measured in simple sunspot groups. Parameters like $\alpha_x$, $J_x$, $\mathcal{R}$, all of which are measurable from photospheric observations, without recourse to models of the coronal field, do not reveal major consistent differences for highly twisted versus weakly twisted flux ropes. Photospheric measurements of these parameters may reveal higher values for \dss, but it is also possible that they may reveal lower values, depending on the instant at which the measurement was taken. Plotting the time evolution of these parameters may also reveal differences, but the results presented here suggest that, with the exceptions of the separation distance between opposite polarities and rotation angle of the opposite polarities, there is no simple dependence of many measurables on the flux rope's initial twist. Even the free energy in the coronal volume, thought to be a proxy for energetic flaring events, does not show an obvious dependence on the initial flux rope twist, despite the much more complicated coronal topology evident in the higher twist simulations. 
{\color{black} In contrast, Figure 8 in \citet{Toriumi11} shows that, for initially weakly twisted flux ropes, there appears to be a dependence of magnetic energy on initial twist. However, the differences in the numerical setups between those simulations and the ones presented here make a direct comparison challenging.  \par 
The lack of dependence of our results on the initial twist arises due to several effects. First, the nature of the kink instability is to reduce the field line twist by converting twist to write \citep{Berger84, Linton96}, so that once a flux rope has kinked, its effective twist, and therefore $\alpha$, is smaller than its initial value, and may turn out to be comparable to that in a lower twist flux rope. Indeed, it is possible that the lower twist flux rope need not, itself, be kink stable for this result to hold. Since it started out with a smaller value of $\alpha$, the lower twist flux rope may, as a result of a kink, experience a smaller decrease in $\alpha$ than the higher twist flux rope, resulting in the two flux ropes having comparable values of $\alpha$. Regardless, it is clear that $\alpha$ is not simply transferred directly from the convection zone to the photosphere, but instead evolves dynamically during the emergence process, explaining its decrease from convection zone to photospheric values. Second, the rapid expansion of the magnetic field once it reaches the coronal level causes a significant decrease in the value of $\alpha$ at the photospheric level, since the expansion of the field into the corona occurs faster than Alfv\'en waves can spread the twist along the field line, creating a gradient in $\alpha$ along the expanding field line. This effect has also been observed in previous high twist simulations \citep{Fan09, Leake13}. Finally, reconnection inside the flux rope itself could destroy a lot of the initial twist structure, perhaps through an inverse cascade of magnetic helicity \citep{Antiochos13,Knizhnik15,Knizhnik17,Knizhnik17b}, leaving a relatively weakly twisted internal structure, forming active regions that have lost much of the information about the structure of the initial flux rope. \par 
The simulations presented here show that the dynamics of emerging highly twisted flux ropes cause copious internal reconnection to occur, as evidenced by the formation of concave up loops and knots during the emergence process. Such complicated topological structures can only be formed, in the absence of an external magnetic field, by internal reconnection. This internal reconnection would likely manifest itself as flaring behavior on the Sun. We conclude that flaring internal to an emerging kinking active region can be strong, independent of the state (or even absence) of external fields around it, in agreement with observations showing that \dss can produce X-class flares within themselves \citep{Zirin73, Zirin87, Wang91, Schmieder94}. 
}

\acknowledgments{
K.J.K.\ was supported for this work by the Chief of Naval Research through the National Research Council. M.G.L.\ was supported for this work by the Chief of Naval Research and by the NASA Living with a Star Program and Heliophysics Supporting Research programs. C.R.D.\ was supported by grants from NASAʼs Living With a Star and Heliophysics Supporting Research programs. K.J.K.\ acknowledges very helpful discussions with J. Leake and L. Tarr. The numerical simulations were performed under a grant of computer time at the Department of Defense High Performance Computing Program. 
}

\newpage
\clearpage

\newpage




\newpage

\newpage

\bibliography{bibliography}
\end{document}